\pdfoutput=1

\newif\ifprintfig
\printfigtrue

\documentclass{emulateapj}
\usepackage{graphicx}
\usepackage{etoolbox}
\usepackage{amsmath}

\newcommand\kpc{{\rm\,kpc}}

\newcommand\Myr{{\rm\,Myr}}

\newcommand\Gyr{{\rm\,Gyr}}

\newcommand\kmsec{{\rm\,km\,s^{-1}}}
\newcommand\kms{\kmsec}

\newcommand\msun{{\rm\,M_\odot}}
\newcommand\Msun{{\rm\,M_\odot}}

\newcommand{\mas}{\textrm{milli-arcseconds}}
\newcommand\clock{\count0=\time \divide\count0 by 60
     \count1=\count0 \multiply\count1 by -60 \advance\count1 by \time
     \number\count0:\ifnum\count1<10{0\number\count1}\else\number\count1\fi}
\newcommand{\thou}{,\!000}
\newcommand{\Teff}{T_{ef\!f}}
\newcommand{\multiaccum}{\texttt{MULTIACCUM}}
\newcommand{\tinytim}{\texttt{TinyTim}}

\newcommand{\filt}[1]{%
    \ifbool{mmode}{\text{#1}}{#1}}
\newcommand{\fw}[1]{\filt{F#1W}}

\shortauthors{Dalcanton et al.}
\shorttitle{Panchromatic Hubble Andromeda Treasury}

\begin{document}

\title{The Panchromatic Hubble Andromeda Treasury}

\author{
Julianne J.~Dalcanton\altaffilmark{1},
Benjamin F.~Williams\altaffilmark{1},
Dustin Lang\altaffilmark{2},
Tod R.~Lauer\altaffilmark{3},
Jason S.~Kalirai\altaffilmark{4},
Anil C.~Seth\altaffilmark{5}, 
Andrew Dolphin\altaffilmark{6}, 
Philip Rosenfield\altaffilmark{1},
Daniel R.~Weisz\altaffilmark{1},
Eric F.~Bell\altaffilmark{7},
Luciana C.~Bianchi\altaffilmark{8},
Martha L.~Boyer\altaffilmark{4},
Nelson Caldwell\altaffilmark{9},
Hui Dong\altaffilmark{3},
Claire E.~Dorman\altaffilmark{10},
Karoline M.~Gilbert\altaffilmark{1,11},
L\'eo Girardi\altaffilmark{12}, 
Stephanie M.~Gogarten\altaffilmark{1},
Karl D.~Gordon\altaffilmark{4},
Puragra Guhathakurta\altaffilmark{10},
Paul W.~Hodge\altaffilmark{1},
Jon A.~Holtzman\altaffilmark{13},
L.~Clifton Johnson\altaffilmark{1},
S{\o}ren S.~Larsen\altaffilmark{14},
Alexia Lewis\altaffilmark{1},
Jason L.~Melbourne\altaffilmark{15},
Knut A.~G.~Olsen\altaffilmark{3},
Hans-Walter Rix\altaffilmark{16},
Keith Rosema\altaffilmark{17},
Abhijit Saha\altaffilmark{3},
Ata Sarajedini\altaffilmark{18},
Evan D.~Skillman\altaffilmark{19},
Krzysztof Z.~Stanek\altaffilmark{20}
}

\altaffiltext{1}{Department of Astronomy, University of Washington, Box 351580, Seattle, WA 98195, USA}
\altaffiltext{2}{Department of Astrophysical Sciences, Princeton University, Princeton, NJ 08544, USA}
\altaffiltext{3}{National Optical Astronomy Observatory, 950 North Cherry Avenue, Tucson, AZ 85719, USA}
\altaffiltext{4}{Space Telescope Science Institute, 3700 San Martin Drive, Baltimore, MD, 21218, USA}
\altaffiltext{5}{Department of Physics \& Astronomy, University of Utah, Salt Lake City, UT 84112, USA}
\altaffiltext{6}{Raytheon Company, 1151 East Hermans Road, Tucson, AZ 85756, USA}
\altaffiltext{7}{Department of Astronomy, University of Michigan, 500 Church St., Ann Arbor, MI 48109, USA}
\altaffiltext{8}{Department of Physics and Astronomy,  Johns Hopkins University, Baltimore, MD 21218, USA}
\altaffiltext{9}{Harvard-Smithsonian Center for Astrophysics, 60 Garden Street Cambridge, MA 02138, USA}
\altaffiltext{10}{University of California Observatories/Lick Observatory, University of California, 1156 High St., Santa Cruz, CA 95064, USA}
\altaffiltext{11}{Hubble Fellow}
\altaffiltext{12}{Osservatorio Astronomico di Padova -- INAF, Vicolo dell'Osservatorio 5, I-35122 Padova, Italy}
\altaffiltext{13}{Department of Astronomy, New Mexico State University, Box 30001-Department 4500, 1320 Frenger Street, Las Cruces, NM 88003, USA}
\altaffiltext{14}{Department of Astrophysics, IMAPP, Radboud University Nijmegen, PO Box 9010, 6500 GL Nijmegen, The Netherlands}
\altaffiltext{15}{Caltech Optical Observatories, Division of Physics, Mathematics and Astronomy, Mail Stop 301-17, California Institute of Technology, Pasadena, CA 91125, USA}
\altaffiltext{16}{Max Planck Institute for Astronomy, Koenigstuhl 17, 69117 Heidelberg, Germany}
\altaffiltext{17}{Random Walk Group, 5209 21st Ave.~N.E., Seattle, WA 98105, USA}
\altaffiltext{18}{Department of Astronomy, University of Florida, Gainesville, FL, 32611, USA}
\altaffiltext{19}{Minnesota Institute for Astrophysics, University of Minnesota, 116 Church Street SE, Minneapolis, MN 55455, USA}
\altaffiltext{20}{Department of Astronomy, The Ohio State University, 140 West 18th Avenue, Columbus OH 43210, USA}

\begin{abstract}
  The Panchromatic Hubble Andromeda Treasury (PHAT) is an on-going
  Hubble Space Telescope (HST) Multicycle Treasury program to image
  $\sim$1/3 of M31's star forming disk in six filters, spanning from
  the ultraviolet (UV) to the near-infrared (NIR).  We use the Wide
  Field Camera 3 (WFC3) and Advanced Camera for Surveys (ACS) to
  resolve the galaxy into millions of individual stars with projected
  radii from 0--20$\kpc$.  The full survey will cover a contiguous 0.5
  square degree area in 828 orbits.  Imaging is being obtained in the
  \fw{275} and \fw{336} filters on the WFC3/UVIS camera, \fw{475} and
  \fw{814} on ACS/WFC, and \fw{110} and \fw{160} on WFC3/IR.  The
  resulting wavelength coverage gives excellent constraints on stellar
  temperature, bolometric luminosity, and extinction for most spectral
  types.  The data produce photometry with a signal-to-noise ratio of
  4 at $m_{\fw{275}}=25.1$, $m_{\fw{336}}=24.9$, $m_{\fw{475}}=27.9$,
  $m_{\fw{814}}=27.1$, $m_{\fw{110}}=25.5$, and $m_{\fw{160}}=24.6$
  for single pointings in the uncrowded outer disk; in the inner disk,
  however, the optical and NIR data are crowding limited, and the
  deepest reliable magnitudes are up to 5 magnitudes brighter.
  Observations are carried out in two orbits per pointing, split
  between WFC3/UVIS and WFC3/IR cameras in primary mode, with ACS/WFC
  run in parallel.  All pointings are dithered to produce
  Nyquist-sampled images in \fw{475}, \fw{814}, and \fw{160}.  We
  describe the observing strategy, photometry, astrometry, and data
  products available for the survey, along with extensive testing of
  photometric stability, crowding errors, spatially-dependent
  photometric biases, and telescope pointing control.  We also report
  on initial fits to the structure of M31's disk, derived from the
  density of red giant branch stars, in a way that is independent of
  assumed mass-to-light ratios and is robust to variations in dust
  extinction.  These fits also show that the 10$\kpc$ ring is not just
  a region of enhanced recent star formation, but is instead a
  dynamical structure containing a significant overdensity of stars
  with ages $>\!1\Gyr$.
\end{abstract}

\vfill

\section{Introduction}  \label{introsec}

Our quest to understand the Universe relies on detailed knowledge of
physical processes that can only be calibrated nearby.  It is
impossible to interpret observations across cosmic time without an
underlying understanding of stellar evolution, star formation, the
initial mass function, the extinction law, and the distance scale, all
of which require detailed studies of individual stars and the
interstellar medium (ISM) on sub-kiloparsec scales.

When the needed studies of stars and gas are carried out in the Milky
Way, they frequently face complications from line-of-sight reddening,
uncertain distances, and background/foreground confusion.  As such, it
is sometimes easier to constrain physical processes in external
galaxies, which are free of the projection effects that can plague
Milky Way studies.  Not only are observations of external galaxies
more straightforward to interpret, but they can also be placed in the
larger context of the surrounding environment (i.e., the ISM,
metallicity, and star formation rate (SFR)).  Galaxies in the Local
Group therefore offer an excellent compromise between being close
enough to resolve relatively faint stars, while being distant enough
to unveil the complex processes that govern star and galaxy evolution
in their full galactic context.

Unfortunately, even the nearest massive galaxies have sufficiently
high stellar surface densities that severe crowding compromises the
detection of fainter, more age-sensitive stellar populations, allowing
only the brightest stars to be studied with typical ground-based
angular resolution in high-surface brightness regions of galaxy disks
\citep[e.g.,][]{massey2006}.  However, with the high angular
resolution available from HST, we have the potential to resolve
millions to billions of stars within the Local Group, grouped into
galaxies with a common distance and foreground extinction.  These
stars, along with their ancestors and descendants (e.g., molecular
clouds, H{\sc ii} regions, variable stars, X-ray binaries, supernova
remnants, etc.), provide transformative tools for strengthening
the foundation on which knowledge of the distant Universe is based.

Within the Local Group, the Andromeda Galaxy (M31) offers the best
proxy for the properties of more distant galaxies.  It is massive
(sampling above the characteristic stellar mass
($3\!-\!5\!\times\!10^{10}\msun$) over which rapid systematic changes
in galaxies' stellar populations and structure occur
\citep[e.g.,][]{kauffmann2003}), hosts spiral structure, and contains
the nearest example of a traditional spheroidal component (outside the
MW).  M31 is also representative of the environments in
which typical stars are found today.  More than half of all stars are
currently found in the disk and bulges of disk-dominated galaxies like
M31 \citep{driver2007}, and more than 3/4 of all stars in the Universe
have metallicities within a factor of two of solar
\citep{gallazzi2008}, comparable to the typical metallicities of stars
in M31.

In addition to its dominant solar-metallicity population of young
stars, M31 also contains significant populations of older super-solar
metallicity stars in its bulge, and sub-solar metallicity stars in its
outskirts \citep[e.g.,][and references therein]{silchenko1998,
  lauer2012, brown2003, worthey2005, kalirai2006, chapman2006,
  fan2008}, making M31 a superb laboratory for constraining stellar
evolution models across at least an order of magnitudue in iron
abundance.  Moreover, because of its high mass, M31 contains
$\gtrsim$90\% of the stars in the Local Group (outside the MW),
making it ideal for generating samples of sufficient size that
Poisson statistics are negligible and even rare phenomena are well
represented.  Finally, the stars in M31 are bright enough to be
accessible spectroscopically, allowing one to augment imaging
observations with spectroscopy, providing measurements of the
kinematics, metallicities, spectral types, and physical parameters of
star clusters and massive main sequence, asymptotic giant branch
(AGB), and red giant branch (RGB) stars.

We have therefore undertaken a new imaging survey of M31's bulge and
disk using HST.  The Panchromatic Hubble Andromeda Treasury (PHAT)
survey is being carried out as a ``Multi-Cycle Treasury'' program to
image a large contiguous area in M31, building upon the existing
ground-based studies that probe M31's most luminous stars \citep[most
recently, ][]{magnier1992, massey2006, mould2004,skrutskie2006}.  The
survey uses HST's new instrumentation to provide spectral coverage
from the UV through the NIR, with which one can effectively measure
the bolometric luminosity, spectral energy distribution, and
morphology of most astrophysical objects.  Broad-band coverage allows
one to constrain the mass, metallicity, and ages of stars, even in the
presence of extinction.  Other objects such as background active
galactic nuclei (AGN), planetary nebulae, X-ray binaries, and
supernova remnants, whose bolometric luminosity peaks outside the
accessible wavelength range, will still have distinctive spectral
energy distributions within the HST filter coverage.  Thus, the broad
wavelength coverage of the survey enables full characterization of
stars, their evolved descendants, and useful background sources.  When
complete, the survey should contain data on more than 100 million
stars, comparable to the number found in the Sloan Digital Sky Survey.

The legacy and scientific value of our survey is rich and diverse.
The survey data can be used to: (1) provide the tightest constraints
to date on the slope of the stellar initial mass function (IMF) above
$\gtrsim\!5\msun$ as a function of environment and metallicity; (2)
provide a rich collection of clusters spanning wide ranges of age and
metallicity, for calibrating models of cluster and stellar evolution
\citep[e.g.,][]{johnson2012}; (3) characterize the history of star
formation as a function of radius and azimuth, revealing the spiral
dynamics, the growth of the galaxy disk and spheroid in action, and
the role of tidal interactions and stellar accretion; (4) create
spatially resolved UV-through-NIR spectral energy distributions of
thousands of previously cataloged X-ray binaries, SN remnants,
Cepheids, planetary nebulae, and Wolf-Rayet stars, allowing full
characterization of these sources; (5) constrain rare phases of
stellar evolution \citep[e.g.,][]{rosenfield2012}; and (6) provide
rich probes of the gas phase and its interaction with star formation,
by using sub-arcsecond extinction mapping as a probe of the molecular
gas, and by comparing parsec-scale gas structures to the recent
history of star formation and stellar mass loss.  In the future, the
survey will provide the fundamental baseline for characterizing
sources in transient surveys (e.g., PanSTARRS, LSST) both by using
direct identification of counterparts, and by associating transients
with the properties of the surrounding stellar population, when the
counterpart is undetected even at HST's depth and resolution.  The
survey will also catalog hundreds of UV-luminous
background AGN, which can be used as absorption line probes of the
detailed physics of the ISM in future UV spectroscopic observations,
and as a reference frame for proper motion studies.

In this paper after giving a brief history of stellar population
studies in M31 (Section~\ref{historysec}), we present the survey
design (Section~\ref{surveysec}) and data reduction
(Section~\ref{datasec}), along with a thorough characterization of the
data quality.  In Section~\ref{resultssec}, we present initial color
magnitude diagrams covering several kiloparsec-scale contiguous
regions, located in the bulge, two major star forming rings, and the
outer disk.  We also present a preliminary analysis of the structure
of M31's disk, based on counts of RGB stars in the NIR.

\section{The History of M31 Resolved Stellar Population Studies}  \label{historysec}

The recognition of the presence of different stellar populations in
M31 began when \citet{hubble1929} noted that the outer parts of the
disk could be resolved into stars, while the central region showed
only unresolved light, despite clear spectroscopic evidence that it
was made up of stars.  When Baade carried out his 20-year study of M31
\citep[see ][]{baade1963}, he used red-sensitive plates to show that
the central regions could indeed be resolved into stars, and had
properties similar to those of Galactic globular clusters. Based on
these M31 studies, Baaded developed the influential concept of
Populations I and II, which formed the foundation of subsequent
population studies. The central areas of M31 were made up of red,
low-luminosity stars (Population II) and the main disk contained
luminous blue stars (Population I). At the time of Baade's work,
theoretical stellar evolution calculations \citep{schwarzschild1965}
were beginning to provide models that nicely fit the kinds of patterns
found observationally by Baade, allowing the empirical facts of
different stellar properties to be turned into quantitative
population histories. Baade concluded that the central bulge of M31
was made up of very old stars, while the bright blue arms were
young. He mapped out the structure of these arms and showed that the
most luminous areas were fragments of arms located at radial distances
of between 8 and $12\kpc$ \citep[see also work
by][]{arp1964}. Subsequent work by many astronomers \citep[see][for
references]{vandenbergh1991,hodge1992} confirmed this pattern, but
showed that the dichotomy of just two populations was too simple.
Population studies of the disk of M31 by \citet{williams2003}, for
example, showed that the populations vary across the disk, implying
different histories of star formation within the broader
classification of Population I.  The structures seen between 8 and
12$\kpc$ are now thought to be a ring of star formation at
$\sim\!10\kpc$ \citep{habing1984,gordon2006}, based in large part on
observations of M31's interstellar medium.

Global population studies were sparse in the later parts of the 20th
century, in spite of major wide-area ground-based CCD surveys of M31's
bright stellar content \citep[e.g.,][]{magnier1992}. Most papers
instead looked at smaller units, e.g., globular clusters
\citep[e.g.,][]{rich1996}, OB associations \citep[e.g.,][]{massey1986,
  haiman1994, hunter1996}, and individual HST fields
\citep[e.g.,][]{rich1995}.  After the turn of the century, large new
catalogs of M31's stars were produced in the optical
\citep{massey2006} and infrared \citep{skrutskie2006,mould2008}, but
much of the more recent work on stellar populations concentrated on
the halo and on extended disk stars \citep[e.g.,][ and many
others]{cuillandre2001, durrell2001, ferguson2001, sarajedini2001,
  rich2004, brown2006, kalirai2006, kalirai2006b, brown2007,
  brown2008, richardson2008, brown2009b, bernard2011}, and the
population of the bulge \citep[e.g.,][]{saglia2010, davidge2001,
  davidge2005, sarajedini2005, stephens2003, olsen2006,
  rosenfield2012}, which holds special interest due to the presence of
a significant super-solar metallicity stellar population.  Analyses of
stellar clusters \citep[e.g.,][]{krienke2007,krienke2008,
  barmby2009,hodge2009,perina2010} and various HST pointings have also
continued in earnest \citep{bellazzini2003}.

These and other studies have confirmed a basic picture where M31 hosts
a clear disk and bulge.  However, there is evidence for more complex
structures in the inner region \citep[including a bar and a boxy
peanut-shaped bulge, in addition to M31's classical
bulge;][]{lindblad1956,stark1977,stark1994,athanassoula2006,beaton2007},
the disk \citep[which shows a change in position angle at
$\sim$18\kpc, most likely due to a warp; e.g.,][]{walterbos1988}, and
the outer disk and halo \citep[most recently,
][]{tempel2010,mcconnachie2009}.  The disk contains ample evidence for
recent star formation, confined largely to major spiral arms or the
$10\kpc$ ring.  Near the ring, M31's current metallicity appears to be
comparable to or higher than that of the Milky Way, with ambiguous
evidence for a gradient in O/H; however, existing metallicity data is
surprisingly sparse, spanning a limited range in radii with large 0.5
dex scatter in the metallicity at any given radius
\citep{rubin1972,dennefeld1981, blair1982,zaritsky1994,galarza1999,
  venn2000,smartt2001, han2001,trundle2002}.  Analyses of older
stellar populations (RGB and PNe) suggests that the disk of M31 hosts
a wide range of stellar metallicities \citep[e.g.,][and references
therein] {worthey2005, jacoby1999, richer1999}.  The metallicity of
older stars in M31's bulge likewise spans a wide range, but seems to
reach super-solar metallicities in the very center \citep[e.g.,][and
references therein]{silchenko1998, lauer2012}.

Somewhat surprisingly, the distance to M31 remained largely
uncontroversial in the last two decades, unlike for the other galaxies
in the Local Group, be it the Large Magellanic Cloud
\citep[e.g.,][]{macri2006,vanleeuwen2007} or M33
\citep[e.g.,][]{bonanos2006,scowcroft2009}.  For example, the Cepheid
distance modulus to M31 of $\mu_0=24.41 \pm 0.08$
(\citealp{freedman1990}; see also \citealp{riess2012}, using data from
the PHAT survey) agrees well with the red clump distance modulus of
$\mu_0=24.47 \pm 0.06$ \citep{stanek1998}, a value numerically
identical to the TRGB-based distance of $\mu_0=24.47 \pm 0.07$
\citep{mcconnachie2005} and very close to the $\mu_0=24.44 \pm 0.12$
measured via a direct method using a detached eclipsing binary
\citep{ribas2005}.  Given such too-good-to-be-true agreement one would
normally suspect a ``bandwagon effect'', but all these papers use
different methods and different zero-point calibrations to derive
their distances. We will therefore adopt an M31 distance modulus of
$\mu_{M31,0}=24.45 \pm 0.05$ (physical distance of $d_{M31}=776 \pm
18\;$kpc) in this paper.

\section{Survey Design}  \label{surveysec}

The optimal survey design for efficiently imaging a large portion of
M31 requires finding solutions to several problems.  These include
determining the best portion of the galaxy to image, the best filters
to use, and the best design of the exposure sequences.  These in turn
require identifying the best way to efficiently tile large areas with
two cameras operating in parallel, while maximizing the photometric
depth and image resolution, given a limited total exposure time for
any location.  In this section we discuss the rationale for the
choices we made to address all of these issues in a way that we hope
has optimized the broad scientific utility of the program.

We start in Section~\ref{coveragesec} with the rationale for the
location and geometry of the PHAT survey area.
Section~\ref{filtersec} gives the scientific rationale for 6-filter
coverage from the UV through the NIR, and describes the specific
filter choices.  Section~\ref{visitsec} explains the exposure strategy
for fitting observations in six filters into two orbits.
Section~\ref{dithersec} shows how the observations are dithered to
optimize spatial resolution and repair detector defects.
Section~\ref{tilingsec} describes how the 2-orbit visits are packed
into an efficient 3$\times$6 mosaic of 18 pointings, building the 23
``bricks'' that tile the survey area.  We close in
Section~\ref{spectroscopysec} with a brief discussion of a coordinated
spectroscopic campaign.

\subsection{Areal Coverage} \label{coveragesec}

We evaluated several possible schemes for mapping M31.  We initially
considered producing a complete map of M31's star forming disk, using
only one orbit per pointing and a reduced number of filters.  However,
meeting the broadest set of scientific goals required more complete
filter coverage, and thus longer exposures at each position.  We
therefore reduced our areal coverage to a generous quadrant of M31.
Since spirals like M31 have fairly regular structure, we judged that a
quadrant would be sufficient to characterize the galaxy while still
covering enough area to allow statistically significant samples of
previously cataloged objects, of star forming regions across the
widest possible range of star formation intensity and metallicity, and
of azimuthal variations in the star formation rate.  We specifically
chose the northeast quadrant, which has the lowest internal
extinction, the largest number of regions with unobscured,
high-intensity star formation, and the least contamination from M32.
However, the tiling is generous enough that a substantial fraction of
the northwest quadrant is covered by the HST imaging as well.

The resulting survey area is a long, roughly rectangular region with a
slight bend in the middle.  The long axis of the survey begins at the
center of the galaxy, covering much of the bulge, and extends to the
northeast following the major axis of the disk out to the last
obvious regions of star formation visible in GALEX imaging \citep[see
maps in ][]{thilker2005}.  The short axis spans the minor axis of the
galaxy, out to a comparable projected radius as the long axis, once
M31's inclination is taken into account.

\begin{figure}
\centerline{
\includegraphics[width=6.1in]{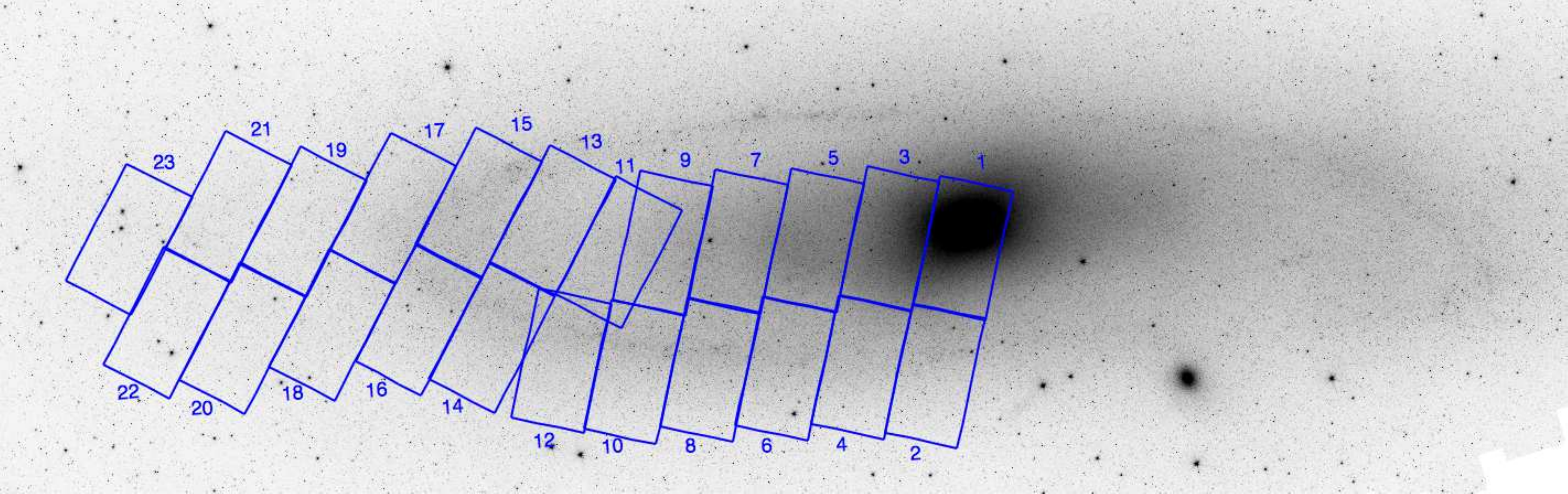}  
}
\centerline{
\includegraphics[width=6.1in]{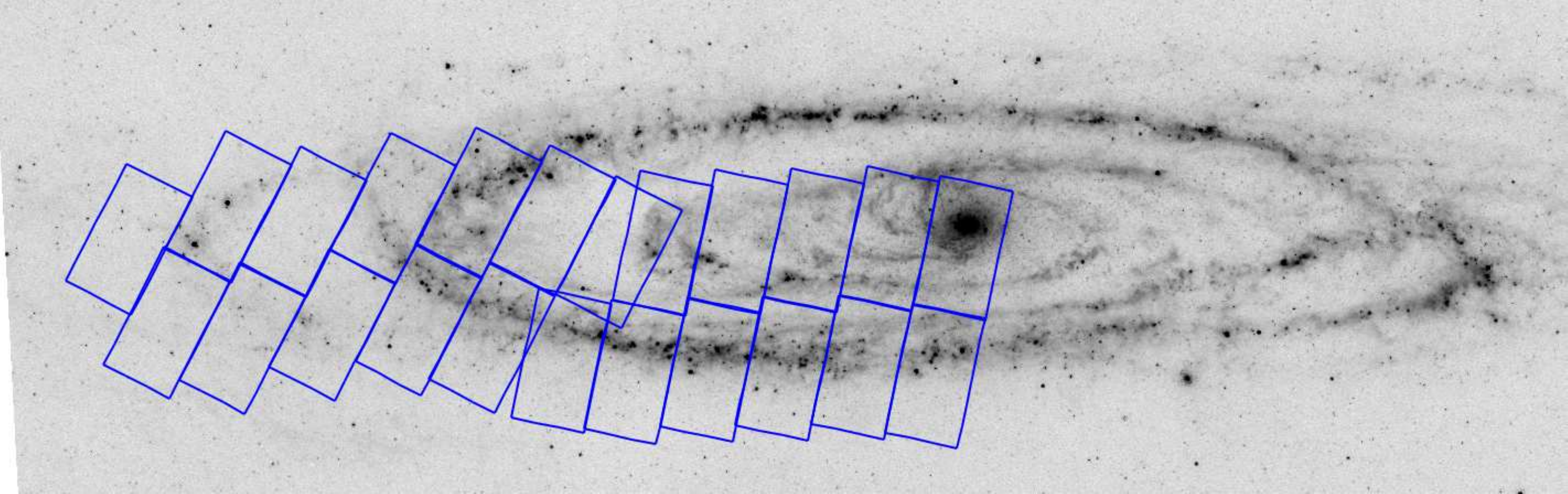}  
}
\centerline{
\includegraphics[width=6.1in]{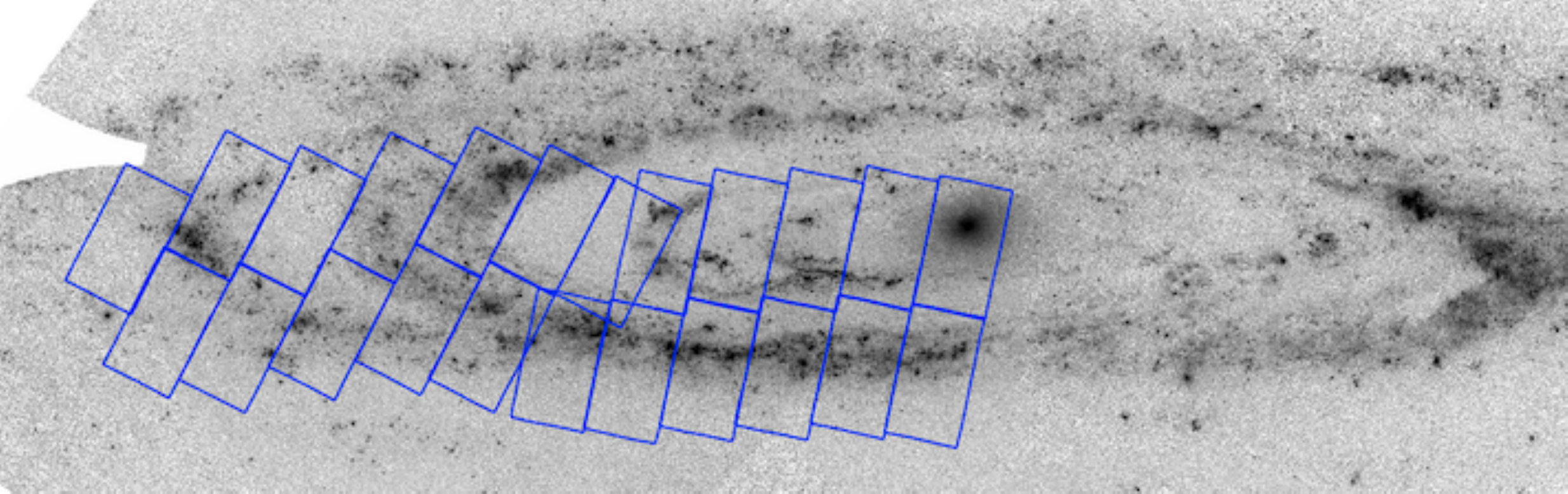}  
}
\centerline{
\includegraphics[width=6.1in]{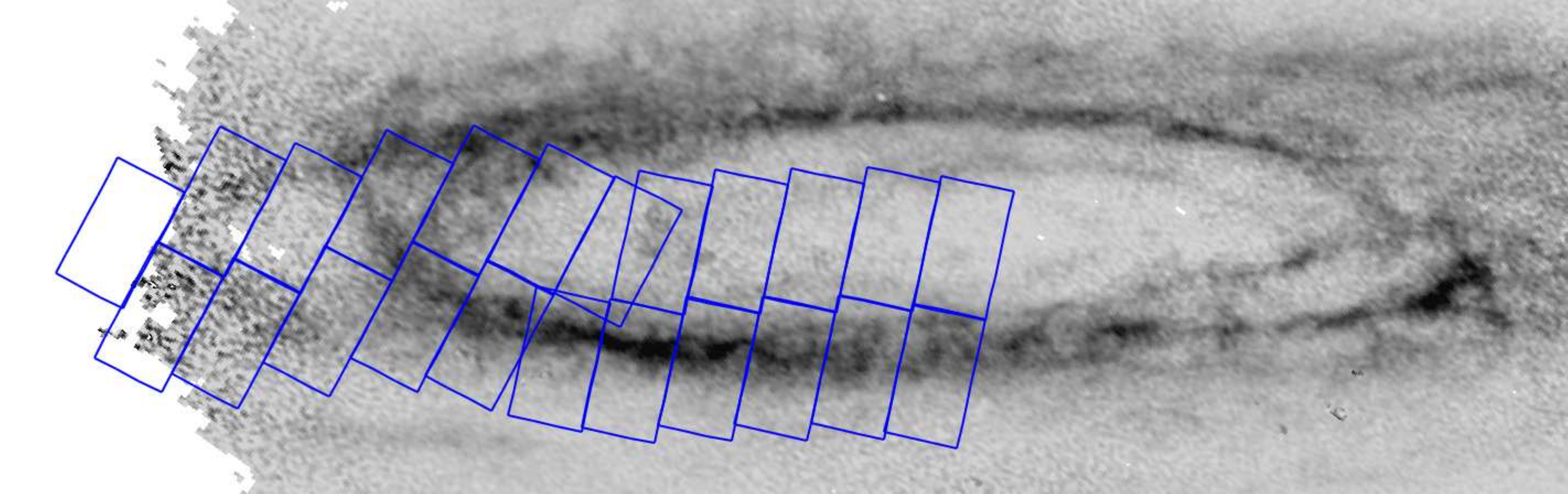}  
}
\caption{Location of PHAT ``bricks'', superimposed on a Spitzer
  3.6$\mu$ image \citep{barmby2006}, 24$\mu$ image \citep{gordon2006},
  GALEX FUV image \citep{thilker2005}, and Westerbork HI image
  \citep{brinks1984}, shown from top to bottom, respectively.  The
  numbering scheme for the bricks is shown in the top panel; odd
  numbers are along the major axis, starting with Brick 1 on the
  bulge.
  \label{footprintfig}}
\end{figure}

In richer detail, the survey area is contiguous and is divided into 23
sub-areas, called ``bricks,'' which solve a number of problems related to
the efficient design and operation of a large multi-year survey; we discuss the
design of the bricks in detail in section~\ref{tilingsec}.  In the present
context they can be considered to be rectangular tiles that ``pave'' the
complete survey area. The bricks are arranged into two
strips that respectively comprise the northwest and southeast halves
of the survey area.  Their detailed orientation and positioning gives the
edges of the survey area a somewhat serrated appearance.
Maps of the brick positions are shown in Figure~\ref{footprintfig},
superimposed on various multi-wavelength images.

The northern strip has 12
bricks extending along the major axis from the bulge to the outer
disk.  The bricks in the southern strip are aligned to complete
coverage of the quadrant.  The naming scheme is such that Brick
1 covers the bulge, with odd numbered bricks extending along the major
axis.  To increase the windows in which observations can be scheduled,
Bricks 1--11 are observed with a slightly different orientation than
Bricks 12--23.  In practice, we use the brick designations as a shorthand
for referring to different regions of interest within the survey.

The survey area samples the diversity of environments within M31.  The
southwest end of the survey encloses much of the bulge, including the
nucleus and the dense stellar population that surrounds it.  The
extent of the survey along both the major and minor axes of M31 is
sufficient to observe the transition of the bulge into the main disk
of M31.  Star formation is probed at many locations within the disk,
the most notable being the 10$\kpc$ ring of strong star formation,
which is tracked throughout the southern edge of the survey. The width
of the survey area completely tracks the roll-off of this arm into the
background disk, and on the southern side of the galaxy, the terminus of
the star forming disk itself, into the surrounding halo.  Weaker arms
of star formation just outside the bulge are sampled, as are
weaker spiral arms well outside of the main ring of star formation.
Between the zones of strong star-formation, the survey samples the
smooth background population of the disk over the entire major axis of
M31.  The northeast end of the survey also captures the termination
of the disk and the transition into the halo.  To a fair
approximation, the bricks that tile the survey area typically sample
one or two of these various features.  Table~\ref{brickobstab}
presents a brief description of the populations present in each brick.

Because of the large number of orbits required by this program, not
all bricks can be observed in a single Cycle.  We have therefore
prioritized the bricks to maximize the possible science output in
early years.  Our Year 1 priority was to complete bricks that sample
the full radial extent of the galaxy along the major axis, focusing on
the bulge and major star forming rings and/or spiral arms (Bricks 1,
9, 15, 21, followed by Bricks 17 and 23).  Year 2 priorities were to
sample the high intensity star forming ring (Bricks 2, 8, 12, 14, and
22), and to increase the sampling of star forming regions on the major
axis (Brick 5).  Year 3 priorities are to complete most observations
of the major star forming ring (Bricks 4, 6, 16, and 18), and to start
building larger contiguous regions in the inner and outer galaxy
(Bricks 3, 19).  Year 4 will be devoted to completing all remaining
areas (Bricks 7, 10, 11, 13, and 20).  The status of observations (as of
Fall 2011) are given in Table~\ref{brickobstab}.

\subsection{Filter Choices} \label{filtersec}

The choice of UV-through-IR filter coverage was driven by a number of
goals.  The need for two optical filters was obvious, given that
optical HST imaging data has proved to be the most efficient route to
deriving the deepest possible color-magnitude diagrams (CMDs) for the
largest number of stars.  Supplementing the optical data with two
additional NIR filters allows one to extend stellar population studies
to dusty regions, and to better constrain the bolometric fluxes of
intrinsically cool stars in important evolutionary phases (AGB stars,
Carbon stars, red core Helium-burning stars).  Adding an additional
two UV filters opens up science that can be done with hot stars, and
in particular permits simultaneous constraints of effective temperature
and extinction, when combined with measurements in optical filters
\citep[e.g.,][]{bianchi2007,romaniello2002,zaritsky1999}.

\begin{figure}
\centerline{
\includegraphics[width=5.5in]{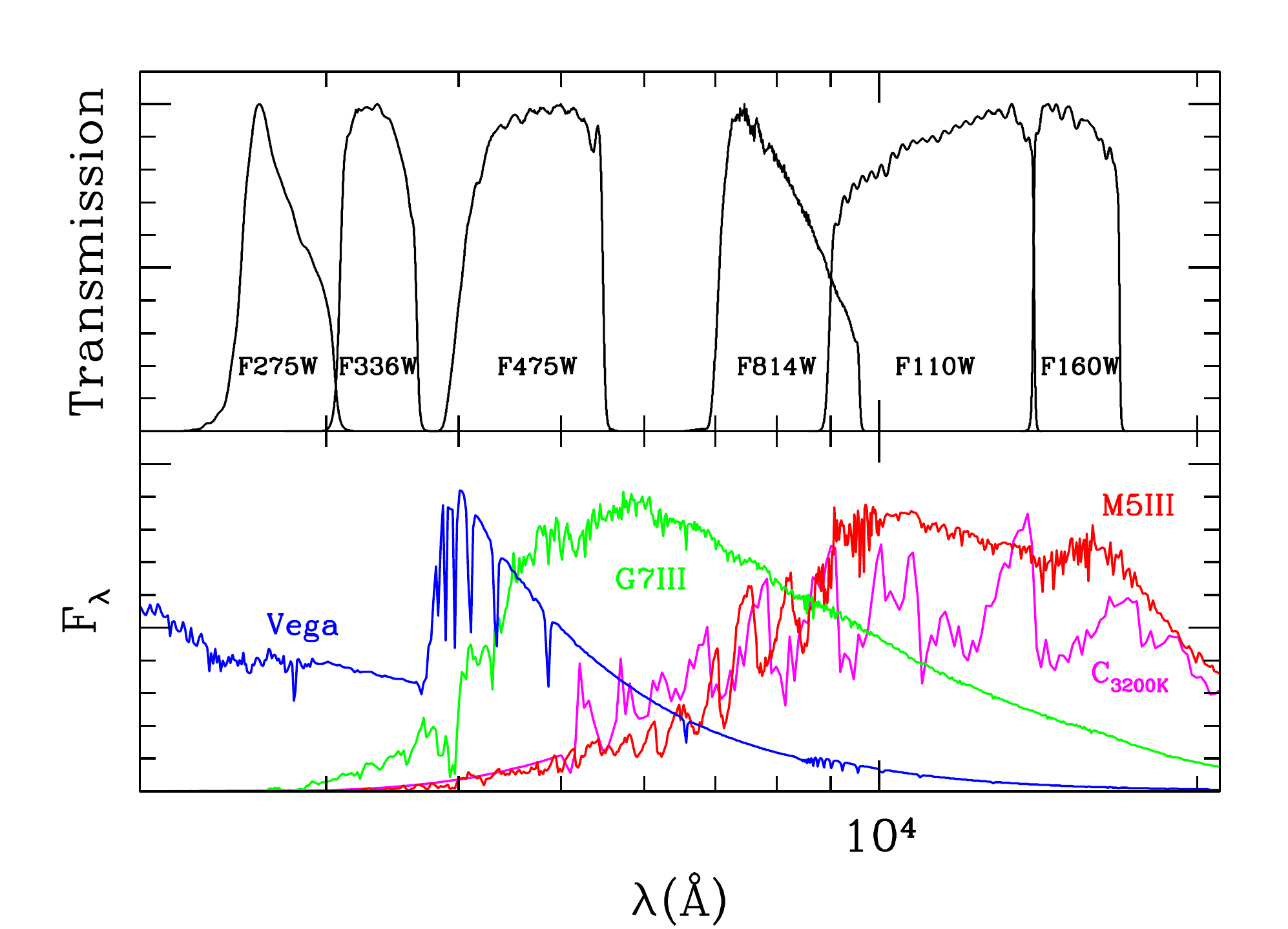}  
}
\caption{The relative transmission of the PHAT filter set (upper
  panel) as a function of wavelength.  The lower panel shows a number
  of stellar spectra for comparison, including a Vega A-star spectrum
  \citep[][ blue]{bohlin2007}, a G7III giant \citep[][
  green]{castelli2004}, a M5III giant \citep[][ red]{fluks1994}, and a
  carbon star with a 3200K atmosphere \citep[][ red]{loidl2001}.
  \label{filterfig}}
\end{figure}

The choice of specific filters, shown in Figure~\ref{filterfig}, was
made as follows:

In the UV, we adopted \fw{336} as the filter giving the highest
throughput.  This filter is also immediately blueward of the Balmer
break, giving the best constraint on its amplitude.  For the bluer UV
filter, we wanted to push to the shortest possible wavelengths to give
the best constraints on stellar temperature for the hottest stars, and
to provide the largest baseline for constraining the extinction law.
However, the bluest wide WFC3/UVIS filters have relatively low
efficiency, and moreover fall within the 2175\AA\ dust extinction
feature, which is known to vary dramatically with environment.  We
therefore adopted \fw{275} as the bluest, high throughput UV filter
that would not be significantly affected by variations in dust
composition.

In the optical, we adopted the \fw{814} filter, which has consistently
offered the highest throughput for stellar populations studies.  The
choice of the bluer optical filter was more problematic, as several
valid choices exist.  The \fw{606} filter offers very high throughput
and has been widely used for previous stellar population studies.
However, it is quite red, providing a limited color baseline in the
optical, and weaker constraints on the amplitude of the Balmer break
(when combined with \fw{336}).  The \fw{555} filter provides a wider
color baseline than \fw{606}, and has also been used for a large number
of HST stellar population studies.  However, it is significantly
narrower, and thus has much lower throughput.  We therefore rejected
these two filters and adopted the \fw{475} filter (approximately
SDSS-$g$).  This filter is as broad as \fw{606}, but is bluer than
both \fw{606} and \fw{555}, providing much better color separation from
\fw{814}.  When combined with \fw{336}, it provides good constraints on
the Balmer break.  The only cost is some loss in depth for
intrinsically red stars (i.e., on the red giant branch).  However,
since much of the disk is crowding limited, this limitation is not
severe at most pointings.  

For the NIR, we use the \fw{110} and \fw{160} filters.  These two
filters provide the highest throughput with the WFC3/IR camera, and
have been used successfully in our previous SNAP survey of stellar
populations in nearby galaxies \citep{dalcanton2012}.  The only
drawback with this combination is the partial overlap of the \fw{110}
and \fw{814} bandpasses.  The only other feasible substitutes for the
\fw{110} filter would have been the \fw{140} filter, which overlaps the
\fw{160} filter, and the \fw{125} filter, which is much narrower than
\fw{110}, and which has less color separation and temperature
sensitivity when paired with \fw{160}.

Almost all of the PHAT filters have been used in calibration
observations of nearby stellar clusters.  The WFC3 Galactic Bulge
Treasury Program \citep[GO-11664;][]{brown2009,brown2010} have taken
calibration observations in \fw{814}, \fw{110}, and \fw{160}.  \fw{336}
observations of the same clusters have been carried out in GO-11729
(PI: Holtzman).  Our survey is executing observations of the same
clusters in combinations of \fw{275} and \fw{475} filters, to complete
the calibrations of our filter set.  Additional calibration in
\fw{336}, \fw{475}, and \fw{814} will be provided by GO-12257 (PI:
Girardi) for intermediate-age Magellanic Cloud clusters.

The adopted filter set should allow us to make strong constraints on
the effective temperature and extinction of the stars in our sample.
With the UV$+$optical filters, we expect to be able to separate
$\Teff$ and $E(B\!-\!V)$ with little degeneracy for both hot stars
\citep[$12\thou\!\lesssim\!\Teff\!\lesssim\!40\thou K$;
][]{massey1995b,romaniello2002,bianchi2007} and cooler stars
\citep[$5500\!-\!6500K$; ][]{zaritsky1999}.  Inspection of reddening-free
diagrams suggests that the optical$+$NIR combination will allow us to
extend $E(B\!-\!V)$ constraints to cooler stars ($\Teff\!<\!5000K$)
as well.

\subsection{Exposure Sequences} \label{visitsec}

The primary aims of our exposure plan are (1) imaging two filters per
camera; (2) achieving Nyquist-sampled images through dithering where
possible; and (3) avoiding saturation of bright sources.  As we
describe below, these goals are balanced against constraints on the
number of images that can be downloaded when running WFC3~and~ACS in
parallel, and on limitations on the duration of an orbit.  Because of
the strains that this program puts on HST's observing schedule, the
exposure sequences must fit within the shortest possible orbit
duration (``{\tt{sched100}}'', i.e., an assumed duration that is
schedulable for 100\% of the orbits).  This constraint maximizes the
schedulability, particularly in the summer observing season
(Section~\ref{tilingsec}) which has a more restrictive scheduling
window.  In the winter observing season, we can relax the orbit length
constraints to ``{\tt{sched60}}'' (a duration that fits within 60\% of
the orbits), giving slightly longer exposures.

Scheduling observations in six filters requires two orbits, with the first
orbit devoted to WFC3/UVIS$+$ACS/WFC and the second to WFC3/IR$+$ACS/WFC; this
ordering minimizes persistence in the WFC3/IR channel
(ISR WFC3 2008-33; McCullough \& Deustua 2010), by allowing more time
for the persistent charge to decay.  We run WFC3 in primary, and ACS
in parallel.\footnote{Because the corrections for differential velocity
  aberration (ISR OSG-CAL-97-06; Cox 1997) by the pointing control
  software are optimized for the primary observations, the ACS
  parallels will not automatically be corrected for this
  milli-arcsecond level effect.  The affine corrections used for
  astrometry (Section~\ref{astrometrysec}) should provide adequate
  corrections, however.}

The strongest constraints on the observing sequence within each orbit
come from the limited buffer space available on-board the HST
instruments.  The ACS buffer can only hold one full-frame ACS/WFC
image, and the WFC3 buffer can hold either two full-frame WFC3/UVIS
images, or three WFC3/IR images (if the number of non-destructive
reads in the latter is reduced to $\sim$10 from the nominal 15
samples) before needing to be dumped.  However, the time to dump the
buffer is substantial ($\sim$340 seconds), which can lead to
significant inefficiencies if the observing sequence does not allow
the buffers to dump in parallel with the observations.  This issue is
particularly severe when running both imaging cameras simultaneously,
because of the high data rates.

\begin{figure}
\centerline{
\includegraphics[width=6.0in]{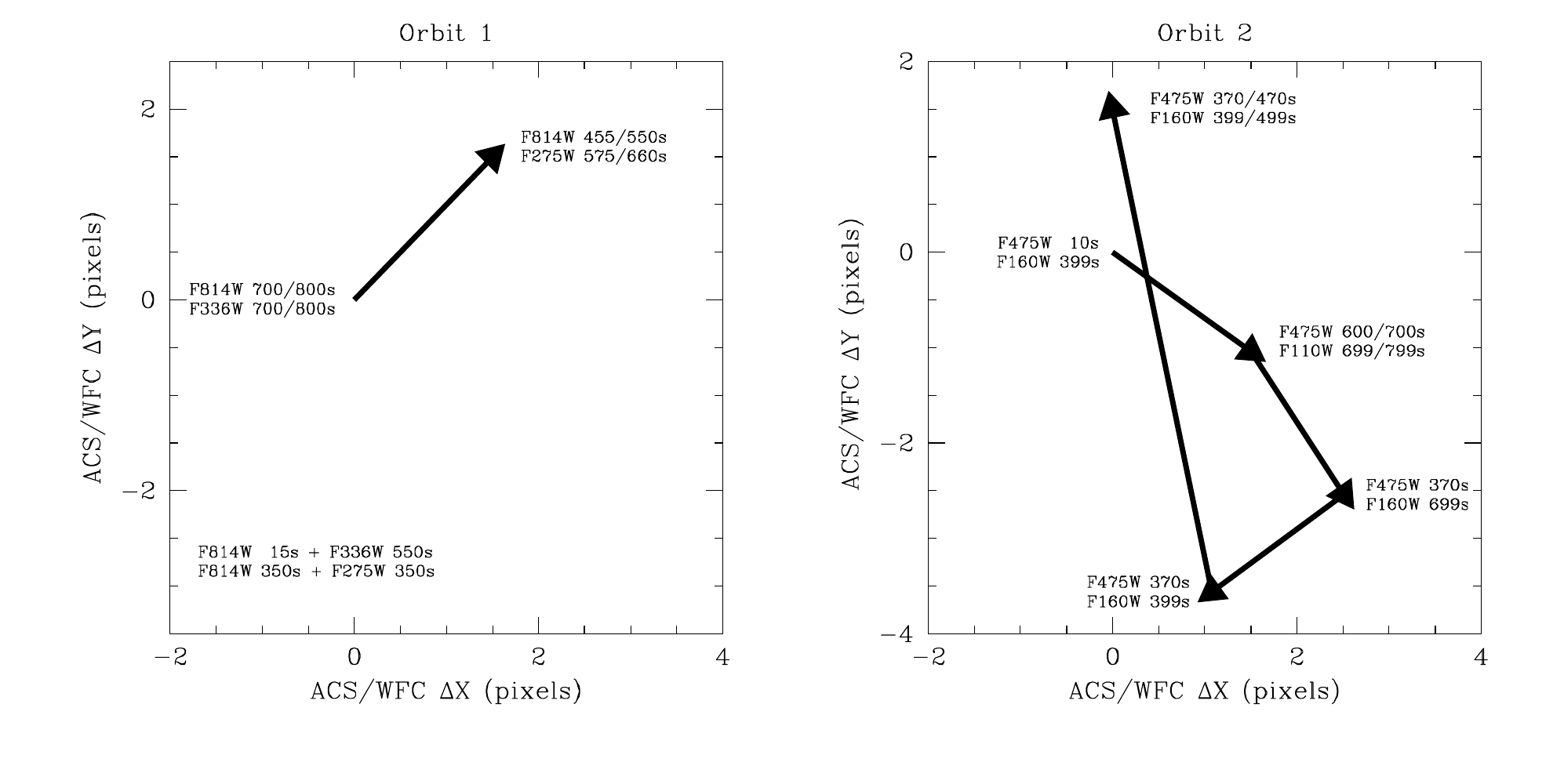}  
}
\caption{Map of the sub-pixel dither and exposure sequence for the two
  orbit visits.  Arrows show the dithers relative to the previous
  exposure.  The text at each vertex indicates the the filter and
  duration of the primary and parallel exposures taken at that
  position.  Exposure times are given for the shorter {\tt{sched100}}
  summer observing sequences and the longer {\tt{sched60}} winter
  sequences (i.e., ``700/800s'' means 700s in summer and 800s in
  winter; see text in Section~\ref{visitsec}).  For Orbit 1, only the
  sub-pixel dither between the third and fourth primary exposures is
  shown, since there is no dither between the first and second
  exposure, and the dither between the second and third is a large
  $\sim$37 pixel dither to span the chip gap.
  \label{dithermapfig}}
\end{figure}

After extensive experimentation, we were able to find exposure
sequences that allow four WFC3/UVIS exposures and four ACS/WFC exposures
in the first orbit, and five WFC3/IR exposures and five ACS/WFC exposures
in the second orbit, with minimal latencies due to buffer dumps.  The
resulting observing sequence also fits in dithers between every pair
of exposures, with the exception of the first two in the
WFC3/UVIS$+$ACS/WFC orbit (see Figure~\ref{dithermapfig}).  This high
observing efficiency comes at the expense of unequal exposure times
for observations in a single filter, and fewer non-destructive reads
in the WFC3/IR observations (to allow the buffer to hold more WFC3/IR
images between dumps).  The sequence is summarized in
Table~\ref{dithertab}.

During the first orbit, the four WFC3/UVIS exposures follow the sequence:
\fw{336} ($550\sec$; $550\sec$), \fw{275} ($350\sec$; $350\sec$), \fw{336}
($700\sec$; $800\sec$), \fw{275} ($575\sec$; $660\sec$), where the
two numbers in parentheses indicate the exposure times in the summer
{\tt{sched100}} and winter {\tt{sched60}} observing seasons,
respectively.  We obtained two exposures in each filter to enable CR
rejection and cover the chip gap; three exposures per filter would have
been preferable, but constraints on buffer dumps limited the observations to
no more than four WFC3/UVIS exposure per orbit.  Note that the majority of the
chip gap is only imaged in one exposure per filter, making CR
rejection more challenging in this region.

The four ACS/WFC exposures in the first orbit are all in \fw{814}, with
exposure times of ($15\sec$; $15\sec$), ($350\sec$; $350\sec$),
($700\sec$; $800\sec$), and ($455\sec$; $550\sec$).  The short
$15\sec$ \fw{814} exposure is included to allow photometry for stars
brighter than $\fw{814}\!\sim\!17.5$.\footnote{There is no equivalent
  short ``guard'' exposure in the UV, since the number of saturated
  stars was expected to be negligible (based on existing ground-based
  $U$-band data from the \citet{massey2006} Local Group Survey), and
  the penalty in exposure time would be large for the unsaturated
  stars (given the small number of possible WFC3/UVIS exposures).  In
  practice, it may be possible to pull out reasonable photometry from
  even the saturated stars (see ISR WFC3 2010-10 by Gilliand et al.~2010 for
  WFC3/UVIS and \citealp{anderson2008} for ACS/WFC).}

 During the second orbit, the five WFC3/IR exposures follow the sequence
\fw{160} (\mbox{\texttt{NSAMP}$=\!9$,} \texttt{STEP200}, $399\sec$;
\mbox{\texttt{NSAMP}$=\!9$}, \texttt{STEP200}, $399\sec$),
\fw{110} (\texttt{NSAMP}$=\!13$, \texttt{STEP100}, $699\sec$;
\mbox{\texttt{NSAMP}$=\!11$,} \texttt{STEP200}, $799\sec$),
\fw{160} (\texttt{NSAMP}$=\!9$, \texttt{STEP200}, $399\sec$;
\texttt{NSAMP}$=\!9$, \texttt{STEP200}, $399\sec$),
\fw{160} (\texttt{NSAMP}$=\!9$, \texttt{STEP200}, $399\sec$;
\texttt{NSAMP}$=\!9$, \texttt{STEP200}, $399\sec$),
\fw{160} (\texttt{NSAMP}$=\!9$, \texttt{STEP200}, $399\sec$;
\texttt{NSAMP}$=\!11$, \texttt{STEP100}, $499\sec$),
where the numbers in
parentheses give the number of samples, the \multiaccum\ exposure
sequence, and resulting exposure time for the summer {\tt{sched100}}
and winter {\tt{sched60}} observing seasons, respectively.  The \texttt{STEP}
sequence was adopted to provide optimal sampling for the wide range of
fluxes expected for observations of stellar populations.  Since the
\multiaccum\ sequence allows for CR rejection in an accumulated
single image, multiple images are not needed.
We therefore chose to use multiple pointings in \fw{160} only, to allow
the construction of a Nyquist-sampled image in this filter, at the
expense of not being able to reject chip defects in \fw{110} (the WFC3/IR
``blobs'' and the ``death star''; ISR WFC3 2010-06, Pirzkal 2010).
This trade-off was preferred, given the lower resolution of the
WFC3/IR channel compared to the other cameras, and the partial
overlap of the \fw{110} and \fw{814} bandpasses.

The five ACS/WFC exposures in the second orbit are all in \fw{475}, with
exposure times of ($10\sec$; $10\sec$), ($600\sec$; $700\sec$),
($370\sec$; $360\sec$), ($370\sec$; $360\sec$), and ($370\sec$;
$470\sec$).  As for \fw{814}, we include a short ``guard'' exposure to
allow photometry of stars brighter than $\fw{475}\!\sim\!18.5$.

The resulting total exposure times for each two-orbit visit are
($925\sec$; $1010\sec$) in \fw{275}, ($1250\sec$; $1350\sec$) in
\fw{336}, ($1720\sec$; $1900\sec$) in \fw{475}, ($1520\sec$; $1715\sec$)
in \fw{814}, ($699\sec$; $799\sec$) in \fw{110}, and ($1596\sec$;
$1696\sec$) in \fw{160}, where the first and second numbers indicate
the exposure times in the summer and winter observing seasons.  The
effective exposure times for the ACS observations will be at least a
factor of two larger due to overlaps (see Figure~\ref{exptimemapfig}),
giving nearly two full orbits of exposure time in both \fw{475} and
\fw{814} outside the chip gap.

\subsection{Dithering Strategy} \label{dithersec}

The exposure sequences were interleaved with small angle maneuvers to
produce dithered images.  Large dithers provide for the repair of
detector defects (hot or bad pixels, missing columns, and so on), as
well as coverage of the segments of M31 that fell into the ACS/WFC
or WFC3/UVIS chip gaps in any single exposure.  Smaller sub-pixel
dithering enables well-sampled (Nyquist) images to be generated from a
set of under-sampled single exposures.  Unfortunately, all of these
tasks were difficult to satisfy simultaneously, given the limited
number of exposures possible, the need to operate WFC3 and ACS in
parallel, and the geometric distortion present in all cameras.  The
distortion causes the angular pixel-scale to vary with
field position in the cameras, such that larger dither steps would
produce a highly variable and generally non-optimal pattern of
sub-pixel steps (once the integral portion of the step is subtracted)
as a function of field location.  Constructing Nyquist-sampled images
thus requires keeping the total size of the dithers small, which
conflicts with the large dithers needed to bridge the CCD-gaps.

With these qualifications, our primary goal was to use dithers to
produce Nyquist-sampled images in as many filters as possible.  We
were able to do this in \fw{475}, \fw{814}, and \fw{160}.  The
\fw{475} and \fw{160} filters each require four dithers such that the
fractional portions of the shifts map out a $2\times2$ pattern of 0.5
pixel steps.  In contrast, only two dither positions are required in
\fw{814} to achieve Nyquist-sampling, given its larger PSF.  Although
the use of a single diagonal dither to produce Nyquist sampling is
less intuitive than traditional $2\times2$ dither pattern, a single
diagonal shift can produce fractional offsets of 0.5 pixels in both
$X$ and $Y$ at the same time, allowing the two images to be interlaced
to produce a new image with a rectilinear sampling-grid that is a
factor $\sqrt{2}$ finer than the native ACS/WFC sampling\footnote{A
  perfect analogue is the interlacing of black and white squares on a
  chess board.  The black and white squares each form a regular grid
  with the sampling interval $\sqrt{2}\times$ larger than the spacing
  of the interlaced squares of the chess board.}.  This approach is
sufficient to ensure adequate sampling for F814W, but not F475W, which
requires $2\times$ finer subsampling.

The dithers used to produce Nyquist-sampling in the three
aforementioned filters are summarized in Table \ref{dithertab}.  The
specific dithers were designed so that where possible the correct
fractional sub-sampling could be achieved in both ACS and WFC3 cameras
simultaneously.  The magnitude of the dithers is actually large enough
so that small defects in the detectors can be stepped over, satisfying
a second goal of the dithering, but not so large that geometric
distortion ruins the accuracy of the shifts.

The design of the dither sequence was helped by the artful interplay
of the location of the guard exposures and filter changes in the
primary and parallel cameras, so that simultaneous sub-sampling
dithers were not always required in both instruments at the same time.
Of the four sub-sampling dithers needed for \fw{475} and \fw{160}, one
dither is ``exact'' for each filter, while three are done
simultaneously in the two filters and provide slightly non-optimal
sub-pixel sampling.  The single \fw{814} sub-sampling dither occurs
between a filter change in WFC3/UVIS and is optimal for that
filter. In general, Nyquist-images can still be readily constructed
from dithers that stray significantly from the optimal sub-pixel steps
\citep{lauer1999a}, a consideration that must be allowed for, even with
an optimally-designed sequence, since the HST Fine Guidance Sensor
(FGS) has limited accuracy in executing the dithers.

The dithers in the ACS \fw{475} and \fw{814} filters are not large
enough to fill in the gap between the two CCDs in that camera.
However, since every region that falls in the chip gap in one brick
will be imaged again in another brick, completely filled coverage with
the ACS can still be achieved.  We thus have chosen to emphasize
Nyquist-sampling in all ACS filters at the expense of depth in the
small portion of the survey that falls into the chip gaps.

Likewise, while the dithers in \fw{160} are large enough to counter
defects of one or two pixels in extent, the shifts are unfortunately
not large enough to completely sample over the ``IR blobs'' or the
``death star'', which have characteristic sizes of 10-15 pixels and 51
pixels, covering a total of $\sim$700 pixels and $\sim$2000 pixels,
respectively (ISR WFC3 2010-06, Pirzkal 2010).  We have found that the
regions of the IR blobs can still produce adequate photometry,
although with slightly higher uncertainties due to the 5-30\%
reduction in sensitivity in the area of the blobs.  The only area then
that completely lacks WFC3/IR data is the 0.2\% of the field covered
by the WFC3/IR ``death star''.  Again, we consider the benefits of
obtaining Nyquist-sampling over most of the survey in \fw{160} to
outweigh the sacrifice of a small portion of the area lost to
unrepaired defects.

In contrast, we could not obtain the four-point dither pattern
required to construct Nyquist-sampled images in either WFC3/UVIS filter.
However, we do not expect this to be a significant scientific
limitation, given that the WFC3/UVIS data are not crowding limited, and that
any star detectable in WFC3/UVIS will also be detected in the
Nyquist-sampled \fw{475} images.  Again, in contrast to the ACS/WFC, we did
include a single large dither of $\sim37$ pixels to cover the chip gap
in both WFC3/UVIS filters, since we have essentially no later
duplicate coverage of the fields with this camera.  

We also could not obtain the set of images needed to achieve
Nyquist-sampling in the WFC3/IR \fw{110} filter, due to the buffer-dump
limits.  However, we expect that every star that is detectable in the
\fw{110} image will also be detectable in either of the Nyquist-sampled
\fw{814} or \fw{160} images.

\begin{figure}
\centerline{
\includegraphics[width=4.5in]{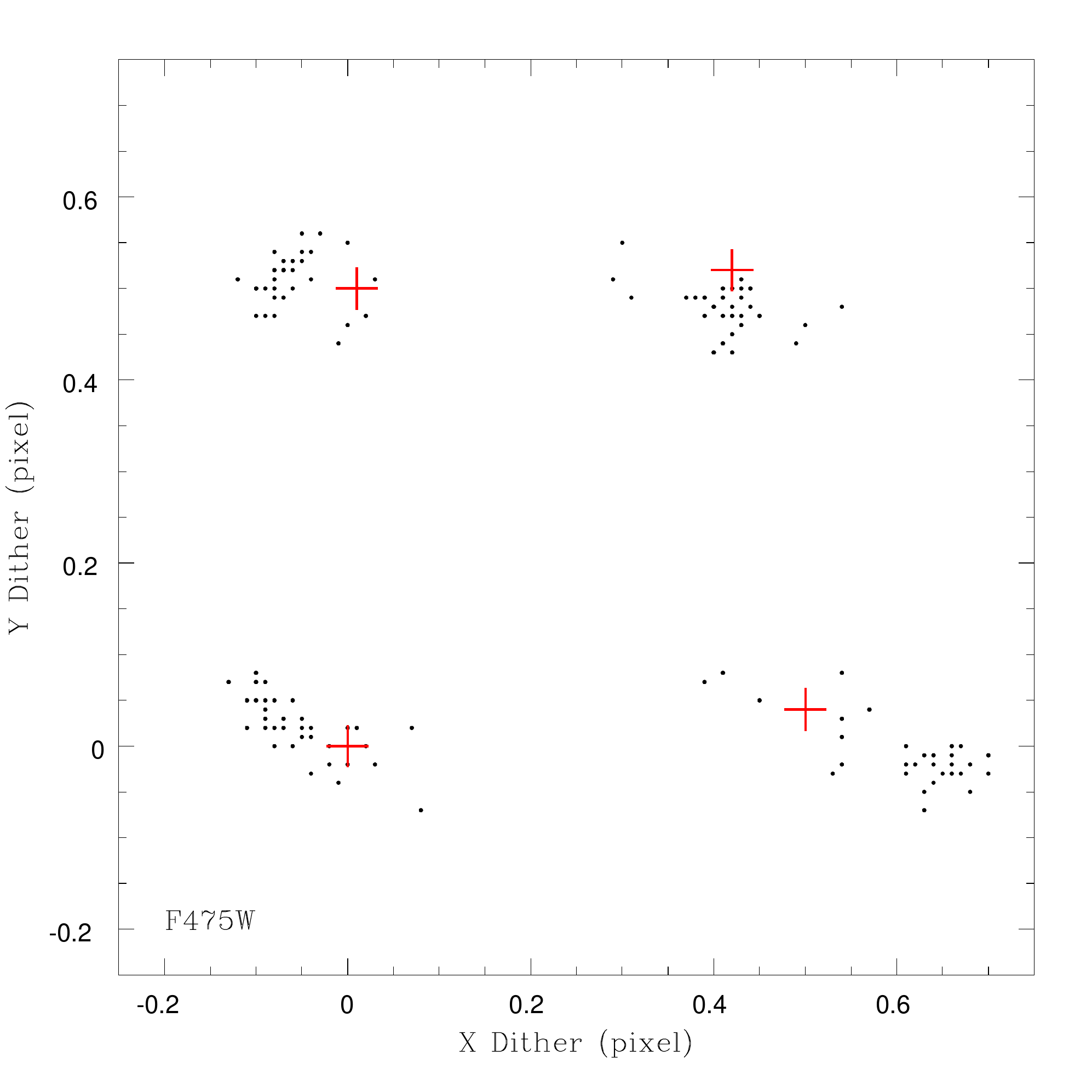}  
}
\caption{The sub-pixel dithers requested (red crosses) and achieved
  (points) are shown for ACS/WFC \fw{475} images obtained during the
  Winter 2010 observing season.  Note that the requested dithers do
  not fall exactly at the optimal $2\times2$ 0.5 pixel locations, as
  the dither sequence operates on both the \fw{475} and \fw{160} filters
  simultaneously, and thus is a compromise between the needs of the
  two filters.  The deviation of the actual dithers from the ideal
  interlace pattern is small enough that algorithms that can
  reconstruct Nyquist-sampled images from non-optimal dithers, such as
  that of \citet{lauer1999a}, should still work satisfactorily.
\label{ditherfig}}
\end{figure}

All dithers are tabulated in Table \ref{dithertab}.  A concern at the
time of the Phase-II preparation of the program was that the accuracy
of the dithers needed to achieve simultaneous Nyquist-sampling in the
\fw{160} exposures in parallel with the \fw{475} exposures would be
difficult to realize in practice.  In fact, due to an unfortunate
$\sqrt{2}$ error in the scale of the dithers, the first season of data
(summer 2010) was not properly Nyquist-sampled.  This error was
corrected in time for all subsequent observations, however, and we
find that the HST FGS is indeed able to perform the sub-pixel dithers
to the level required.  Figure \ref{ditherfig} shows the dithers
realized in \fw{475} during the Dec 2010 -- Feb 2011 period of
observation.  This filter makes the strongest demands for dither
accuracy of the three filters in which we will construct
Nyquist-sampled images, but nearly all dithers fell within 0.15 pixel
of the targeted offsets, which is sufficient to achieve the desired
degree of sub-sampling \citep{lauer1999a}.

\subsection{Tiling Strategy} \label{tilingsec}

We use the basic exposure sequence above (two orbits per
pointing, one with WFC3/IR and one with WFC3/UVIS in primary, and ACS/WFC
in parallel) as the foundation for a highly efficient tiling scheme.

The primary tiling is based on WFC3/IR, which has the smallest FOV of
all three imaging cameras.  We arrange the WFC3/IR pointings in a
3$\times$6 grid, with $<$5$\arcsec$ overlap among the WFC3/IR FOVs;
due to distortion, the overlaps range from $\sim\!1.5\arcsec$ to
$\sim\!5\arcsec$ for adjacent pointings.  The WFC3/UVIS exposures use
a small offset of 1.655 pixels in {\tt{X}} and 2.96 pixels in {\tt{Y}}
with respect to the standard {\tt{UVIS-CENTER}} aperture, such that
the same pointing center is maintained with respect to the
{\tt{IR-FIX}} aperture used to define the WFC3/IR pointings.  The
overlap between adjacent WFC3/UVIS exposures is $\sim\,0.5-1\arcmin$, due
to the camera's larger FOV.

We cover the 3$\times$6 ``brick'' with ACS by observing the two
3$\times$3 halves of the brick with two orientations, 180$^\circ$ apart.
Due to telescope constraints, observations in these two orientations
are taken $\sim$6~months apart, in two ``seasons''.  In the first
observing season, the observations produce a contiguous 3$\times$3
half-brick of overlapping WFC3 observations, and an adjacent
3$\times$3 half-brick of overlapping ACS observations.  Due to the
large ACS FOV, most points within the survey region have ACS data from
two to four separate visits.  In the second observing season, the
telescope is rotated by $180^\circ$, and completes the 3$\times$3 WFC3
pointings on the half-brick covered previously with ACS.  The switch
in orientation leads the parallel ACS pointings to simultaneously
cover the 3$\times$3 WFC3 pointings taken during the previous
orientation.

The resulting 3$\times$6 brick has complete areal coverage in all
three cameras, producing a contiguous area of
$12\arcmin\!\times\!6.5\arcmin$ in 36 orbits; exposure maps of the
bricks are shown in Figure~\ref{exptimemapfig}.  Each pointing in the
brick follows a consistent naming scheme, such that Field 1 is
the pointing in the northeast corner of the brick, and Field 18
is the pointing in the southwest.  The full naming scheme for each
target position is of the form ``{\tt{M31-B\#\#-F\#\#-XXX}}'', where
``\texttt{\#\#}'' represents a two-digit number; the number after
{\tt{B}} indicates the brick number, the number after
{\tt{F}} indicates the field number within the brick,
and the {\tt{XXX}} indicates the camera
on the WFC3 or ACS instruments ({\tt{UVIS}}, {\tt{WFC}}, {\tt{IR}}).
Note that parallels are named according to the area they overlap,
rather than for the location of the primary; thus, ACS images labeled
as ``B01-F04'' overlap WFC3 images of the same name.

The default orientation of the brick was set to allow the observations
to be maximally schedulable in both $180^\circ$ orientations.  The
optimal orientations had large numbers of schedulable days in the
summer (peaking in July) and the winter (peaking in January).  While a
single orientation for the whole survey would be preferable for
producing minimal overlaps between adjacent bricks, upon consultation
with STScI it was decided to adopt two default orientations for the
survey.  This choice allowed slightly longer scheduling windows for
the observations, reducing their impact on the scheduling of other
programs.  The 11 bricks closest to the center of the galaxy (Bricks
1-10, and Brick 12) are observed with {\tt{ORIENT}} set to 69 (winter)
or 249 (summer).  The remainder of the bricks in the outer galaxy are
observed with {\tt{ORIENT}} set to 54 (winter) or 234 (summer).  Each
brick is assigned to a unique proposal ID number (PID).  The
approximate corners of the NIR footprint of the bricks and their PIDs
are given in Table~\ref{brickcornertab}.

\subsection{Spectroscopy} \label{spectroscopysec}

The PHAT imaging described here is being augmented with extensive
spectroscopy.  Individual stars in M31 are bright enough be observed
spectroscopically, allowing us to measure the kinematics,
metallicities, spectral types, and physical parameters of star
clusters and massive main sequence, asymptotic giant branch (AGB), and
red giant branch (RGB) stars.  

The majority of PHAT spectroscopy to date has been carried out with
DEIMOS on Keck ($R\!=\!6000$, $\sim$6400--9100\AA), using a similar
observing set-up as \citet{gilbert2006}, \citet{kalirai2006}, and
\citet{guhathakurta2006}.  A total of 21 slitmasks covering much of
the HST footprint have already been observed, and have produced
$\sim$5000 radial velocities accurate to $\sigma_{v}\!=\!5\kms$ down
to $I_o\!=\!22$ ($\sim$2$^m$ fainter than the TRGB); preliminary
reductions are described in \citet{dorman2012}.  These spectra are
being used primarily for kinematic decomposition of different M31
sub-populations.  Future analyses will constrain stellar metallicities
and spectral types from the Ca triplet and other absorption features.
These data also can be used to estimate ionized gas metallicities from
H$\alpha$/[NII]/[SII] emission lines.

Additional spectroscopic programs are underway, with the goals of: (1)
measuring accurate HII region abundances using weak line methods; (2)
measuring spectral types of hot stars using medium resolution
spectroscopy at $<6000$\AA; (3) measuring the masses and metallicities
of stellar clusters.  When possible, these various spectroscopic
programs also target serendipitous ``interesting'' sources from the
PHAT survey, including PNe, AGN candidates, and X-ray counterparts.
We are also extending the star cluster and planetary nebula survey of
\citet{caldwell2009,caldwell2011} to PHAT targets, with more than 200
new spectra obtained thus far.  These various spectroscopic programs
will be described in future papers.

\section{Data} \label{datasec}

We now describe the present state of the PHAT observations and
pipeline for image processing, stellar photometry, and astrometry.  We
also include assessments of the current data quality.  Because this
project is still actively acquiring data, we expect future data
releases to incorporate on-going improvements in data processing and
in the resulting photometric catalogs.  We note planned upgrades and
processing revisions throughout this section, for cases where substantial
development work is already underway.

\subsection{Observations} \label{observationsec}

Figure~\ref{footprintfig} and Table~\ref{brickobstab} summarize the
observations discussed in this paper.  We list bricks, their PIDs, and
the range of dates over which data were taken for each orientation
(i.e., the winter observations which populate WFC3 observations of the
eastern halves of the bricks, and the summer observations which
populate WFC3 observations of the western halves).  Observations for
the program began July 12, 2010, and complete 6-filter coverage is
available for Bricks 1, 9, 15, 17, and 21 as of Fall 2011.  Brick 23
is nearly complete, but two fields are being re-observed due to
guiding failures (Section~\ref{problemsec}).  Our primary scheduling
priority is to finish bricks, and we therefore anticipate that
observations of the current ``half-bricks'' (2, 5, 8, 12, 14, 16, 18,
and 22) will be completed in the winter 2011/2012 observing season.
Observations are currently scheduled for the first halves of Bricks 4
and 6, as well.

Figure~\ref{brickwidergbfig} shows false-color images (generated by
Zolt Levay of STScI) of Bricks 1, 9, 15, and 21, which were the first
four bricks completed in the program.
Figure~\ref{brickwidergbzoomfig} shows a small portion of a
false-color image for Brick 9, and reveals the rich information that
can be seen at full HST resolution.

\subsubsection{Known Problems}  \label{problemsec}

In a data set this large, there are bound to be glitches in the
observations.  Here we briefly point out known issues with some of the 
observations

There were several problems affecting observations of Brick 23 (PID
GO-12070).  The Fine Guidance Sensors (FGS) failed to acquire guide
stars during Visits 03 and 13, making data from these observations
unusable.  As a result, Brick 23 is currently missing high quality
WFC3 observations for Fields 03 and 13, and ACS observations for Field
16; the ACS parallels in Field 06 appear to be usable.  
Data will be taken for these fields in the winter 2011/2012
observing season. In addition, one WFC3/UVIS
exposure for target M31-B23-F12-UVIS was not completely read off the
HST recorder before being overwritten.  As a result, half of one line
is missing from the image.

There are also currently missing observations in Brick 16 (PID 12106).
In one of our requested orientations, there is no guide star available
for the pointing that covers Field 17 with WFC3 and Field 14 with ACS.
We therefore have to observe these two regions with the other
orientation, and will have to point at each independently.  The
additional orbits needed for these observations have been ``borrowed''
from the interface between the two brick orientations, where fields in
Brick 13 have a large degree of overlap with Brick 15.  We anticipate
that data will be taken for these fields in the winter 2011/2012
observing season.

An FGS guide star lock failure also affected one observation in Brick
1 (Visit 3, Orbit 1, of GO-12058 on 2010-12-17).  This target was
subsequently re-observed on 2011-01-03.

In addition to the above observational difficulties (which should be
remedied in upcoming observations), visual inspection of the images
shows that a small portion of them are slightly affected by a number
of different artifacts, most of which are well-known features of
previous HST imaging observations (see HLA ISR 2008-01 by
M.~Stankeiwics, S.~Gonzaga, and B.~Whitmore for ACS, and ISR WFC3
2011-16 by P.~McCullough for WFC3/UVIS).  Strongly over-exposed stars
in ACS/WFC and WFC3/UVIS, for example, exhibit ``bleeding-charge''
tails, which are common to all images obtained with CCDs.  At the same
time, bright stars also generate a number of scattered-light artifacts
in ACS/WFC and WFC3/UVIS, such as weak out-of-focus pupil reflections
or ``ghosts'', which often manifest as pairs of highly-elliptical
rings of light, somewhat resembling a pair of reading spectacles.
Bright stars also generate more diffuse or complex ``flares'' of
light, known as ``dragon's breath'' to users of these instruments.
Both the dragon's breath and the ``spectacles'' can occur at angles of
a few arcminutes away from the bright stars, which may not be even
present in the image.  These artifacts should have minimal impact on
our photometry, given that they affect only a trivial fraction of the
survey area.  Moreover, because our photometry uses local sky
measurements, these additive effects usually have little impact on our
photometry, beyond increased noise from the elevated background.

While scattered light artifacts will be repeated in all images within
a given exposure sequence, there are also transient events that
typically affect only one image in the sequence.  Surprisingly common
events are long trails due to ``space debris'' passing through the
field during an exposure. In some cases the trail may affect both CCDs
in ACS/WFC or WFC3/UVIS.  These are most likely small particles in
orbit around the Earth --- in many cases the object generating the
trail is clearly out of focus, but a point source will come into focus
only when it is $\gtrsim2\times10^4$ km away from the telescope. Again we
have done no special processing to remove these trails or the spurious
sources that they may produce; such processing is unnecessary for
ACS, where all but the faintest trails will be flagged as cosmic rays,
but additional image flagging could potentially be beneficial for
WFC3/UVIS, where there are only 2 exposures per filter.  The trails
are typically extremely straight and uniform, and thus catalog-level
contamination might be described with a fairly simple model.  At the
other end of the scale, we have also detected a few asteroids, which
may be seen at different locations in several exposures within a
sequence.

The other transient scattered light artifact seen in our data is the
illumination of $\sim$1/3 of the WFC3/IR channel when pointed near a
bright Earth limb.  This effect has been described and characterized
in \citet{dalcanton2012}, and is likewise thought to have very little
impact on the photometry.

Several exposures exhibit strong cosmic-ray ``clusters,'' in which
several dozen cosmic ray events form a tight cluster, superficially
resembling a small star cluster.  These events can be corrected
with standard cosmic-ray repair algorithms; however, it may be worth
noting that most of the data of the image affected may be lost over
the extent of the cluster.

Lastly, we note that errors in the transmission of the images to
the ground in rare cases has resulted in the loss of a small number of
pixels in a few images.  The lost pixels typically occur in a small
contiguous segment of a few hundred pixels within a single row on the
CCD imagers (we have not experienced any lost data with WFC3/IR).  In
all cases the lost data has been flagged in automatic reduction
of the images, both in the data-quality image, and as anomalous
values in the flattened image itself.

\subsection{Image Processing}  \label{processingsec}

The first goal of our data processing is the production of a
homogeneous catalog of 6-band photometry for all of the stars detected
in our survey area.  Obtaining such a catalog requires well-calibrated
images that are cleaned of cosmic rays (CRs). 

The ACS camera has been well-studied and is well-calibrated in
general.  However, the camera has had issues with the bias level since the
replacement of key electronic components during Servicing Mission 4.
For WFC3, the calibration has been steadily evolving over the course
of our survey.  These issues have required us to go beyond the standard
pipeline image processing when bias-correcting, flat-fielding, and
flagging CRs in our imaging data.  We now describe the image
processing and CR rejection for the ACS and WFC3 cameras.

\subsubsection{ACS}

The ACS data were processed starting with the {\tt{*.flt}} images from
the HST archive.  These images were flat-fielded by the OPUS pipeline.
As our data come in every 6 months, multiple versions of the OPUS
pipeline were applied to different portions of our data.  Depending on
the observation date, versions 2010\_2, 2010\_3, and 2011\_1e were
used to generate the {\tt{*.flt}} images.  Since the ACS calibrations
have been stable for several years, any changes are likely to be
slight and are unlikely to affect the photometry.

Due to issues with the ACS bias level (ISR ACS 2011-05, Grogin et
al.~2011), the pipeline-processed images suffer from several problems.
The first major problem is that the ACS bias level shows a clear
striping pattern, whose row-to-row fluctuations vary from image to
image.  These bias level variations are only apparent in images where
the sky level was low; in such cases the row-to-row changes in the
bias level have a larger amplitude than the typical sky noise.  The
second notable problem is that, in rows where the default pipeline
bias subtraction was poor, the initial data quality extensions from
STScI flag a high percentage of pixels as being affected by cosmic
rays.  We reset these CR-flagged pixels to allow us to determine the
CR-affected pixels independently, after correcting for the bias
problems.

To correct for the bias striping, we used a destriping algorithm
developed by Norman Grogin ({\tt{csc2\_kill.py}}, ISR ACS 2011-05).
This algorithm attempts to correct images for striping, but is only
helpful in cases where the bias striping can be accurately measured
above the sky noise and where the bias striping corrections are not
inadvertently correcting for real structure in the background sky
(say, in the bulge, where there are large gradients in the unresolved
light).  Therefore, after initially applying the de-striping algorithm, we
then test the results to be sure that the changes to the image are
at the level appropriate for bias-stripe correction.  To do so, we
plot histograms of the difference between the final and corrected
pixel values in one CCD column.  If this distribution has an amplitude
and width characteristic of the bias striping as described in the ISR,
we replace the original {\tt{*.flt}} image with the de-striped
version.  Otherwise the exposure is deemed to be unaffected by the
striping, and the original exposure is unchanged.  Roughly 50\% of
the exposures were kept as-is.

After bias striping is corrected (if needed), we flag CRs as follows.
For our longer exposures, the exposure times are similar enough to
allow for reliable rejections based on high outliers from the median
of the image stack.  Therefore, all exposures longer than 50 seconds
were put through the PyRAF routines {\tt{tweakshifts}} and
{\tt{multidrizzle}}, which flags CR-affected pixels using this
median-image technique.  We found that CR-rejection was far too
aggressive when the short ``guard'' exposures were included in the
image stack.  We therefore handle CR rejection in the short exposures
independently.  All ACS exposures shorter than 50 seconds are instead
put through the IDL routine {\tt{lacosmic}} \citep{vandokkum2001},
which uses Laplacian edge-detection to flag pixels that show sharp
edges associated with CRs, and which proved to be effective for
removing obvious CRs from short single exposures.  Once these steps
are completed, the ACS data are ready to enter our photometry
pipeline.

\subsubsection{WFC3}

While our ACS imaging required only a few minor changes to the
pipeline-processed images, WFC3 is a sufficiently new instrument that its
entire calibration has been in a state of flux during our survey,
making homogeneous calibration more of a challenge.  Furthermore, our
WFC3/UVIS data were obtained with just two exposures per filter, making
CR-masking particularly difficult.

The WFC3 data were calibrated starting with the raw images,
because the UVIS and IR flats and distortion calibrations are
continuing to evolve.  All of our data were processed in PyRAF with
{\tt{calwf3}} in STSDAS version 3.12.  We used a single set of flat
fields for all the images used is this release
({\tt{f110w\_lpflt.fits}}, {\tt{f160w\_lpflt.fits}},
{\tt{f275w\_lpflt.fits}}, and {\tt{f336w\_lpflt.fits}}, with header
dates of 2008-12-09, 2008-12-09, 2009-04-24, and 2009-03-31,
respectively).  For consistency, we also used a single set of
distortion correction tables ({\tt{210710\_uvis\_idc.fits}},
{\tt{t20100519\_ir\_idc.fits}}, and {\tt{u7n18501j\_idc.fits}} for
WFC3/UVIS, WFC3/IR, and ACS, respectively).  It is known, however, that the
distortion has changed as a function of time, for ACS at least (ISR
ACS 2007-08); in future releases, we will be solving for
time-dependent distortion corrections internal to the PHAT data set,
as described below in Section~\ref{astrometrysec}.  We anticipate
updating all calibration images before each major rerun of the full
data set.

After flat-fielding, we flagged CRs in the calibrated WFC3 images.
The calibrated WFC3/IR images are essentially free of CRs, as expected
due to the many non-destructive reads taken during data collection.
However, the WFC3/UVIS images, which contain only two exposures in
each filter, were plagued by CRs.  We attempt to mitigate the CR
effects by running all WFC3/UVIS exposures though the IDL routine
{\tt{lacosmic}} \citep{vandokkum2001}, as was done for the short ACS
guard exposures.  We also process the images through the PyRAF
routines {\tt{tweakshifts}} and {\tt{multidrizzle}} using the
{\tt{minmed}} algorithm to flag CR-affected pixels.

Unfortunately, even after these techniques are applied, the WFC3/UVIS data
still contain some obvious CRs.  More aggressive CR rejection was found
to eliminate central pixels of stars, and thus we cannot pursue more
aggressive CR rejection at the image-processing level.  Instead of
risking degradation of photometry for real stars, we cull CR
artifacts from our photometric catalogs, since they have poor fits to
the PSF model and are anomalously ``sharp''.  We are therefore able to
cleanly remove CR-affected photometric measurements in our
post-processing, as will be detailed in our description of photometry.
Thus, while the residual CRs result in less attractive looking images,
they do not affect our final photometry catalogs significantly.  We
plan to continue to experiment with ways to generate cleaner WFC3/UVIS
images as the survey continues.

After calibration, all WFC3 and ACS images are masked using the data
quality (DQ) extensions of the {\tt{*.flt}} images, which are created
during image processing.  The only modification was to accept data in
the small IR blobs (data quality flag 512), as described in
Section~\ref{dithersec}.  All images are then multiplied by the
appropriate pixel area map to allow for accurate photometry to be
performed on the original calibrated and undrizzled images.

\subsection{Measuring Resolved Stellar Photometry}

We measure stellar photometry on stacks of the final calibrated and CR
flagged images.  The PHAT images are extremely crowded, making point
spread function (PSF) fitting the only viable technique for producing
accurate photometry.  This technique requires a well-measured PSF
model, calibration of that model against aperture photometry, and an
algorithm for fitting the PSF and local background to all sources.

All of our photometry was produced by the software package DOLPHOT~1.2
\citep[][ and many unpublished updates]{dolphin02}.  This software
iteratively identifies peaks and uses the PSF models from
\tinytim\ \citep{krist1995,hook2008} to simultaneously fit the
sky and PSF to every peak within a stack of images, for multiple
filters taken with a single camera.  Minor corrections for differences
between the PSF model and the true PSF in each exposure are calculated
by DOLPHOT using neighbor-subtracted bright stars in the field,
primarily to account for changes in the telescope focus.  Since each
exposure is fit simultaneously, the original HST PSF is left intact to
provide the highest possible quality PSF fits, making these
corrections as small as possible, typically of order 1-5\% (see
Section~\ref{psfcorrsec}).

The exact processing steps employed by DOLPHOT are described in detail
in \citet{dolphin2000}.  To summarize, DOLPHOT uses the
following steps to determine the star list and stellar photometry from
a stack of aligned images for a single camera:
\begin{itemize}
\item An iterative search for stars in the image stack.
\item Iterative PSF-fitting photometry of individual stars, using
  images with stellar neighbors subtracted.
\item PSF adjustment using bright, relatively isolated stars on
  neighbor-subtracted images.
\item A second round of iterative PSF-fitting photometry on neighbor-subtracted
  images, using the refined PSF.
\item Calculation and application of aperture corrections derived from
  bright, relatively isolated stars on neighbor-subtracted images.
\item Computation of individual VEGAMAG instrumental magnitudes and
  uncertainties.  
\item Application of CTE corrections to the ACS magnitudes; WFC3/UVIS
  CTE corrections will be included when they become available from
  STScI.
\item Computation of combined per-filter VEGAMAG magnitudes from all
  measurements that are unsaturated and contain enough valid pixels
  near the core (i.e., do not fall on a cosmic ray or bad column).
\end{itemize}

Because DOLPHOT works on the uncombined image stack, precise alignment
of the images is crucial for producing reliable PSF photometry.  Any
errors in image alignment would lead directly to systematic
photometric errors, due to the PSF peak being misaligned with those of
the stars.  DOLPHOT solves for the relative alignment of images by
matching bright stars, starting from an initial estimate of the shifts
provided by the user.  The distribution of alignment values taken from
the matched stars typically shows a clear central peak, with a width
of order $\sim$0.1 pixels for the star-to-star alignments within a
single image.  The resulting image-to-image alignment is therefore
good to $\ll$0.1 pixels. The PHAT processing pipeline flags cases
where the distribution of alignment values lacks a central peak or
has a width of a pixel or more; these rare cases are then subject to
further by-hand intervention.

We adopted DOLPHOT parameters (Table~\ref{dolphottab}) that would give
high completeness in highly-crowded regions and minimize systematics
in each camera, and that also could be used for fields with a wide
range of stellar density (thus yielding homogeneous photometry from
the bulge to the outer disk).  We selected these initial parameters
based on past experience with stellar population surveys
\citep{dalcanton2009,dalcanton2012}, and on a coarse sampling of
parameter space.  The resulting parameters appear to have accomplished
our goals, as they yield reliable results throughout the survey area
according to our fake star tests (Section~\ref{dataqualitysec} below).
However, given that we have identified a few issues affecting
photometry at the few percent level (described below), we expect to
optimize the processing parameters further as the survey progresses.
As such, we are currently experimenting with a larger grid of
potential DOLPHOT parameter sets to quantitatively determine the
optimal set for precision and completeness.  These parameters will
likely be updated in our next data release, with details of any
changes being described in subsequent papers and in supporting
materials provided with the data distribution via the Multi-Mission
Archive at Space Telescope (MAST).

The key parameters for producing high quality photometry have proved
to be {\tt{Rchi}} and {\tt{Fitsky}}. {\tt{Rchi}} is the radius inside
of which the centroid of the PSF is fit.  This value is set to be
quite small (about one full-width at half maximum of the PSF) to avoid
it being affected by poorly-subtracted neighbors.  Subsequent
star-matching between cameras and astrometric alignment also uses this
same value of {\tt{Rchi}}.  {\tt{Fitsky}} sets how the local
background level is measured; we have adopted {\tt{Fitsky}}$=3$,
which forces simultaneous fits to the PSF and the background level
inside of the fitting radius.  This strategy is optimal for very
crowded fields, but is also computationally expensive.  For this first
pass through our data we conservatively applied this algorithm for sky
measurement.

Once the PSF-fitted magnitudes of all stars have been measured, the
values are corrected to a $0.5\arcsec$ aperture, equivalent to using an
aperture correction measured using neighbor-subtracted bright stars on
the image (see Section~\ref{apcorrsec}).  

The ACS/WFC and WFC3/UVIS images both suffer from charge transfer
efficiency (CTE) problems, such that not all charge is read out of a
pixel, leaving residual charge that is instead read out in subsequent
rows.  This effect leads to a small bleed trail of charge that is most
severe for low electron levels and for rows near the beginning of the
readout (i.e., near the chip gap for the ACS/WFC and WFC3/UVIS
packages).  The trail of charge is made up of flux that should have
been included in the main body of the star, and thus photometry of
stars affected by CTE will be systematically faint, and positions of
the affected stars will be slightly shifted in the $Y$-direction.

Luckily, CTE is sufficiently well-behaved that it is possible to
derive reasonable corrections for its effects.  The fluxes for ACS
measurements are corrected for CTE using the prescription of Chiaberge
(ISR ACS 2009-01).  Typical corrections range from 0 to 0.2 magnitudes
in Brick 01 where the background is high (hundreds of counts per
pixel) to 0 to 0.5 magnitudes in Brick 21 where the background is low
(tens of counts per pixel). $Y$-coordinates closer to the chip gap
have larger corrections, as do fainter stars, which leads to the broad
range in correction values.

We have also examined the possibility of using the image-based CTE
corrections of Anderson \& Bedin (ISR ACS 2010-03), which correct images
for CTE loss prior to running our pipeline.  Unfortunately, the
procedure in its current form has the characteristic that all peaks,
including noise spikes in the background, are treated as stars.  This
has the result of magnifying background noise peaks and CRs above our
detection threshold, creating enormous numbers of very faint false
detections, most notably near the chip gap.  Given this complication,
we have chosen to retain the catalog-based CTE corrections for their
superior noise properties at faint magnitudes.

Currently, there are no CTE corrections available for WFC3/UVIS.
Given that CTE is clearly evident in the WFC3/UVIS images (manifesting
as obvious trails in $Y$ for sources near the chip gap), our WFC3/UVIS
photometry is undoubtedly biased to fainter fluxes for sources close
to the chip gap.  We unfortunately cannot constrain the size of the
effect directly with our data, since our tiling scheme is such that
overlapping WFC3/UVIS observations leave stars at comparable $Y$
chip positions (see Figure~\ref{exptimemapfig}).  We will be
incorporating WFC3/UVIS CTE corrections into our photometry as they
become available.

Finally, the fluxes are transformed to infinite aperture and converted
to Vega magnitudes using the zeropoints dated 15-July-2008 in the ACS
Users Handbook and ISRs WFC3 2009-30 and WFC3 2009-31 (J. Kalirai et
al.~2009) for WFC3.  These zeropoints will be updated should new
values be published before our next run through the full data set.

The photometry for this first pass through the PHAT survey data was
performed on the data from each field and each camera separately.
Therefore any depth that may be gained from overlapping data from
neighboring fields is not included in the current photometry.  This
added depth will be of only modest benefit in the crowding-limited
portions of the survey, but will be significant in other portions.  We
anticipate that our second data release will contain simultaneous
photometry on full image stacks, for all cameras.

Before discussing the production of the final photometric catalogs
(Section~\ref{catalogsec}), we briefly discuss the amplitudes of the
PSF and aperture corrections.

\subsubsection{Amplitude of PSF Corrections}  \label{psfcorrsec}

The \tinytim\ point-spread function models used for PSF
photometry are unfortunately not perfect.  For ACS, they are known to
vary with time (see Figures 8 and 9 in ISR ACS 2006-01 by J.~Anderson
and I.~King) and to have small errors with
position \citep[e.g.,][]{jee2007}.  For WFC3, the errors are likely to
be even larger, because no post-flight updates to the \tinytim\ models
have been implemented at this time.

Due to these temporal variations in the PSF, as well as possible systematic
errors in the \tinytim\ PSFs themselves, DOLPHOT computes a PSF
adjustment image that is added to each PSF.  This calculation is done
midway through the iterative photometry process, once a reasonably
final star list has been reached. The process involves analysis of
residuals near bright, relatively isolated stars in images in which
all detected stars have been subtracted.  Although an entire
difference image (actual minus \tinytim\ model) is calculated and
applied to the photometry, only the value of this image in the central
pixel is reported. A positive value indicates that the data are
sharper than the model; a negative value indicates the reverse.

Figure~\ref{psfcorrfig} shows the adopted PSF correction for the
central pixel, for every image taken prior to Fall of 2011, with the
exception of the short ``guard'' exposures in \fw{475} and \fw{814}.
The amplitude of the typical PSF correction is relatively small
($<$0.05 in the UV, and $<$0.01 in the optical and NIR), and has no
strong dependence on the local stellar density (i.e., Brick number) or
the time of observation (not shown).  Table~\ref{photcorrtab} lists
the median PSF correction and the semi-interquartile width, as well as
the number of stars used to calculate the PSF correction.  The scatter
in the measured corrections increases for the bluer cameras, most
likely due to the smaller number of high signal-to-noise sources in
the WFC3/UVIS images.

The sign and amplitude of the central pixel PSF corrections are a
first order indication of the errors in the PSF model.  For the WFC3/UVIS
data, the PSF corrections are substantial and are consistently
negative, indicating the the current \tinytim\ model for the WFC3/UVIS
PSF is slightly broader than the true data (i.e., the PSF correction
must put extra flux back into the central pixel).  The PSF corrections
for ACS and WFC3/IR are a factor of three smaller, suggesting that the
\tinytim\ models for these cameras are in better agreement with the
data, at least within the very core of the PSF.  The agreement
outside the central pixel, however, is unconstrained by these initial
tests.

\subsubsection {Amplitude of Aperture Corrections}  \label{apcorrsec}

To calculate a star's magnitude, DOLPHOT adjusts the flux in the
corrected \tinytim\ PSF to minimize residuals in the image stack.  The model
PSFs are scaled to unity within an aperture of $3\arcsec$ in radius,
such that the internal magnitudes within DOLPHOT are calibrated as if
the zero points were for $3\arcsec$ apertures.  However, especially in
crowded fields, we find it more practical to compute aperture
corrections within a smaller aperture of $0.5\arcsec$ in radius, and
then apply encircled energy corrections to calibrate our photometry
using the published zero points (which are for infinite aperture).
DOLPHOT therefore corrects its internal PSF magnitude to an aperture
magnitude calculated within a $0.5\arcsec$ radius.  DOLPHOT calculates
these aperture corrections by identifying $\sim$200 bright, reasonably
isolated stars in each image, computing aperture photometry on the
neighbor-subtracted residual image, and then measuring the differences
between these aperture magnitudes and the stars' PSF magnitudes.  These
measurements are then combined in a weighted average with outlier
rejection to derive a single aperture correction which is applied to
the PSF magnitudes of all stars in the image, for a given filter.

Figure~\ref{apcorrfig} shows the aperture corrections for every image
taken prior to Fall of 2011, excluding the short ``guard'' exposures
in \fw{475} and \fw{814}; Table~\ref{photcorrtab} lists the median
aperture correction and the semi-interquartile width for each filter,
as well as the median number of stars used to calculate the aperture
correction.  The number of stars is typically high ($\sim$200) for the
crowded optical and NIR data.  The UV fields, however, frequently
suffer from a paucity of bright sources, leading to much smaller
numbers of suitable stars and larger image-to-image scatter; this
problem is most severe in Bricks 22 and 23 in the outer disk, and
Brick 1 in the bulge.  The Brick 1 bulge fields also suffer from
reduced numbers of suitable stars in the optical and NIR, though not
to the same degree as in the UV.  The image-to-image scatter tends to
be systematically larger for bluer filters, due to the lower number of
high signal-to-noise stars.

Table~\ref{photcorrtab} shows that the mean aperture correction for a
given filter is small ($<0.06$~mag for all but $F160W$), and has a
typical image-to-image scatter of less than $0.02$~mag across the
entire data set.  However, inspection of Figure~\ref{apcorrfig}
indicates that the aperture corrections do have a modest systematic
variation with brick number.  We found no time-dependence in the
amplitude of the aperture correction, suggesting that the observed
variation with brick number is likely driven by the systematic
decrease in the stellar density and/or background level with increasing
radius, since larger brick numbers and even brick numbers have larger
projected radii.  The effect is most noticeable in Brick 1, which has
the largest range of stellar densities internal to the brick, due to
the presence of the center of the bulge in Field 10.

In contrast to what is seen in Figure~\ref{apcorrfig}, the aperture
correction should in principle be a constant that solely corrects for
the difference in the encircled energy between $0.5\arcsec$ and
$3\arcsec$, assuming perfect PSF models and measurements.  The fact that
the aperture correction depends somewhat on the local level of
crowding indicates that at least one of the key assumptions has been
violated.  In other words, the data in Figure~\ref{apcorrfig} suggest
that there is a slight bias in the measurement of the aperture
correction calculated for bright stars, which may result from some
combination of errors in the PSF model, background-dependent biases in
the initial PSF magnitude, or subtle problems in constructing the
empirical PSF used to measure the aperture correction.  As such, it is
not clear whether the observed crowding-dependence in the aperture
correction is introducing systematic errors or correcting for them.

At the present time, the $\sim$0.02~mag scatter must be included in
the error budget of the resulting photometry, with the understanding
that there may be small systematic offsets in photometry measured in
different regions of M31, particularly in Brick 1 for the NIR.
However, the systematic variations in the aperture correction are
typically smaller than empirical systematic errors derived from repeat
measurements (shown below in Figure~\ref{systematicfig}), suggesting
the aperture corrections are not dominating the systematic error
budget.  They are also much smaller than the typical random
uncertainties due to photon counting (Figure~\ref{photerrfig}, below).
We are currently investigating the origin of this bias, and expect to
have it and other small systematic uncertainties resolved before the
second data release.

\subsection{Producing Photometric Catalogs}  \label{catalogsec}

The lowest level data products produced by DOLPHOT are
``{\tt{*.phot}}'' files, which include the photometry for every
detected source (both optimized for the entire image stack, as well as
for each individual image), along with several metrics for the quality
of each photometric measurement.  These catalogs offer the greatest
amount of information, but are frequently much larger than needed for
most applications.  We therefore generate additional smaller catalogs
restricted only to stellar sources with good photometric quality.

The metrics we have found most useful for generating
high-completeness, 
\linebreak[4] low-contamination catalogs are measures of the
``sharpness'' and ``crowding'', which are quantified by DOLPHOT and
returned as the parameters ``{\tt{sharp}}'' and ``{\tt{crowd}}'' for
each star.  Sharpness quantifies the degree to which an object is more
or less centrally peaked than the PSF \citep[see eqn.~14 of
][]{dolphin2000}.  Very centrally peaked sources (positive sharpness)
are likely CRs, as they do not have the wings associated with true
point sources.  Very diffuse sources (negative sharpness) are likely
background galaxies or unresolved groups of stars.  Crowding
quantifies the degree to which the photometry of the source may have
been affected by neighboring sources.  The value for {\tt{crowd}} is
determined by calculating the difference between the magnitude
measured for the star before and after the subtraction of neighboring
sources.  Large numbers of bright neighboring sources therefore lead
to a large value for the crowding parameter.

Using these metrics, we cull lower-quality measurements from our
catalogs to produce color-magnitude diagrams showing well-defined
features.  The cuts were done at two levels for each camera,
generating two catalogs with different restrictions on data quality.
The first cuts the photometry at the most basic level, to ensure that
all stars in the catalog have a high signal-to-noise detection in at
least one filter.  The resulting ``{\tt{st}}'' catalog contains only
objects that either were detected with signal-to-noise S/N$>$4 in both
filters or were detected in one filter at S/N$>$4 with
{\tt{sharp}}$^2<$0.1; the additional sharpness cut is used for the
one-filter detections to suppress CRs.  These cuts admit a very high
percentage of all measurements but tend to remove any residual CRs
from the catalogs.

The second cuts are designed to produce sharp-featured CMDs with
well-measured colors, by removing highly crowded stars, stars with
poor PSF fits and stars with very low S/N in either filter.  These
cuts were made by keeping only measurements with S/N$>$4 in both
filters for a given camera, {\tt{crowd}}$_1$ + {\tt{crowd}}$_2$ $<$
1.0 (0.48 for WFC3/IR), and ({\tt{sharp}}$_1$ + {\tt{sharp}}$_2$)$^2$
$<$ 0.075 (0.12 for WFC3/IR).  These cuts were determined by
experimenting with a series of possible cuts, looking at the CMD of
the rejected objects, and choosing the cuts that removed the largest
number of stars without generating strong features in CMDs made from
the rejected stars.  These {\tt{gst}} (``good star'') catalogs
contain only stars with the highest quality photometry, but suffer
from more incompleteness in crowded regions, such as dense stellar
clusters.  They are also limited in depth by the requirement that
stars have good signal-to-noise in both filters for a given camera.
In cases where the two filters have very different depths for typical
stellar colors (such as for \fw{275} and \fw{336}), the {\tt{gst}}
catalogs will be missing many high quality detections in the deeper
filter.  Thus, when combining data from across cameras (say, when
generating a $\fw{336}-\fw{475}$ CMD), it may be preferable to revert to
the {\tt{st}} catalogs to avoid this restriction, but then to apply
joint crowding cuts on the detections from the different cameras.

For PHAT data releases, the {\tt{st}} and {\tt{gst}} photometry
catalogs are packaged into binary FITS tables, and distributed via
MAST. 

\subsection{Artificial Star Tests}   \label{artificialstarsec}

We assess the completeness and uncertainty as a function of color and
magnitude of the {\tt{st}} and {\tt{gst}} catalogs using extensive
artificial (``fake'') star tests.  We insert a series of 10$^5$ fake
stars into each field with the appropriate PSF, for each camera
independently.  This strategy produces 5.4 million fake star tests per
brick.  Each fake star is added individually to the image stack (i.e.,
one artificial star at a time), and the photometry is rerun in a
sub-region around the star to determine if the star was found, and if
so, to determine what its measured magnitude was; this method is
efficient, and does not run the risk of artificially changing the
local crowding. The measured properties of the fake stars are then put
through our quality cuts to determine which stars would be included in
the final {\tt{st}} and {\tt{gst}} catalogs.

The properties of the fake stars are chosen such that 50\% of the
stars randomly sample the entire observed range of color and
magnitude, and 50\% sample the actual color-magnitude distribution
measured for the individual field.  Spatially, the stars are
distributed randomly throughout each field.  While they therefore will
broadly sample the photometric quality of the field, they will not
necessarily be representative of the photometry in stellar clusters,
which have higher than average local densities, but which cover only a
tiny fraction of the area.

Taking a coarse grid in color and magnitude, we calculate the
completeness in each box of the grid and interpolate to find the 50\%
completeness points.  The limiting magnitude in the red filter is then
the 50\% completeness magnitude in the red band, measured at a color
that is blue enough to ensure that non-detection in the red filter was
the cause for low completeness (i.e., blueward of the corner at the faint
end of the color-magnitude diagram). The blue band limit is then the
50\% completeness magnitude in the blue band, measured at a color that
is red enough to ensure that non-detection in the blue filter was the
cause for low completeness (redward of the corner at the faint end of
the color-magnitude diagram).  These 50\% completeness values are then
tabulated as our magnitude limits in each field with the current
processing version.

\subsubsection{Survey Depth}   \label{depthsec}

Based on the 50\% completeness limits, the faintest limiting
magnitudes in the PHAT survey are found to be roughly 25.1, 24.8,
27.9, 27.1, 25.0, and 24.0, in the \fw{275}, \fw{336}, \fw{475},
\fw{814}, \fw{110}, and \fw{160} filters, respectively.  For solar
metallicity main sequence stars, these depths correspond roughly to
$2.6\Msun$, $2.8\Msun$, $1.5\Msun$, $1.5\Msun$, $3.5\Msun$, and
$6\Msun$, respectively, for zero foreground extinction.  However,
these depths are a strong function of local stellar density for the
optical and NIR, as we now show.

\subsubsection{Crowding and Depth vs Radius}   \label{depthradiussec}

For isolated stars, the depth of photometry is a constant set
by the uncertainty in inferring a star's flux from a limited number
of photons.  At the surface density and distance of M31, however, one
cannot assume that stars are isolated, or that their primary
photometric uncertainties are due to photon counting.  In M31's disk,
the number of stars that fall above the limiting magnitude is
sufficiently high that the PSFs of the individual stars will overlap.
While crowded field photometry can compensate somewhat when detecting
and measuring photometry for partially overlapping stars, at some
point the level of overlap is sufficiently large that faint stars are
completely undetectable.  The crowding also adds additional
photometric uncertainties to the detected stars, since the reported
photometry for a single star now depends on subtracting flux from
neighboring stars as well.  In these regimes, the limiting magnitude
is no longer set by Poisson counting of electrons in the CCD, but
instead is ``crowding-limited''.  In crowding-limited data, the depth
of the observations increases little with increased exposure time
(although the photometry of the detected stars does improve), and only
observations with higher angular resolution, or of lower surface
density regions can significantly improve the limiting magnitudes. In
M31, both the ACS and the WFC3/IR observations will be crowding
limited over much of the disk, with the crowding limit being brighter
in the higher surface brightness central regions.

We can use the 50\% completeness limits to determine the radial
dependence of the survey depth of the survey.  We assign a radius to
each field by calculating the median right ascension and declination
of the detected stars, and then solving for the deprojected major axis
length of an ellipse passing through that position (assuming an
inclined disk with $i=70^\circ$, a position angle of $43^\circ$, and a
center at $\alpha(2000)=10.68473^\circ$ and
$\delta(2000)=41.26905^\circ$; we discuss the choice of inclination
and position angle further in Section~\ref{m31structuresec} below).  We
also calculate the mean number surface density of stars with S/N$>$4
(i.e., stars per arcsecond$^2$) in the {\tt{st}} and the culled,
high-quality {\tt{gst}} catalogs for each field.

We plot the radial dependence of the 50\% completeness magnitude limit
in Figure~\ref{maglimradiusfig}. As expected, we see a strong radial
dependence in the completeness limit, in spite of all fields having nearly
identical exposure times. In both the optical and the NIR, there is a
brighter magnitude limit in regions with higher stellar densities.
The effective magnitude limit of the ACS filters is more than 4
magnitudes brighter in the bulge than in the diffuse outer disk.  The
variation in the WFC3/IR channel is even larger, being $\sim$5.5
magnitudes deeper in the outer disk than in the bulge; given the
larger pixels of the WFC3/IR camera, it is not surprising that the NIR
observations are the most severely affected by crowding.

In contrast, there is little radial variation in the 50\% completeness
limit in the WFC3/UVIS data, with only the slightest hint of an upturn in
the central bulge field.  The limiting magnitude of the WFC3/UVIS data are
therefore set by photon-counting, not crowding.  Longer WFC3/UVIS
observations could therefore be expected to yield significantly deeper
CMDs, unlike for the PHAT WFC3/IR and ACS/WFC CMDs, which are the
deepest that are currently possible at these radii in M31.  We note
that the observed radial variation agrees extremely well with
predictions initially made using \citet{olsen2003}'s crowding simulations
(modified for the PHAT observing parameters) in advance of the start
of survey operations.

We can see the impact of stellar density on photometric depth more
directly in Figure~\ref{maglimdensityfig}, where we plot the
correlation between the 50\% completeness limit and the stellar
density of all stars detected with S/N$>$4 in a given filter (open
circles, from the {\tt{st}} catalogs), and for the subset of stars
with reliable photometry and S/N$>$4 in both of the filters observed
with a single camera (solid circles, from the {\tt{gst}} catalogs).
At high stellar densities, the photometric quality cuts significantly
reduce the numbers of stars in the {\tt{gst}} catalog compared to
the less stringent {\tt{st}} catalog, by roughly a factor of three.
These cuts have a particularly large impact on the \fw{336} filter,
because of the requirement that the {\tt{gst}} catalog also have
S/N$>$4 for the shallower \fw{275} observations.

For photon-limited observations, we expect the limiting magnitude in
Figure~\ref{maglimdensityfig} to be constant for a fixed exposure
time, and to have no dependence on the stellar surface density.  We
see this expected behavior in the WFC3/UVIS filters, where there is no
correlation between the completeness limit and the stellar density.

The relationship between stellar density and the magnitude limit is
far more complicated for the ACS data, which is crowding-limited over
most of the disk.  At the lowest stellar densities shown, there is no
obvious relationship between the completeness limit and the stellar
surface density, indicating that some of the ACS data has reached a
low enough surface density ($\lesssim\!6\,{\rm{stars\,arcsecond}}^{-2}$) that
the photometry becomes photon-limited.  These limiting magnitudes are
roughly 27.9 in \fw{475} and 27.1 in \fw{814} for the ACS exposures taken
at a single pointing.\footnote{The overlapping ACS observations from
  adjacent pointings will increase the depths in these photon-limited
  regions.}  At higher stellar surface densities, the limiting
magnitude of the ACS data is systematically brighter, as the crowding
begins to affect the ability to detect faint stars (the {\tt{st}}
points), or to produce reliable photometry (the {\tt{gst}} points).
Instead, the completeness limit becomes pinned to whichever depth
produces $\sim\!14\,{\rm{stars\,arcsecond}}^{-2}$ with reliable photometry
(i.e., in regions with intrinsically high stellar surface densities,
the density of bright stars is sufficiently high that the maximum
surface density can be reached with only the brightest stars in the
luminosity function, leading to a bright limiting magnitude).

The NIR data in Figure~\ref{maglimdensityfig} is similar to the
optical ACS data in many respects.  However, it lacks the turnover at
low stellar surface densities to a single 50\% completeness magnitude,
indicating that the WFC3/IR data is crowding-limited throughout the
disk, as was expected from our initial simulations.  The surface
density of reliably measured stars is essentially constant at
$\sim\!2\,{\rm{stars\,arcsecond}}^{-2}$ throughout the galaxy.  The small
ripples in the stellar surface density at $\sim$24.3$\,$mag in \fw{110}
and $\sim$23.2$\,$mag in \fw{160} are due to the increasing numbers of
stars at the magnitude of the well-populated red clump, and the bright
ripples at 19 and 20th magnitude (in \fw{110} and \fw{160},
respectively) are due the increasing numbers of stars at the NIR tip
of the red giant branch (TRGB).  Hints of a bump caused by the red clump is
also visible in the ACS data (at $\sim$24.8$\,$mag in \fw{814}), but it
is less pronounced, because the red clump is somewhat blurred by dust
in the optical, making it a less pronounced feature in the luminosity
function.

Figure~\ref{maglimdensityfig} also shows the substantial penalty that
results from the larger WFC3/IR pixel scale.  At the crowding limit,
the ACS camera produces reliable photometry for $\sim$7 times as many
stars per arcsecond$^2$ than the WFC3/IR channel.  Deeper NIR data in M31
will only be possible with a telescope$+$camera system with smaller pixels
and higher angular resolution.

\newcommand{\masunit}{\mas}

\subsection{Astrometry}  \label{astrometrysec}
Producing accurate 6-filter photometry for crowded-field photometry
requires exquisite sub-pixel alignment between overlapping images.  We
have made careful astrometric solutions to place our PHAT data on a
universal reference frame with an absolute error of $100\,\masunit$
(dominated by errors in the ground-based reference catalog) and
relative errors of $10\,\masunit$ between stars observed in the three
cameras.  The astrometric alignment is a multi-stage process in which
individual exposures within a field are aligned to each other, 
overlapping regions within bricks are used to tie the fields
together, and finally bricks are aligned to a ground-based
absolute astrometric reference frame.  We describe each of these steps
below.

\subsubsection{Alignment of individual exposures taken with one camera
  during one visit:}
As discussed briefly above, the individual exposures in an image stack
are aligned within the DOLPHOT processing module.  DOLPHOT identifies
bright stars within each exposure, and then uses the geometric
distortion correction to calculate relative shifts to apply to each
exposure, based upon initial guesses generated from the astrometry in
the header.  Typically hundreds to thousands of stars are used in the
alignment, with $\sim$300, 525, 850, 800, 6000, 6250 average alignment
stars for the \fw{275}, \fw{336}, \fw{475}, \fw{814}, \fw{110}, and \fw{160}
filters, respectively; the number of alignment stars used for the NIR
filters varies dramatically with radius, from $\sim$2000 in the outer
disk to 20,000 in the inner bulge.  The resulting corrections to the
(relative) header-based astrometry are typically of the order of 1-2
pixels or less.  The only exceptions found in the data so far are for
the ACS images in Brick 21, Field 2, which have relative shifts that
differ substantially from what would be inferred from the headers;
data quality in this field is otherwise unexceptional.  For WFC3/UVIS and
ACS/WFC, each chip is aligned independently.  The RMS of the shifts
for pairs of alignment stars are typically $\sim$0.2 pixels for
WFC3/UVIS and ACS/WFC, and $0.1$ pixels for WFC3/IR. 

\subsubsection{Aligning all images for a single camera:}
The photometric catalogs initially generated by DOLPHOT contain
distortion-corrected astrometry, tied to the astrometry of a reference
frame generated with {\tt{MULTIDRIZZLE}}.  These reference frames, in
general, are translated and rotated with respect to a global reference
frame (due to errors in guide star catalog positions) and also have
shear, scale, and distortion (due to residuals from the pipeline
distortion corrections).  We aim to remove these errors by aligning
the individual catalogs with each other and with an astrometric
reference catalog tied to a global reference frame.  For aligning the
astrometry we use the {\tt{gst}} catalogs for ACS/WFC and WFC3/IR,
since they contain fewer spurious stars, and the {\tt{st}} catalogs
for WFC3/UVIS, with cuts on the \fw{336} detections only (S/N$>$4 and
{\tt{crowd}} $\le\!0.25$) to avoid eliminating the large number of
valid \fw{336}-detected stars that were undetected in the shallow
\fw{275} imaging.  We split the ACS/WFC catalogs along the CCD gap and
handle the two CCDs separately.

We begin by coarsely aligning the individual catalogs in celestial
coordinates, allowing only translational corrections.  For the large ACS/WFC
catalogs, we cut to bright ($\fw{475}\!<\!24$) stars for speed.  Given a
set of spatially overlapping catalogs, we search each pair of catalogs for pairs
of stars (one from each catalog) within $2\arcsec$ of each other, with
this limiting separation set by the expected astrometric uncertainty
of the cameras relative to the guide stars.  This large matching
radius is necessary to handle the typical offsets we find in our data.
If the affine (rotation, shear and scale) and distortion corrections
are small, then the distribution of positional differences between the
pairs of stars will include a peak (due to true matches) and a
background (due to false matches).  We also observe a ring of lower
match density around the peak due to DOLPHOT's closeness-of-sources
limit.  The position of the peak in $(\Delta
\textrm{RA},\Delta \textrm{Dec})$ space is the mean positional offset
between the two catalogs, and the width of the peak is due to
measurement errors plus affine and distortion errors.  We fit for
$(\Delta \textrm{RA},\Delta \textrm{Dec})$ using a Gaussian peak plus
flat background model using the Expectation Maximization algorithm
(EM; \citealp{dempster1977}).  Once we have measured the offsets between
each pair of overlapping catalogs, we solve a least-squares equation
to find a consistent set of translational corrections that will
simultaneously minimize the distance between matched pairs of stars
for the entire set of catalogs.

\subsubsection{Tying photometric catalogs to a global astrometric frame:}
After applying the initial translational corrections described above,
the contiguous catalogs for each camera are roughly aligned
internally.  Next, we align the cameras with each other and with a
global reference frame.  This step is required for eventual
comparisons with other multi-wavelength data sets.

We have defined a dense network of global astrometric standards using
archival $i^{\prime}$ data from MegaCam CFHT (courtesy of Jean-Charles
Cuillandre).  The CFHT processed images were subjected to iterative
PSF photometry.  The photometry in the resulting astrometric catalog
extends down to $i^\prime\!=\!21.5$ and contains 1.7 million sources.
For comparison, the \citet{massey2006} Local Group Survey (LGS)
catalog contains one-seventh as many sources over the same region.
The positions of the brighter, non-saturated stars were matched to
2MASS \citep{skrutskie2006}---which itself is tied to
Tycho-2 \citep{hog2000}---to produce a well-constrained astrometric
solution for the entire region.  The dispersion for matches between
CFHT sources with $i^{\prime} < 18$ and 2MASS is about $0.2$
arcseconds.

We begin the global alignment by roughly aligning the ACS/WFC catalogs
to the CFHT-based reference catalog.  We first merge the ACS/WFC
catalogs, removing one star from each pair within $30\,\mas$ of each
other, and cutting to $\fw{814} < 21.5$ to match the depth of the
reference catalog.  Then, as before, we find stars in the ACS/WFC and
CFHT catalogs within $1\arcsec$ of each other, fit a Gaussian peak
plus background model, and use the peak weights to fit a least-squares
affine correction (translation, rotation, scale, and shear) to the
ACS/WFC catalogs.  Similarly, we roughly align the WFC3/UVIS and
WFC3/IR catalogs to the ACS/WFC catalogs.  

Finally, we repeat the alignment process, this time allowing each
individual catalog an affine correction to bring it into alignment
with its peers and with the reference catalog.  We do this, as before,
by finding nearby pairs (both between pairs of catalogs, and between
the catalogs and the reference), fitting a Gaussian peak plus
background model to properly weight the matches, and computing a
least-squares set of affine corrections.\footnote{We note that we have
  more than ample signal to perform these fits, in spite of the large
  number of parameters needed for the field-by-field affine
  transformations.  The least-squares problem when aligning a set of
  18 ACS/WFC catalogs to the CFHT-based reference typically has $\sim
  10^7$ matches, but only $216$ parameters (6 per catalog).}

We characterize the accuracy of the photometry as the width of the
Gaussian peak fit to the distribution of all offsets between pairs of
matched stars.  The final alignment of the ACS/WFC catalog to the
CFHT reference frame has a 1-$\sigma$ accuracy of $\sim45\,\mas$ per
coordinate in the outer disk, increasing to $\sim60\,\mas$ in the
bulge. The internal alignments within the WFC3/UVIS, ACS/WFC, and
WFC3/IR catalogs have 1-$\sigma$ accuracies of $\sim8.5\,\mas$,
$\sim6\,\mas$, and $\sim6\,\mas$ respectively.  The relative alignments of
the WFC3/UVIS and WFC3/IR catalogs to ACS/WFC have 1-$\sigma$
accuracies of $\sim6\,\mas$ and $\sim3.7\,\mas$ respectively.

Even with the inclusion of affine corrections, we see clear structured
residuals of up to $\sim5\,\mas$ when matching different cameras to
each other.  These residuals have a consistent structure from
field-to-field within a brick, and are almost certainly caused by
errors in the geometric distortion correction.  In future releases, we
expect to refine the above procedure to include corrections to this
geometric distortion.

\subsubsection{Merging photometric catalogs for all cameras:}
For some applications it is helpful to have catalogs of matched
multi-camera (6-filter) photometry.  This step involves merging
measurements made with three different cameras, with multiple
overlapping pointings, in highly crowded fields.  As a result, this
step can only be done after all catalogs have been tied together
astrometrically to a high degree of accuracy.  We generate merged,
six-filter catalogs as follows.  We first remove duplicates from
catalogs of overlapping fields, for each camera.  First, we identify
and group any ACS/WFC stars that appear within $50\,\mas$ of each
other in multiple catalogs. We then replace them with a single ACS
measurement assigned to the mean of the measured positions and
magnitudes in the overlapping catalogs.  We do the same for WFC3/UVIS
and WFC3/IR, resulting in a single brick-wide catalog for each
camera, with duplicates removed.  We then generate the multi-camera
photometric catalog by matching the merged ACS/WFC stars with the
merged WFC3/UVIS and WFC3/IR stars, using a matching radius of
$50\,\mas$.  Stars are not required to be matched in multiple cameras,
such that a source with no counterparts within $50\,\mas$ will still
be included in the merged catalog.  

We note that deblending sources near the detection limit in highly
crowded fields does not always produce a unique result, leading to
slight variations in the positions and assigned magnitudes of
deblended stars.  As such, source matching becomes less reliable at
faint magnitudes.  In future releases, we will be doing source
identification and photometry simultaneously across all overlapping
images in all three cameras, which should eliminate this problem.

\subsubsection{Propagating astrometry back into the photometric catalogs:}
The result of our astrometric alignment process is an affine
transformation for each camera in each field.  We apply these affine
transformations to the FITS World Coordinate System (WCS) headers of
the input {\tt{*.flt}} images, along with the small relative shifts
that DOLPHOT calculated during alignment of the image stacks.  We
modify the WCS {\tt{CRVAL}} headers to apply the shifts, and the
{\tt{CD}} matrix elements to apply the remaining affine terms.  We
also propagate the revised astrometry back to the individual
photometric catalogs.

\subsection{Data Quality}  \label{dataqualitysec}

We have taken a number of different steps to characterize the current
quality of the photometry.  In the following sections we discuss the
raw photometric uncertainties reported by DOLPHOT
(Section~\ref{photerrsec}), the uncertainties and biases due to
crowding, as assessed by artificial star tests
(Section~\ref{fakeerrsec}), and uncertainties due to systematic
errors, as measured from repeatability tests in overlapping images
(Section~\ref{repeaterrsec}).  All results in this section are for
individual fields in the current data release, which do not take
advantage of additional depth due to overlapping adjacent fields, or
upcoming improvements in the flat fields, distortion corrections, or
PSF models.  We therefore expect that some of these uncertainties will
be reduced in upcoming versions of the PHAT photometry.

\subsubsection{DOLPHOT-reported Uncertainties} \label{photerrsec}

DOLPHOT reports photometric uncertainties for each star as part of its
standard output.  These uncertainties are based on the Poisson
variation in the flux from the star, from the sky, and from
neighboring stars whose flux has been subtracted.  The value of the
uncertainty (in magnitudes) can be found in the {\tt{MAG1\_ERR}} and
{\tt{MAG2\_ERR}} entries of the {\tt{gst}} and {\tt{st}}
photometric catalogs.

In Figure~\ref{photerrfig} we plot the median photometric uncertainty
as a function of magnitude, for all 6 filters, in Field 9 of Bricks 1,
9, 15, and 21 (upper left to lower right).  The median uncertainty is
given in magnitudes, and was calculated for stars with S/N$>$4 in the
named filter, using data from the {\tt{st}} catalogs.  The median
uncertainty and magnitude was calculated for groups of 25, 400, and
200 stars, for the WFC3/UVIS, ACS/WFC, and WFC3/IR cameras,
respectively.  We have not included stars that only appear in a subset
of the images (i.e., those that are near the edges of an image or near
a chip gap); such stars have much higher reported photometric
uncertainties, as expected, but are not representative of the bulk of
the photometric catalog.  We have also excluded stars that touch the
edge of an image, or that have saturated pixels (i.e.,
{\tt{FLAG}}$>$0).

The median photometric uncertainty in Figure~\ref{photerrfig}
correlates strongly with magnitude, as expected.  However, the slope
of the correlation is different for each camera, due largely to the
changing importance of the sky brightness.  In the UV, the uncertainty
is dominated by the Poisson variation in the flux from the stars
themselves, since the PHAT WFC3/UVIS images are essentially uncrowded and
the background of unresolved flux is quite dark.  In the optical and
NIR, the situation is quite different, due to a high level of crowding
and brighter backgrounds.  The correlation between photometric
uncertainty and magnitude therefore follows a different slope than in
the UV.  Moreover, the exact relation changes with radius as well, due
to the darker background and lower stellar density at large radii.

We note that the photometric uncertainties shown in
Figure~\ref{photerrfig} are not always the dominant source of error.
For faint stars, flux from unresolved stars below the detection limit
can significantly bias the resulting magnitudes.  For bright stars,
systematic uncertainties in the flat field, CTE corrections, or PSF
model can dominate over the formally small photometric error.  As we
show below in Section~\ref{repeaterrsec}, these systematic errors have
typical amplitudes of 0.02--0.04 magnitudes for bright stars; we
include a horizontal line in Figure~\ref{photerrfig} to indicate the
approximate magnitude range where systematic uncertainties will become
important.

\subsubsection{Crowding Uncertainties and Biases from Artificial Stars} \label{fakeerrsec}

The accuracy of stellar photometry can be strongly affected by
crowding in regions of high stellar density.  In such cases stars
that are too faint to be detected individually can blend with
other stars to rise above the detection limit.  Thus, many of the
faint stars in the photometric catalog have been biased to brighter
magnitudes by their neighbors.  High crowding also makes it difficult
to accurately measure the local background, because all of the pixels
outside a star's photometric aperture are filled with other stars.
Finally, crowding tends to increase the photometric uncertainty,
because the measured flux depends on the accuracy with which the flux
from all surrounding stars can be subtracted from the image.

The only way to accurately assess the effect of crowding is through
extensive artificial stars tests (described above in
Section~\ref{artificialstarsec}).  These tests allow us to measure the
difference between the true and recovered magnitude, as a function of
the star's color, magnitude, and position.  

In Figures~\ref{fakestarsuvfig}--\ref{fakestarsirfig}, we show a
series of three-panel plots containing the cumulative distributions of
the magnitude difference (left panels), the fractional flux difference
scaled by the flux uncertainty (center panels), and the color
difference (right panels) as a function of magnitude (where redder
colors and thicker lines indicate fainter stars), for the UV
(Figure~\ref{fakestarsuvfig}), optical (Figure~\ref{fakestarsacsfig}),
and NIR (Figure~\ref{fakestarsirfig}), using Field 9 of Bricks 1, 9,
15, and 21 (top to bottom, respectively).  We only include data for
Brick 1 in the UV analysis; the WFC3/UVIS data is uncrowded at all
radii outside the very central regions of the bulge, and thus the
artificial star tests yield identical results at all radii.  All
quantities are given as the measured value minus the true value for
the inserted star.  Photometry only includes stars from the {\tt{gst}}
catalogs.

Figure~\ref{fakestarsuvfig} shows that the effect of crowding is
minimal for the \fw{275} and \fw{336} filters.  The differences between
the input and output magnitudes and/or colors increase for fainter
stars (left and right panels, respectively).  However, if one scales
the flux difference by the photometric uncertainty reported by DOLPHOT
(center panel), it is clear that the distribution of flux differences
is essentially Gaussian, with a width set by the photon-counting
uncertainties alone.  The only sign of crowding effects is an
extremely slight tail towards brighter fluxes at faint magnitudes in
the deeper \fw{336} filter (thick red line in the central panel of the
bottom row).    

The effects of crowding are more severe in the optical, as can been
seen in Figure~\ref{fakestarsacsfig}. Unlike in the UV, the
distributions of flux differences (center panels) do not follow the
Gaussian distribution expected for uncertainties dominated by
photon-counting (with the possible exception of the brightest stars in
the least crowded fields).  The distributions are instead skewed, such
that stars are preferentially detected as brighter than their true
magnitude.  The skewing yields a significantly increased probability
that a star will be reported as being brighter than
expected for photon-counting uncertainties (i.e., compare the fraction
of stars that are more than 5$\sigma$ brighter to either the Gaussian
expectation, or to the fraction that are 5$\sigma$ fainter).  However,
the median magnitude difference is not highly biased outside of the
bulge, and the median flux errors are typically biased by less than
the expected photometric error.  The effects of crowding are
noticeably worse in the inner regions of the galaxy, where the surface
density is highest.  Crowding also has more
of an effect in the \fw{814} filter, which has a larger PSF, and which
is deeper than \fw{475}, and thus has a higher density of resolved
stars.

The crowding errors are largest in the NIR
(Figure~\ref{fakestarsirfig}), which suffers from worse resolution
than either ACS or WFC3/UVIS, due to bigger pixels and an
intrinsically broader PSF at long wavelengths.  As a result, the
effects of crowding are severe at all radii, even the outer disk.
These effects can be clearly seen in the distributions of magnitude
errors, which are dramatically wider than expected from photon
counting along (middle panels), and which show clear skewing (left
panels).

Surprisingly, the NIR photometry is biased towards {\emph{fainter}}
magnitudes for the median star.  The bias is slight in \fw{110}, but
noticeable in \fw{160}, where the crowding is most severe.  This
difference between the two filters then produces a slight blueward
shift in the median \fw{110}$-$\fw{160} color ($<$0.05 mag in all but
the faintest bin).  In all cases, the systematic bias in the median
magnitude or color is significantly smaller than the width of the
error distribution, such that the error properties will be dominated
by random rather than systematic errors in most applications.  The
systematic bias is also of the same order as both the intrinsic errors
in repeat measurements (see Section~\ref{repeaterrsec} below) and the
likely uncertainties in the current zero points.  As such, while the
biases in Figure~\ref{fakestarsirfig} are undesirable, they are
unlikely to be the dominant source of systematic errors, except
perhaps in the densest regions in the bulge.  Unfortunately, we do not
currently have a convincing explanation for this offset at this point.
We are continuing to investigate, and expect to have the issue
resolved for the second data release.

\subsubsection{Repeatability and Systematic Uncertainties} \label{repeaterrsec}

There are a number of position-dependent uncertainties that are not
reflected in the DOLPHOT-reported uncertainties or the artificial star
tests, but that will contribute to magnitude differences when the same star
is measured in different parts of a single chip.  These include errors
in the flat-field, the position-dependent \tinytim\  model for the PSF,
the CTE correction, and the image-to-image DOLPHOT
aperture corrections.  These will manifest in the distribution of
magnitude differences as a constant systematic error that adds
linearly with the random photometric error.  We can test for these
effects via the repeatability of our photometry when comparing the
magnitudes of individual stars in overlapping observations.

We use the PHAT astrometry to identify all pairs of matched stars from
the {\tt{gst}} catalogs whose positional offsets give them a better
than 75\% chance of being a true match, based on the observed
distribution of astrometric offsets.  These matched pairs sample the
entire area of the ACS/WFC, due to the large amount of overlap between
adjacent fields.  In the WFC3/UVIS camera, the matched pairs cover
more than 50\% of the chips, but do not sample a square ``hole'' in
the center of the field of view.  In the WFC3/IR channel, the matched
pairs only sample the very edges of the field of view
($\sim3$--$5\arcsec$), due to their limited overlap (see
Figure~\ref{exptimemapfig}).  Thus, one should be aware that only the
results for ACS/WFC can be seen as truly representative of the entire
field of view, whereas for the WFC3/IR channel, we are only probing
the extreme edges of the chip, where many of the systematic effects
are likely to be at their worst.

In Figure~\ref{overlapmagdifffig} we plot the cumulative distribution
of observed magnitude differences between repeated observations of
stars within limited magnitude ranges (i.e., bins of 0.5 magnitude,
with bluer lines indicating brighter magnitude ranges).  We also plot
the 1$\sigma$ widths of the distributions as a function of the median
magnitude in each bin (black line), derived from the observed
inter-quartile range of the observed distribution, assuming a Gaussian
distribution.  These widths are compared with the expected width of
the distribution, based solely on photon counting statistics (blue
line).  The quadrature difference between the observed and expected
width (red line) is then a measure of the systematic error (although
an imperfect one, since systematic errors are unlikely to be strictly
Gaussian).  These measurements are shown for Brick 15, but the
results are representative of what we see in other disk fields with
very different crowding levels.

Figure~\ref{overlapmagdifffig} shows several clear trends.  First, the
distribution of magnitude differences becomes systematically wider at
faint magnitudes.  However, comparing the observed to the expected
width (black versus blue line) shows that the majority of this trend
is driven by the larger variation in photon counting for small fluxes.
Second, at bright magnitudes, the observed magnitude variations are
much larger than expected from the formal photometric errors, by
factors of 10 in some cases.  This difference is a clear signal of
systematic errors (red line).  However, the systematic errors are
small ($\lesssim0.05$ mag) at bright magnitudes, where it is easiest to
assess their amplitudes.  Finally, photometric errors begin to
dominate at faint magnitudes in most cases.  The magnitude where
photometric magnitudes begin to dominate depends on the camera,
filter, and degree of crowding.  For example, the WFC3/IR observations are
crowding-limited, and thus can only detect stars much brighter than
the true photometric limit of the observations
(see Section~\ref{depthradiussec}).  Thus, the WFC3/IR detected stars
tend to have lower average photometric uncertainties, making the
systematic uncertainties proportionally more important.

We have calculated the width of the observed distribution of color
differences in a similar fashion.  Because spatial-dependent biases in
flat fields and PSF models are often correlated between filters,
systematic biases in colors may in fact be smaller than those for
magnitudes alone.\footnote{We see this effect for crowding errors as
  well (Section~\ref{fakeerrsec}).}  In
Figure~\ref{systematicfig} we compare the inferred level of systematic
bias in color (black line) to those in the magnitudes of individual
filters (blue and red lines, for the bluer and redder filters used for
an individual camera).  The systematic errors in colors are indeed
smaller than those for individual magnitudes, by up to a factor of two
in the NIR.  They have typical amplitudes of only 0.02 mag, rising to
$\sim$0.04 mag in the faintest bins.  They also appear to have a
weaker dependence on the brightness of a star, and are thus more
constant with magnitude.  

We can use stars that are measured in multiple images to assess
whether the systematic biases in Figure~\ref{systematicfig} have
spatial structure, as one would expect for flat-fielding errors or
spatially-dependent errors in the PSF model.  For ACS, we have the
benefit of highly overlapping images, such that an individual star
appears in multiple quadrants of the ACS chips, and such that every
position on each chip has overlapping data.  We use these overlaps to
assess whether there are locations in the ACS FOV where stars are
consistently measured to be brighter or fainter.  We first measure the
magnitude difference $\delta m$ for each pair of overlapping images
for each star in the matched astrometric catalogs.  If we assume that
this magnitude difference results entirely from systematic errors,
then we can map out, in pixel coordinates, the magnitude corrections
required to minimize all the $\delta m$ residuals.  We do this by
spatially binning the $\delta m$ values based on the pixel positions
of the two measurements of each star.  Weighting using the geometric
mean of the errors in the two magnitude measurements, and rejecting
$>3 \sigma$ outliers, we find the maximum-likelihood magnitude
corrections on a $100\times100$ grid in pixel space.

The resulting maps for Bricks 15 and 21 are shown in
Figure~\ref{magdiffmapfig} (top and bottom row, respectively) for
\fw{475} (left) and \fw{814} (right).  If there were no
spatially-varying systematics, the $\delta m$ values in each bin would
be randomly distributed around zero and our correction map would have
small random values.  Instead, we find coherent spatial patterns, at
the $\sim0.03$ mag level, that are consistent between bricks.  The
amplitude of the residual structure is clearly worse for \fw{475} than
for \fw{814}.  We believe that this difference most likely reflects
the better large scale flat fields that are available for \fw{814},
which was calibrated against ground-based globular cluster
observations (ISR ACS 2002-008 by J.~Mack et al.; ISR ACS 2006-06 by
R.~Gilliland et al.).  Even in \fw{814}, however, there are small
$\lesssim0.01$ mag systematics associated with the chip gap; hints of
these also have been reported in the analysis in ISR ACS 2003-10 by
R.~van der Marel (see Figure 2).  The maps in
Figure~\ref{magdiffmapfig} also indicate larger systematic errors
where the boundaries of individual images lie in the
distortion-corrected multi-drizzled reference images; stars that fall
on these image edges are more likely to have unreliable photometry,
and thus it is not clear whether the offsets at those positions
reflects flat-fielding errors, or the difficulty in doing reliable
photometry near the edge of a chip.  Even the largest offsets,
however, are still less than 5\%.

In future releases we expect to incorporate new large scale flats
based upon maps like those in Figure~\ref{magdiffmapfig}.
Unfortunately, we cannot generate equivalent maps for WFC3, due to the
lack of overlapping data that covers the whole field of view for each
chip.

\section{Results}  \label{resultssec}

\subsection{Color Magnitude Diagrams}  \label{cmdsec}

The stellar populations revealed by the PHAT imaging are best
understood through CMDs produced from the photometric catalogs.  In
this section we present CMDs for a range of radii within M31.  To
facilitate interpretation of these data in the following discussion,
we include in Figure~\ref{fakeCMDfig} artificial CMDs generated for
the PHAT filter set, showing the positions of stars as function of
their age and metallicity (for $A_V=0$ and a constant star formation
rate, assuming photometric errors and limits based on artificial star
tests from Brick 23).  We also show the expected Galactic foreground
populations in Figure~\ref{fakeFGfig}, calculated with TRILEGAL's
default Milky Way model \citep{girardi2005} (assuming no photometric
errors, but applying depth cuts to approximate those in
Figure~\ref{fakeCMDfig}).

In what follows, we show CMDs for stars in Bricks 1, 9, 15, and 21,
drawn from the {\tt{gst}} catalogs.  For reference, Bricks 9, 15,
and 21 are sampling a series of M31's star forming rings at
progressively larger radii, whereas Brick 1 is almost entirely
dominated by the bulge (Figure~\ref{footprintfig}).  The stellar
surface density varies dramatically among these regions, with higher
numbered bricks having systematically lower stellar surface densities,
and thus deeper CMDs in the optical and NIR, due to reduced crowding
(Figure~\ref{maglimradiusfig}).  There are also significant variations
in the stellar density within Brick 1 itself, leading to a highly
non-uniform limiting magnitudes within this single brick.

\subsubsection{Brick-Wide CMDs}

In Figures~\ref{hessbrickuvfig}--\ref{hessbrickirfig}, we show Hess
diagrams for the UV, optical, and NIR, for stars in Bricks 1, 9, 15,
and 21 (top left to bottom right).  

The UV disk CMDs (Figure~\ref{hessbrickuvfig}) are dominated by a
clear diagonal sequence of main sequence stars, extending down to
$\fw{336}\sim24.5$ (corresponding to a $\sim\!3\msun$ star at the
distance of M31).  These young stars have ages of $\lesssim500\Myr$
(Figure~\ref{fakeCMDfig}; top right), assuming roughly solar
metallicity for the young stellar populations.  The main sequence
starts to become more vertical brighter than $\fw{336}\sim19.5$
($\gtrsim15\msun$ for solar metallicity main sequence stars) and has a
color of $\fw{275}-\fw{336}\sim-0.45$ along the blue edge of the sequence.
This color is slightly redder than the color of $\fw{275}-\fw{336}\sim-0.6$
seen in the unreddened simulations in Figure~\ref{fakeCMDfig}, due to
extinction from a Galactic foreground and dust associated with M31's
star forming regions.

The observed UV CMDs show a notable absence of stars redward of the
main sequence, beyond the modest numbers expected from the Galactic
foreground (Figure~\ref{fakeFGfig}).  As can be seen in
Figure~\ref{fakeCMDfig}, even modestly sub-solar stellar populations
can be expected to host a dramatic second sequence of blue core Helium
burning stars, in addition to the main sequence.  Additional
simulations (not shown) suggest that there should be a clear sequence
of blue core Helium burning stars present for metallicities as high as
[Fe/H]$=-0.25$.  The absence of such a sequence suggests that even the
outer disk of M31 has close to solar metallicity at the present day.

In contrast to the main sequence dominated CMDs for Bricks 9, 15, and
21, the bulge population in Brick~1 shows a radically different UV CMD
morphology.  Although a smattering of main sequence stars are present,
the most noticeable sequence is much bluer and more vertical, and
dominates at colors of $\fw{275}-\fw{336}\sim-0.5$, fainter than
$\fw{336}\sim20$.  This sequence is made up of hot post-horizontal
branch stars that had anomalously high mass loss rates on the RGB,
which then altered what would otherwise be normal AGB evolution.  The
properties of this population are discussed extensively in
\citet{rosenfield2012} for the PHAT dataset.

The optical disk CMDs in Figure~\ref{hessbrickoptfig} shows a much
wider range of stellar ages than the UV (see Figure~\ref{fakeCMDfig}).
Outside of the bulge (i.e., Bricks 9, 15, and 21), there is an obvious
main sequence, extending nearly vertically at $\fw{475}-\fw{814}\sim0$.  The
faintest detectable stars in the main sequence are found in Brick 21,
and have typical masses of $\sim1.5\Msun$, assuming solar metallicity.
At redder colors, the stellar populations are dominated by a prominent
red giant branch, punctuated by a well-populated red clump at
$\fw{814}\sim24.5$ and a noticeable AGB bump at $\fw{814}\sim23.5$
\citep{gallart1998}.  The RGB, red clump, and AGB bump are all smeared
out diagonally by differential reddening. This effect is particularly
severe in Brick 15, which covers the main star-forming ring.  The RGB
terminates at bright magnitudes, with the tip of the red giant branch
(TRGB) falling at $\fw{814}\sim20.5$ for bluer RGB stars, and
$\fw{814}\sim21.5$ for redder RGB stars.  There is a significant
population of AGB stars brightwards of the TRGB, with the bulk being
no more than 1.5 magnitude brighter than the TRGB.

There are also subtle, less populated features that can be discerned
in the optical CMDs.  First is a hint of a vertical sequence emerging
brightward of the red clump and bluewards of the RGB.  This feature is
most likely due to core helium burning stars, which have
characteristic ages of less than several hundred Myrs \citep[see, for
example, Fig.~1 of ][]{mcquinn2011}.  These stars typically appear as
two narrow sequences in lower metallicity dwarf galaxies \citep[see
example CMDs in][]{dalcanton2009}, but in M31 the two sequences are
redder and less distinct, due both to higher metallicity and
differential reddening.  As a result, the ``sequence'' manifests more
as a subtle ``edge'' at $\fw{475}-\fw{814}\sim1.5$ just above the red
clump, corresponding to the blue edge of the core Helium burning
evolution (see the solar metallicity panels of
Figure~\ref{fakeCMDfig}).  The brighter stars in the core Helium
burning sequences (i.e., classical supergiants) are difficult to
separate from the Galactic foreground, which overlaps their positions
in the optical CMD (Figure~\ref{fakeFGfig}; see also
\citet{drout2009}), making it challenging to trace this feature to
brighter magnitudes.

In addition to the core Helium burning sequences, there is a more
subtle secondary overdensity of stars extending 0.5 mag faintward from
the red clump, which is most evident in Brick 21 because of the lower
level of crowding and better contrast in the CMD. This feature
corresponds to the ``secondary clump'' \citep{girardi1999} caused by
He-burning stars which are just massive enough to have ignited He in
non-degenerate conditions, at ages of 1~Gyr.

The optical bulge CMD in Brick 1 shows some similarities with the disk
CMDs.  The bulge region is dominated by a very strong red giant branch
with a clear TRGB, and a substantial population of AGB stars
populating the region brighter than the TRGB.  The RGB is also very
broad, which is most likely due to a significant spread in stellar
metallicity, given the lack of substantial differential reddening in
the bulge \citep[e.g.,][]{montalto2009}.  The crowding limit varies
dramatically within different regions of Brick 1, however, such that
the faint red clump is only detectable in the fields that are far from
the center of the bulge.

The bluer optical populations in Brick 1 differ substantially from
those seen in the disk fields, in two major ways.  First, there is a
noticeable diagonal sequence extending blueward from the RGB at
$\fw{814}\sim24$, down to fainter magnitudes.  In
Figure~\ref{bulgetrackfig} we overlay the post-AGB (P-AGB), post-early
AGB (PE-AGB), AGB manqu\'e, and zero age horizontal branch
evolutionary tracks that were used to analyze the UV populations in
\citet{rosenfield2012}.\footnote{PE-AGB and AGB manqu\'e stars are
  stars which lost too much mass on the RGB to undergo normal AGB
  evolution.}  The dashed portion of the cyan line indicates a very
fast interval of AGB manqu\'e evolution.  Based on these tracks, it
seems most likely that the diagonal sequence is made up primarily of
blue horizontal branch stars (which are long-lived), though with a
possible small contribution of AGB manqu\'e.  However, an exact
attribution will depend on fuller modeling that takes evolutionary
lifetimes and likely reddening into account.  We note that the stars
in the diagonal sequence are primarily found far from the center of
the bulge, but it is not yet clear if this trend is due only to the
increased depth at large radii, or reflects a real population
gradient; this population is being analyzed more fully in
\citet{williams2012}.  

The second major difference between the bulge and disk's blue
populations is the behavior of the blue edge of the optical CMDs.  In
the disk regions, the upper blue edge is defined by the nearly
vertical main sequence.  In Brick 1, however, there are few massive
main sequence stars.  Instead, the blue edge is curved to redder
colors at bright magnitudes.  Comparing to the tracks in
Figure~\ref{bulgetrackfig}, it appears that the blue edge in the bulge
is defined by a combination of post-AGB and post-early AGB stars, with
a possible contribution of main sequence stars at fainter magnitudes.

Figure~\ref{hessbrickirfig} shows the NIR color magnitude diagrams for
all four bricks.  The stars in these CMDs fall primarily on the older,
redder RGB and AGB sequences, with an additional small contribution
from the brightest main sequence stars.  The RGB is quite narrow in
all the disk fields, and a clear TRGB is present at $\fw{160}\sim18.3$.
However, differential reddening is still apparent, even at these
longer wavelengths.  The effect is most easily seen in Brick 15, where
large amounts of dust lead the RGB to appear doubled, due to stars
seen in front of and behind dust in the mid-plane of the galaxy.  In
the crowded bulge data, the RGB is broadened and does not extend more
than $\sim$3 magnitudes below the TRGB.  In the outer disk, however,
the CMDs are sufficiently deep that we see both a well-populated red
clump and AGB bump.  These features are extended to redder, fainter
colors, suggesting that there is significant differential reddening,
even at these large radii.  Red core helium burning stars are present
as well, but as discussed in \citet{dalcanton2012}, they are difficult
to distinguish from the main RGB and AGB sequences in this filter
combination.

The PHAT bricks are sufficiently large to contain a sizable number of
foreground MW stars. These appear more numerous in the NIR CMDs, and
appear as a narrow vertical feature at $\fw{110}-\fw{160}\sim0.7$ mag. Such
a feature is evident in all our NIR CMDs. Simulations with TRILEGAL
\citep{girardi2005} reproduce well the shape and stellar density of
this feature (Figure~\ref{fakeFGfig}).

\subsection{Global Structure of the M31 Disk}  \label{m31structuresec}

Our resolved stellar data enable us to study the spatial distribution
of selected stellar populations.  Here we present initial results on
the spatial distribution of the abundant metal-rich ([Fe/H] $>-0.7$) RGB
stars with ages $\gtrsim1$~Gyr.  Previous studies of the structure of
M31's inner disk have relied on integrated light measurements of the
combined stellar population.  Recent work by \citet{courteau2011} uses
$I$ band and Spitzer 3.6~$\mu$m integrated light to fit the inner
profile of M31.  However, integrated light is an imperfect tracer of stellar
mass, mixing stellar populations with varying mass-to-light ratios.
While the NIR wavelengths used in recent work minimize the effects of
dust, variations in $M/L$ ratio at NIR wavelengths from young
populations are still large due to contribution from supergiant and
TP-AGB stars \citep[e.g.,][]{melbourne2012}.  By using star counts rather
than integrated light we can reduce the contamination of these younger
populations and directly study the structure of an older, more
dynamically mixed population.

We select bright RGB stars using the NIR \fw{110} and \fw{160}
filters, as this enables detection of even highly reddened sources.
Figure~\ref{rgbselectionfig} shows a NIR CMD of stars in Bricks 8 and
9, with our metal-rich RGB selection region overlaid as a blue
polygon.  The blue edge of this region was selected to contain old RGB
stars ($\sim$10~Gyr) up to $\sim$1 magnitude below the RGB tip with
metallicities [Fe/H] $> -0.7$.  Younger RGB stars also fall within the
box, with the solar metallicity 1.5~Gyr isochrone falling along the
blue edge of the selection region.  The redward extension of this box
follows the reddening vector, and contains stars with $E(B-V) \lesssim
3$.  Only the brightest RGB stars were considered so that our
selection box remains well above the completeness limit over all
regions except the bulge.  The RGB selection box excludes younger
populations such as unreddened red supergiants and TP-AGB stars.

After selecting the stars based on their position in the CMD, we
calculated the densities of these stars in $30\arcsec \times
30\arcsec$ bins over all the regions with available WFC3/IR data.  The
RGB stellar densities are shown in the top panel of
Figure~\ref{densitymapfig} and range from $>$1
$\textrm{stars\,arcsecond}^{-2}$ in the bulge to $\sim0.01
\,\textrm{stars\,csecond}^{-2}$ in Brick 23.  As would be
expected from an old population, the RGB density distribution is quite
smooth, despite the outer fields covering several regions with strong
star formation or high dust extinction.  The axes in
Figure~\ref{densitymapfig} are rotated to follow the major and minor
axes assuming a position angle of 35$^\circ$ \citep{paturel2003}.

We fit the 2-D RGB density of the stellar disk to an inclined
exponential disk model after applying a small correction for dust
extinction discussed below.  We do not attempt to fit the bulge, due
to incompleteness, and exclude the central region from our fits.
Initial fits to the full data yield a disk scale length of 6.1~kpc.
However, this fit is clearly biased by the presence of an overdensity
of RGB stars between 9 and 12.5 kpc, resulting in a model that
significantly over-predicts the number of stars in Brick 21--23
(beyond 1 degree along the major axis).  Excluding the region between
9.2 and 12.5 kpc, we obtain much better fits to the outer part of the
galaxy, and a scale length of 5.3~kpc, in excellent agreement with the
5.3$\pm$0.5~kpc found by \citet{courteau2011}.  The best-fit position
angle of the disk is 43.2$^\circ$, with an inclination of 70$^\circ$.
This inclination is somewhat lower than the 74$^\circ$ inclination
found by \citet{courteau2011} and the canonical value of 77$^\circ$
assumed in earlier work \citep[e.g.,][]{walterbos1988}, perhaps due to
our sampling an intrinsically thicker distribution of stars or our
incomplete spatial coverage.  The position angle agrees well with that
quoted by \citeauthor{seigar2008} (\citeyear{seigar2008}; $45^\circ$)
and is slightly rotated from the value given in
\citeauthor{tempel2010} (\citeyear{tempel2010}; $38.1^\circ$).  The
best-fit model and residuals are shown in the middle and bottom panels
of Figure~\ref{densitymapfig}.  The areas fitted are represented
reasonably well by a single exponential disk: the reduced $\chi^2$ of
this fit is 1.7, assuming Poisson statistics for the RGB star counts.
In contrast, inclusion of the 10~kpc ring region results in a best-fit
reduced $\chi^2$ of 3.7.

Examination of the residual map in the bottom panel of
Figure~\ref{densitymapfig} shows a clear and significant overdensity
between 9 and $12.5\kpc$ along the major axis, with $\sim$40\% more
stars than expected in our best-fit exponential model.  This
overdensity is associated with the ``$10\kpc$ ring'' seen prominently
in UV emission \citep{thilker2005} and in the mid-IR
\citep{barmby2006}.  This ring hosts ample recent star formation, and
visual inspection of the PHAT data shows the expected population of
young main sequence stars in this region.  However, since the stars
selected for density mapping have ages $>$1~Gyr, the overdensity
observed here implies that the $10\kpc$ ring, originally identified
through evidence for recent star formation
\citep{habing1984,gordon2006}, is overdense in more than just recently
formed stars.

A potential complication with this interpretation is that the higher
extinction associated with this region could redden additional stars
into our selection box.  In Figure~\ref{rgbselectionfig}, the green
box indicates areas of the CMD from which stars could be reddened into
the selection box.  This region includes primarily metal-poor RGB
stars, AGB stars, and a small number of young supergiants.  In the
fitted region, the median number of stars in this box (referred to
hereafter as $C_{obs}$) is 27\% of the number of stars in our
selection box (hereafter $R_{obs}$).  The upper envelope in
$C_{obs}/R_{obs}$ is about 40\%.  Thus replicating the $\sim$40\%
overdensity seen in the 10~kpc ring would require all of these stars
to be reddened into the metal-rich RGB box.  This scenario is
implausible, especially given that the RGB overdensity extends over a
wide range of dust extinction.  We can attempt to use the observed
$C_{obs}/R_{obs}$ to make a correction for the effects of dust
extinction on $R_{obs}$.  Assuming an intrinsic $C/R$ ratio of 40\%,
we can calculate $R_{intrinsic}$ based on the observed
$C_{obs}/R_{obs}$:

\begin{equation}
R_{intrinsic} = R_{obs} \frac{1+C_{obs}/R_{obs}}{1.4}
\end{equation}

This correction does create a significantly smoother RGB density
distribution, reducing the scatter in the fit residuals by $\sim$20\%
both in the fitted regions and in the 10~kpc ring.  This reduction in
residual scatter suggests that this correction does help account for the
effect of dust extinction on our number counts and therefore we have
included it in our fits discussed above and shown in
Figure~\ref{densitymapfig}.  However, the correction does not lessen the
significance of the 10~kpc ring overdensity, at least in part because
this region has a wide range of $C_{obs}/R_{obs}$.  It therefore
appears that the observed overdensity does in fact reflect a
significant increase in the number metal-rich RGB stars at radii
between 9 and 12.5 kpc along the major axis.

In summary, we find disk scale lengths in agreement with previous
estimates and find a strong overdensity ($\sim$40\% over the expected
density) of RGB stars affiliated with the ``10~kpc ring''.  These fits
suggest that this structure is long-lived, and not simply an
enhancement of recent star formation due to a spiral arm.  A much more
detailed analysis of the M31 structure as a function of stellar
population will be possible once the full data set is in hand.

\section{Conclusions}   \label{conclusionsec}

We have presented the survey design, observations, and data products
for the Panchromatic Hubble Andromeda Treasury.  We have demonstrated
the superb data quality offered by this new legacy of HST
observations.  Given that the optical and NIR data are crowding-limited
over essentially all of M31's disk, the data at these wavelengths
cannot be surpassed in depth until higher resolution facilities become
available.  Our hope is that the combination of wide wavelength
coverage, photometric quality, and broad areal coverage will make this
data set a rich scientific resource for decades to come.

\acknowledgements 

\begin{figure}
\centerline{
\includegraphics[width=2.25in]{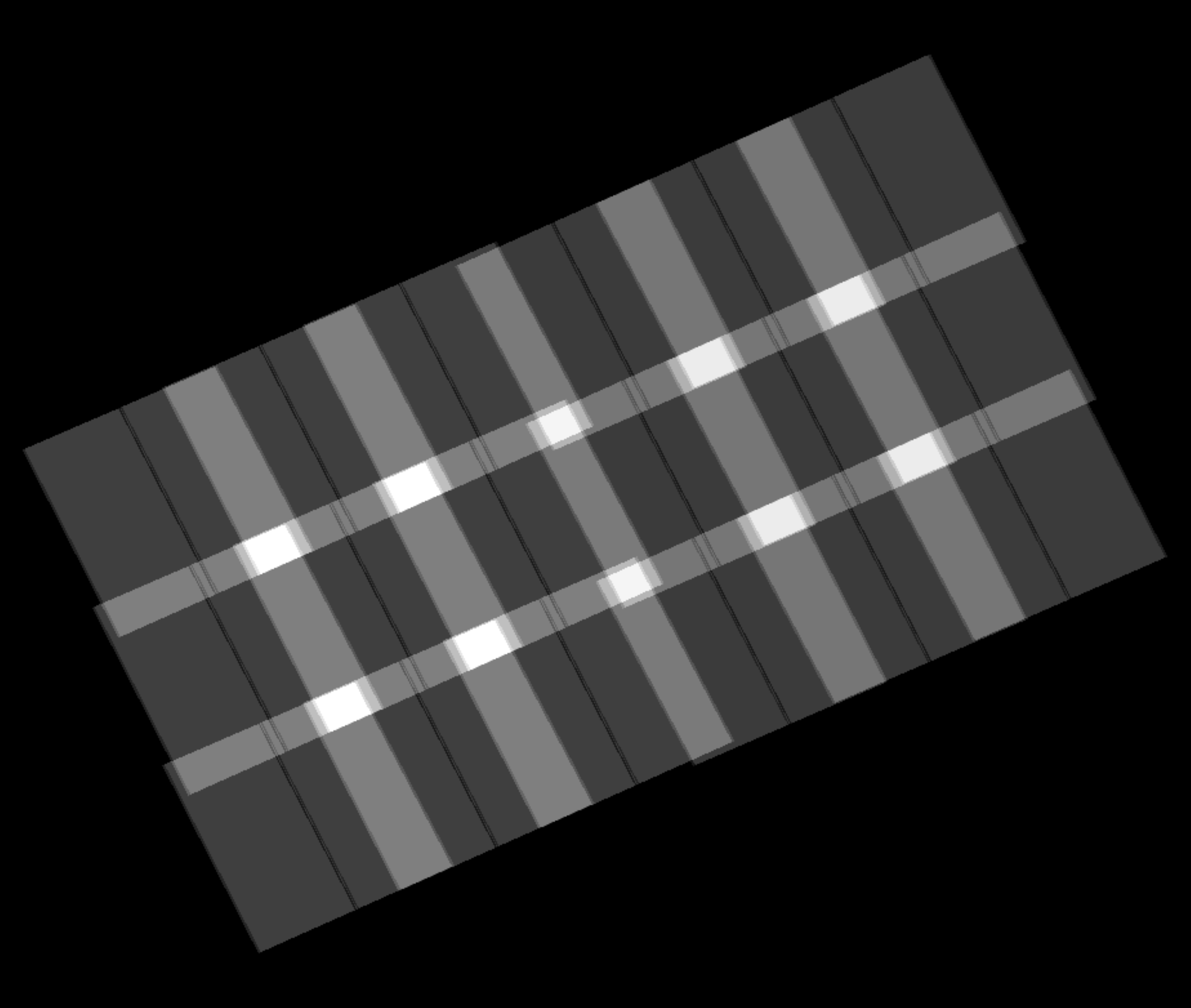}  
\includegraphics[width=2.25in]{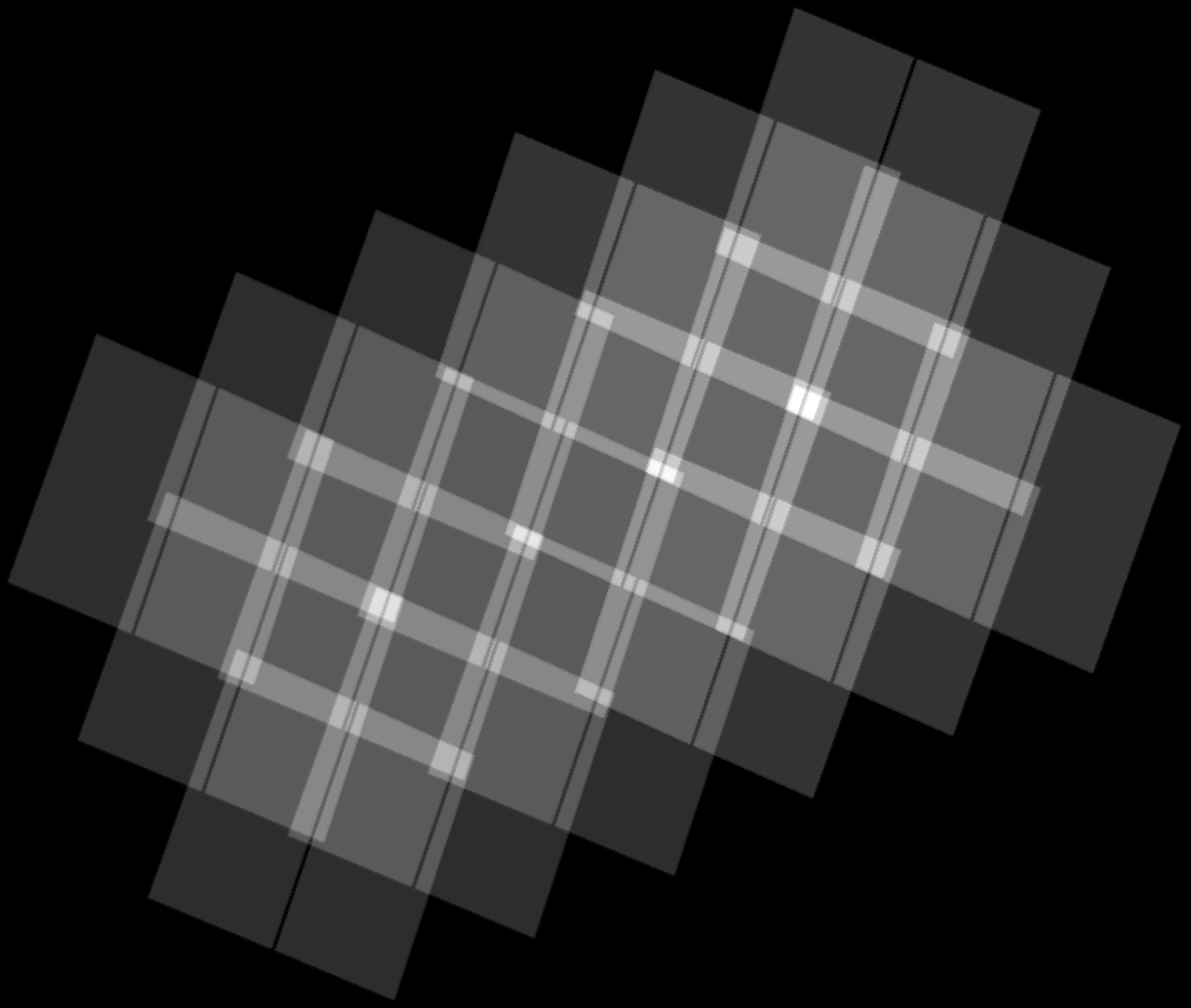}  
\includegraphics[width=2.25in]{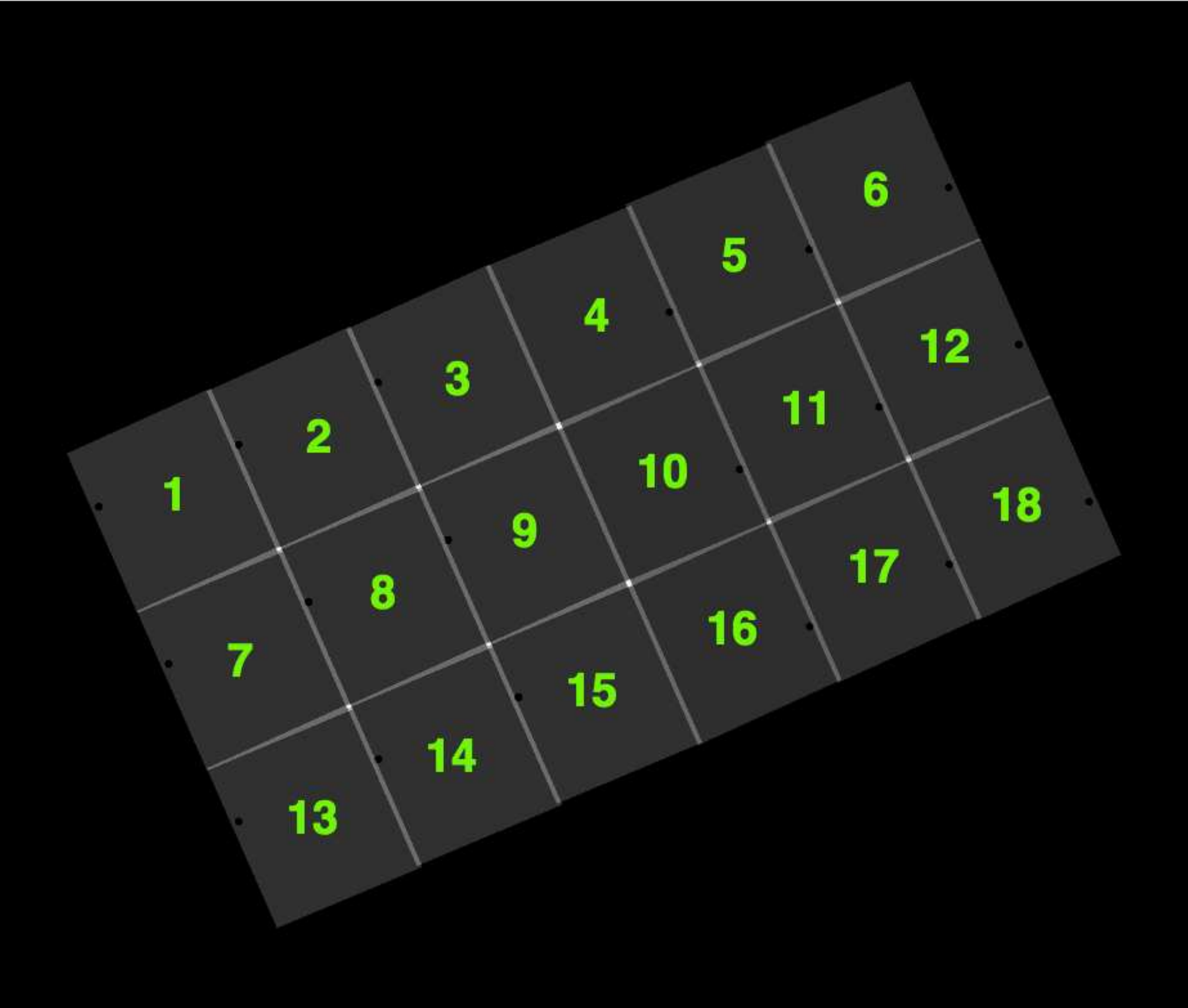}  
}
\caption{Maps of the relative exposure times across a brick for the
  WFC3/UVIS (left), ACS/WFC (middle), and WFC3/IR (right) tilings. The
  labeling of PHAT ``fields'' within a brick is shown superimposed on the
  WFC3/IR tiling. The left and right 3$\times$3 half-bricks are taken
  with orientations 180$^\circ$ apart, as can be seen by the relative
  position of the dark circle in the WFC3/IR exposure map. The WFC3/UVIS
  exposures are dithered to cover the chip gap.  The WFC chip gap in ACS
  is not dithered within a single orbit, but is covered by overlapping 
  exposures.  
  \label{exptimemapfig}}
\end{figure}

\begin{figure}
\centerline{
\includegraphics[width=3.55in]{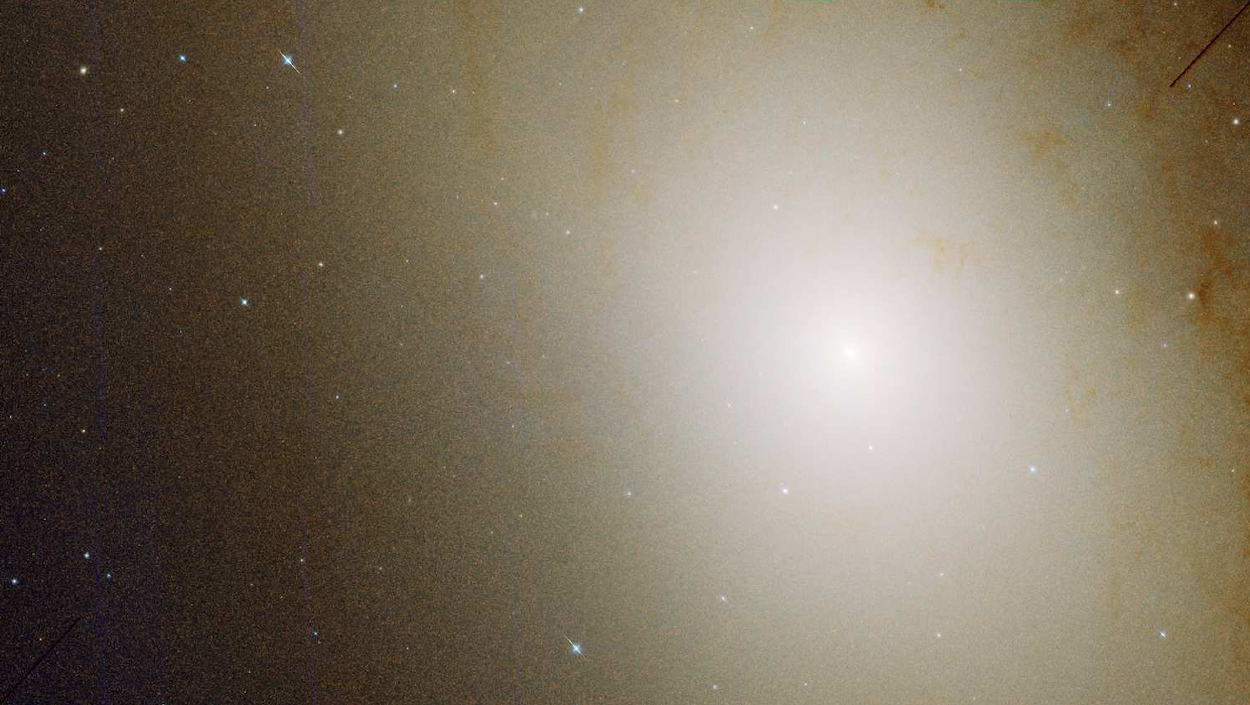}  
\includegraphics[width=3.55in]{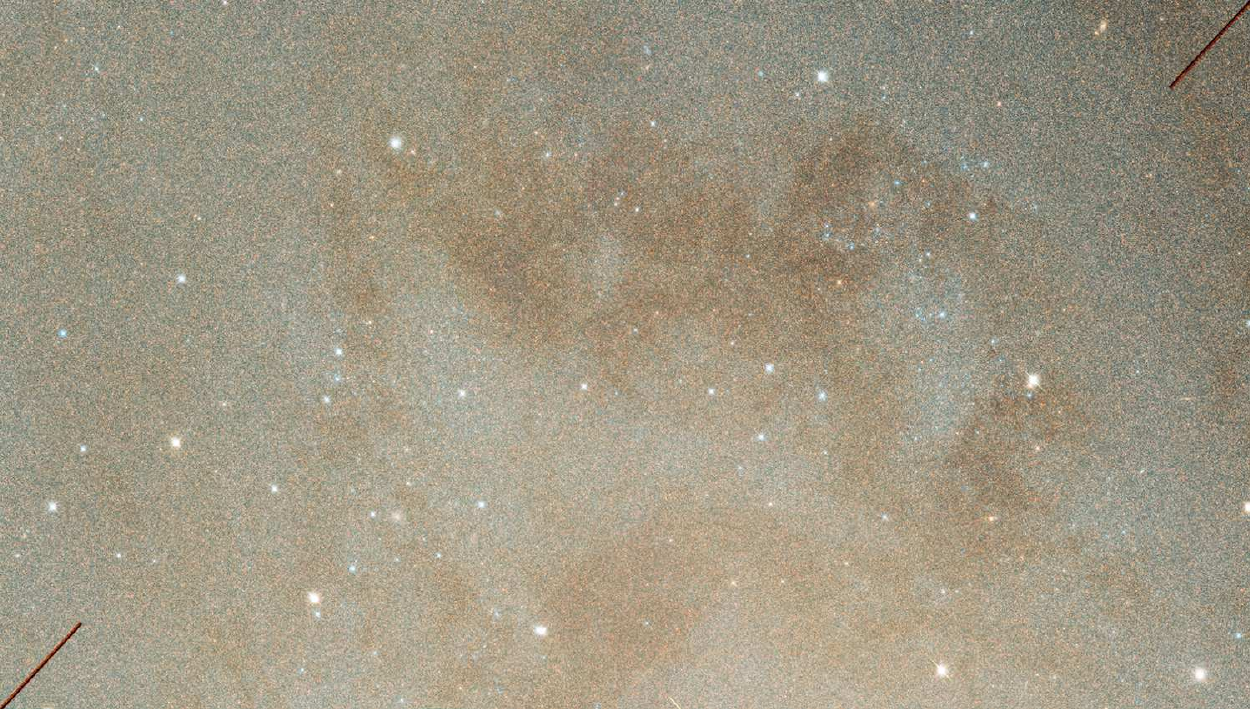}  
}
\vskip0.25cm
\centerline{
\includegraphics[width=3.55in]{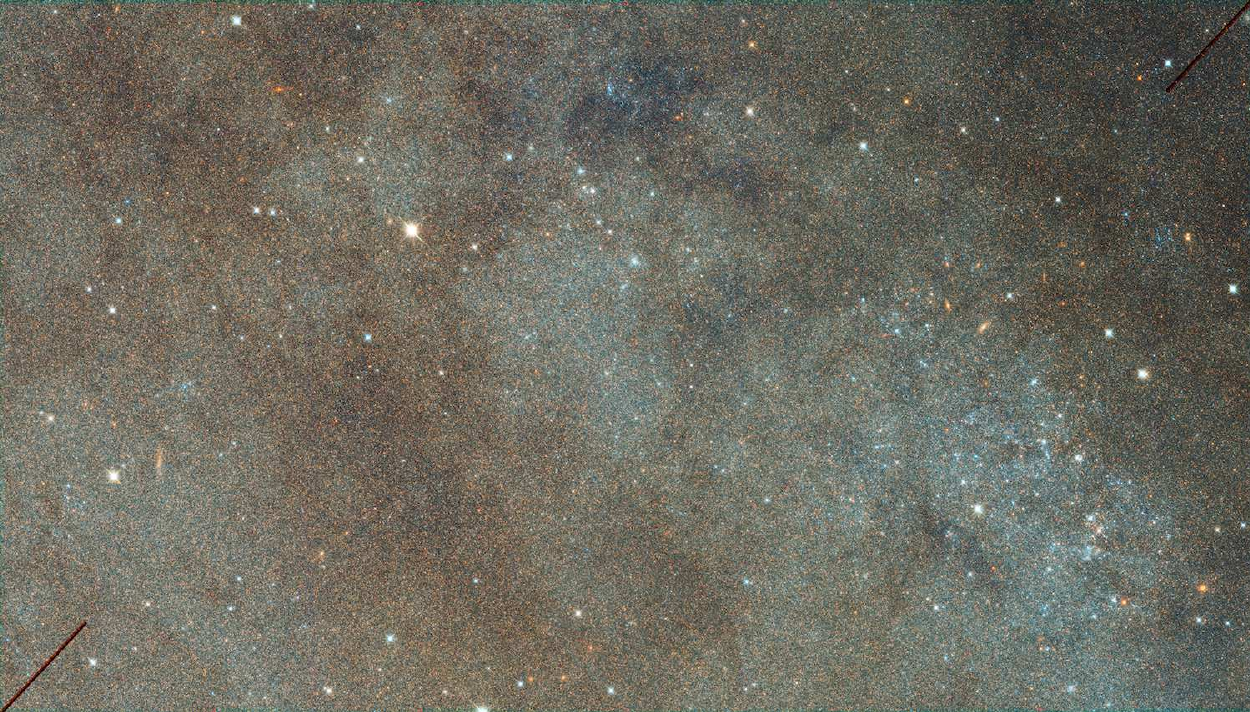}  
\includegraphics[width=3.55in]{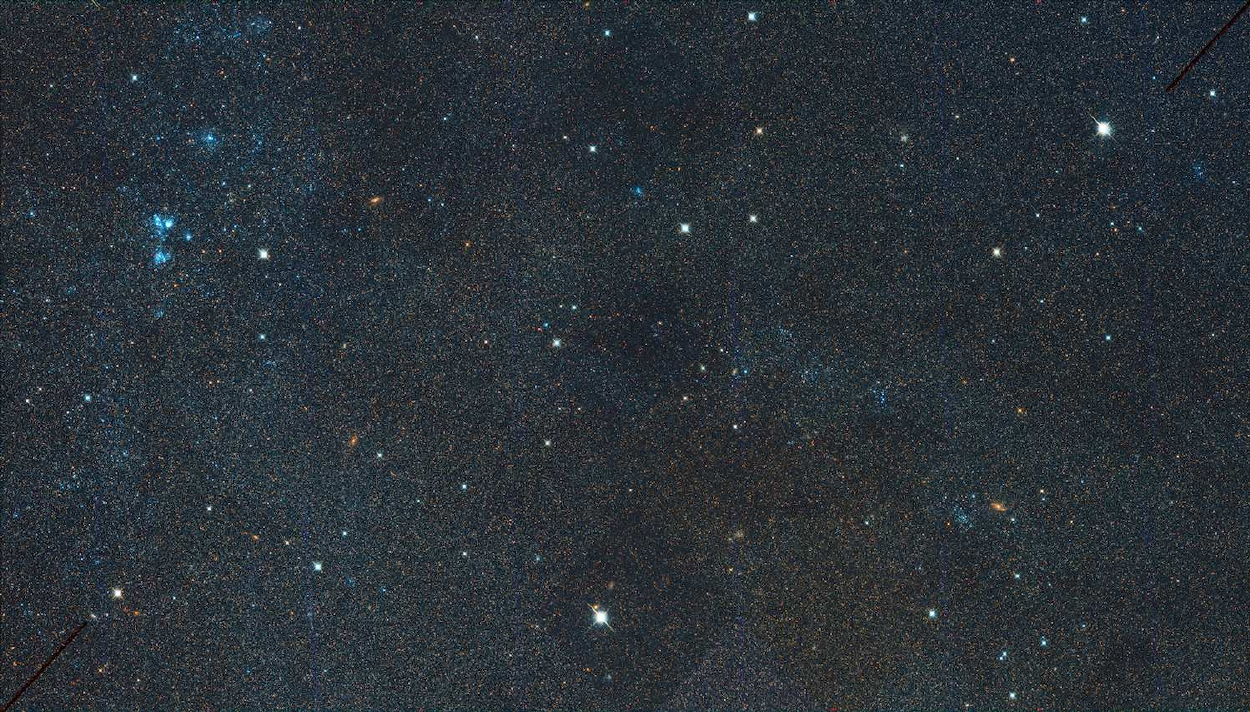}  
}
\caption{Four-color composite ($\fw{336}+\fw{475}+\fw{814}+\fw{160}$)
  images of Bricks 1, 9, 15, and 21 (upper left to lower right,
  respectively).  The flux scaling is identical for Bricks 9, 15, and
  21, to better show the radial change in the stellar density.  Dark
  diagonal lines in the upper right and lower left corners are
  produced by the ACS/WFC chip gap (see Figure~\ref{exptimemapfig}).
  Images were created by Zolt Levay (STScI). \label{brickwidergbfig}}
\end{figure}

\begin{figure}
\centerline{
\includegraphics[width=6.5in]{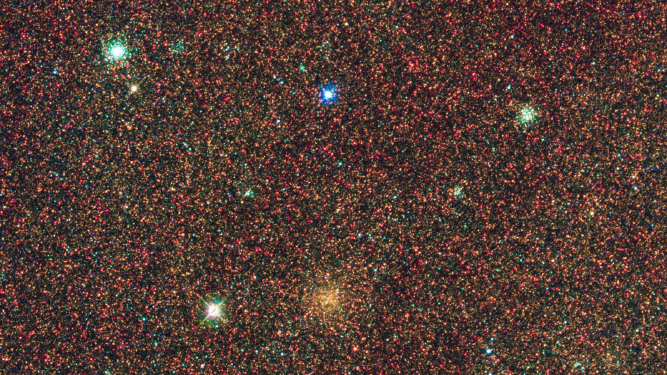}  
}
\caption{Close-up of a false-color image of Brick 9, showing the rich
  detail available at full HST resolution.  Courtesy Zolt Levay
  (STScI).
  \label{brickwidergbzoomfig}}
\end{figure}

The authors are very happy to acknowledge discussions with Jay
Anderson, Tom Brown, Suzanne Hawley, and Alessandro Bressan.  We are
grateful to Stefano Casertano for sharing the code to produce the
exposure time maps in Figure~\ref{exptimemapfig}, Alessandro Bressan
for providing the tracks plotted in Figure~\ref{bulgetrackfig},
Jean-Charles Cuillandre for allowing us to use his CFHT imaging to tie
our data to a global astrometric frame, to Pauline Barmby for
providing us the Spitzer IRAC images, and the anonymous referee for an
extremely knowledgeable and prompt report.  The project has received
superb support from personnel at the Space Telescope Science
Institute, including Alison Vick, Ken Sembach, Neill Reid, and the ACS
and WFC3 instrument teams.  Zolt Levay is particularly thanked for the
beautiful visualizations found in Figure~\ref{brickwidergbfig}, which
far surpassed anything we were able to generate on our own. Stan
Vlcek, Sarah Garner, and Pat Taylor at UW have been instrumental in
helping with logistics for the project.  This work was supported by
the Space Telescope Science Institute through GO-12055.
L.G.~acknowledges support from contract ASI-INAF I/009/10/0.


\bibliographystyle{apj}  


\bigskip
\vfill
\begin{deluxetable}{ccccccl}
\tabletypesize{\scriptsize}
\tablecolumns{9}
\tablewidth{0pt}
\tablecaption{Timing of Brick Observations}
\tablehead{\colhead{Brick}     & \colhead{PID}     &
           \multicolumn{2}{c}{WFC3 East \& ACS West} &
           \multicolumn{2}{c}{WFC3 West \& ACS East} &
           \colhead{M31 Components Covered}\\
           \colhead{}         & \colhead{}     &
           \colhead{Start}    & \colhead{End} &
           \colhead{Start}    & \colhead{End} &
           \colhead{}   }
\startdata
\phn1 & 12058  & 2010-12-14  & 2010-12-26\tablenotemark{a}  & 2010-07-21  & 2010-07-25 & Bulge, including nucleus   \\
\phn2 & 12073  &             &             & 2011-06-30  & 2011-08-03   & Outer bulge, NE star-forming arm  \\
\phn3 & 12109  &             &             &             &              & Outer bulge, inner disk, inner star-forming arms  \\
\phn4 & 12107  &             &             &             &              & NE star-forming arm  \\
\phn5 & 12074  &             &             & 2011-07-13  & 2011-08-03   & Inner-disk star formation  \\
\phn6 & 12105  &             &             &             &              & NE star-forming arm  \\
\phn7 & 12113  &             &             &             &              & Inner-disk star formation  \\
\phn8 & 12075  &             &             & 2011-07-26  & 2011-08-07   & NE star-forming arm; intense star formation; outer disk  \\
\phn9 & 12057  & 2010-12-30  & 2011-01-01  & 2010-07-12  & 2010-07-16   & Inner-disk star formation; significant overlap with brick 11  \\
10    & 12111  &             &             &             &              & NE star-forming arm; outer disk  \\
11    & 12115  &             &             &             &              & Smooth inter-arm disk; significant overlap with brick 9  \\
12    & 12071  &             &             & 2011-07-01  & 2011-07-13   & NE star-forming arm; outer disk  \\
13    & 12114  &             &             &             &              & Smooth, mainly inactive inter-arm disk  \\
14    & 12072  &             &             & 2011-07-22  & 2011-08-12   & NE star-forming arms; outer arm star formation  \\
15    & 12056  & 2011-01-10  & 2011-01-25  & 2010-08-07  & 2010-08-15   & Richest sample of main-arm star formation  \\
16    & 12106  &             &             & 2011-08-29  & 2011-09-03\tablenotemark{b}   & Main and outer-arm star formation \\
17    & 12059  & 2010-12-29  & 2011-01-09  & 2011-07-26  & 2011-08-06   & End of main star-forming ring  \\
18    & 12108  &             &             & 2011-08-21  & 2011-08-28   & Outer-disk star formation  \\
19    & 12110  &             &             &             &              & Inactive region in outer disk \\
20    & 12112  &             &             &             &              & Some outer-disk star formation; transition to halo \\
21    & 12055  & 2011-01-10  & 2011-01-15  & 2010-07-26  & 2010-07-28   & Outer-disk region of strong star formation  \\
22    & 12076  &             &             & 2011-08-14  & 2011-08-17   & Outside of disk; transition to halo  \\
23    & 12070  & 2011-01-26  & 2011-02-09\tablenotemark{b}  & 2011-08-08  & 2011-08-14   & Outside of disk; transition to halo \\
\enddata
\tablenotetext{a}{One orbit covering B01-F03/F06 for ACS$F814W$/UVIS
  were completed in 2011-01-03, after a guide star failure during the
  previous year's observing season.}
\tablenotetext{a}{One pointing has not yet been observed due to lack
  of guide stars (B16-F14/F17 for ACS/WFC3) or currently has invalid
  data due to failed guiding (B21-F16 for ACS); these observations are
  expected to be finished in the winter 2011/2012 observing season.}
\label{brickobstab}
\end{deluxetable}


\begin{deluxetable}{ccccccl}
\tabletypesize{\footnotesize}
\tablecolumns{7}
\tablecaption{Two-orbit Exposure Sequence}
\tablehead{\colhead{ACS}&\colhead{Exp}
&\colhead{$\Delta x,\Delta y$}&\colhead{WFC3}&\colhead{Exp}&\colhead{$\Delta x,\Delta y$}&\colhead{ } \\
\colhead{Filter}&\colhead{(s)}
&\colhead{(pixels)}&\colhead{Filter}&\colhead{(s)}&\colhead{(pixels)}&\colhead{Notes}}
\startdata
\fw{814}&     15&                        &\fw{336}&550& &\fw{814} guard \\
\fw{814}&    350&                        &\fw{275}&350& & \\
\multicolumn{7}{c}{{\emph{Large 37 Pixel Dither in $\Delta Y$ to Cover UVIS Chip Gap}}}\\
\fw{814}&700/800&$0.00,\ \phantom{-}0.00$&\fw{336}&700/800& & \\
\fw{814}&455/550&$1.50,\ \phantom{-}1.50$&\fw{275}&575/660& &ACS Nyquist dither \\
\hline\\
\fw{475}&     10&                        &\fw{160}&399&$0.00,\ \phantom{-}0.00$&\fw{475} guard \\
\fw{475}&600/700&$\phantom{-}0.00,\ \phantom{-}0.00$&\fw{110}&699/799& &ACS Nyquist dither \\
\fw{475}&    370&$\phantom{-}1.01,\ -1.50$&\fw{160}&699&$1.52,\ \phantom{-}1.42$&ACS$+$WFC3 Nyquist dither \\ 
\fw{475}&    370&$\phantom{-}0.48,\ -2.52$&\fw{160}&399&$1.94,\ \phantom{-}1.56$&ACS$+$WFC3 Nyquist dither \\ 
\fw{475}&370/470&$-1.50,\ \phantom{-}1.00$&\fw{160}&399/499&$1.50,\ -0.01$&ACS$+$WFC3 Nyquist dither \\
\enddata
\label{dithertab}
\tablecomments{The \fw{814} and WFC/UVIS exposures are obtained in the first
orbit.  The \fw{475} and WFC/IR exposures are obtained in the second orbit.
The dithers take place prior to the nominal exposure for which they are
listed.  Dithers are in pixels for the given camera.  These are only specified
for the sets of images for which dithers are designed to achieve the
proper sub-pixel sampling needed to generate Nyquist-sampled images.
The origin of the dithers is specified by the first image in the sequence.
Several exposures have two values for the exposure time.  The longer times
are for images obtained with the ``winter'' sequence, which allows for longer
orbital windows.}
\end{deluxetable}

\clearpage
\begin{deluxetable}{cccccccccc}
\tabletypesize{\scriptsize}
\tablecolumns{9}
\tablewidth{0pt}
\tablecaption{Approximate Corners of Bricks\tablenotemark{a}}
\tablehead{\colhead{Brick}     & \colhead{PID}     &
           \colhead{RA$_1$}    & \colhead{Dec$_1$} &
           \colhead{RA$_2$}    & \colhead{Dec$_2$} &
           \colhead{RA$_3$}    & \colhead{Dec$_3$} &
           \colhead{RA$_4$}    & \colhead{Dec$_4$} }
\startdata
\phn1 & 12058  & 10.87969 & 41.26532 &  10.63671 & 41.34631 &  10.57610 & 41.24387 &  10.81892 & 41.16290 \\
\phn2 & 12073  & 11.12029 & 41.18549 &  10.87790 & 41.26699 &  10.81699 & 41.16468 &  11.05921 & 41.08321 \\
\phn3 & 12109  & 10.95216 & 41.36298 &  10.70890 & 41.44413 &  10.64809 & 41.34173 &  10.89118 & 41.26061 \\
\phn4 & 12107  & 11.19357 & 41.28269 &  10.95092 & 41.36434 &  10.88980 & 41.26207 &  11.13230 & 41.18044 \\
\phn5 & 12074  & 11.04482 & 41.45377 &  10.80133 & 41.53511 &  10.74029 & 41.43276 &  10.98361 & 41.35144 \\
\phn6 & 12105  & 11.28646 & 41.37401 &  11.04358 & 41.45586 &  10.98223 & 41.35363 &  11.22495 & 41.27182 \\
\phn7 & 12113  & 11.13774 & 41.54510 &  10.89403 & 41.62664 &  10.83276 & 41.52434 &  11.07630 & 41.44284 \\
\phn8 & 12075  & 11.37848 & 41.46474 &  11.13539 & 41.54679 &  11.07381 & 41.44462 &  11.31675 & 41.36260 \\
\phn9 & 12057  & 11.22980 & 41.63532 &  10.98587 & 41.71706 &  10.92436 & 41.61481 &  11.16813 & 41.53310 \\
10    & 12111  & 11.47410 & 41.55495 &  11.23079 & 41.63719 &  11.16898 & 41.53508 &  11.41213 & 41.45287 \\
11    & 12115  & 11.28791 & 41.71003 &  11.02353 & 41.74143 &  10.99982 & 41.63066 &  11.26404 & 41.59933 \\
12    & 12071  & 11.54277 & 41.65341 &  11.29918 & 41.73581 &  11.23717 & 41.63373 &  11.48060 & 41.55137 \\
13    & 12114  & 11.31717 & 41.81950 &  11.05228 & 41.85054 &  11.02883 & 41.73975 &  11.29355 & 41.70886 \\
14    & 12072  & 11.57892 & 41.78777 &  11.31430 & 41.81941 &  11.29041 & 41.70867 &  11.55487 & 41.67718 \\
15    & 12056  & 11.37644 & 41.92498 &  11.11115 & 41.95616 &  11.08756 & 41.84538 &  11.35268 & 41.81435 \\
16    & 12106  & 11.63914 & 41.89310 &  11.37413 & 41.92490 &  11.35010 & 41.81416 &  11.61495 & 41.78253 \\
17    & 12059  & 11.48855 & 42.02423 &  11.22291 & 42.05568 &  11.19909 & 41.94492 &  11.46456 & 41.91363 \\
18    & 12108  & 11.75272 & 41.99292 &  11.48736 & 42.02497 &  11.46310 & 41.91427 &  11.72830 & 41.88238 \\
19    & 12110  & 11.62005 & 42.12072 &  11.35409 & 42.15246 &  11.33001 & 42.04174 &  11.59581 & 42.01015 \\
20    & 12112  & 11.88344 & 42.08995 &  11.61776 & 42.12230 &  11.59324 & 42.01163 &  11.85876 & 41.97944 \\
21    & 12055  & 11.68571 & 42.22514 &  11.41935 & 42.25704 &  11.39512 & 42.14633 &  11.66131 & 42.11460 \\
22    & 12076  & 11.95062 & 42.19503 &  11.68455 & 42.22754 &  11.65987 & 42.11690 &  11.92579 & 42.08456 \\
23    & 12070  & 11.86064 & 42.31678 &  11.59401 & 42.34909 &  11.56944 & 42.23843 &  11.83591 & 42.20629 \\
\enddata
\label{brickcornertab}
\tablenotetext{a}{All coordinates given in J2000.  Corners are given
  for the approximate limits of the WFC3/IR coverage; WFC3/UVIS and
  ACS/WFC images will extend beyond these limits (see
  Figure~\ref{exptimemapfig}).}
\end{deluxetable}


\clearpage
\begin{deluxetable}{llccc}
\tablecolumns{5}
\tabletypesize{\scriptsize}
\tablewidth{0pt}
\tablecaption{Dolphot Processing Parameters}
\tablehead{\colhead{Description}     & \colhead{Parameter}     &
           \colhead{WFC3/UVIS}    & \colhead{ACS/WFC} &
           \colhead{WFC3/IR}    }
\startdata
Photometry Aperture Size                            & {\tt{RAper}}       & 8      & 8     &    5      \\
Inner Sky Radius                                    & {\tt{RSky0}}       & 15     & 15    &	8      \\     
Outer Sky Radius                                    & {\tt{Rsky1}}       & 35     & 35    &	20     \\     
$\chi$-statistic Aperture Size                      & {\tt{Rchi}}        & 3.0    & 2.0   &    1.5    \\      
Photometry Type                                     & {\tt{PSFPhot}}     & 1      & 1     &	1      \\   
Fit Sky?                                            & {\tt{FitSky}}      & 3      & 3     &	3      \\    
Spacing For Sky Measurement                         & {\tt{SkipSky}}     & 2      & 2     &	2      \\   
Sigma Clipping For Sky                              & {\tt{SkySig}}      & 2.25   & 2.25  &	2.25   \\    
Second Pass Finding Stars                           & {\tt{SecondPass}}  & 25     & 25    &	25     \\
Searching Algorithm                                 & {\tt{SearchMode}}  & 1      & 1     &	1      \\
Sigma Detection Threshold                           & {\tt{SigFind}}     & 3.0    & 3.0   &	3.0    \\   
Multiple for Quick-and-dirty Photometry             & {\tt{SigFindMult}} & 0.85   & 0.85  &	0.85   \\
Sigma Output Threshold                              & {\tt{SigFinal}}    & 3.5    & 3.5   &	3.5    \\  
Maximum Iterations                                  & {\tt{MaxIT}}       & 25     & 25    &	25     \\     
Noise Multiple in {\tt{imgadd}}                     & {\tt{NoiseMult}}   & 0.10   & 0.10  &	0.10   \\ 
Fraction of Saturate Limit                          & {\tt{FSat}}        & 0.999  & 0.999 &	0.999  \\      
Use Saturated Cores?                                & {\tt{FlagMask}}    & 4      & 4     &	4      \\  
Find/Make Aperture Corrections?                     & {\tt{ApCor}}       & 1      & 1     &	1      \\     
Force Type 1/2?                                     & {\tt{Force1}}      & 1      & 1     &	1      \\    
Align Images?                                       & {\tt{Align}}       & 2      & 2     &	2      \\     
Allow Cross Terms in Alignment?                     & {\tt{Rotate}}      & 1      & 1     &	1      \\    
Centroid Box Size                                   & {\tt{RCentroid}}   & 1      & 1     &	1      \\ 
Search Step for Position Iterations                 & {\tt{PosStep}}     & 0.1    & 0.1   &	0.1    \\   
Maximum Single-Step in Position Iterations          & {\tt{dPosMax}}     & 2.5    & 2.5   &	2.5    \\   
Minimum Separation for Two Stars for Cleaning       & {\tt{RCombine}}    & 1.415  & 1.415 &	1.415  \\  
PSF Size                                            & {\tt{RPSF}}        & 10     & 10    &	10     \\      
Minimum S/N for PSF Parameter Fits                  & {\tt{SigPSF}}      & 3.0    & 3.0   &	3.0    \\    
Make PSF Residual Image?                            & {\tt{PSFres}}      & 1      & 1     &	1      \\    
Coordinate Offset                                   & {\tt{Psfoff}}      & 0.0    & 0.0   &	0.0    \\    
WFC3 PSF library                                    & {\tt{UseWFC3}}     & 1      & -     &	1      \\   
ACS PSF library                                     & {\tt{UseACS}}      & -      & 1     &	-      \\   
\enddata
\label{dolphottab}
\end{deluxetable}

\begin{deluxetable}{lcccccc}
\tablecolumns{7}
\tablewidth{0pt}
\tablecaption{PSF \& Aperture Corrections}
\tablehead{\colhead{Camera} & \colhead{Filter}& \colhead{Chip}& \colhead{$n_{psf}$}& \colhead{PSF Correction}  & \colhead{$n_{ap}$} & \colhead{Aperture Correction}}
\startdata
WFC3/UVIS & \fw{275} & 1   & 102  &           -0.0271$\pm$0.0047     & 147    & 0.0584$\pm$0.0176 \\
WFC3/UVIS & \fw{275} & 2   &\phn93&           -0.0307$\pm$0.0044     & 125    & 0.0561$\pm$0.0167 \\
WFC3/UVIS & \fw{336} & 1   & 115  &           -0.0267$\pm$0.0049     & 199    & 0.0188$\pm$0.0154 \\
WFC3/UVIS & \fw{336} & 2   & 111  &           -0.0304$\pm$0.0048     & 199    & 0.0187$\pm$0.0142 \\
ACS/WFC   & \fw{475} & 1   & 204  & \phantom{-}0.0051$\pm$0.0027     & 198    & 0.0048$\pm$0.0082 \\
ACS/WFC   & \fw{475} & 2   & 202  & \phantom{-}0.0064$\pm$0.0021     & 198    & 0.0059$\pm$0.0083 \\
ACS/WFC   & \fw{814} & 1   & 231  & \phantom{-}0.0091$\pm$0.0013     & 200    & 0.0159$\pm$0.0057 \\
ACS/WFC   & \fw{814} & 2   & 233  & \phantom{-}0.0057$\pm$0.0010     & 200    & 0.0161$\pm$0.0053 \\
WFC3/IR   & \fw{110} &     & 241  &           -0.0059$\pm$0.0014     & 199    & 0.0573$\pm$0.0057 \\
WFC3/IR   & \fw{160} &     & 241  & \phantom{-}0.0062$\pm$0.0015     & 200    & 0.0784$\pm$0.0064 \\
\enddata
\label{photcorrtab}
\tablecomments{$n_{psf}$ and $n_{ap}$ are the median numbers of stars
  used to calculate the PSF and aperture corrections, respectively.
  PSF and aperture corrections are listed as the median $\pm$ the
  semi-interquartile range.  All statistics have excluded data in
  Brick 1. The statistics for the short guard exposures in ACS have
  also been excluded.}
\end{deluxetable}


\clearpage
\vfill
\clearpage

\vfill

\vfill
\clearpage

\vfill

\vfill
\clearpage

\begin{figure}
\centerline{
\includegraphics[width=6.0in]{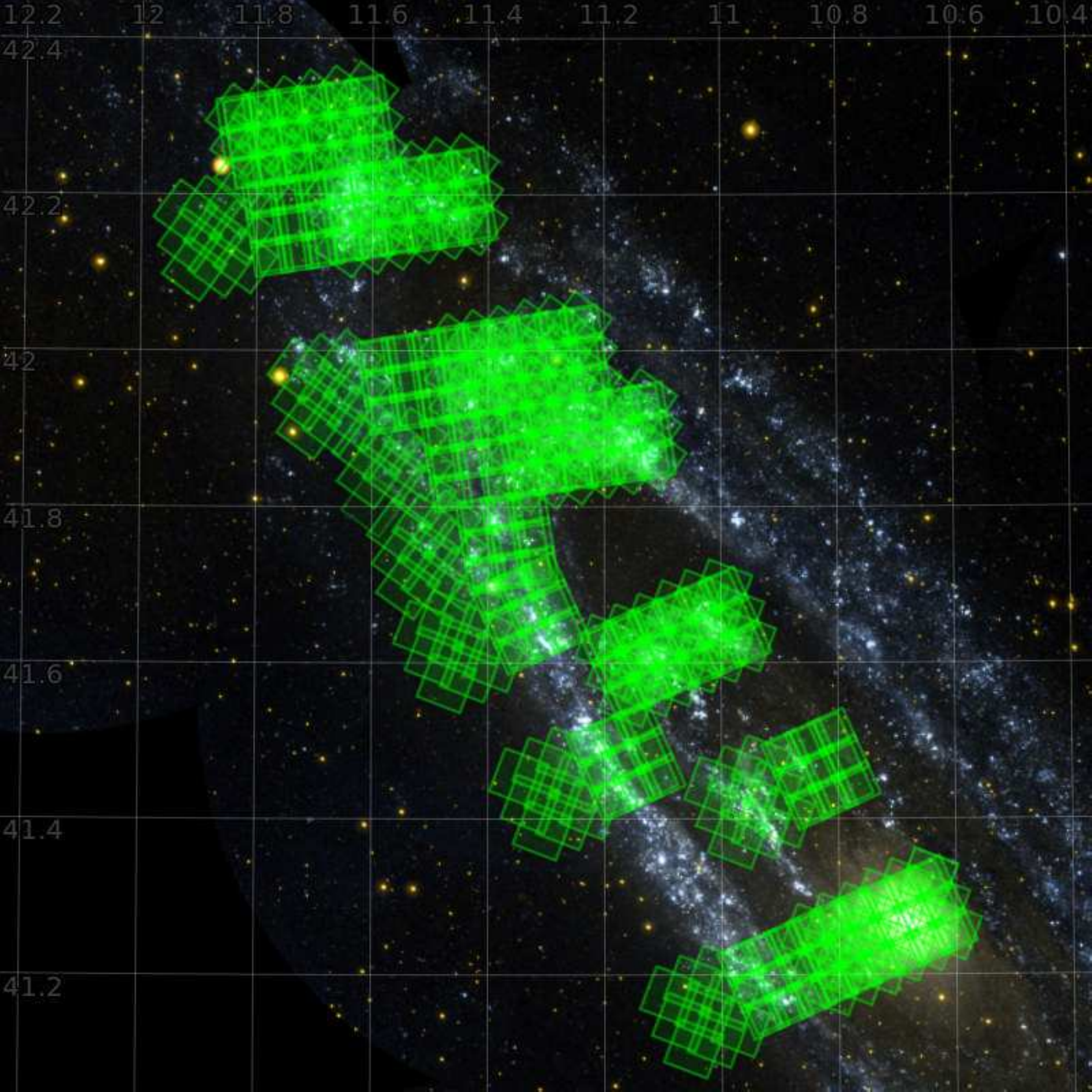}  
}
\caption{Current PHAT coverage, as of Fall 2011, superimposed on a
  GALEX FUV$+$NUV image.  Bricks 1, 9, 15, 17, 21, and 23 are largely
  complete, as are the first halves of Bricks 2, 5, 8, 12, 14, 16, 18,
  and 22.  There are missing data in Brick 16 (Fields 14~and~17) and
  Brick 23 (Fields 13~and~16) that will be filled in during future
  observation cycles.
  \label{coveragefig}}
\end{figure}
\vfill
\clearpage

\begin{figure}
\centerline{
\includegraphics[width=6.0in]{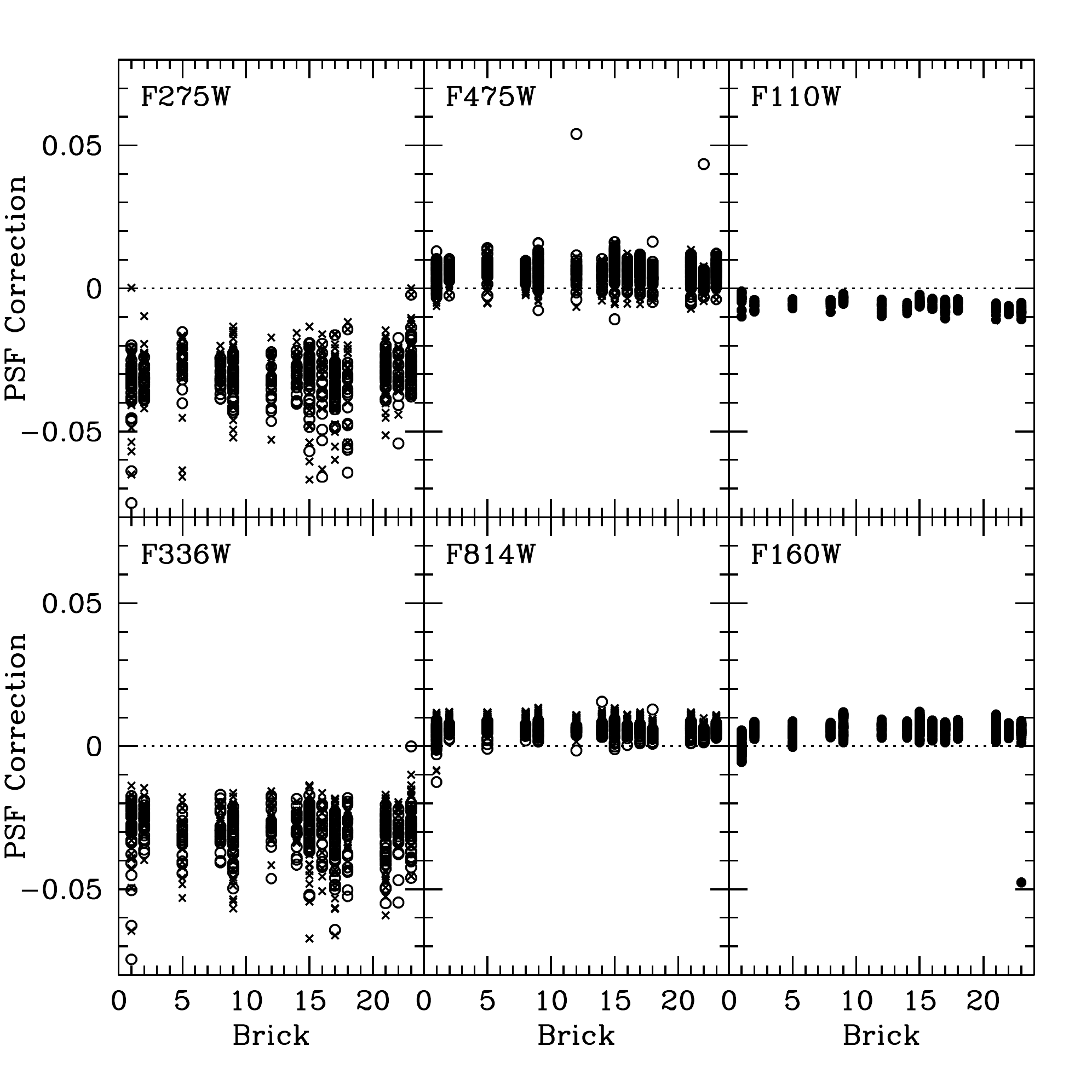}  
}
\caption{PSF corrections in magnitudes for all images in the survey to date,
  excluding the short guard exposures in \fw{475} and \fw{814}.  For the
  WFC3/UVIS and ACS/WFC data, crosses and open circles indicate the corrections
  for chips 1 and 2, respectively. Median values are given in
  Table~\ref{photcorrtab}.  The PSF corrections show no dependence on
  time or local stellar density.
\label{psfcorrfig}}
\end{figure}

\begin{figure}
\centerline{
\includegraphics[width=6.0in]{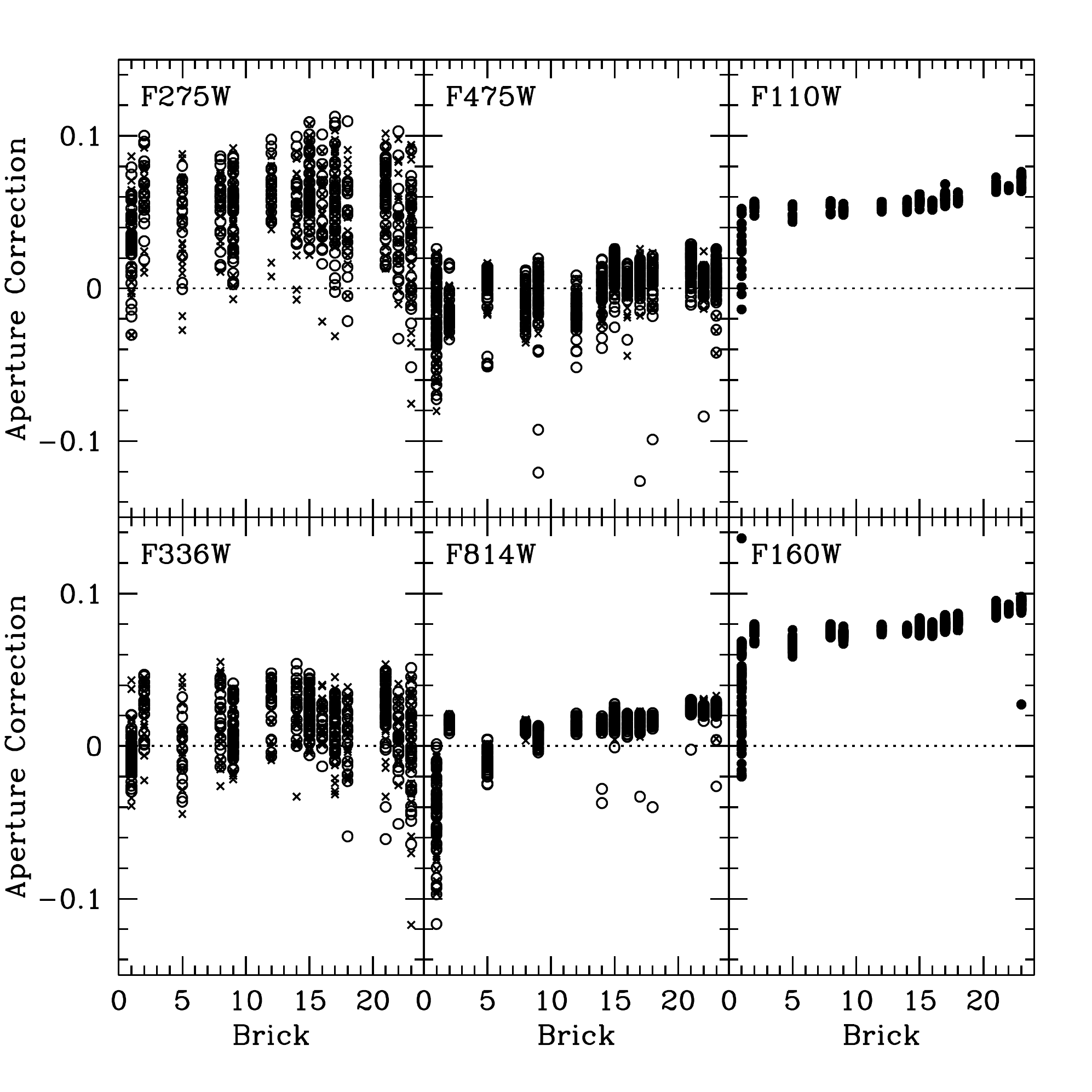}  
}
\caption{Aperture corrections in magnitudes for all images in the survey to date,
  excluding the short guard exposures in \fw{475} and \fw{814}.  For the
  WFC3/UVIS and ACS/WFC data, crosses and open circles indicate the corrections
  for chips 1 and 2, respectively.  The aperture corrections vary with
  brick number, due to variations in the local stellar density, which
  decreases for increasing brick number, and for even-numbered bricks.
  The scatter tends to be systematically larger for bluer filters.
  The scatter is largest in Brick 1, which has the largest range of
  stellar densities internal to the brick, due to the presence of the
  center of the bulge in Field 10.  Median values are given in
  Table~\ref{photcorrtab}, excluding both the guard exposures and
  images from Brick 1.
  \label{apcorrfig}}
\end{figure}

\begin{figure}
\centerline{
\includegraphics[width=4.0in]{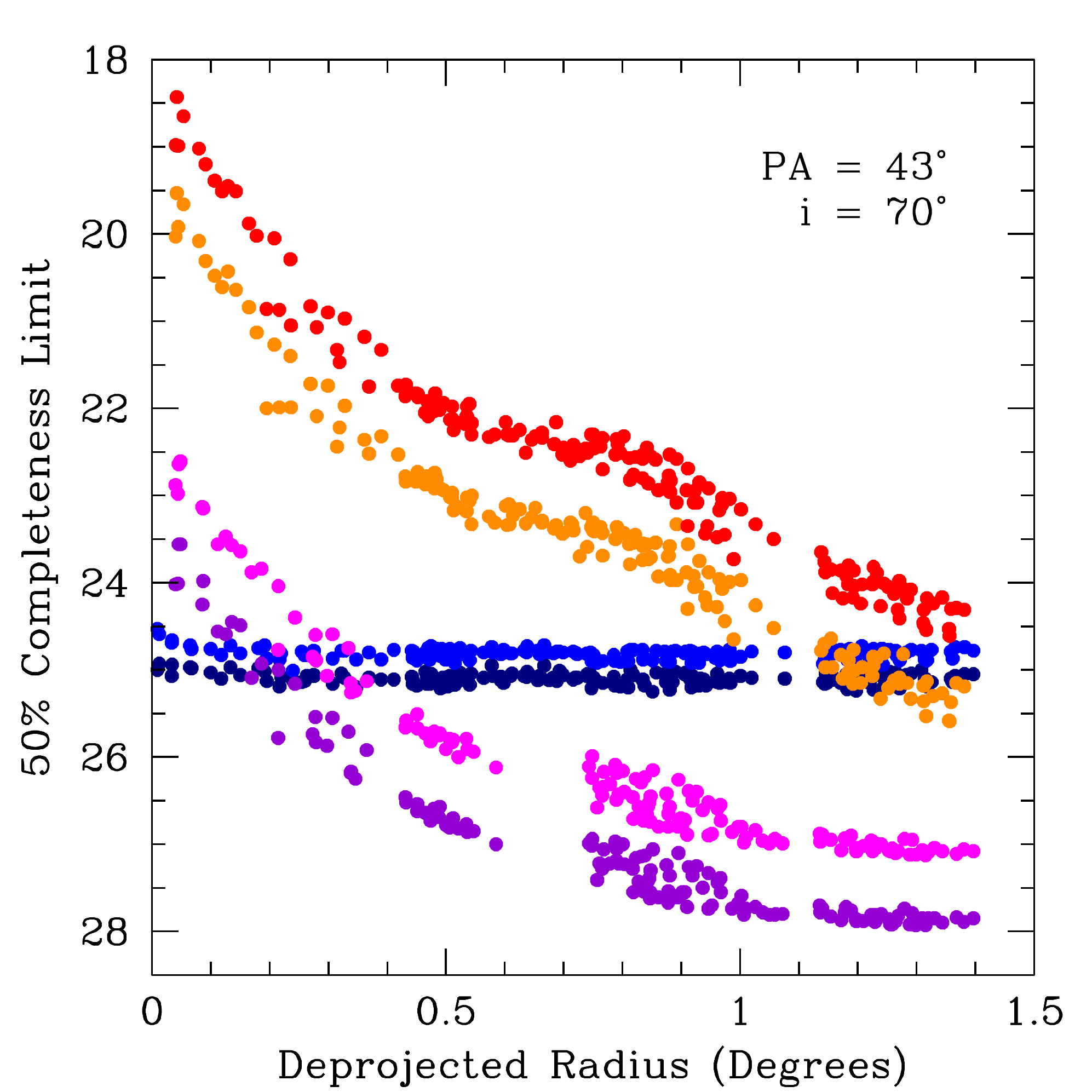}  
}
\caption{The magnitude of the 50\% completeness limit, plotted as a
  function of radius for the high photometric quality {\tt{gst}}
  catalog detections, for all fields observed with the WFC3/UVIS
  (horizontal blue points), ACS/WFC (lowest two purple sequences) and
  WFC3/IR (highest red sequences) cameras.  The two filters in each
  camera are color-coded as in Figure~\ref{maglimdensityfig}. The UV
  observations are photon-limited across the disk, with the exception
  of the very innermost regions of the central bulge field, and
  therefore have an essentially constant depth across the survey area.
  The optical observations are crowding-limited over most of the disk,
  and thus the depth of the ACS observations is a strong function of
  radius.  The optical data does become photon-limited at a projected
  radius of $\sim$0.9$^\circ$, beyond which the depth of the
  observations becomes constant.  The NIR observations are
  crowding-limited across the entire observed area, such that they are
  $\sim$5.5 magnitudes deeper in the uncrowded outer regions than in the
  bulge, in spite of having identical exposure times at all radii. 
  \label{maglimradiusfig}}
\end{figure}
\vfill

\begin{figure}
\centerline{
\includegraphics[width=2.15in]{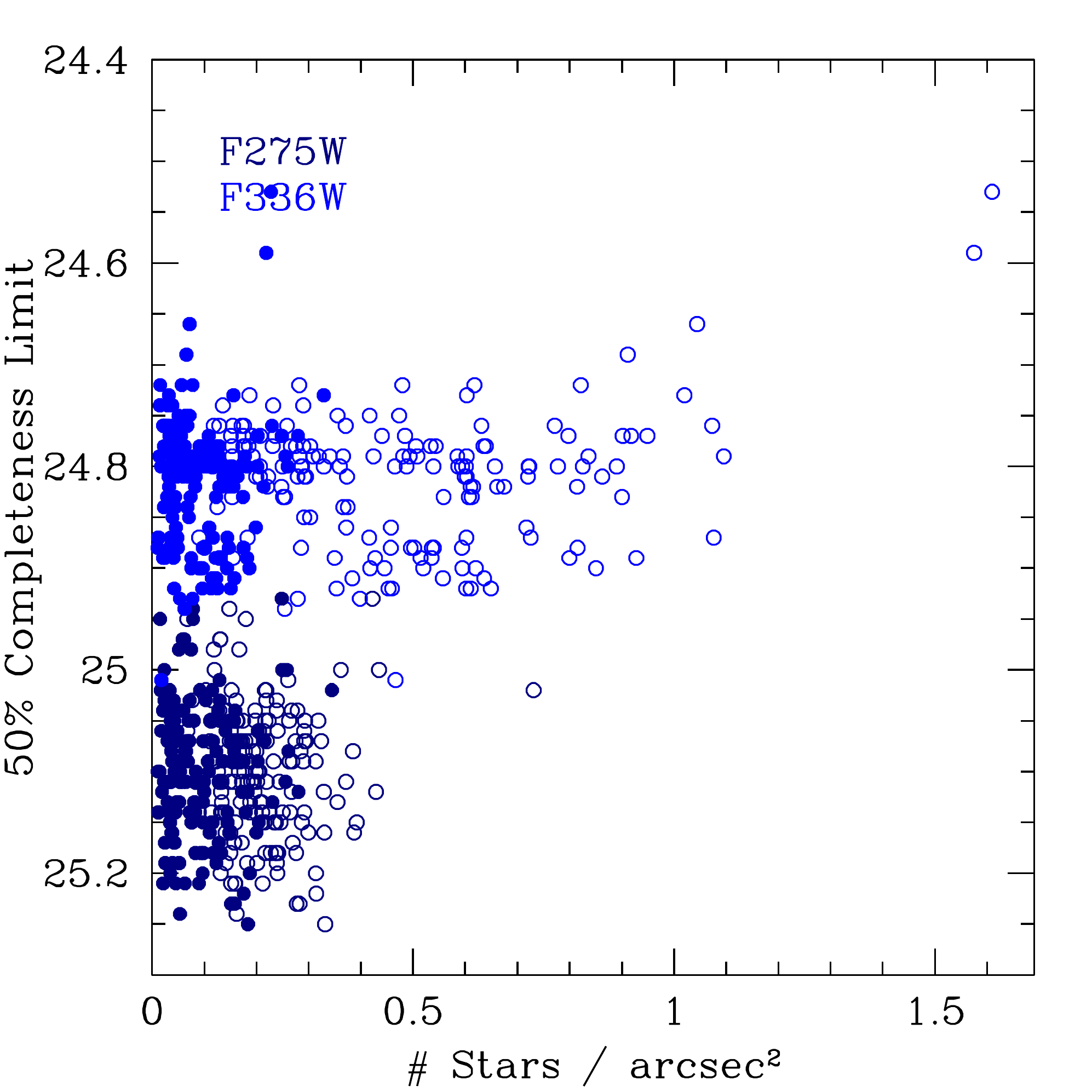}  
\includegraphics[width=2.15in]{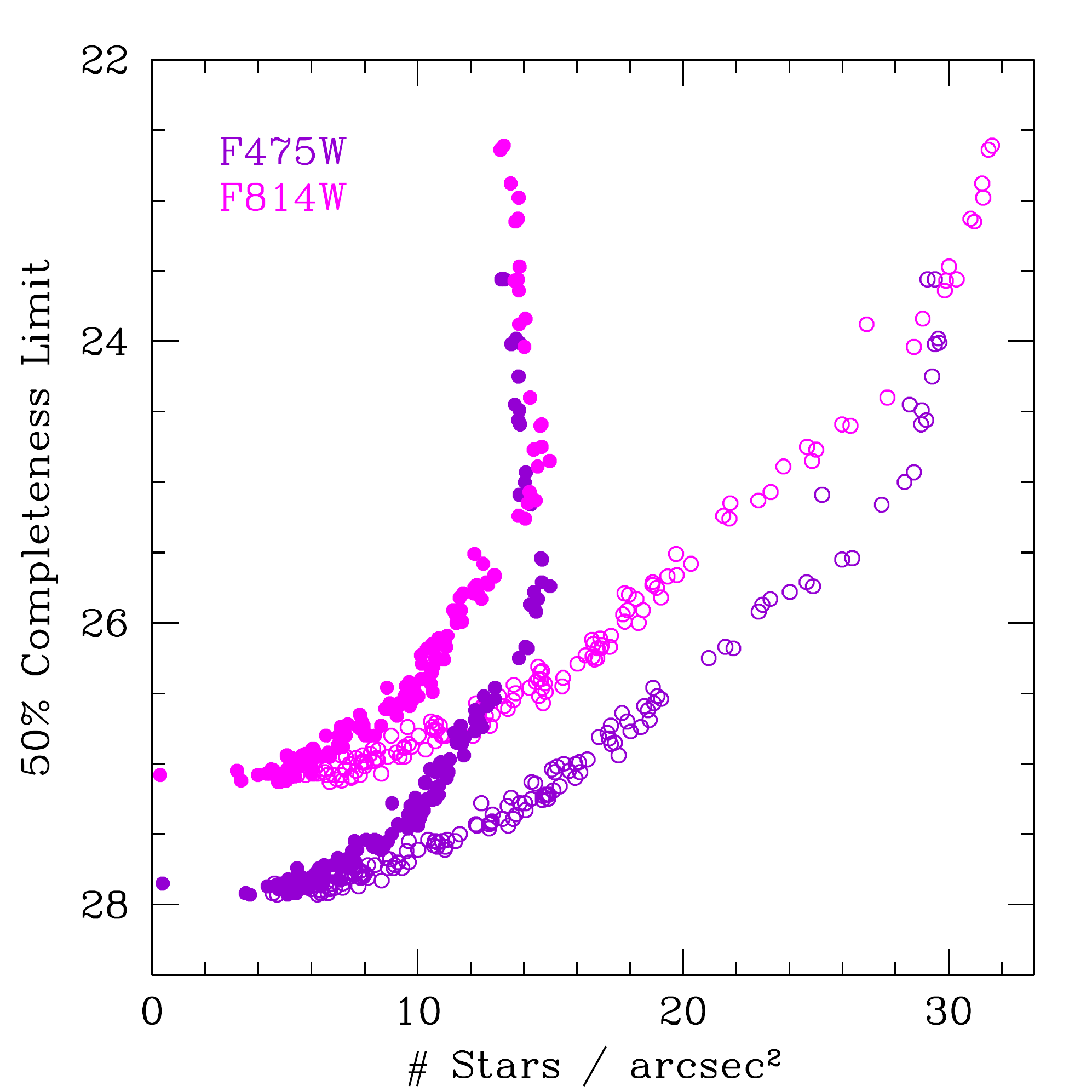}  
\includegraphics[width=2.15in]{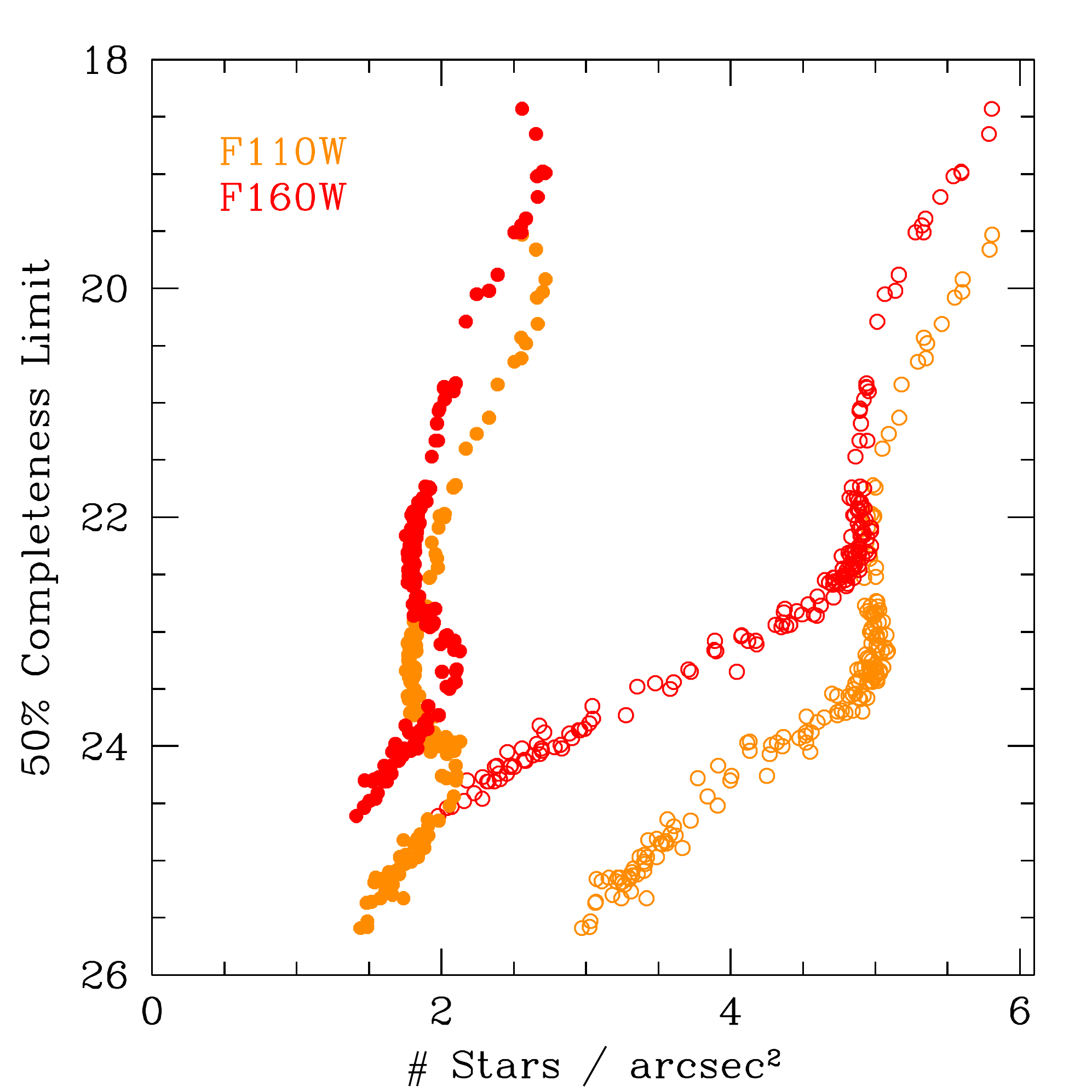}  
}
\caption{The magnitude of the 50\% completeness limit as a function of
  the stellar surface density for the {\tt{st}} catalog detections
  (S/N$>$4 in a single filter; open circles), and the high photometric
  quality {\tt{gst}} catalog detections (S/N$>$4 in both filters for a
  given camera, with additional stringent quality cuts; solid
  circles), for all individual fields observed with the WFC3/UVIS
  (left), ACS/WFC (center) and WFC3/IR (right) cameras by the end of
  the summer 2011 observing season (Table~\ref{brickobstab}).  The two
  filters in each camera are color-coded as indicated in the
  individual panels.  At low stellar surface densities, the limiting
  magnitude becomes constant, indicating that the depth is
  photon-limited and is independent of stellar density (seen in the
  UVIS and WFC plots). At high enough stellar surface densities,
  however, the number of detected (open circles) and/or well-measured
  (filled circles) stars becomes essentially constant, with the
  limiting magnitude being set by the point in the integrated
  luminosity function at which that maximum number of detectable stars
  is reached (seen in the IR and WFC plots); slight deviations from
  the constant stellar density are due to rapid changes in the stellar
  luminosity function at the TRGB and the red clump. The UV data (left
  panel) are clearly photon-limited (i.e., distributed horizontally on
  the plot), and have two different limiting magnitudes, because of
  the longer exposure times used in the winter observing season.  The
  optical data (central panel) are photon-limited at stellar surface
  densities less than
  $\lesssim\!6\,\textrm{stars}\,\textrm{arcsecond}^{-2}$, but are
  crowding-limited at most positions.  The NIR data (right panel)
  never reach the theoretical photon-counting limit, and are
  crowding-limited everywhere.  The central and right hand plots also
  show how the fraction of stars with well-measured photometry
  increases as the stellar density drops and the images become less
  crowded.  \label{maglimdensityfig}}
\end{figure}
\vfill
\clearpage

\begin{figure}
\centerline{
\includegraphics[width=3.25in]{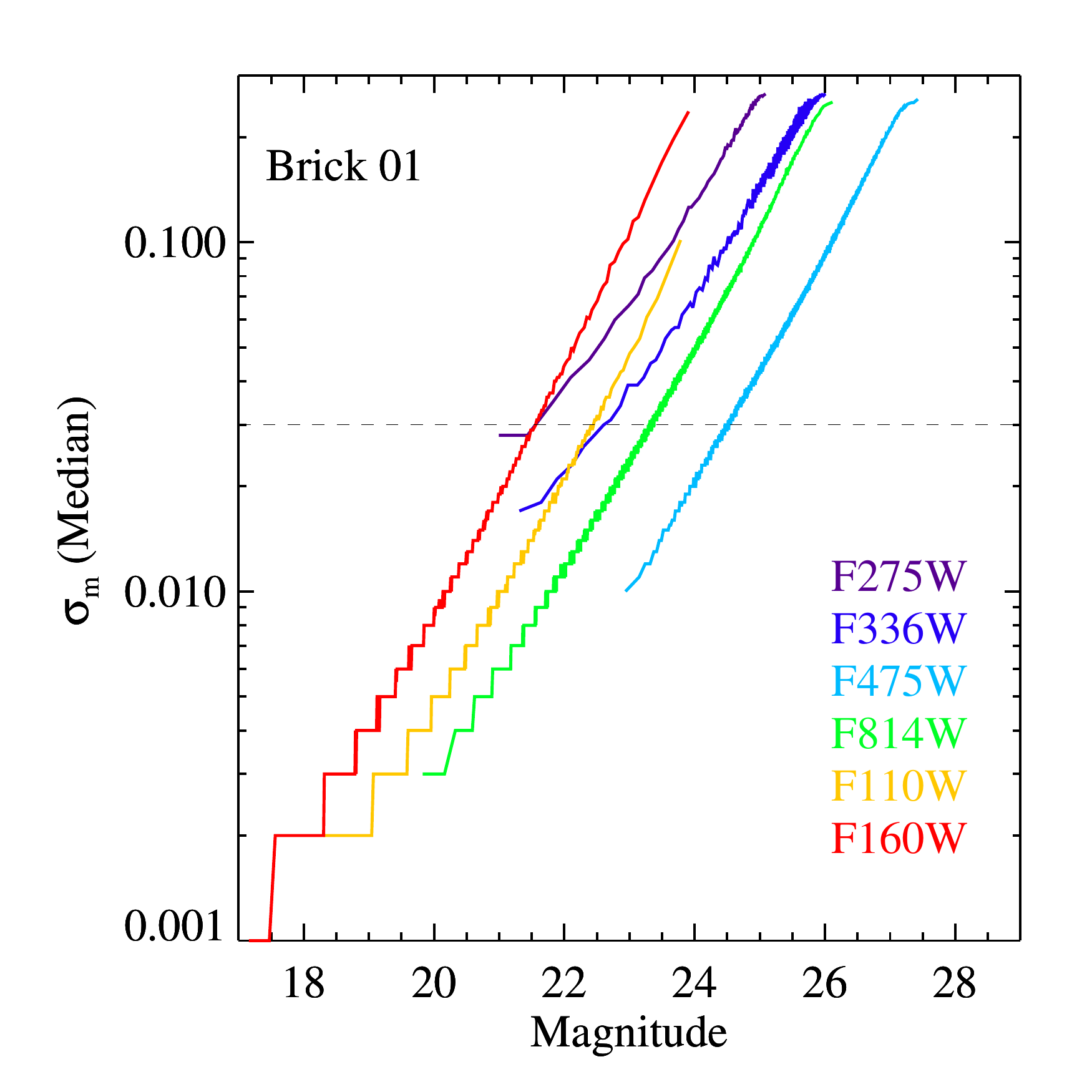}  
\includegraphics[width=3.25in]{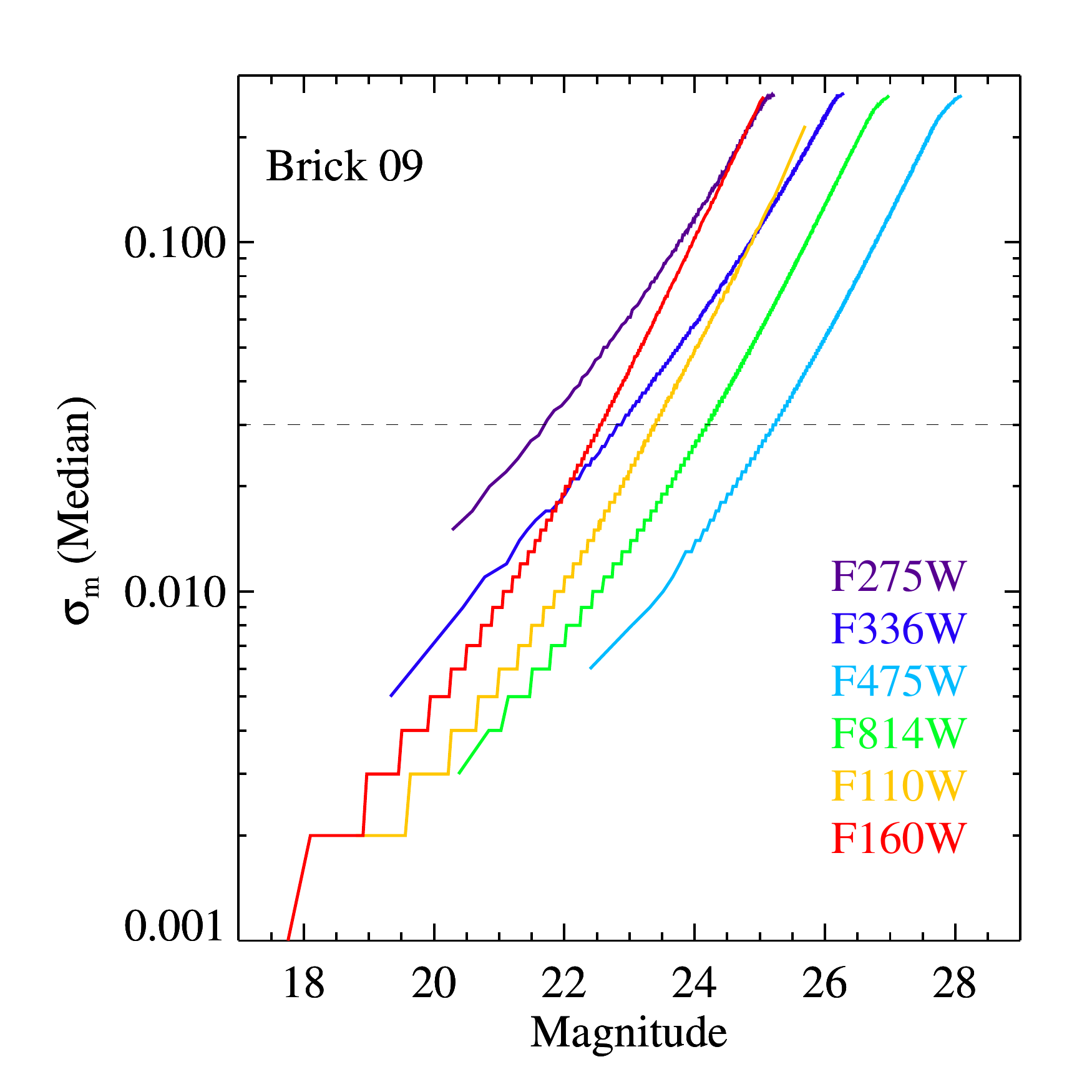}  
}
\centerline{
\includegraphics[width=3.25in]{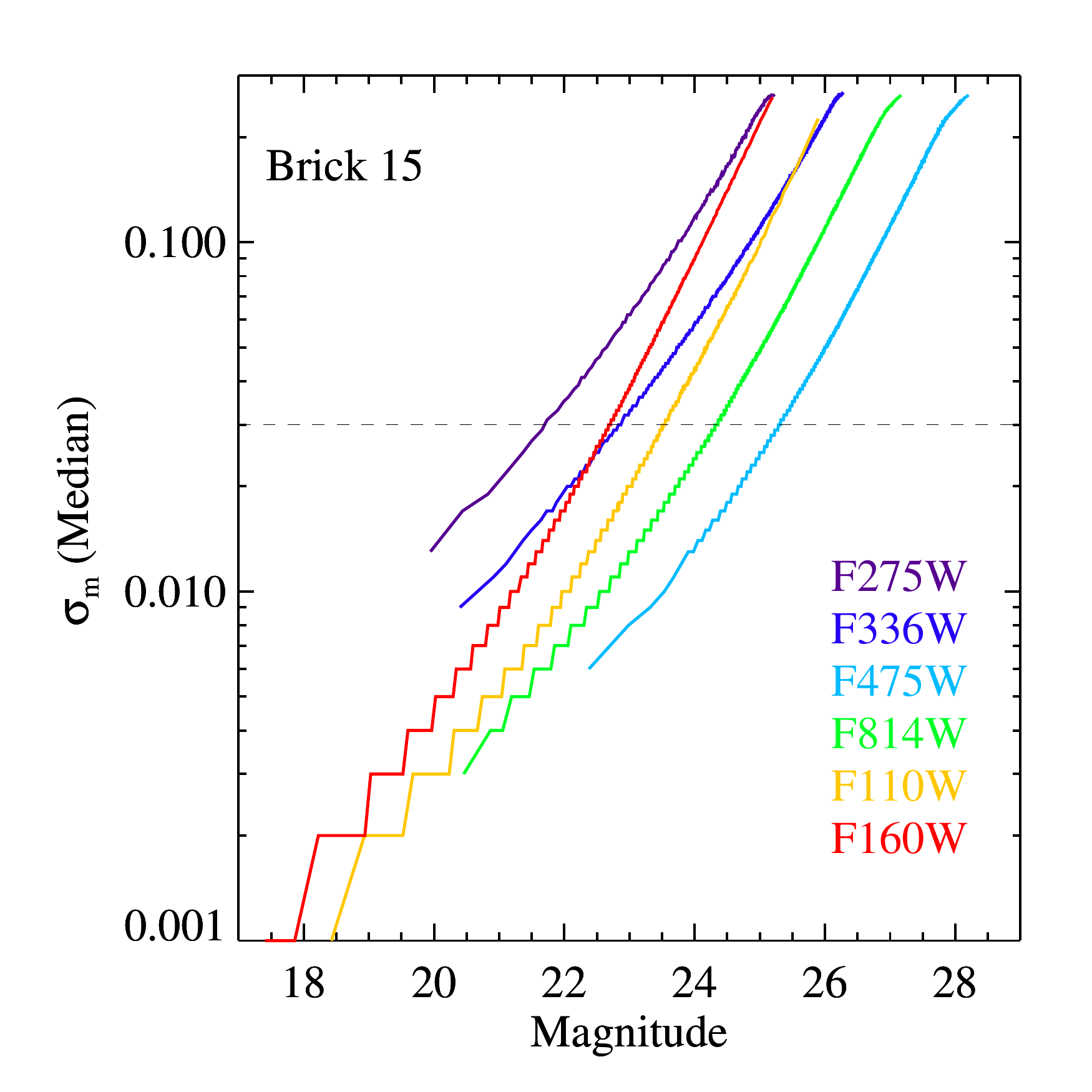}  
\includegraphics[width=3.25in]{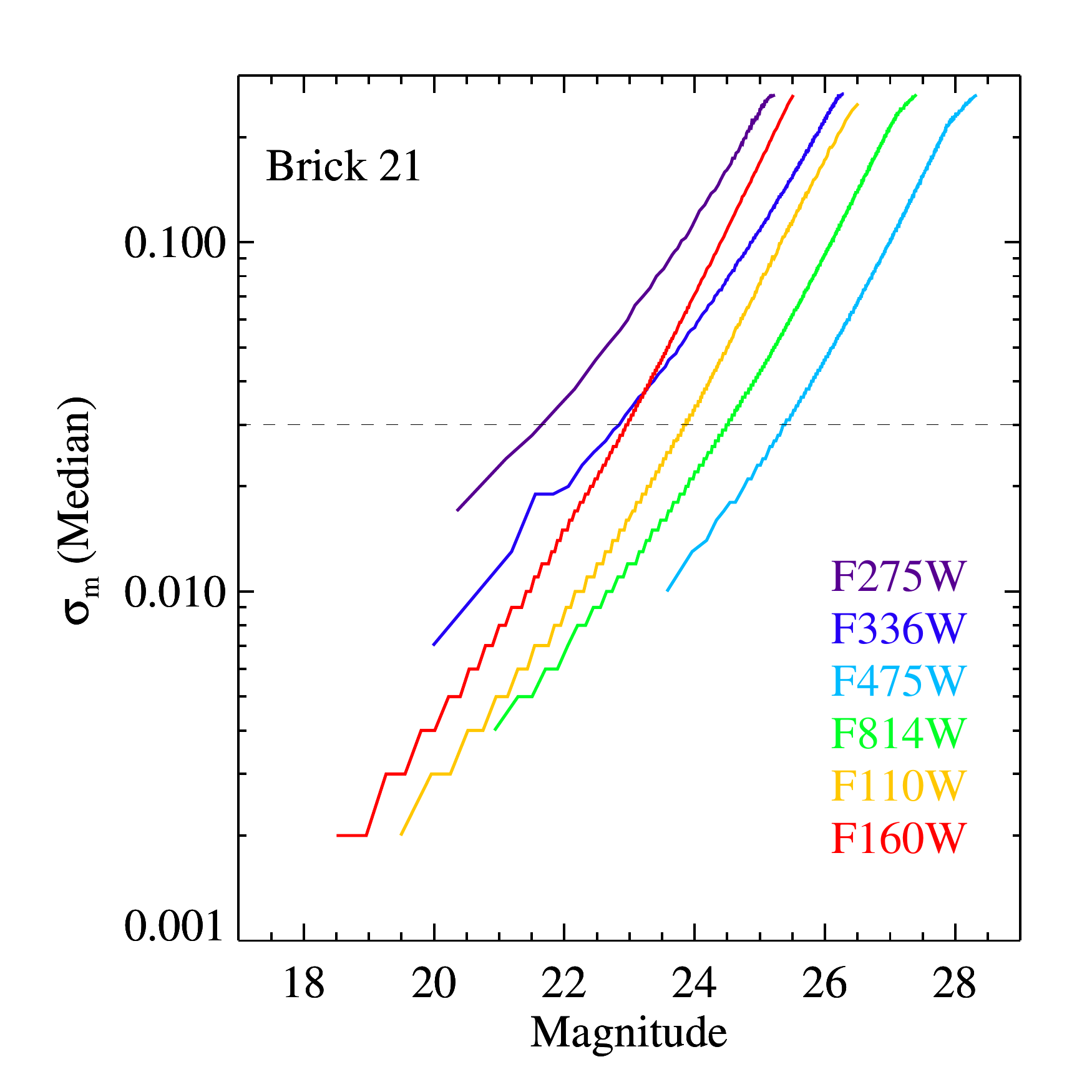}  
}
\caption{Median photometric errors reported by DOLPHOT in the six PHAT
  filters, for Bricks 1, 9, 15, and 21 (upper left to lower right).
  Medians are calculated for groups of 25, 400, or 200 stars, for the
  UV, optical, and NIR channels, respectively.  Photometry is for stars
  with S/N$>$4 in a given filter, restricted to stars with
  well-measured photometry ({\tt{FLAG=0}}) that do not fall near the
  chip gap or the edges of the image. Note that the formal photometric
  uncertainties underestimate the true uncertainty, which is
  frequently dominated by crowding errors at the faint end
  (Figures~\ref{fakestarsuvfig}-\ref{fakestarsirfig}), and by
  systematic errors in the bright end; the horizontal dashed line
  indicates a characteristic amplitude of systematic uncertainties
  (Figure~\ref{systematicfig}).  The steps at bright magnitudes are
  due to the limited number of significant figures used for the
  DOLPHOT photometric uncertainties.
  \label{photerrfig}}
\end{figure}
\vfill
\clearpage

\begin{figure}
\centerline{
\includegraphics[width=5.75in]{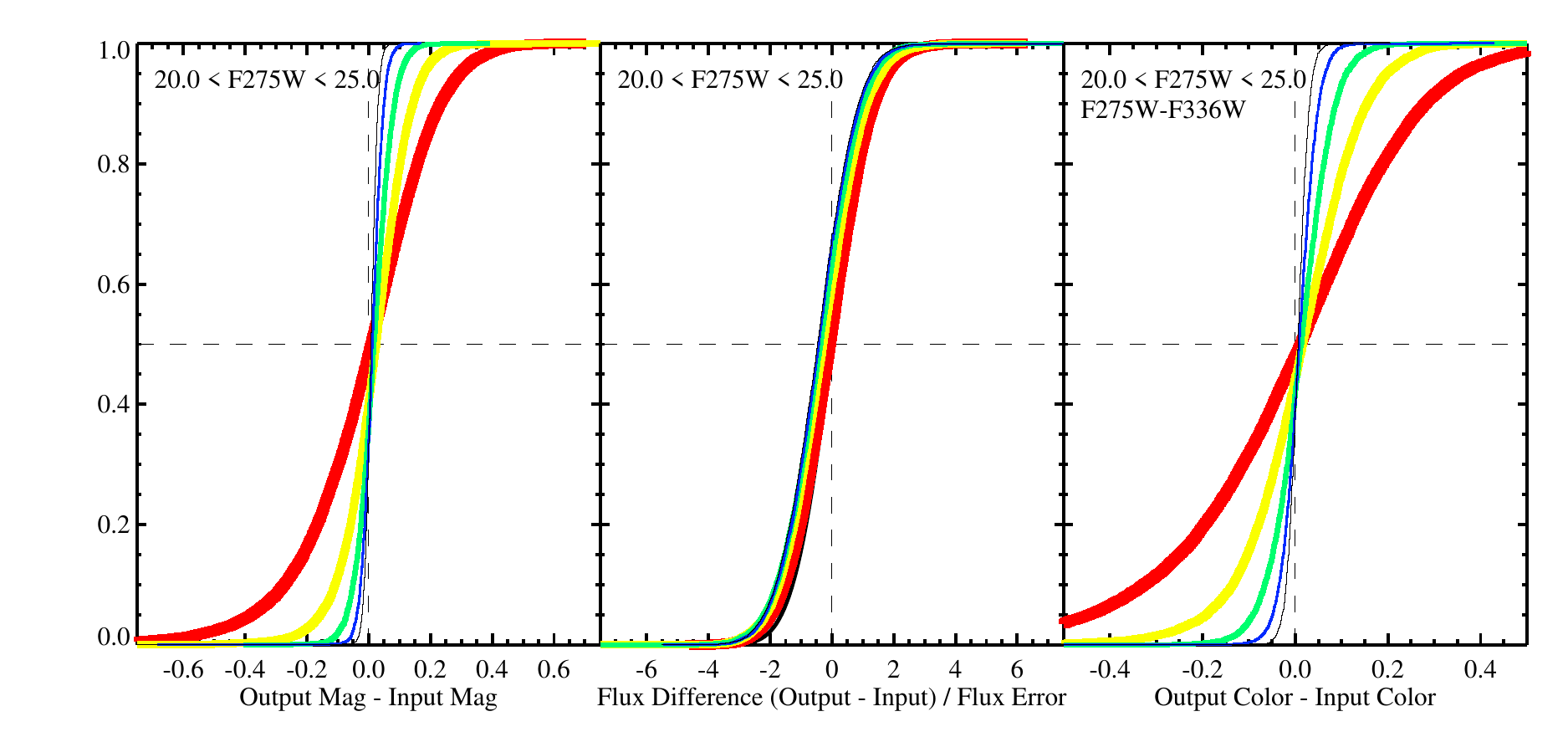}  
}
\centerline{
\includegraphics[width=5.75in]{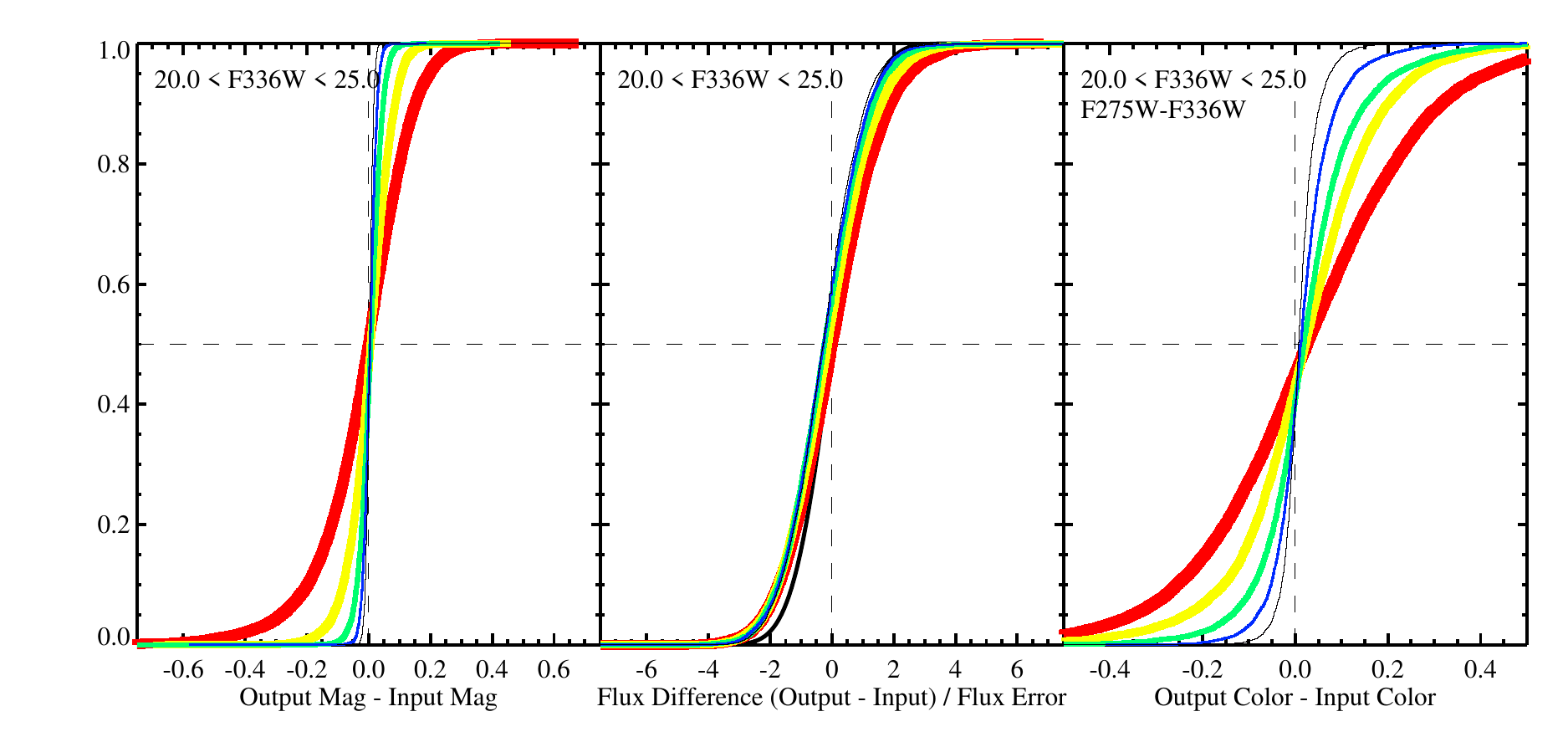}  
}
\caption{Comparisons between the input and output properties of
  artificial stars inserted into WFC3/UVIS images in Brick 1, Field 9,
  for \fw{275} (top row) and \fw{336} (bottom row).  Panels show the
  cumulative distribution of: (left) the difference between the
  recovered and inserted magnitude, such that negative numbers
  indicates stars that are recovered brighter than their true
  magnitude; (center) the fractional difference in flux, scaled by the
  photometric uncertainty reported by DOLPHOT such that positive
  numbers indicate stars that are recovered brighter than their true
  magnitude; and (right) the difference between the recovered and the
  input color of the artificial stars, such that positive values
  indicate stars that are recovered redder than their true color.
  Each curve shows the distribution for artificial stars with
  recovered magnitudes in a 1 magnitude wide range, spanning the total
  magnitude interval specified on each panel, color coded such that
  fainter stars are plotted with redder colors and thicker lines
  (i.e., the top row analyzes bins starting with stars having
  $20<m_{\fw{275},out}<21$ (thin purple line), down to
  $24<m_{\fw{275},out}<25$ (thick red line)).  The center panel also
  includes a black line showing the expectation for a Gaussian with a
  width set equal to the median photometric error of stars in the bin.
  The UV data follow the Gaussian expectation well, because of the
  lack of crowding in the UV data everywhere outside the very center
  of the bulge.
  \label{fakestarsuvfig}}
\end{figure}
\vfill
\clearpage

\begin{figure}
\centerline{
\includegraphics[width=3in]{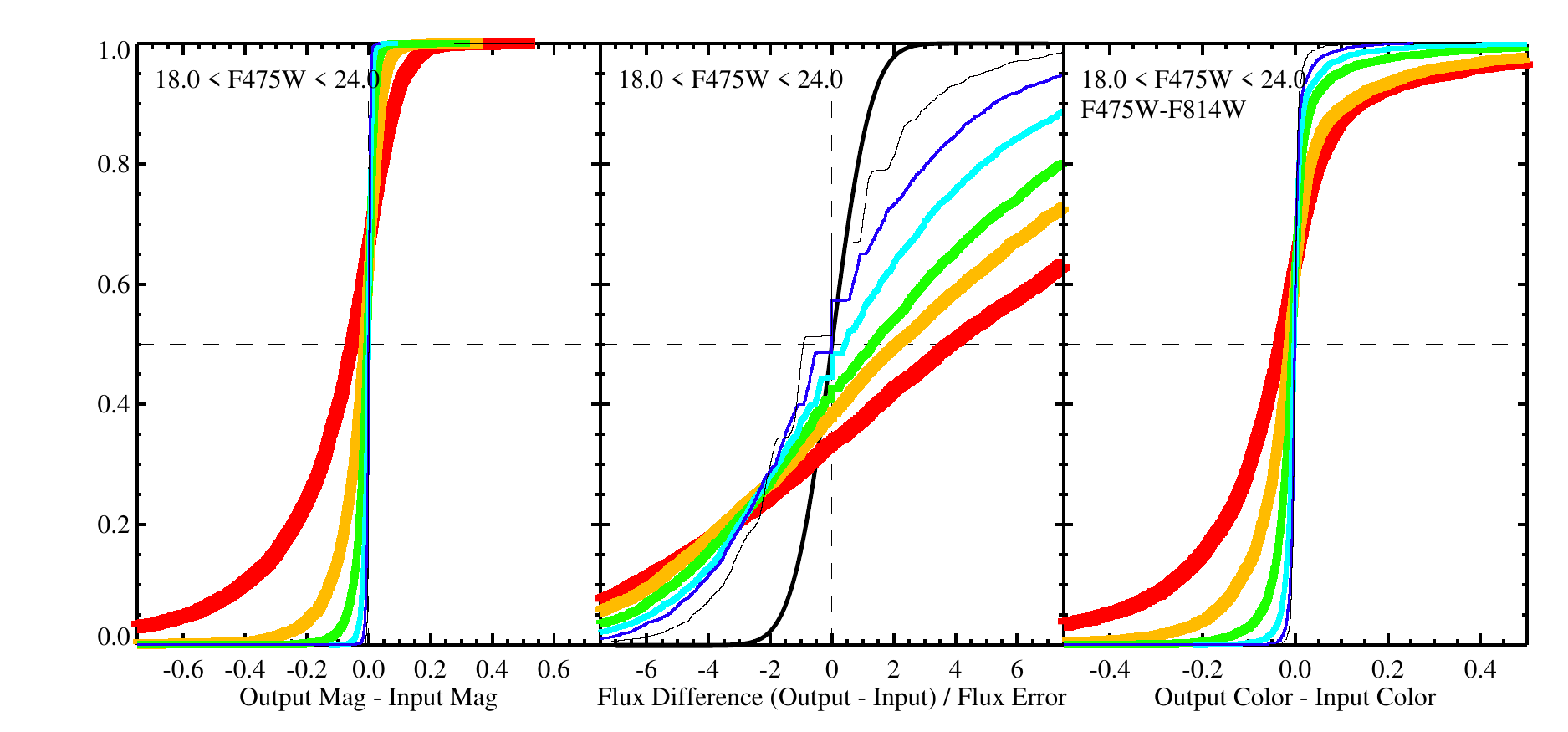}  
\includegraphics[width=3in]{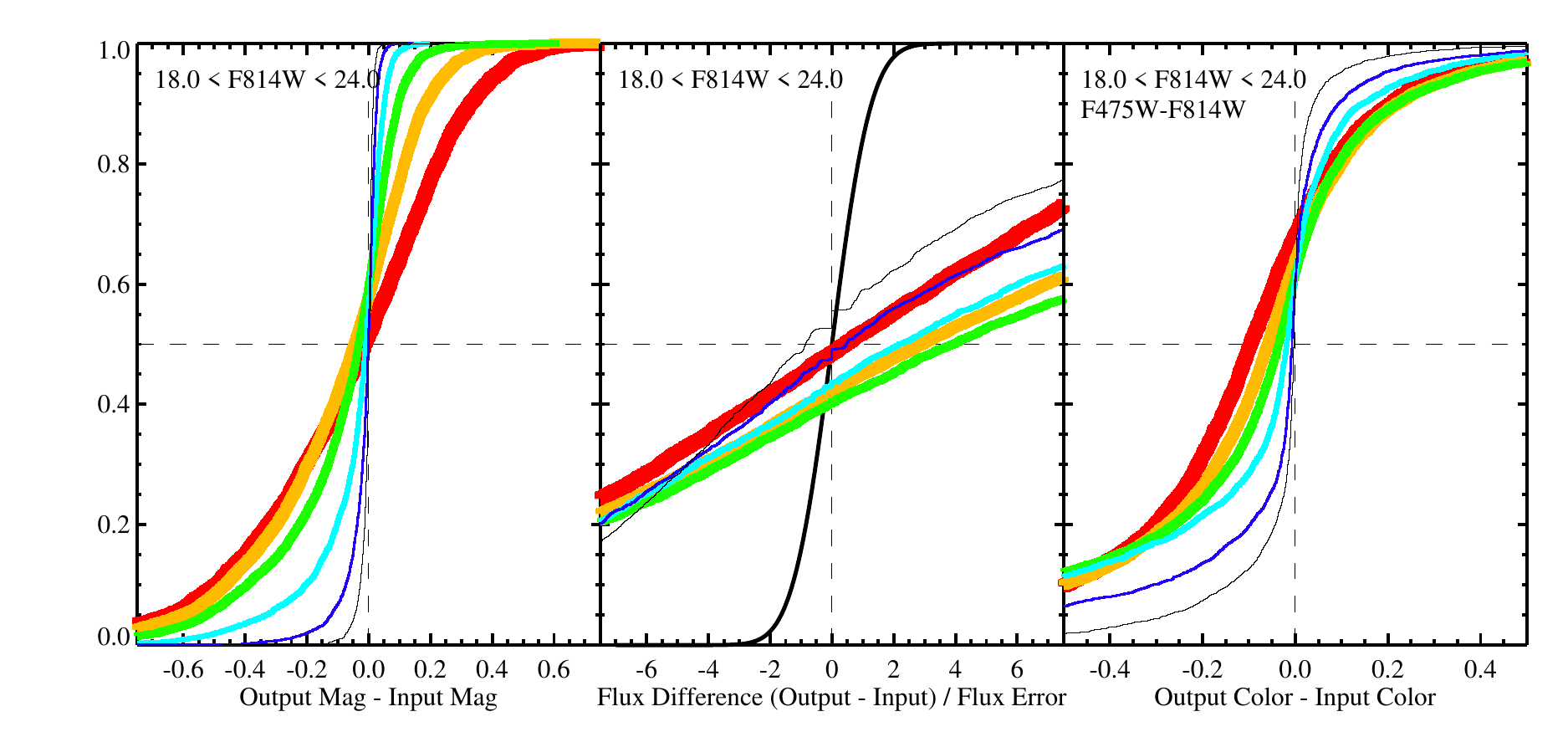}  
}
\centerline{
\includegraphics[width=3in]{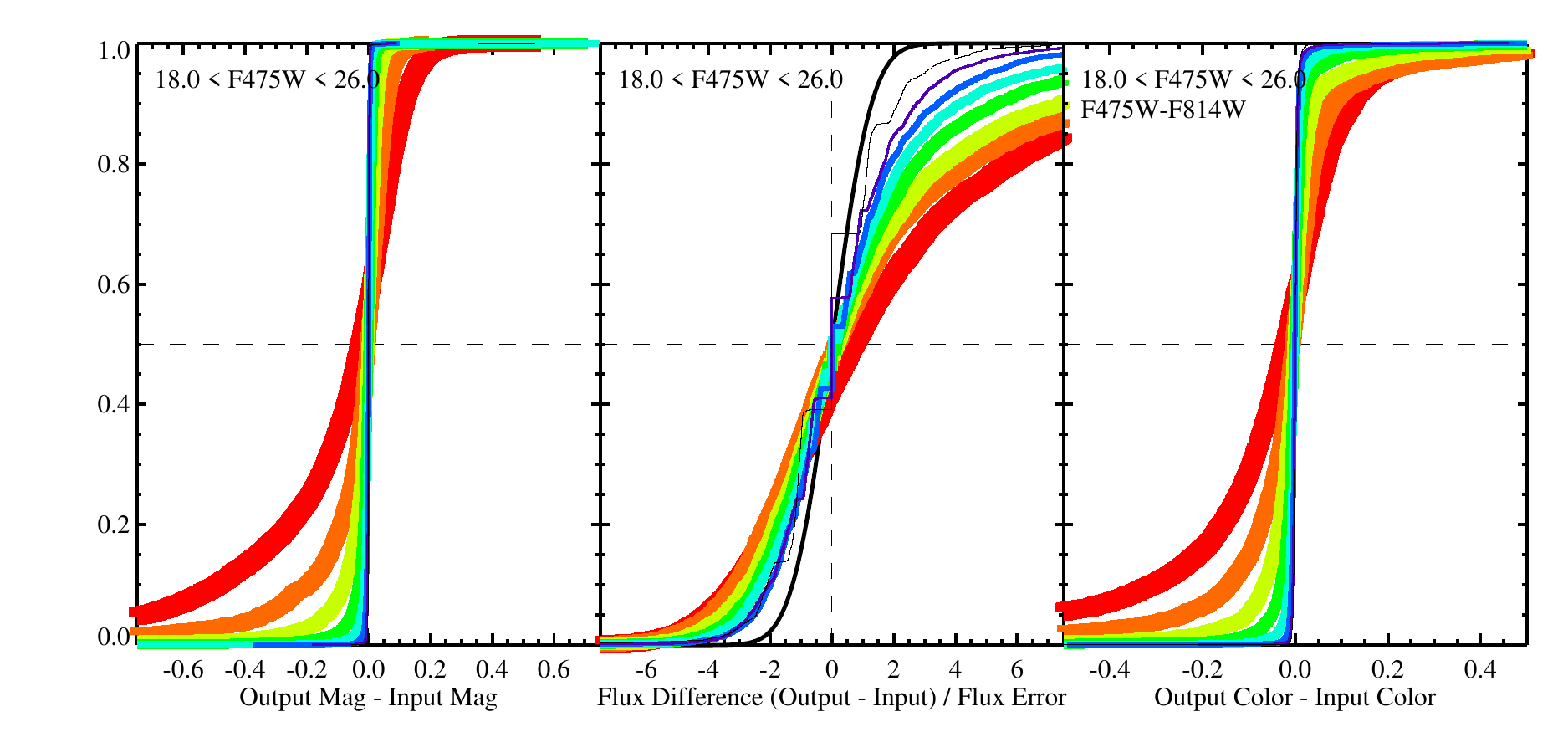}  
\includegraphics[width=3in]{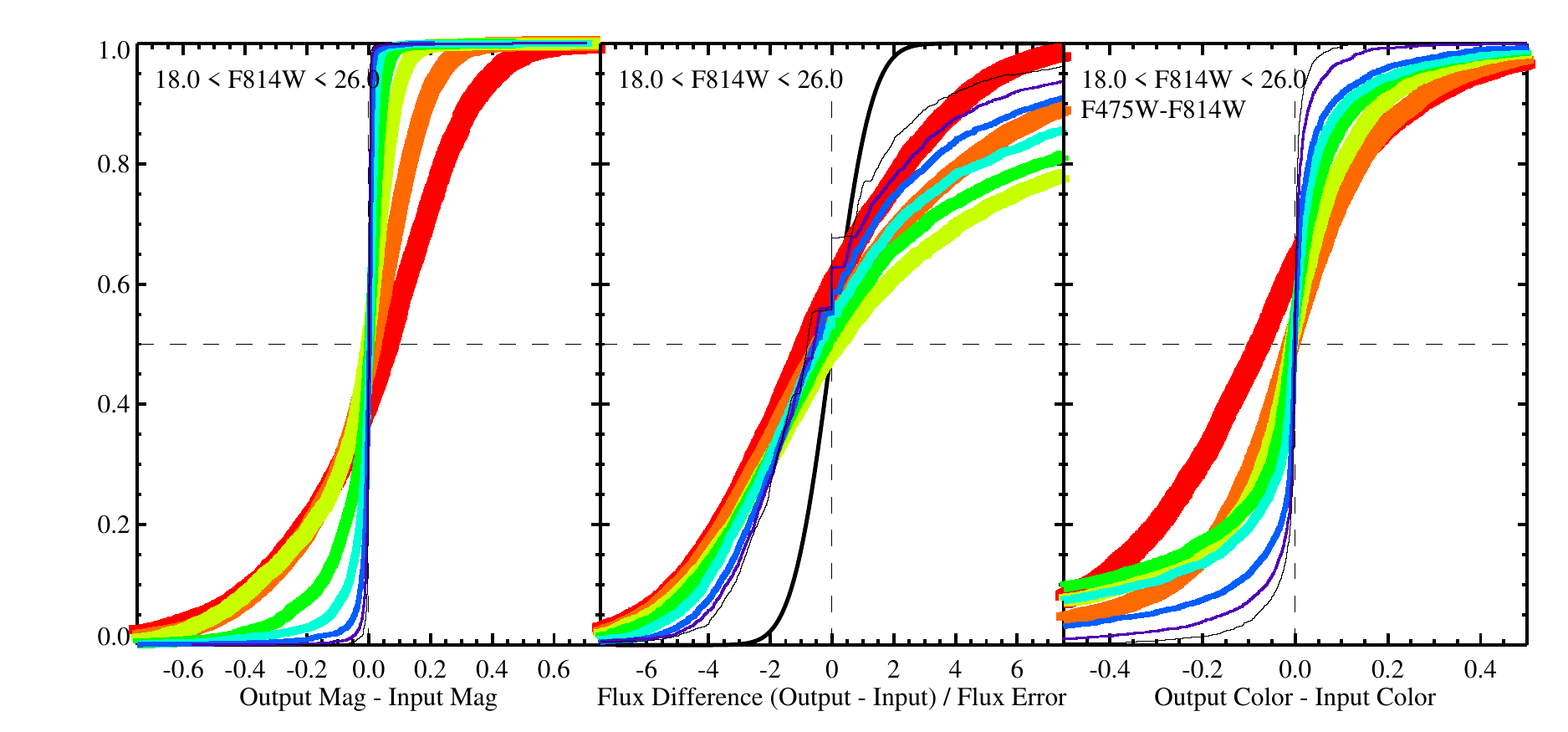}  
}
\centerline{
\includegraphics[width=3in]{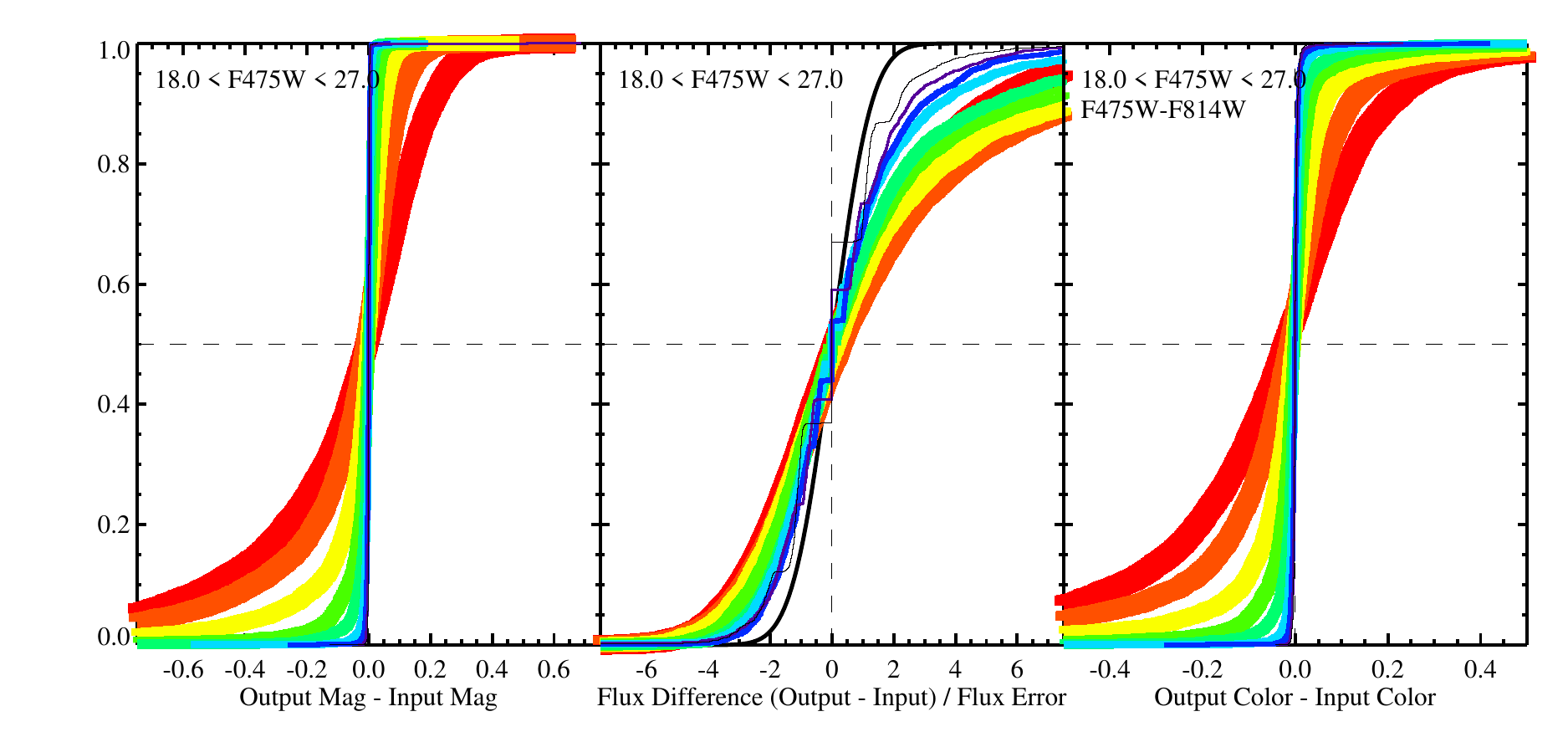}  
\includegraphics[width=3in]{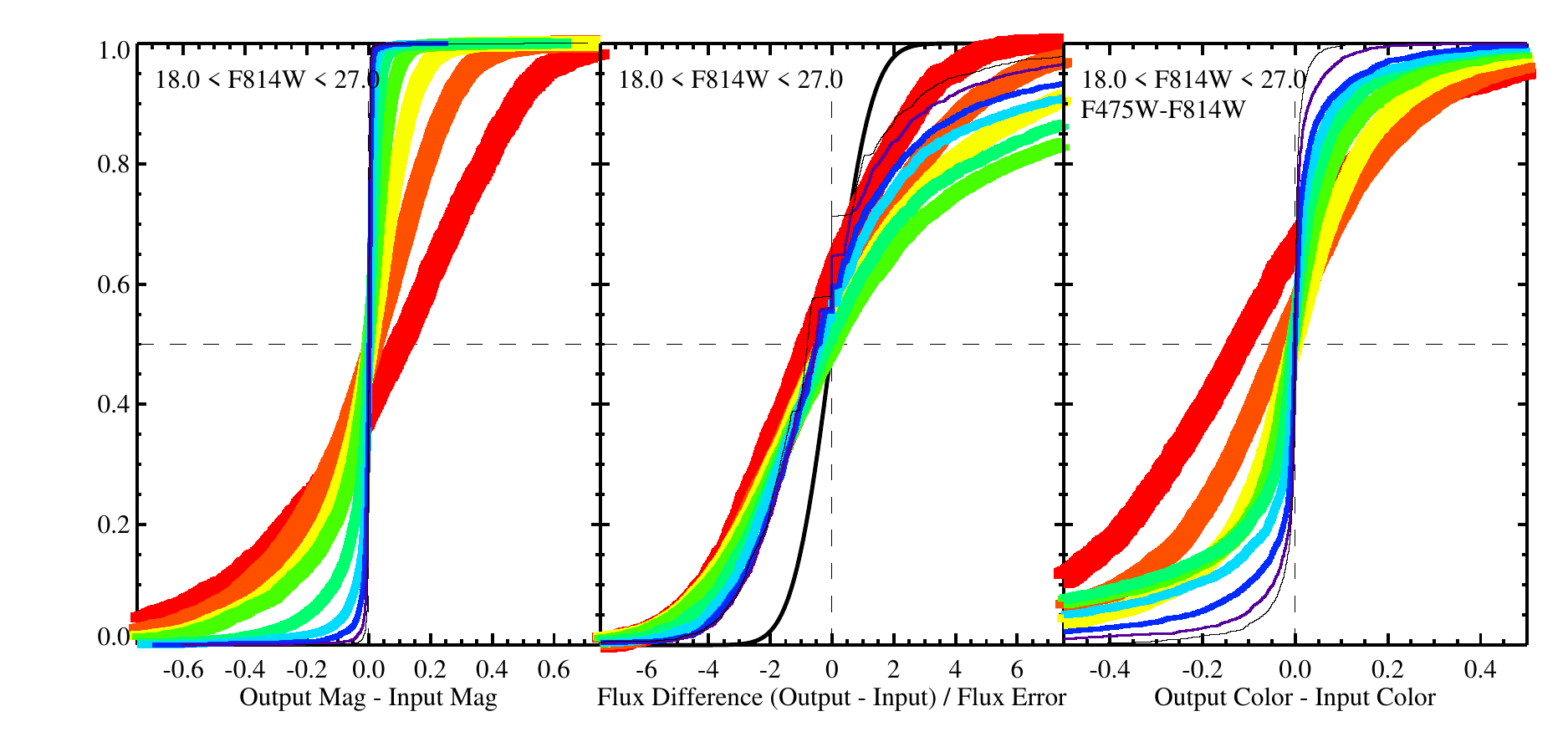}  
}
\centerline{
\includegraphics[width=3in]{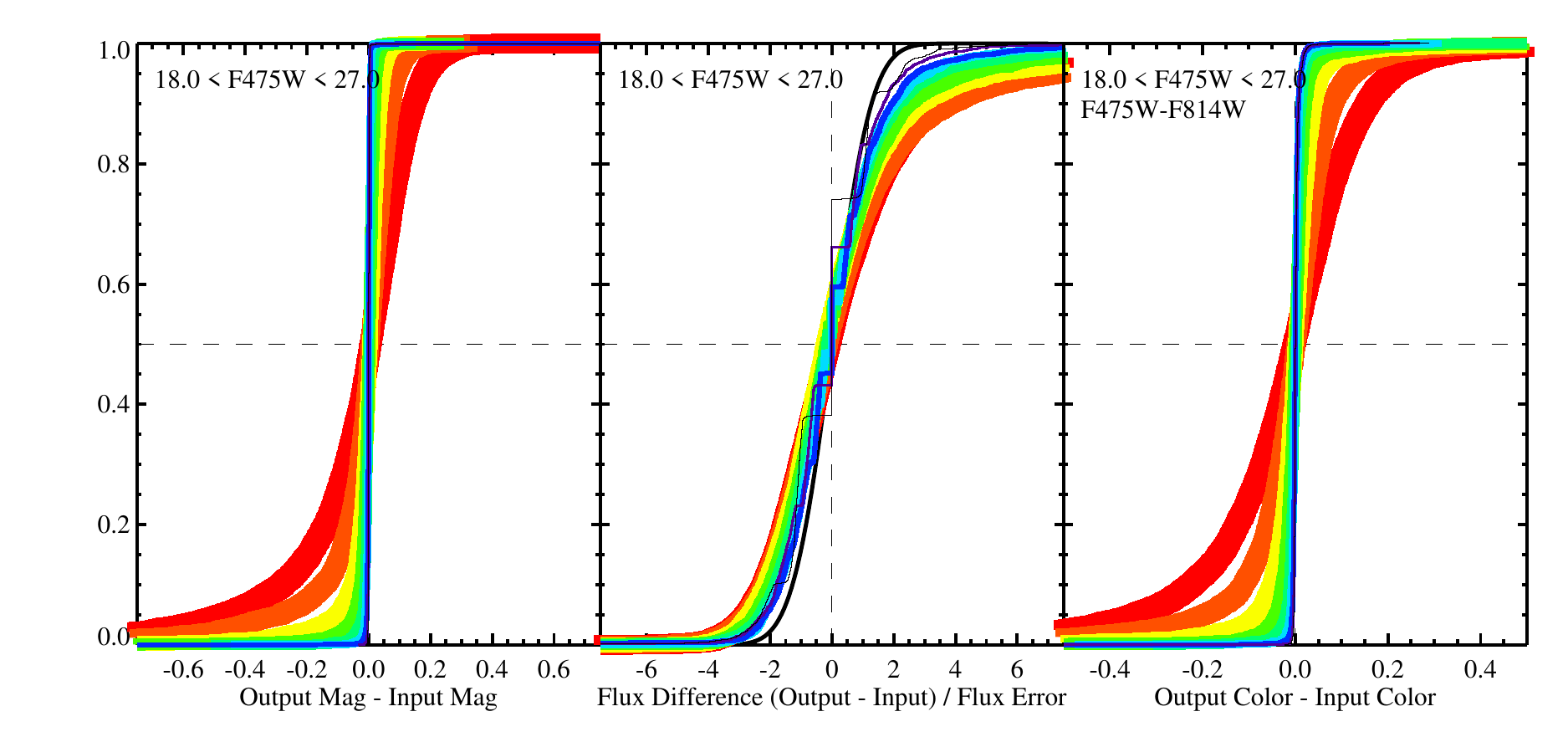}  
\includegraphics[width=3in]{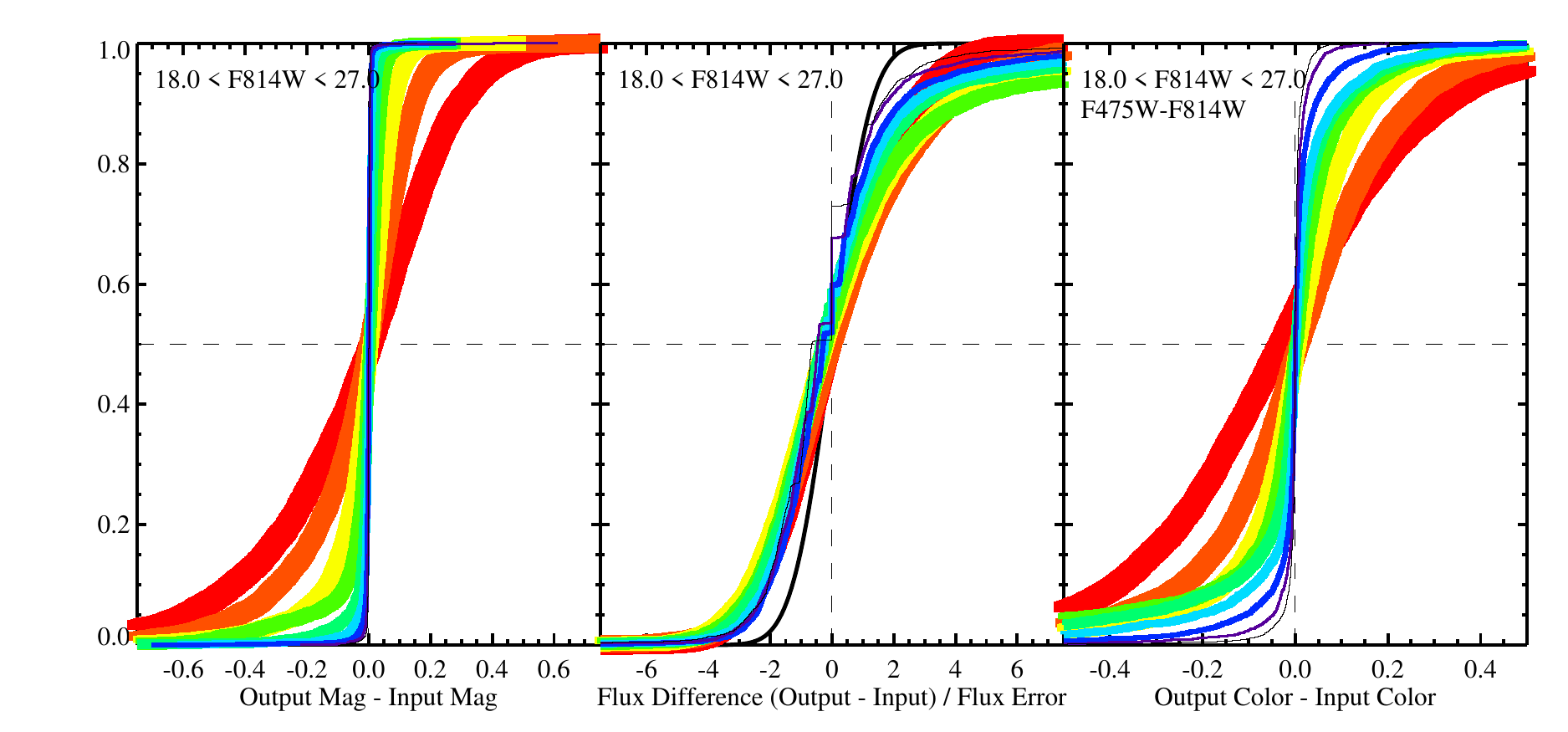}  
}
\caption{Comparisons between the input and output properties of
  artificial stars inserted into ACS/WFC images in Field 9 of Bricks
  1, 9, 15, and 23 (top to bottom rows, respectively), for \fw{475}
  (left column) and \fw{814} (right column).  Panels show the
  cumulative distribution of: (left) the difference between the
  recovered and inserted magnitude, such that negative numbers
  indicates stars that are recovered brighter than their true
  magnitude; (center) the fractional difference in flux, scaled by the
  photometric uncertainty reported by DOLPHOT such that positive
  numbers indicate stars that are recovered brighter than their true
  magnitude; and (right) the difference between the recovered and the
  input color of the artificial stars, such that positive values
  indicate stars that are recovered redder than their true color.
  Line types indicate different bins in recovered magnitude, as
  described in Figure~\ref{fakestarsuvfig}.  Unlike the UV data, the
  optical data depart dramatically from a Gaussian due to the effects
  of crowding, in all but the outermost regions (see middle panel of
  each 3-panel plot).  There is a tail of stars recovered with a flux that
  is significantly larger than their true flux, due to contributions
  from nearby stars.  This bias can be several times larger than the
  DOLPHOT photometric error for faint stars in highly crowded regions.
  \label{fakestarsacsfig}}
\end{figure}
\vfill
\clearpage

\begin{figure}
\centerline{
\includegraphics[width=3in]{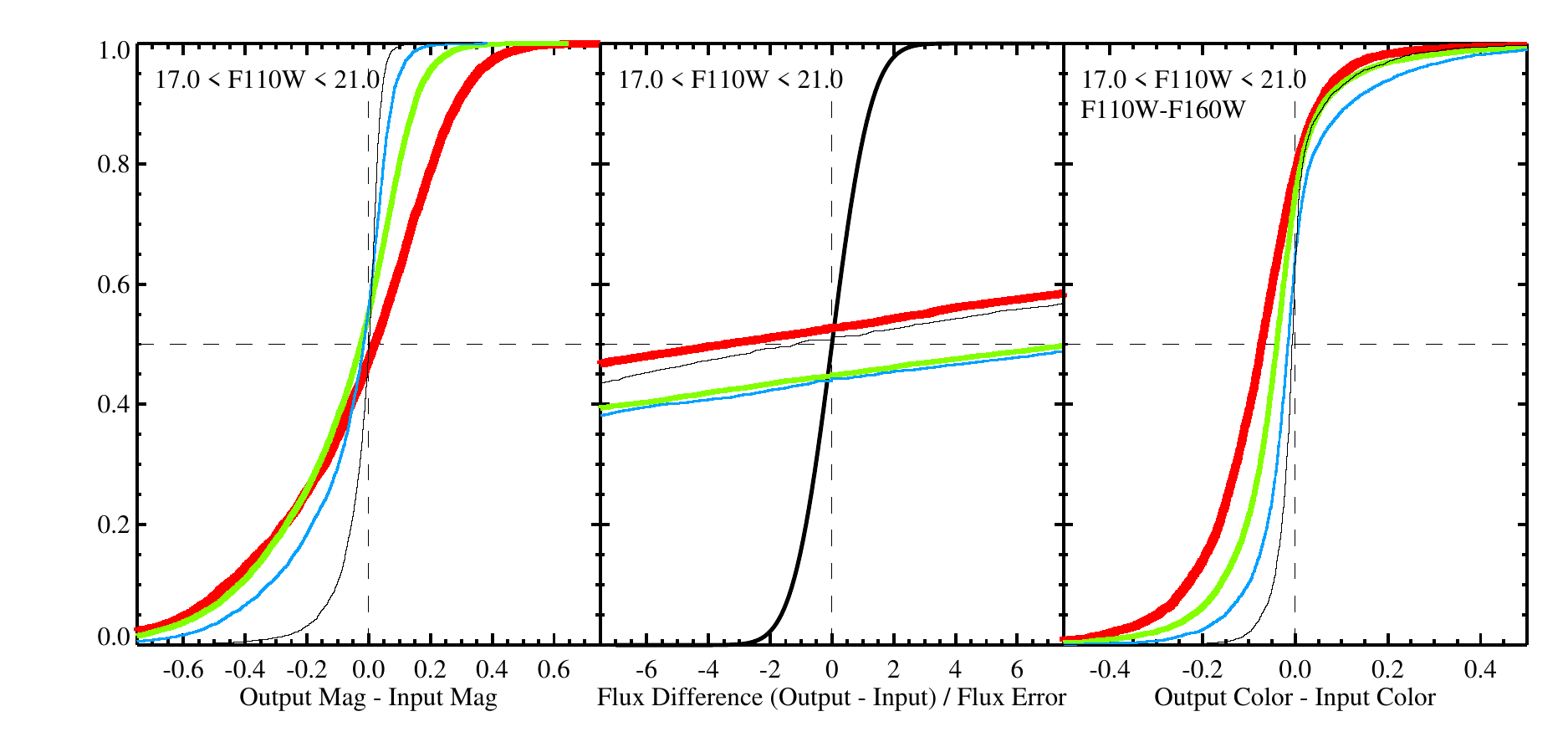}  
\includegraphics[width=3in]{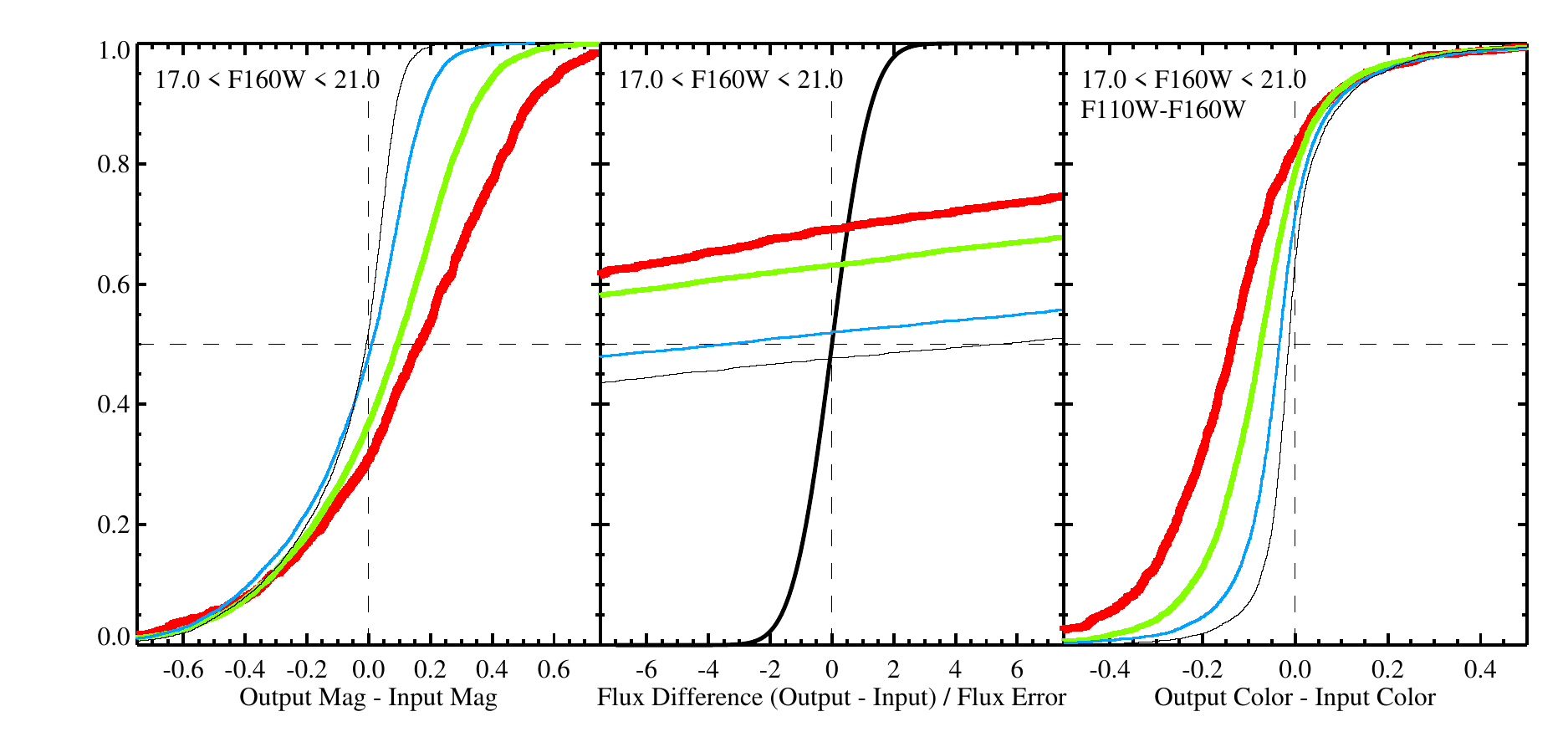}  
}
\centerline{
\includegraphics[width=3in]{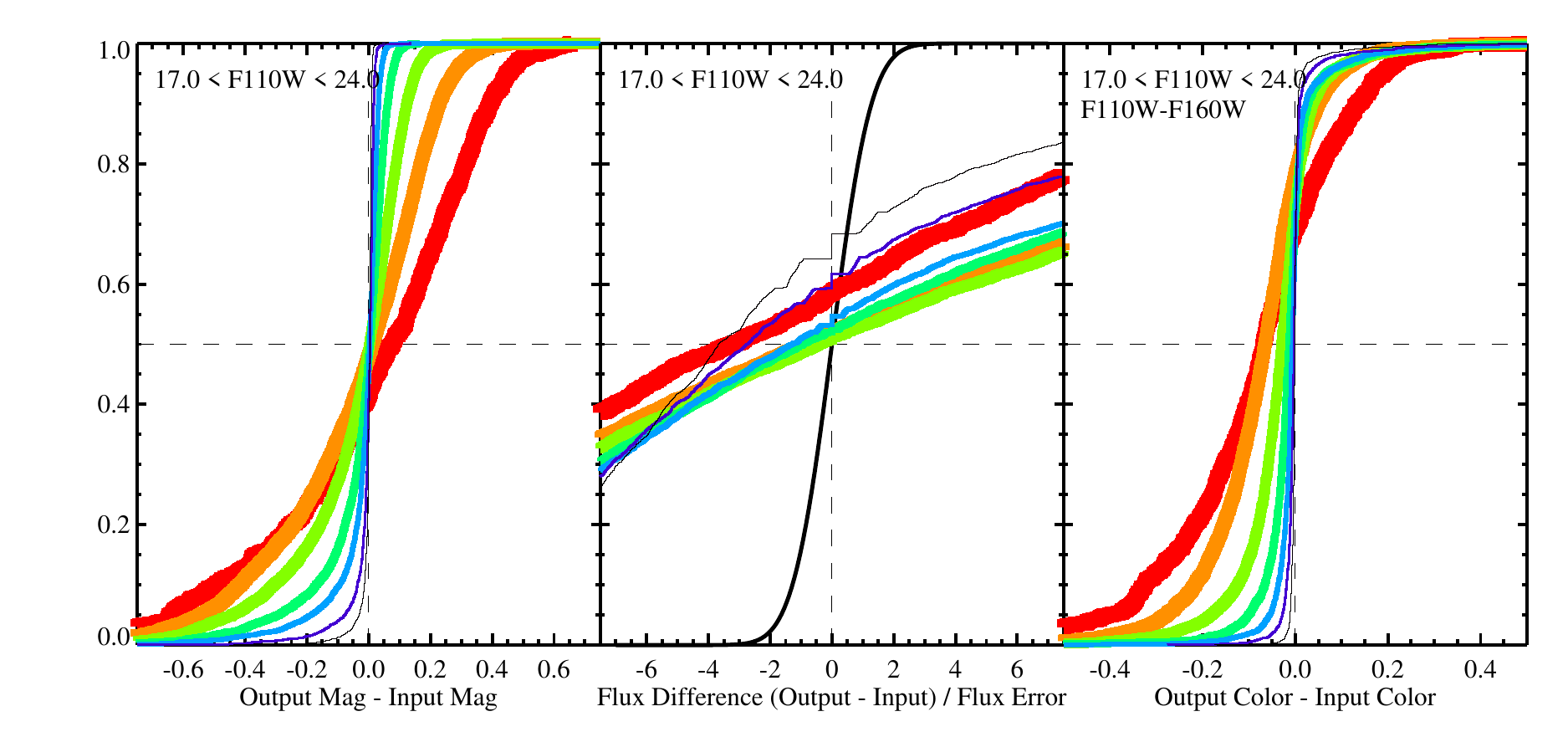}  
\includegraphics[width=3in]{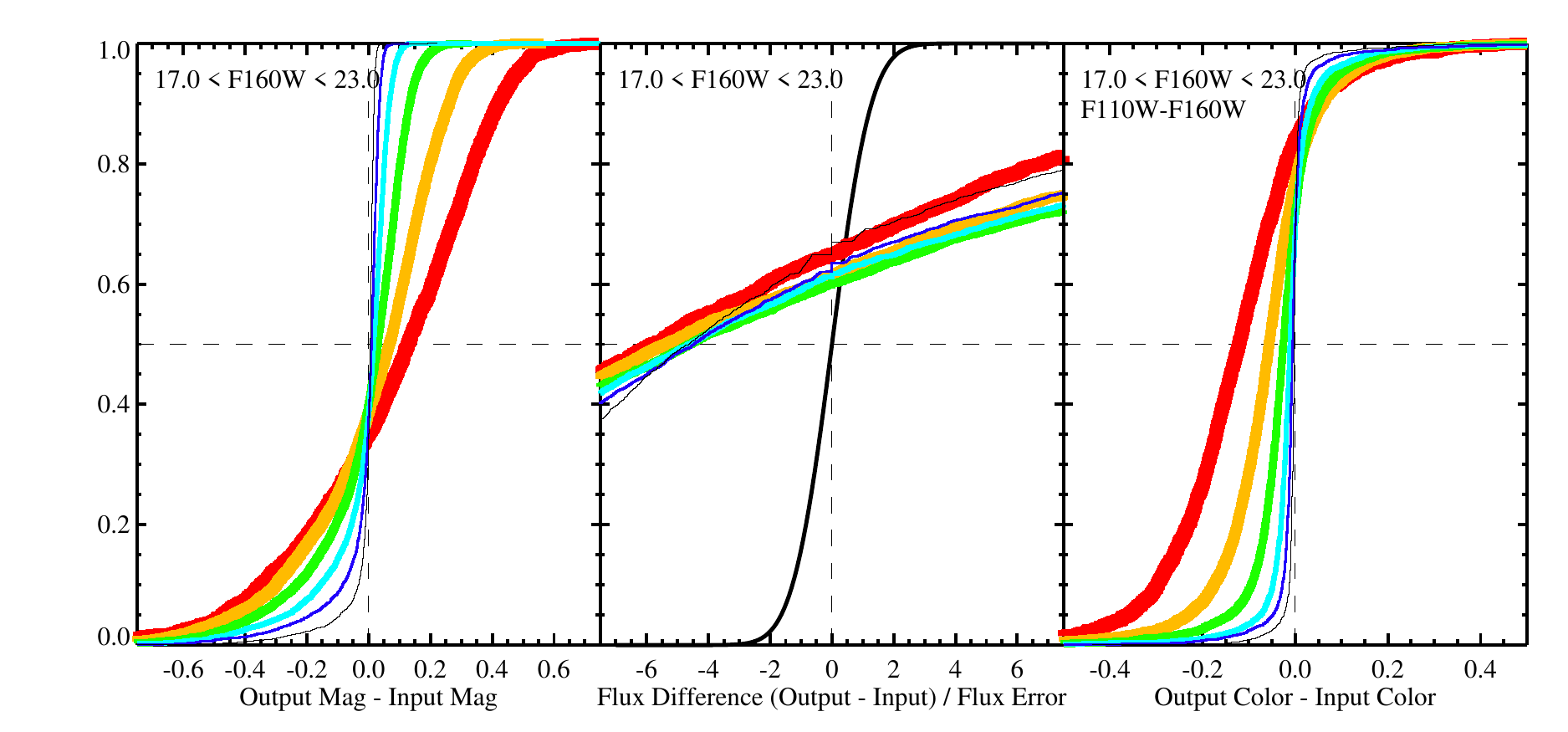}  
}
\centerline{
\includegraphics[width=3in]{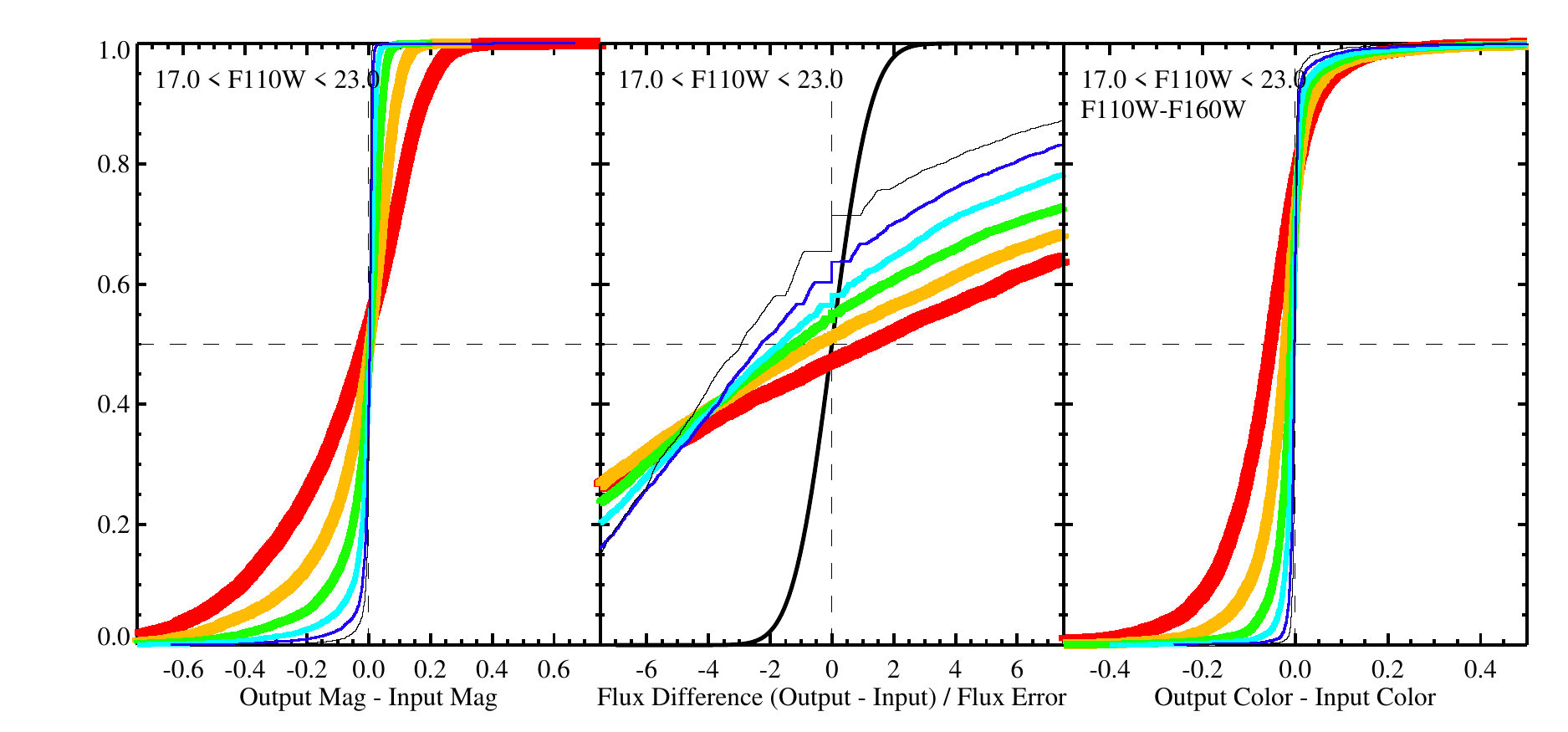}  
\includegraphics[width=3in]{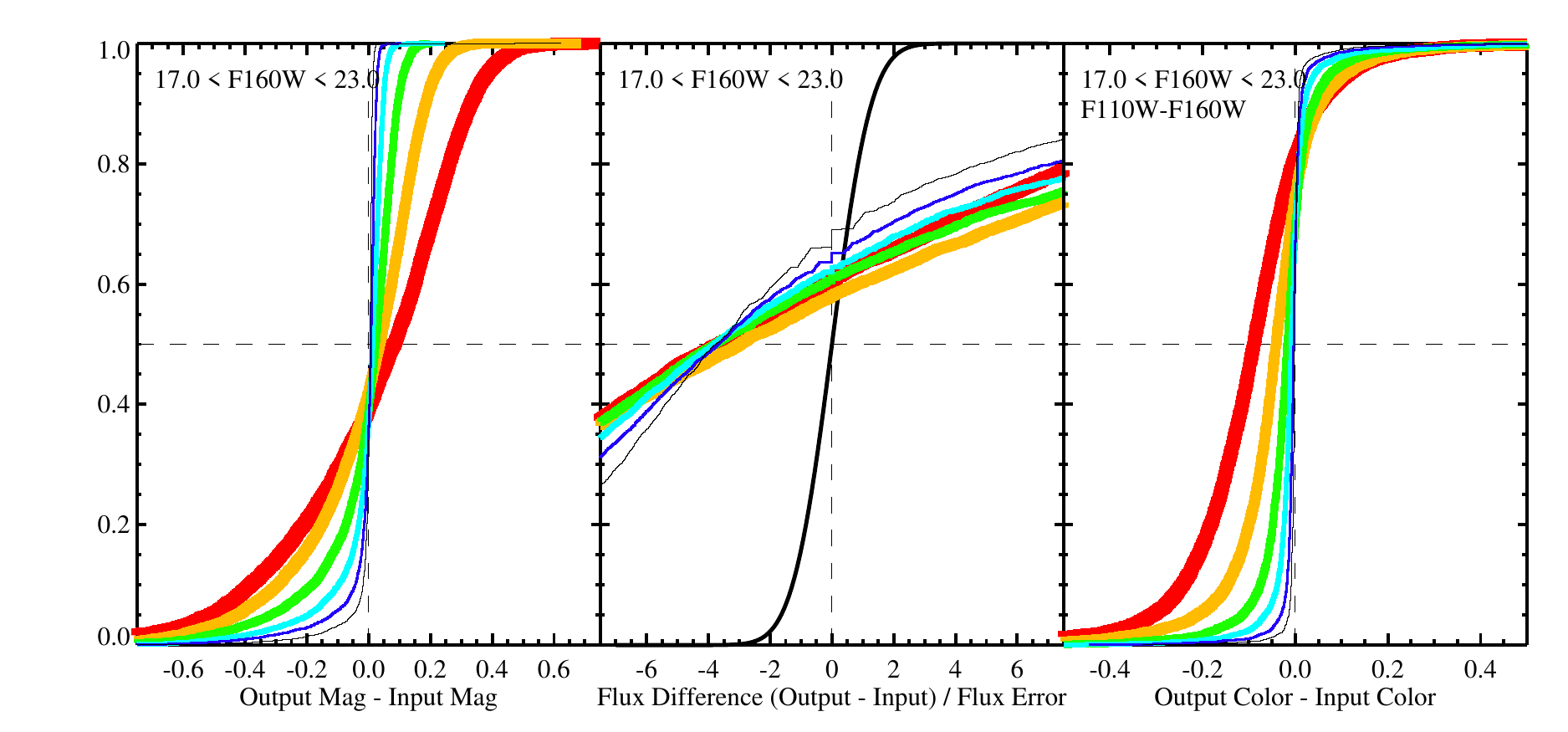}  
}
\centerline{
\includegraphics[width=3in]{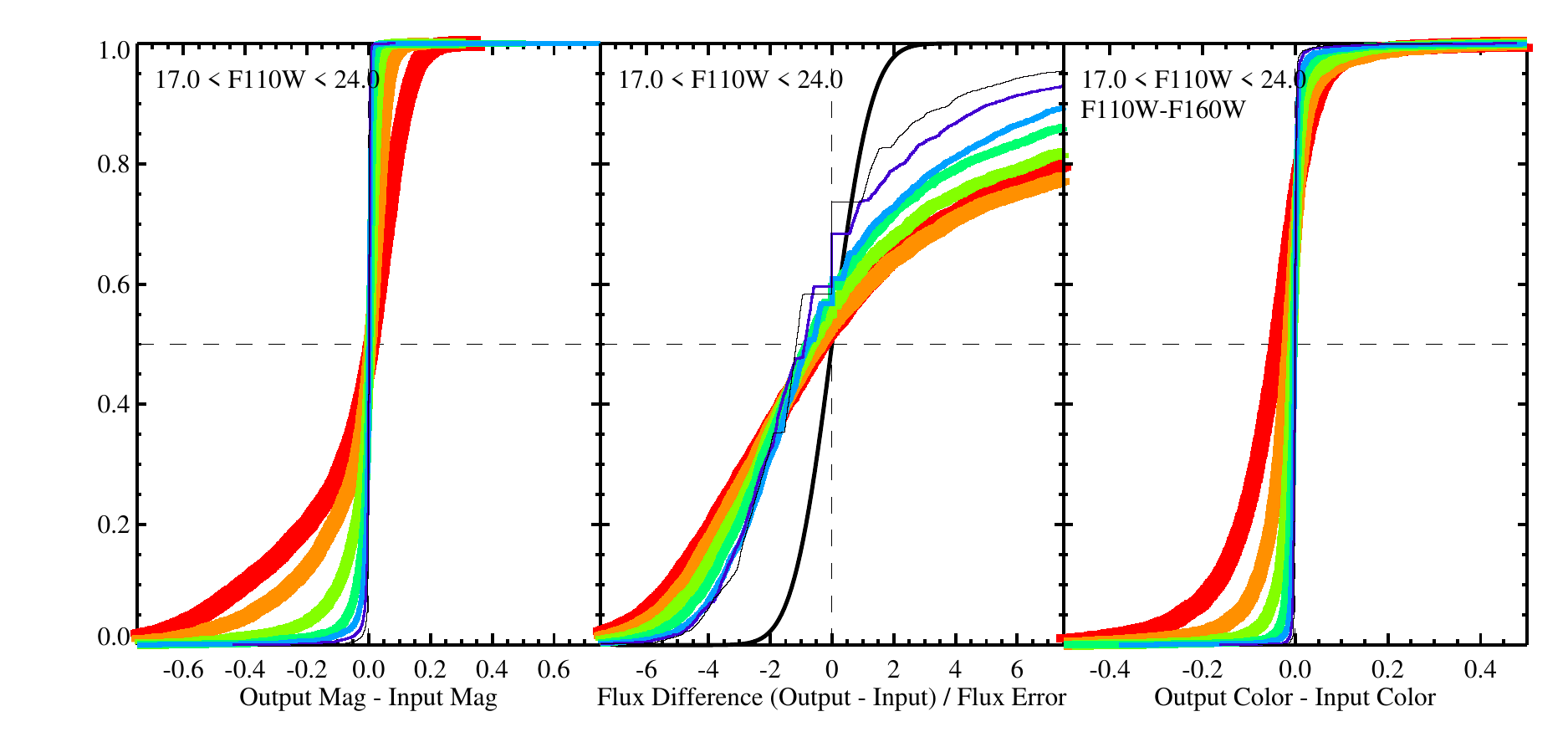}  
\includegraphics[width=3in]{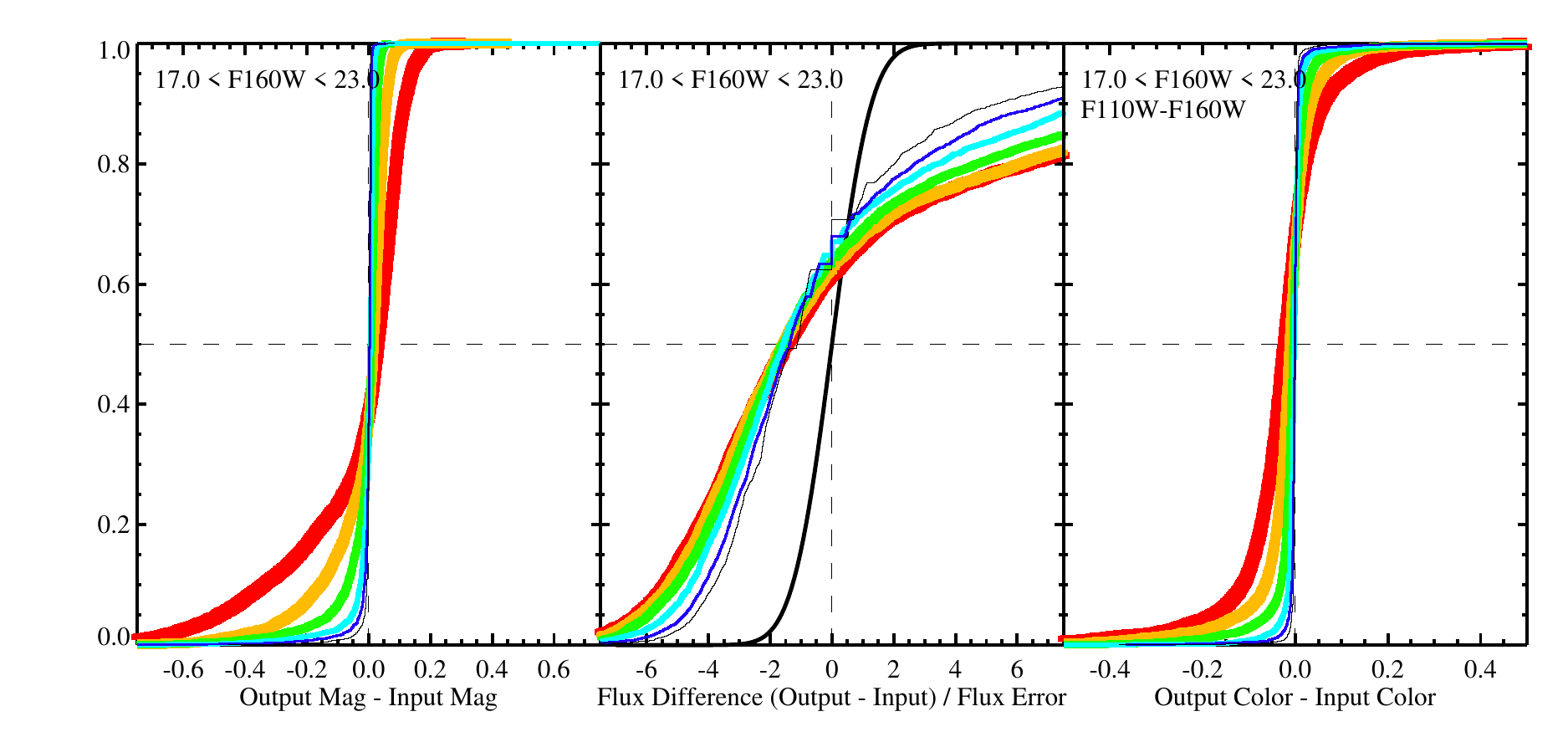}  
}
\caption{Comparisons between the input and output properties of
  artificial stars inserted into WFC3/IR images in Field 9 of Bricks
  1, 9, 15, and 23 (top to bottom rows, respectively), for \fw{110}
  (left column) and \fw{160} (right column).  Panels show the
  cumulative distribution of: (left) the difference between the
  recovered and inserted magnitude, such that negative numbers
  indicates stars that are recovered brighter than their true
  magnitude; (center) the fractional difference in flux, scaled by the
  photometric uncertainty reported by DOLPHOT such that positive
  numbers indicate stars that are recovered brighter than their true
  magnitude; and (right) the difference between the recovered and the
  input color of the artificial stars, such that positive values
  indicate stars that are recovered redder than their true color.
  Line types indicate different bins in recovered magnitude, as
  described in Figure~\ref{fakestarsuvfig}.  The WFC3/IR data depart
  dramatically from a Gaussian due to the effects of crowding,
  everywhere in the disk (see middle panel of each 3-panel plot).  The
  effect is much worse than in the ACS data, due to the larger WFC3/IR
  pixel size.  Moreover, the median star is recovered with a flux that
  is significantly {\emph{fainter}} than its true flux, although
  crowding does induce the expected tail to brighter fluxes.
  \label{fakestarsirfig}}
\end{figure}
\vfill
\clearpage

\begin{figure}
\centerline{
\includegraphics[width=2.25in]{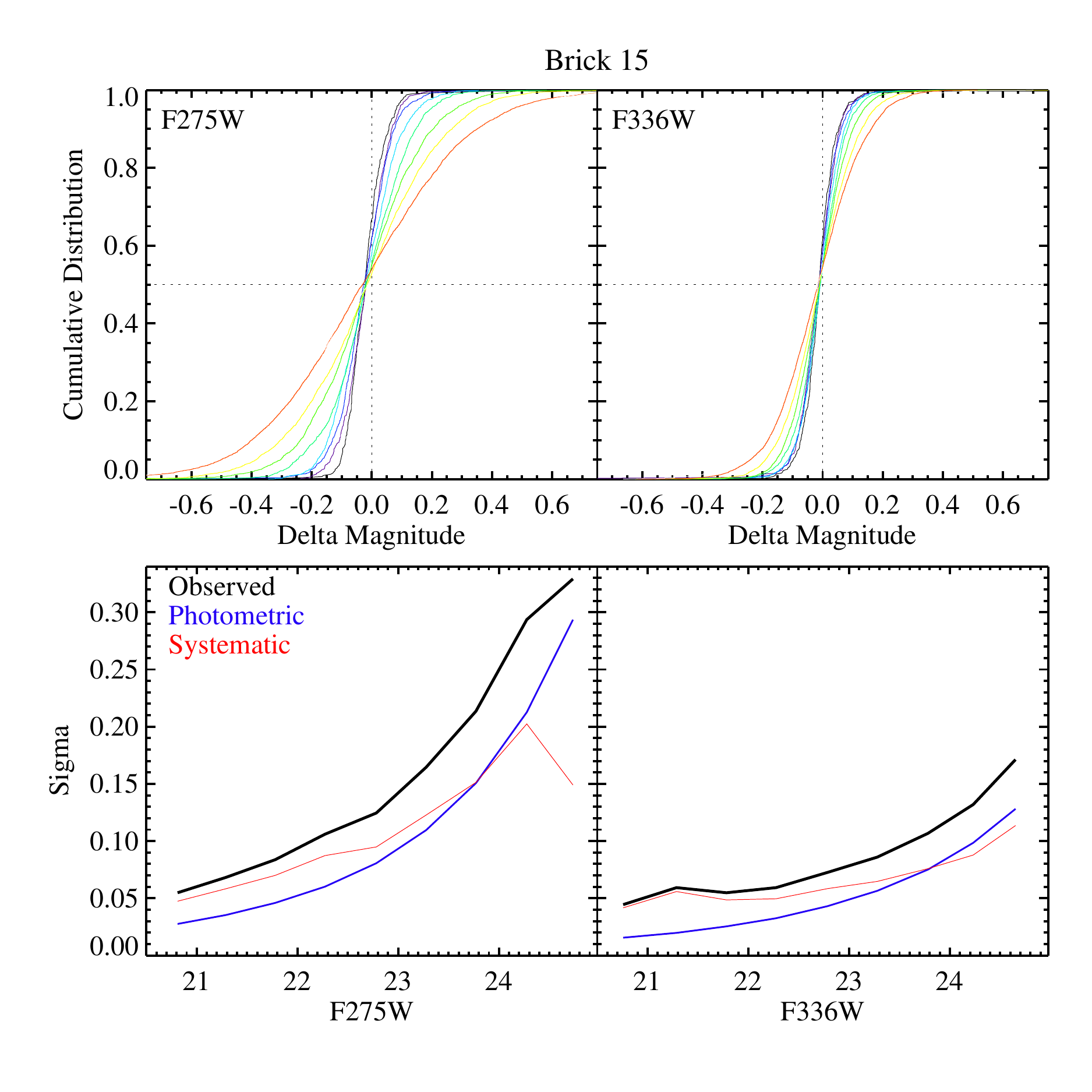}  
\includegraphics[width=2.25in]{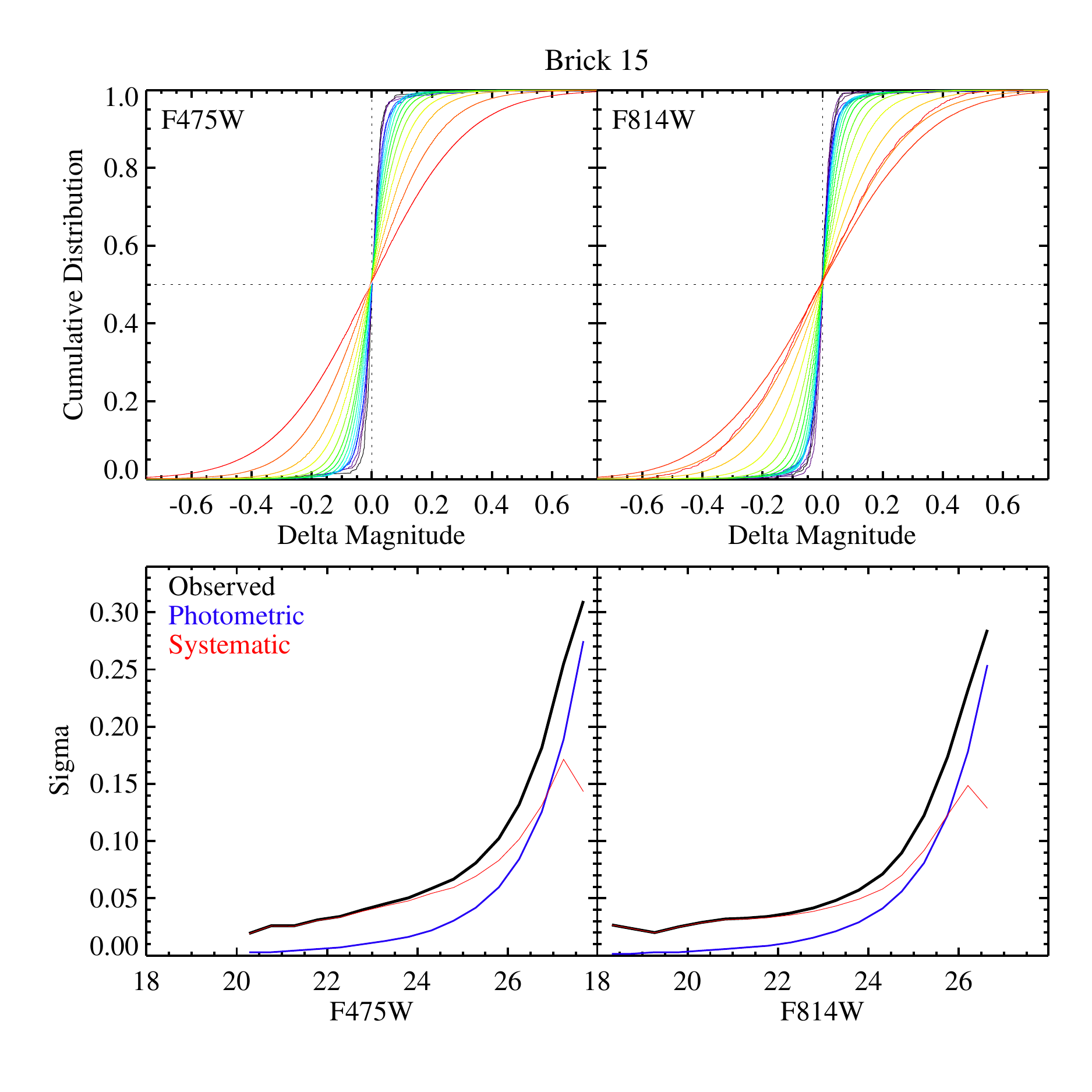}  
\includegraphics[width=2.25in]{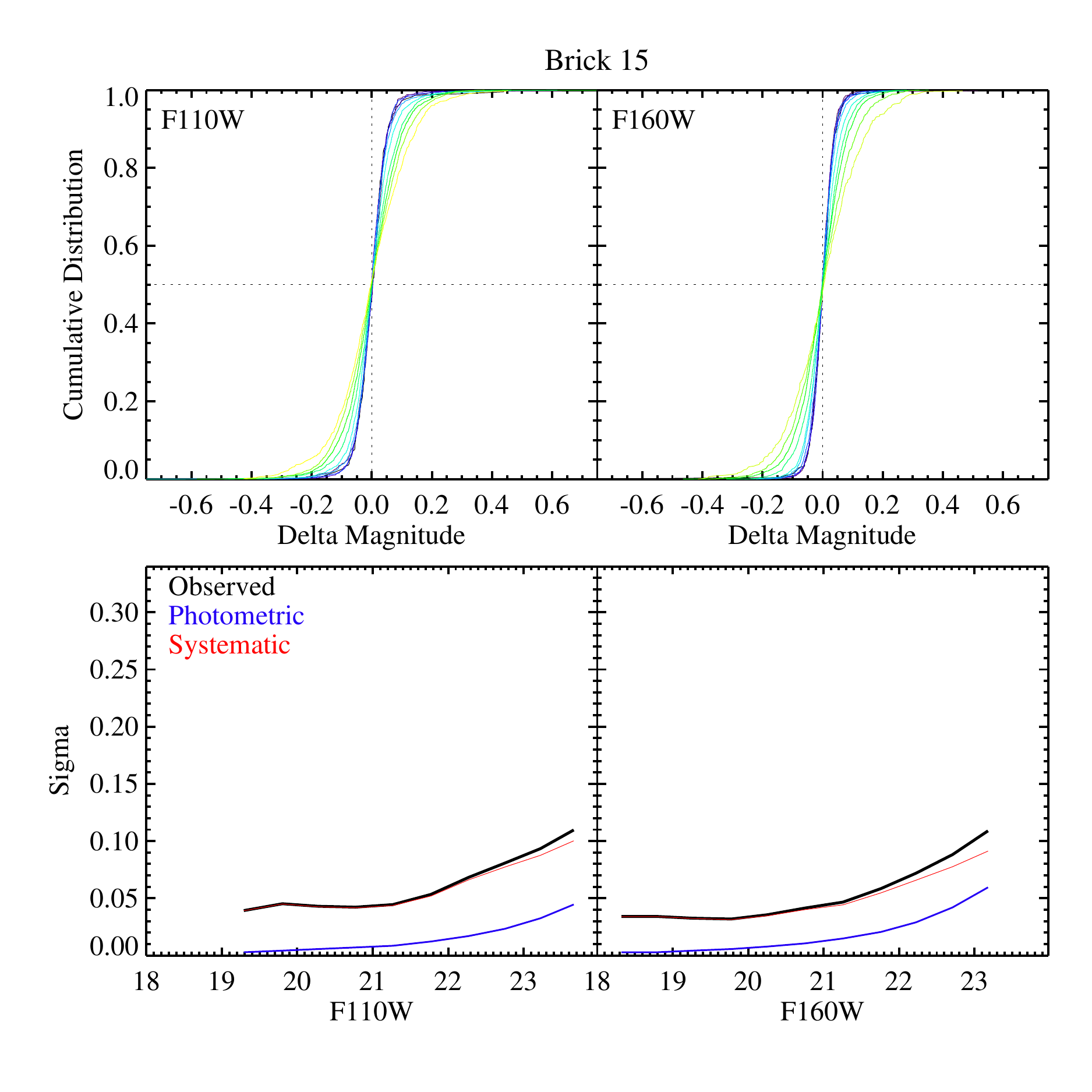}  
}
\caption{Top Row: Cumulative distributions of magnitude differences
  between repeated measurements of stars for Brick 15, binned into
  groups of stars with comparable magnitude (0.5 mag bin widths, with
  redder colors indicating groups of fainter stars), for WFC3/UVIS
  (left), ACS/WFC (center), WFC3/IR (right).  Bottom row: 1$\sigma$
  widths of the observed distributions of magnitude differences (black
  line, as inferred from the interquartile width to avoid skew from
  variable stars), as a function of the median stellar magnitude in a
  bin for Brick 15.  These widths are compared to the expected width
  from photometric uncertainties alone (blue line).  The quadrature
  difference between them is an estimate of the systematic error,
  plotted as a red line.  Systematic errors dominate the observed
  magnitude differences at bright magnitudes, and are larger at faint
  magnitudes, but still are typically small.  The
  behavior of the magnitude differences is similar in other disk
  fields.  Because the curves are based on pairs of overlapping stars,
  the plotted magnitude differences do not characterize the behavior
  of the full chip area for WFC3's UVIS and IR channels, which have
  limited spatial overlap.
  \label{overlapmagdifffig}}
\end{figure}
\vfill

\begin{figure}
\centerline{
\includegraphics[width=6.5in]{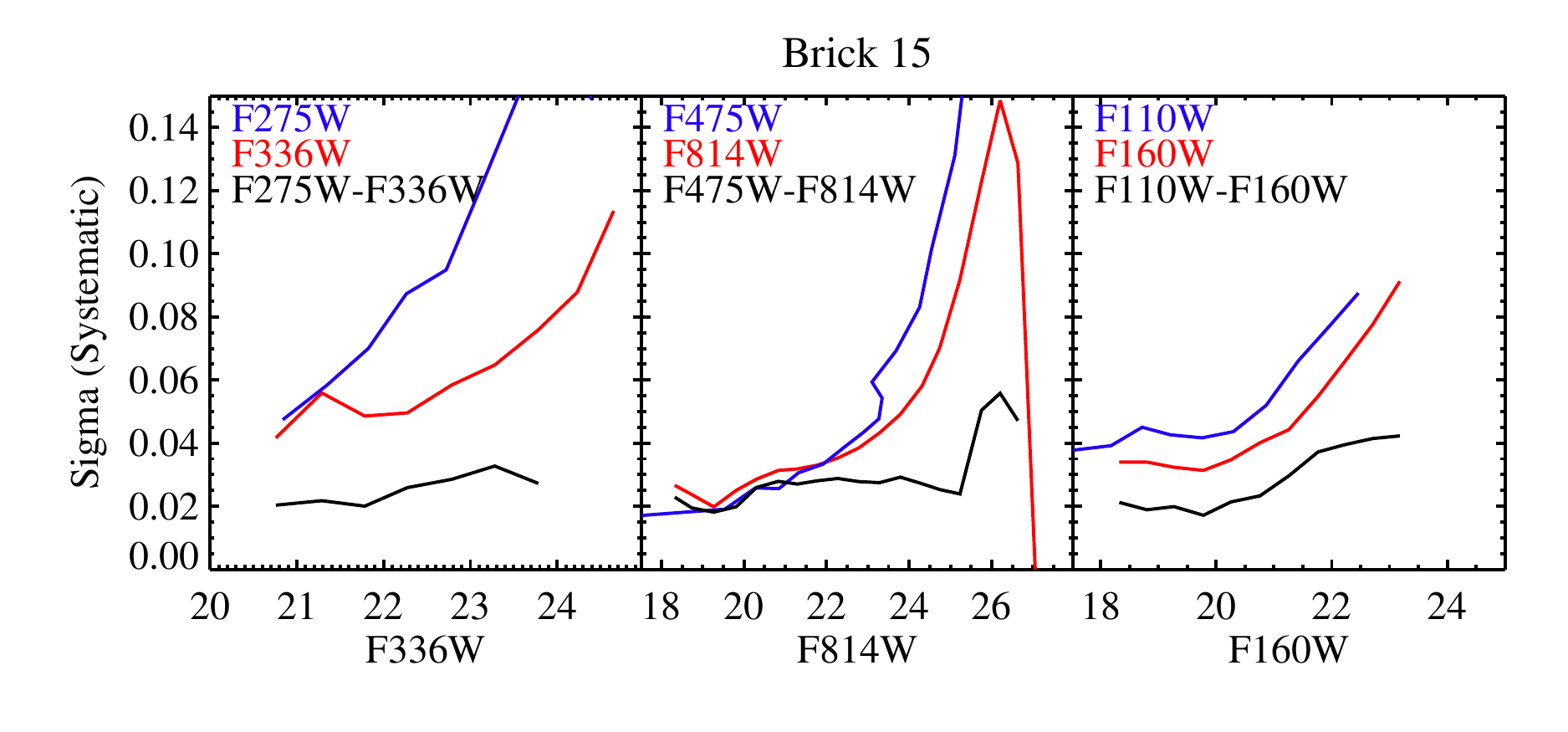}  
}
\caption{Inferred systematic errors as a function of magnitude for
  WFC3/UVIS (left), ACS/WFC (center), WFC3/IR (right), measured in
  Brick 15.  The blue and red lines indicate the systematic error in
  magnitude measurements for the bluer and redder filter in a single
  camera, respectively (see Figure~\ref{overlapmagdifffig}).  The
  black line shows the systematic error in the color.  The systematic
  uncertainties are typically smaller in color than in magnitude,
  because many of the positionally dependent uncertainties (flat
  fields, PSF models) are likely to correlate between the two filters,
  and thus partially cancel out.  The behavior of the magnitude
  differences is similar in other disk fields.  Because the curves are based on
  pairs of overlapping stars, the plotted magnitude
  differences do not characterize the behavior of the full chip area
  for WFC3's UVIS and IR channels, which have limited spatial overlap
  (see Figure~\ref{exptimemapfig}).
\label{systematicfig}}
\end{figure}
\vfill
\clearpage

\begin{figure}
\centerline{
\includegraphics[width=3.25in]{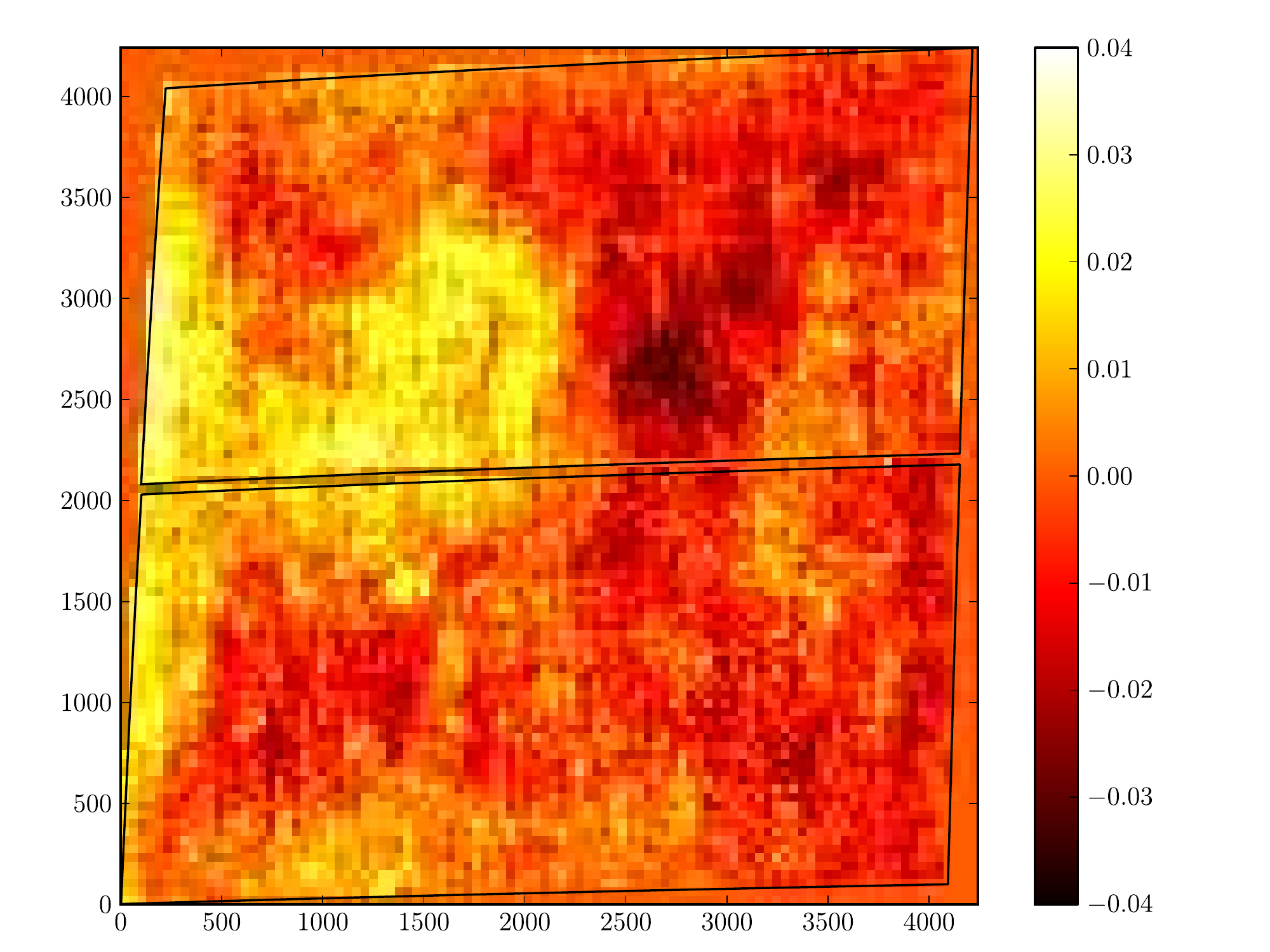}
\includegraphics[width=3.25in]{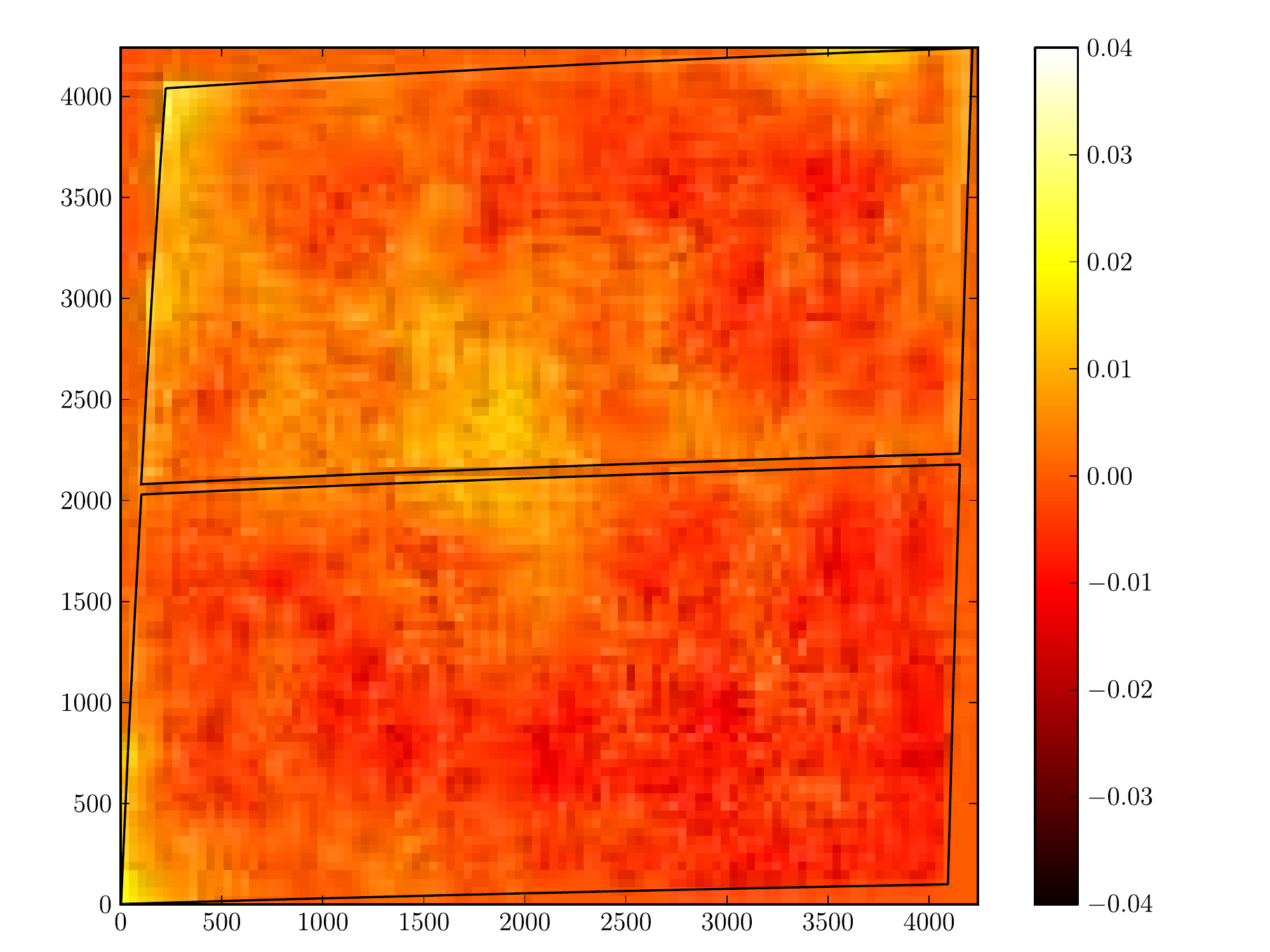}
}
\centerline{
\includegraphics[width=3.25in]{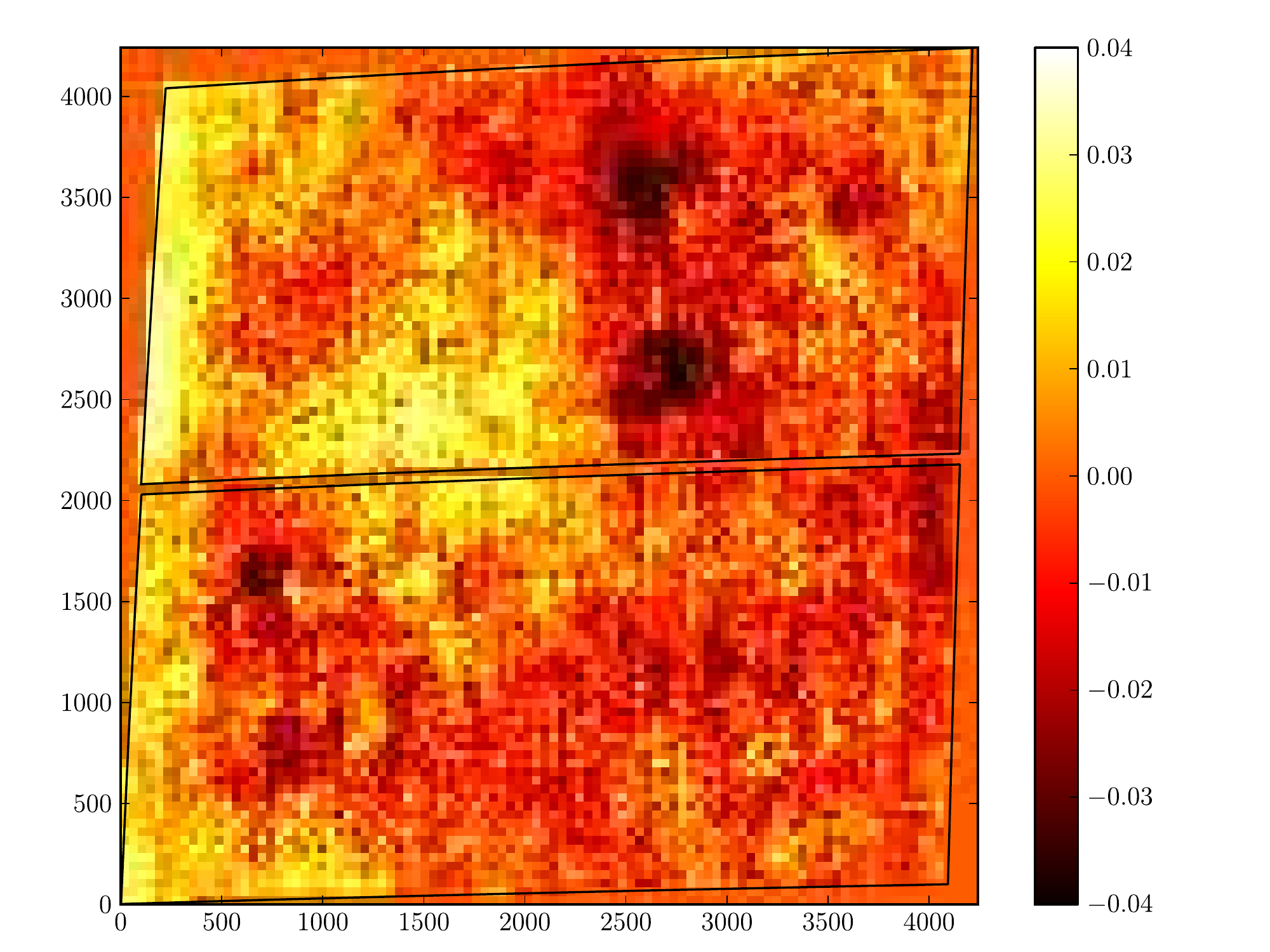}
\includegraphics[width=3.25in]{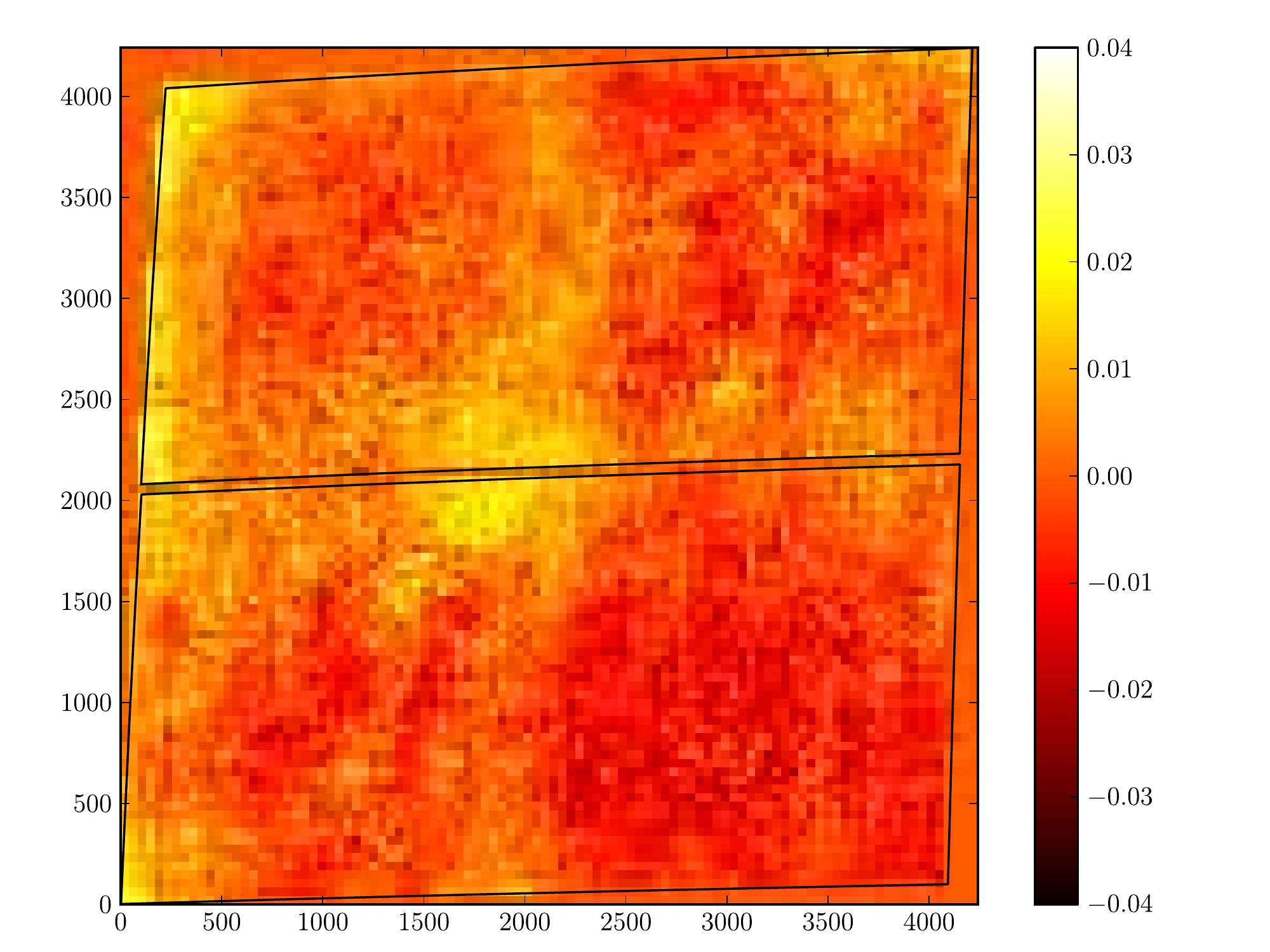}
}
\caption{Map of magnitude corrections as a function of position on the
  ACS reference image for \fw{475} (left column) and \fw{814} (right
  column), and Bricks 15 and 21 (top and bottom rows, respectively).
  The color bar on the right refers to the correction, in magnitudes,
  required to minimize the differences between our measurements for
  stars that appear in multiple images.  The map of \fw{814} offsets is
  quite smooth.  In contrast, the \fw{475} filter shows strong
  positionally-dependent variations that are consistent between the
  two bricks.  These are likely to result from the lack of a large
  scale flat field correction for the \fw{475} filter in the standard
  STScI flat field.\label{magdiffmapfig}}
  \end{figure}
\vfill
\clearpage

\begin{figure}
\centerline{
\includegraphics[width=2.25in]{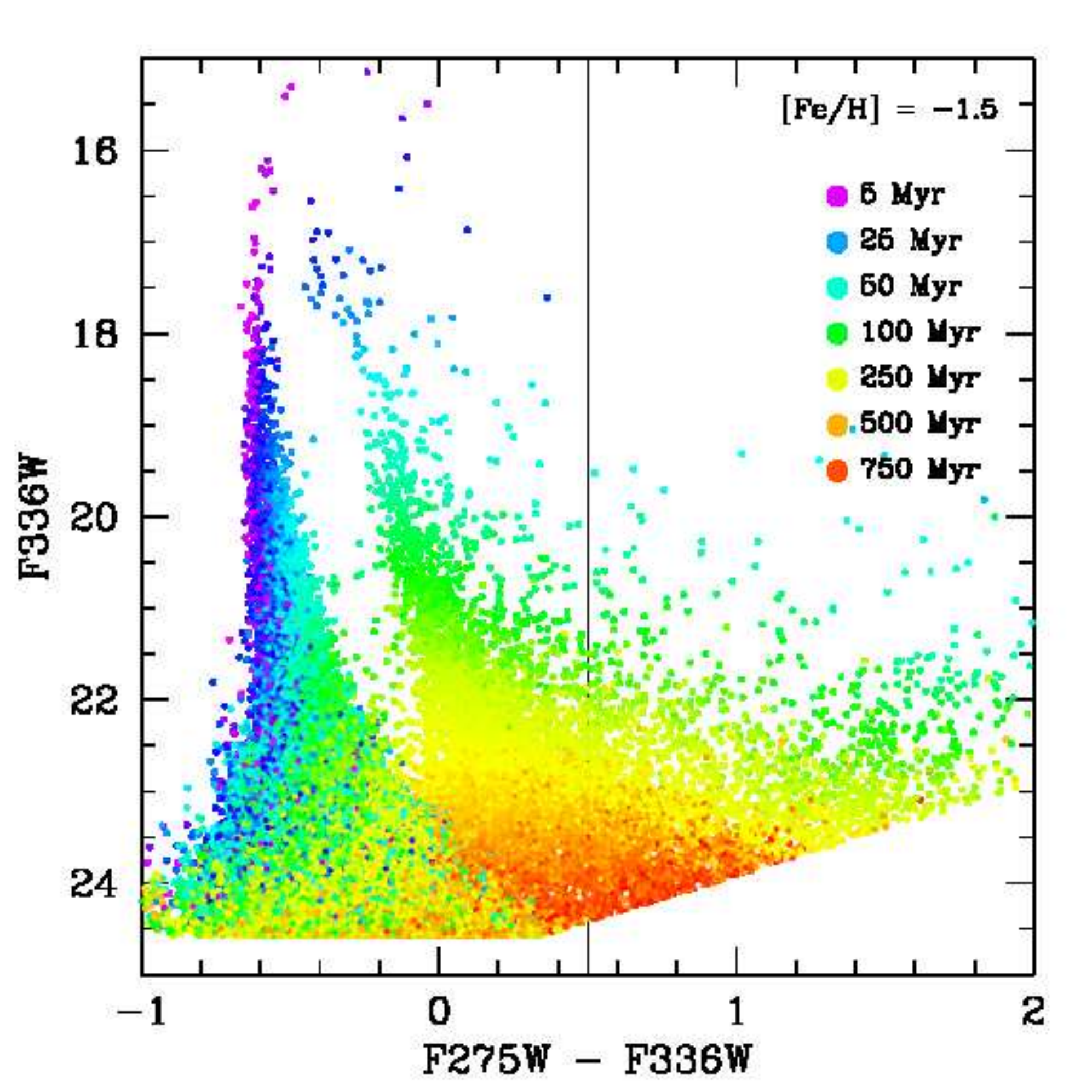}  
\includegraphics[width=2.25in]{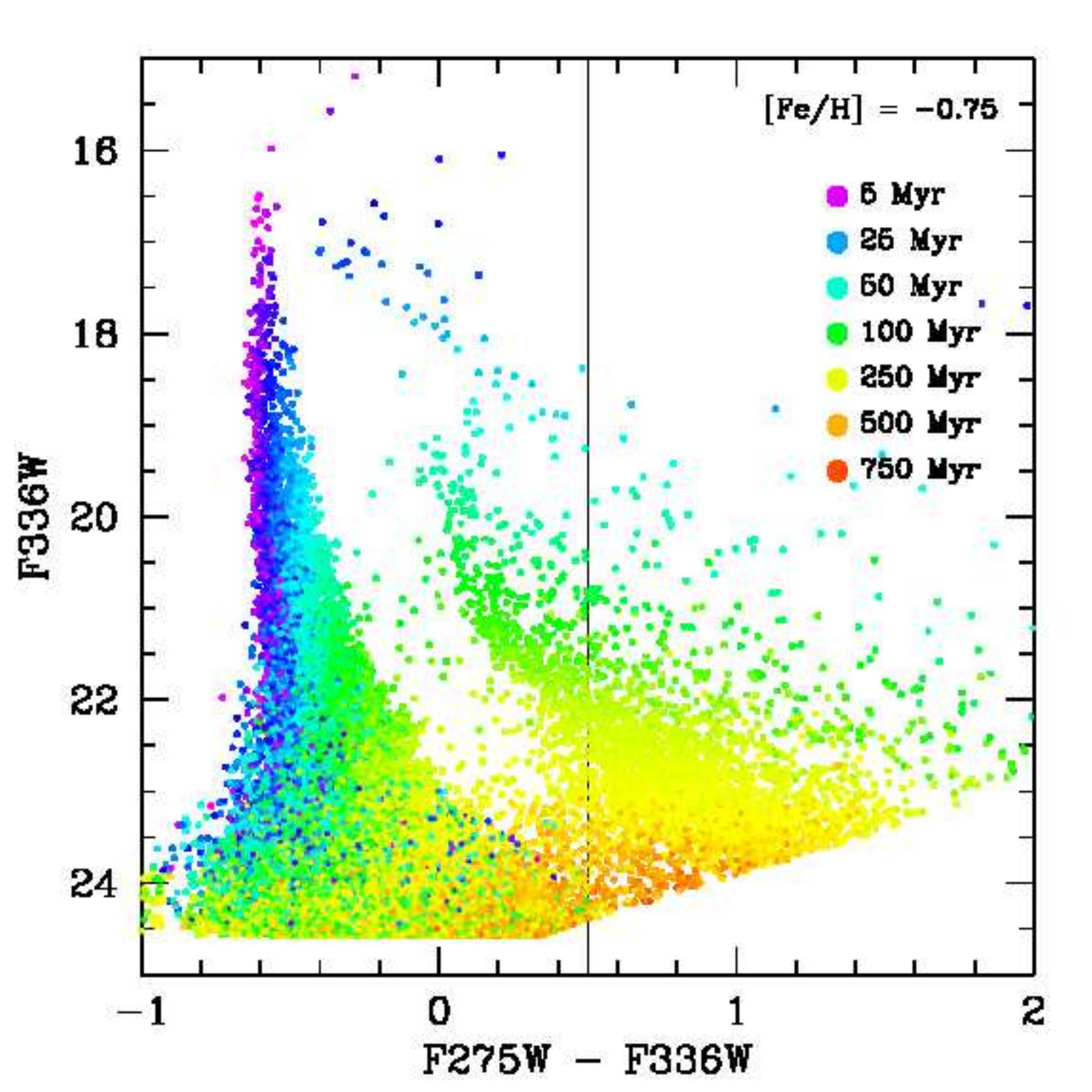}  
\includegraphics[width=2.25in]{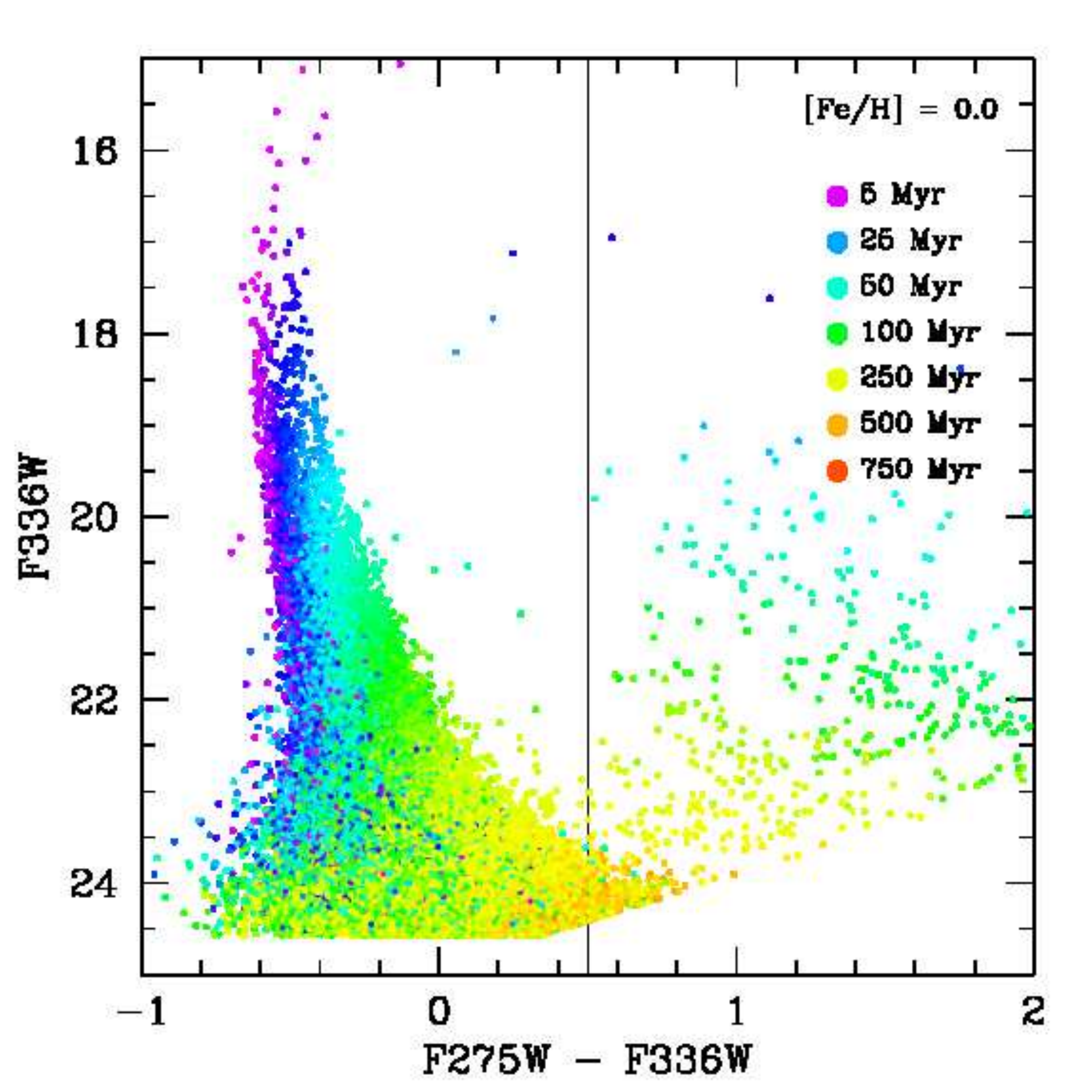}  
}
\centerline{
\includegraphics[width=2.25in]{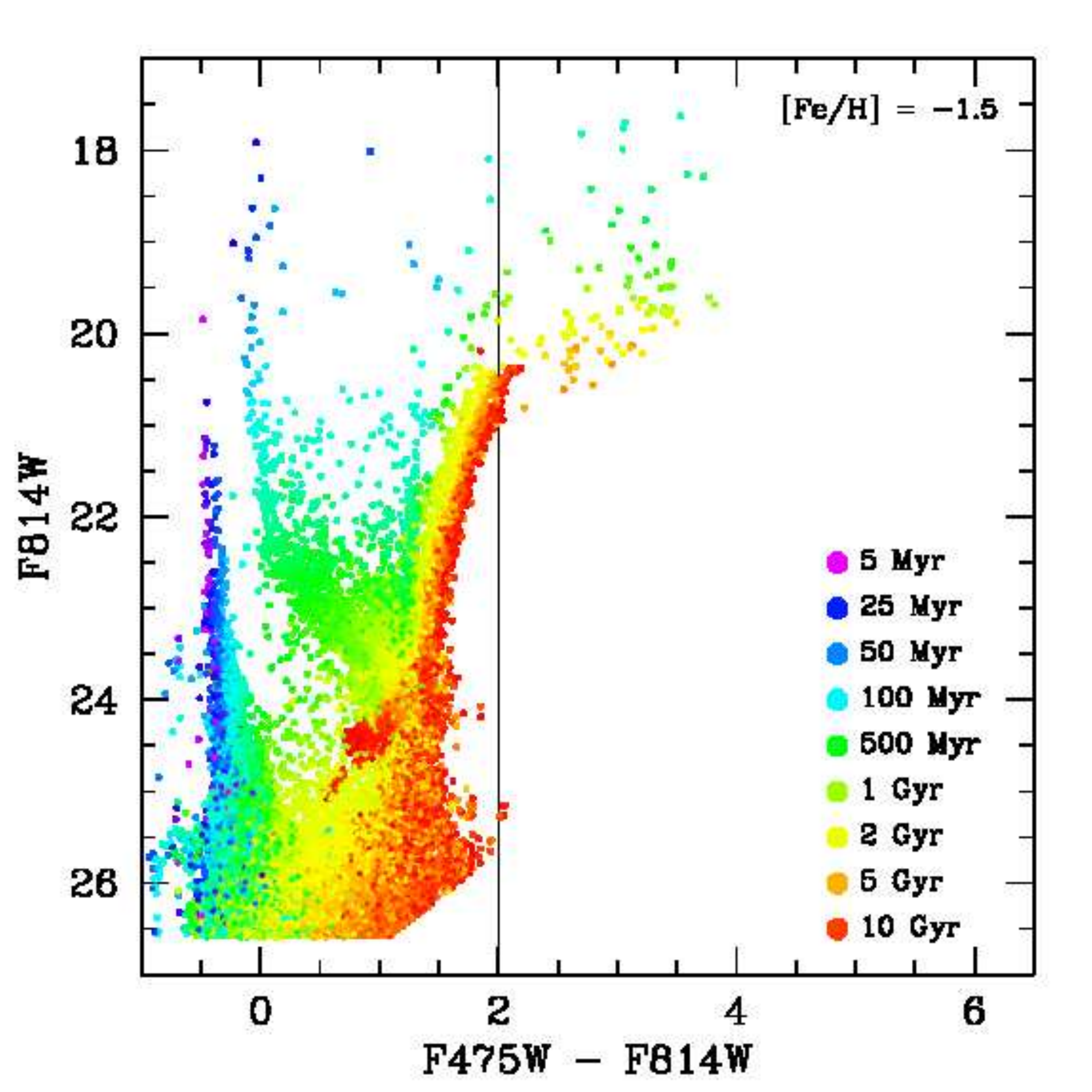}  
\includegraphics[width=2.25in]{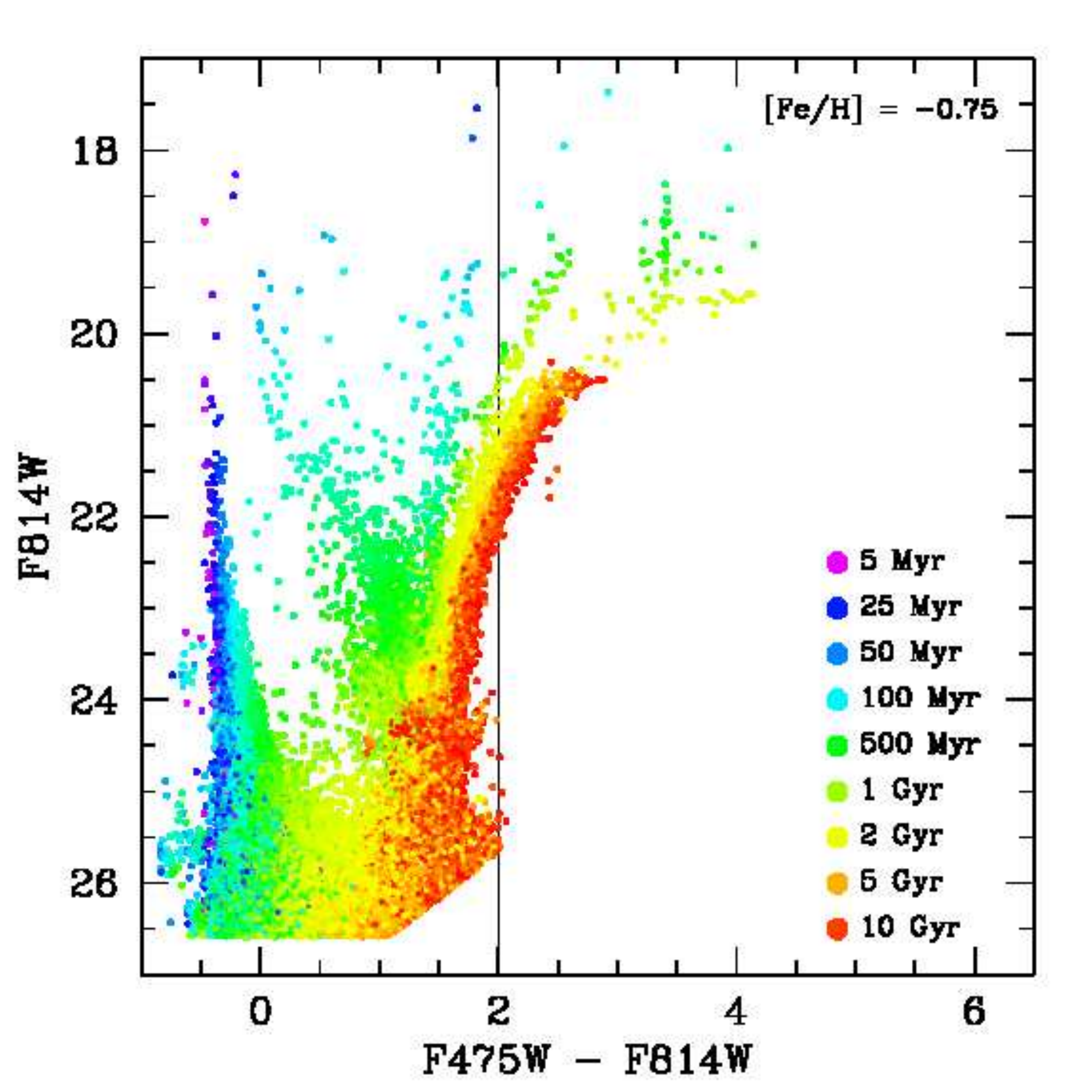}  
\includegraphics[width=2.25in]{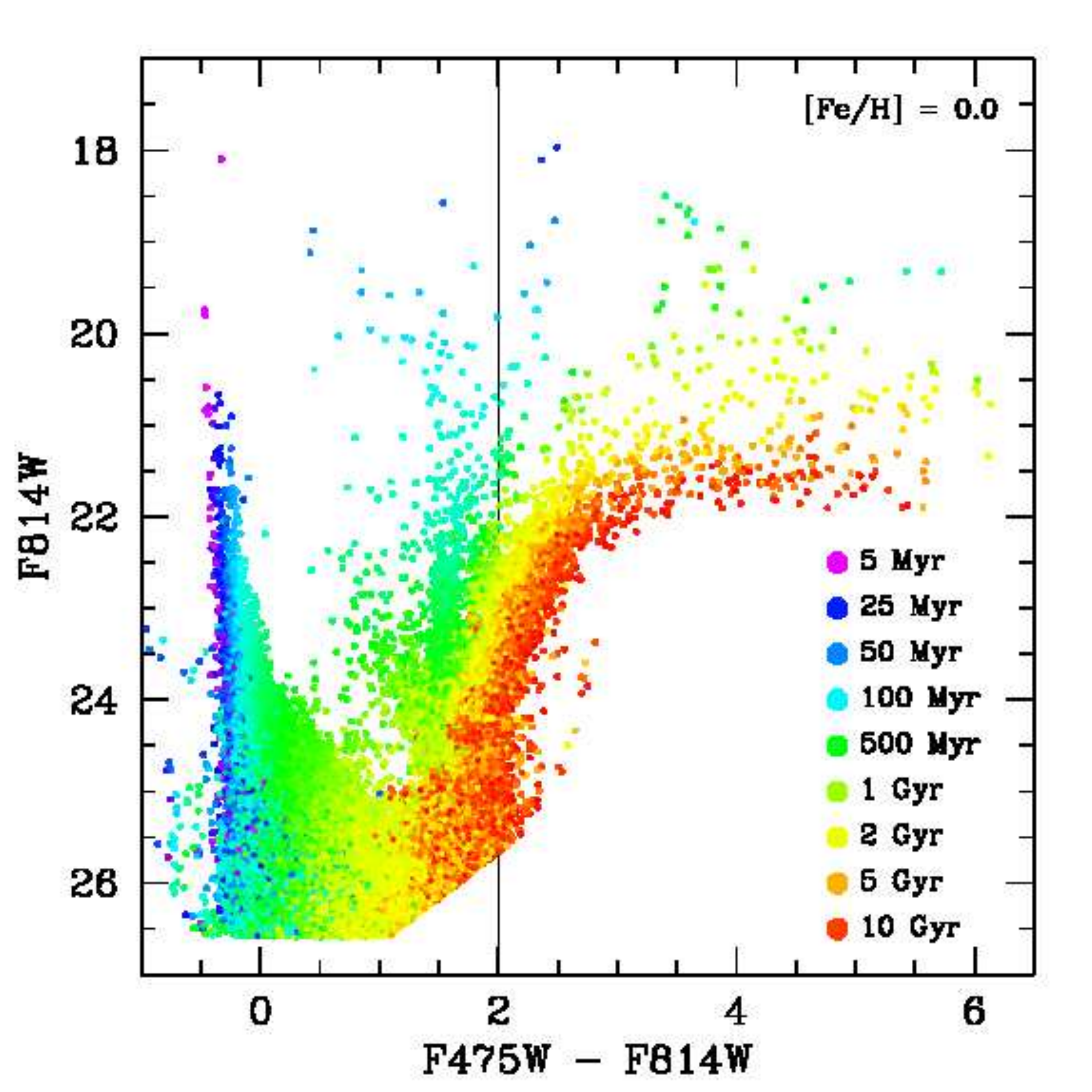}  
}
\centerline{
\includegraphics[width=2.25in]{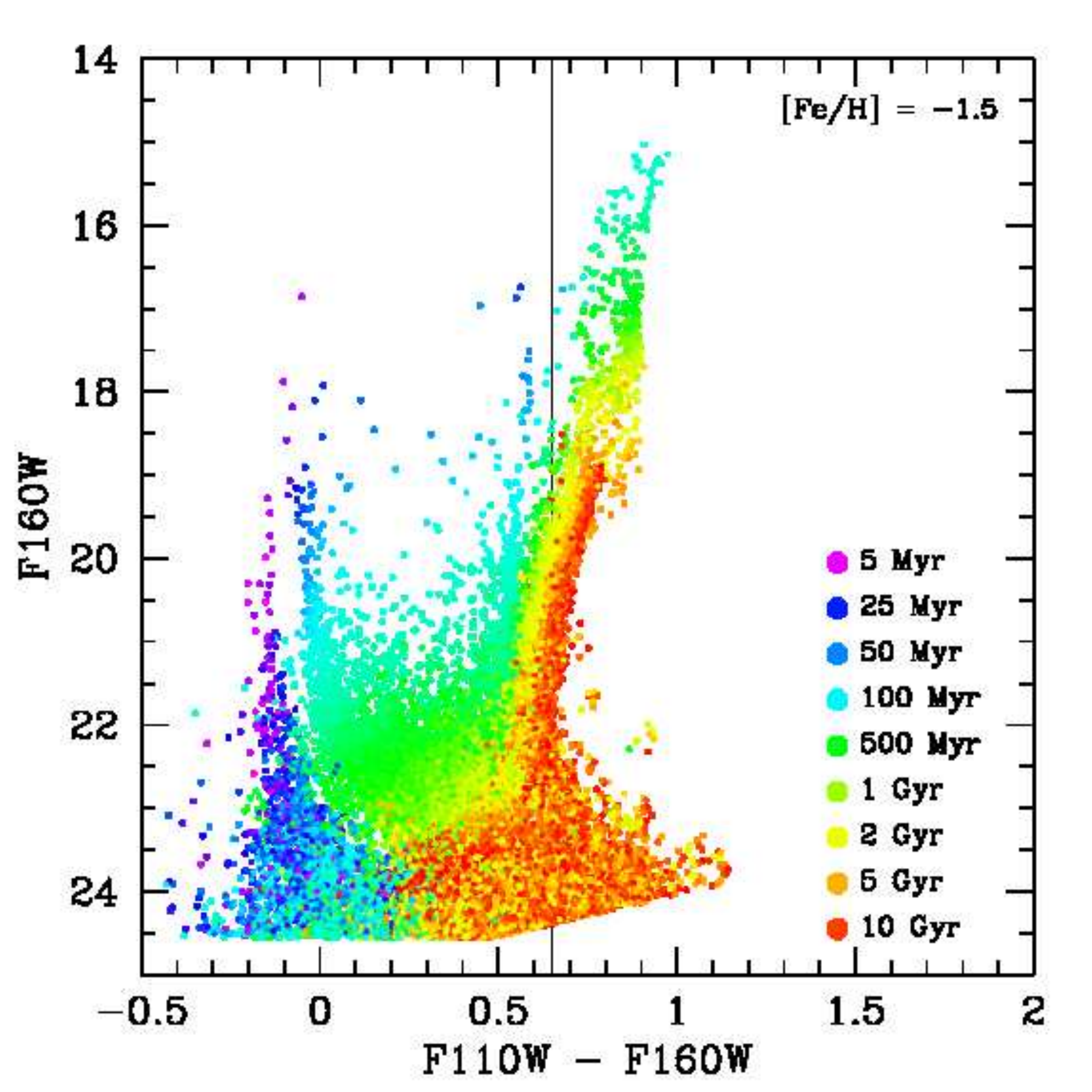}  
\includegraphics[width=2.25in]{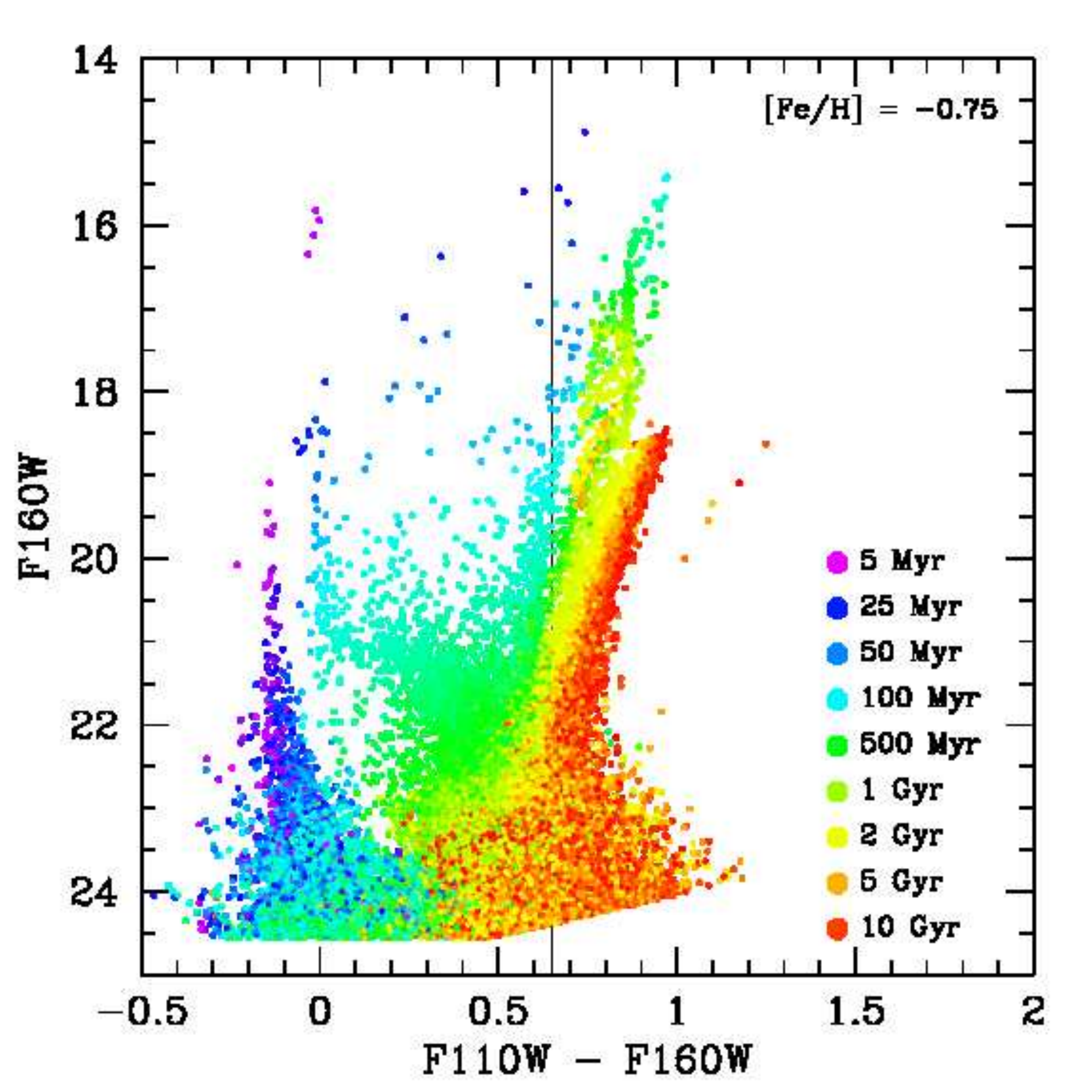}  
\includegraphics[width=2.25in]{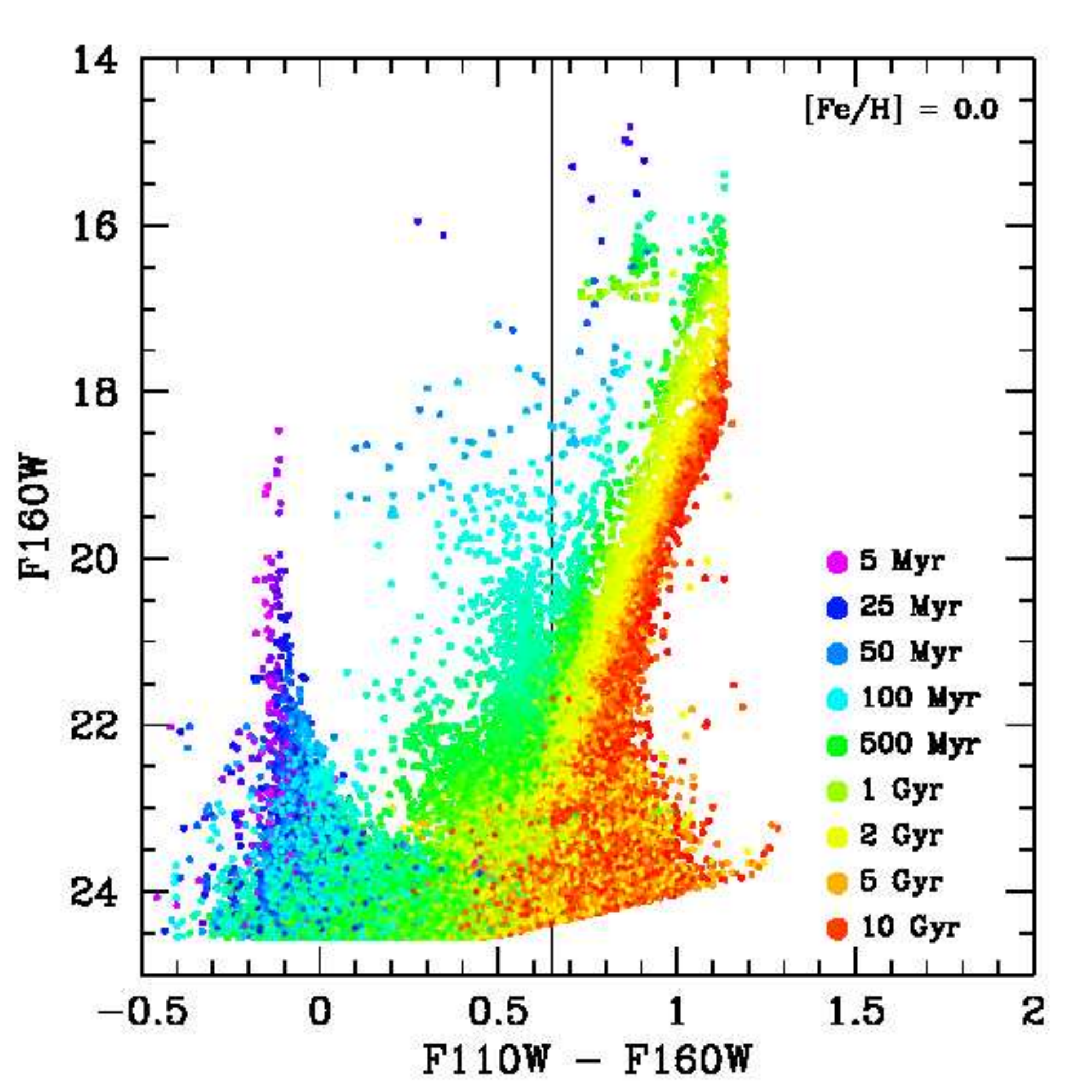}  
}
\caption{\footnotesize{Simulated CMDs for WFC3/UVIS (top row), ACS/WFC
    (middle row), and WFC3/IR (bottom row), for three different
    metallicities ([Fe/H] $=\!-1.5$, $-0.75$, and 0.0; left, middle,
    and right columns, respectively), with points color coded by age.
    The simulations assume a constant star formation rate, no
    foreground or internal extinction, and photometric errors and
    completeness derived from artificial star tests for outer disk
    fields in Brick 23.  The solar metallicity CMDs are most likely to
    be characteristic of the young stellar populations throughout the
    disk, and the older stellar populations in the bulge.  The
    intermediate metallicity ([Fe/H] $=\!-0.75$) CMDs are likely to
    be representative of the older thin and thick disk stars
    \citep[e.g.,][]{collins2011}, and the low metallicity CMDs are
    likely to be characteristic of a potential stellar halo population
    \citep{chapman2006,kalirai2006}.  Age color-coding is the same for
    all plots in single row, but varies from camera to camera.
    Vertical lines are plotted at a constant color for each camera,
    and are shown for reference, to allow comparisons between
    different metallicities and with foreground populations shown in
    Figure~\ref{fakeFGfig}.}
\label{fakeCMDfig}}
\end{figure}
\vfill

\begin{figure}
\centerline{
\includegraphics[width=2.25in]{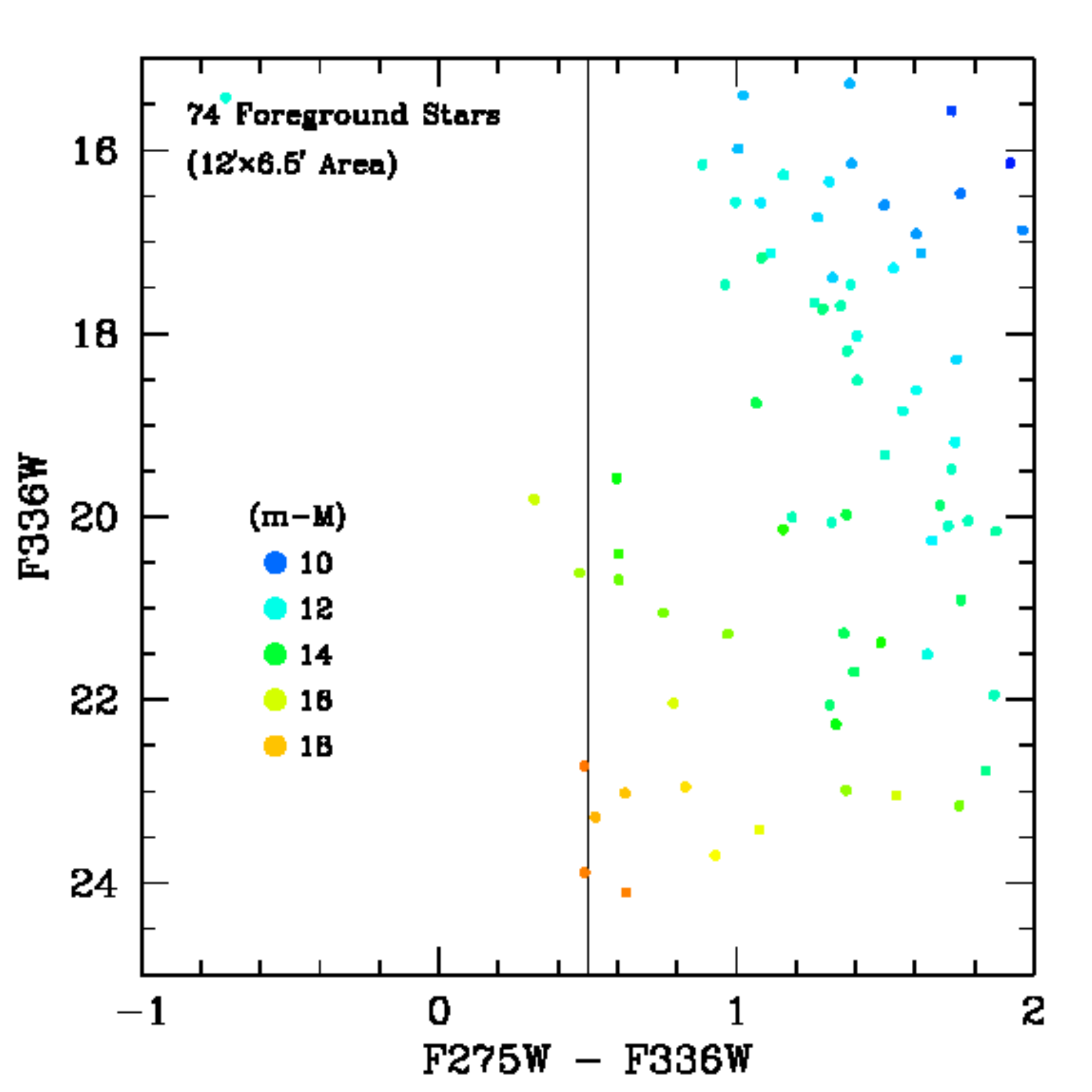}  
\includegraphics[width=2.25in]{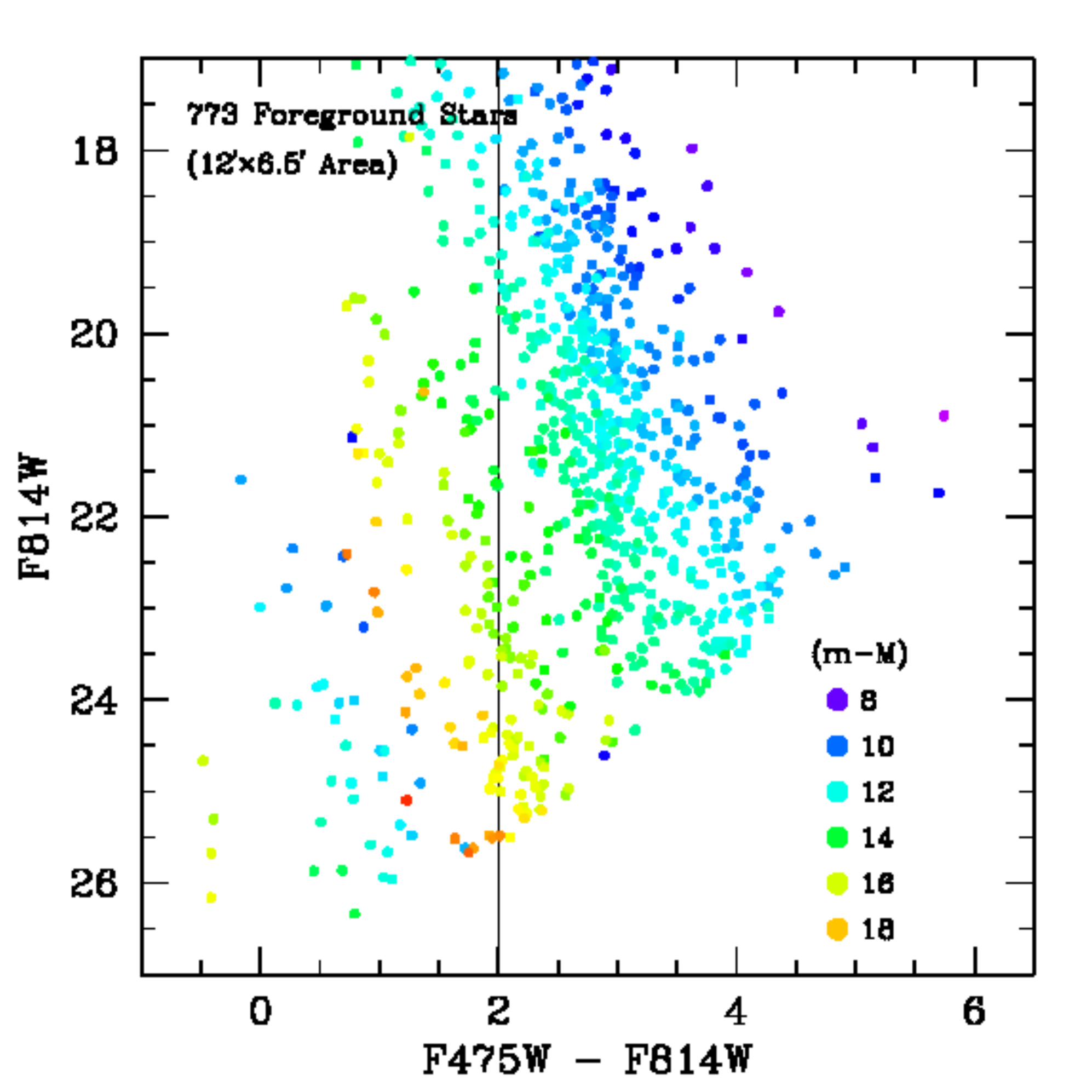}  
\includegraphics[width=2.25in]{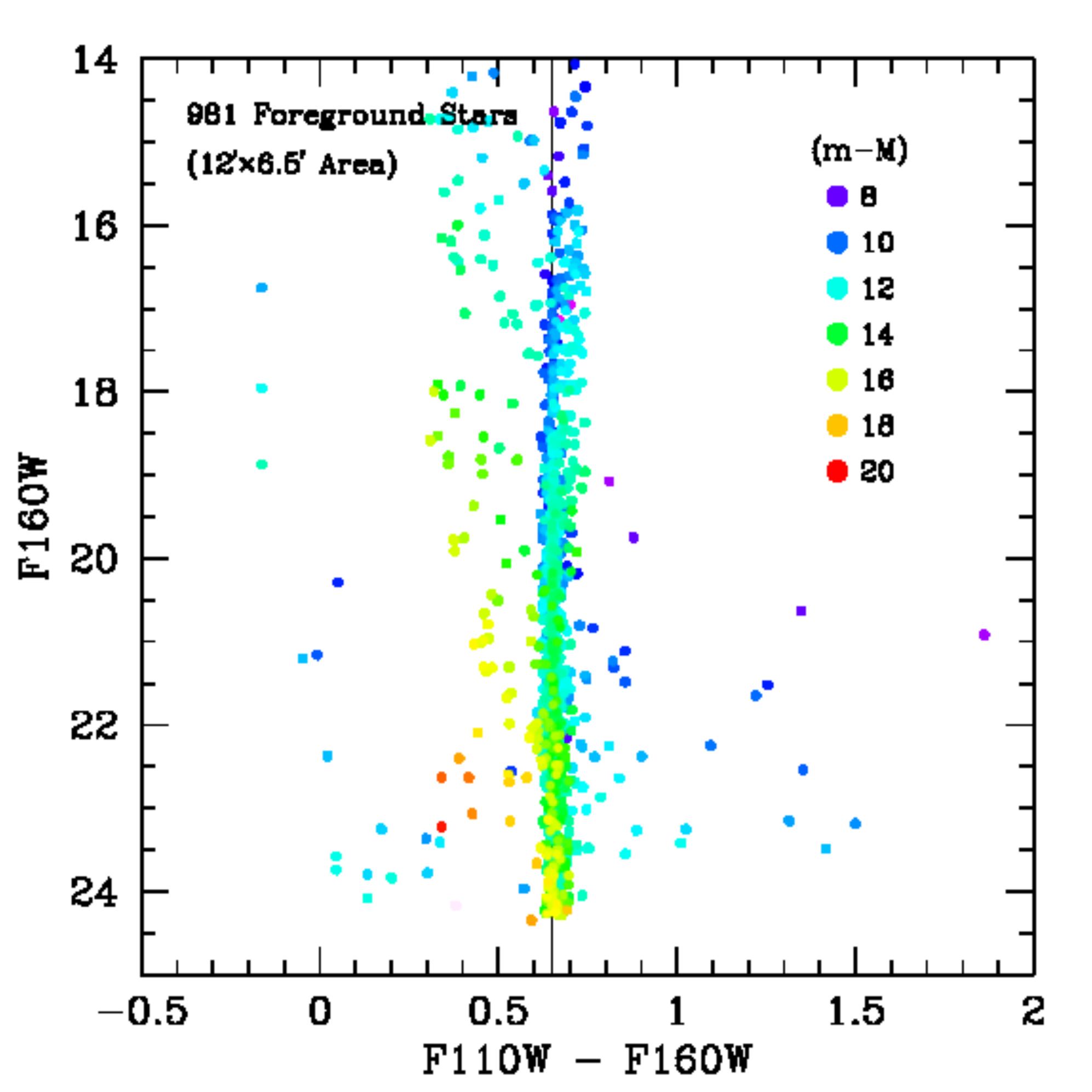}  
}
\caption{Simulated CMDs for expected foreground populations in
  WFC3/UVIS (left), ACS/WFC (center), and WFC3/IR (right), calculated
  from TRILEGAL's default Milky Way model \citep{girardi2005}.  CMDs
  assume a default area of $12\arcmin\times6.5\arcmin$, corresponding
  to the 3$\times$6 WFC3/IR footprint of a single brick; the number of
  stars in each CMD is indicated in the plots.  Stars in each panel
  are color coded by their distance modulus $(m-M)$.  The color and
  magnitude ranges are identical to those in Figure~\ref{fakeCMDfig}.
  Vertical lines are shown for reference, to allow comparisons with
  simulated stellar populations shown in Figure~\ref{fakeCMDfig}.
\label{fakeFGfig}}
\end{figure}
\vfill
\clearpage

\begin{figure}
\centerline{
\includegraphics[width=3.25in]{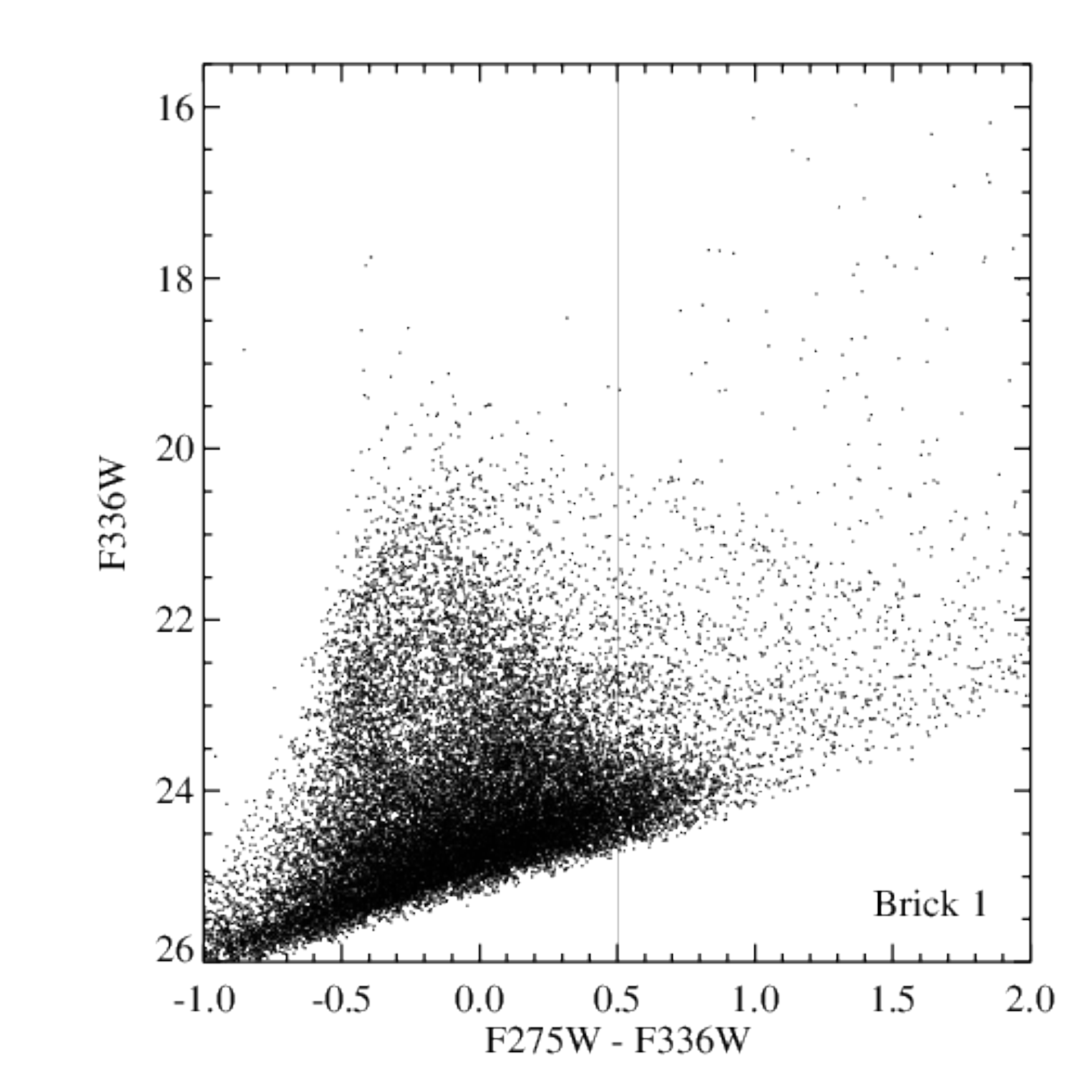}  
\includegraphics[width=3.25in]{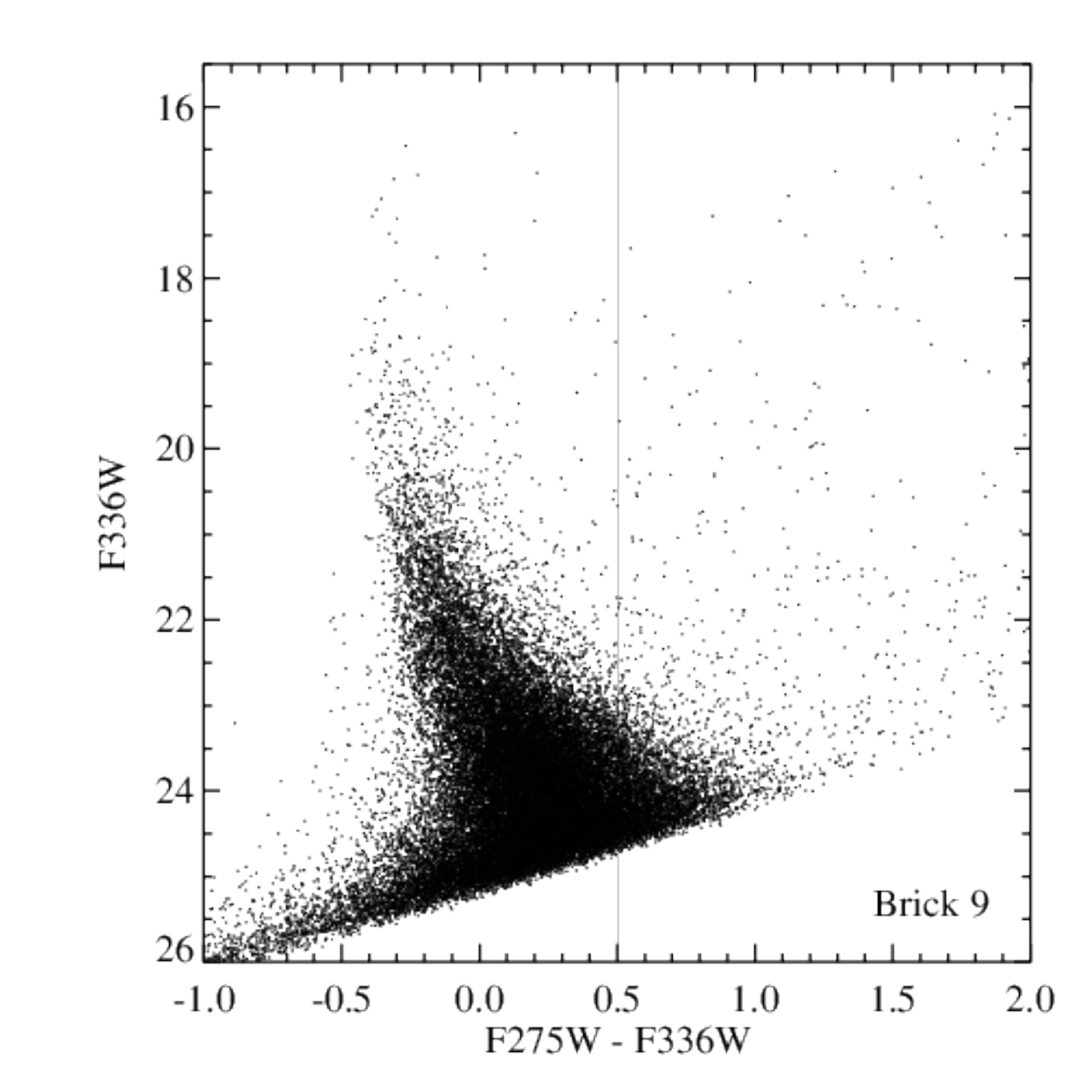}  
}
\centerline{
\includegraphics[width=3.25in]{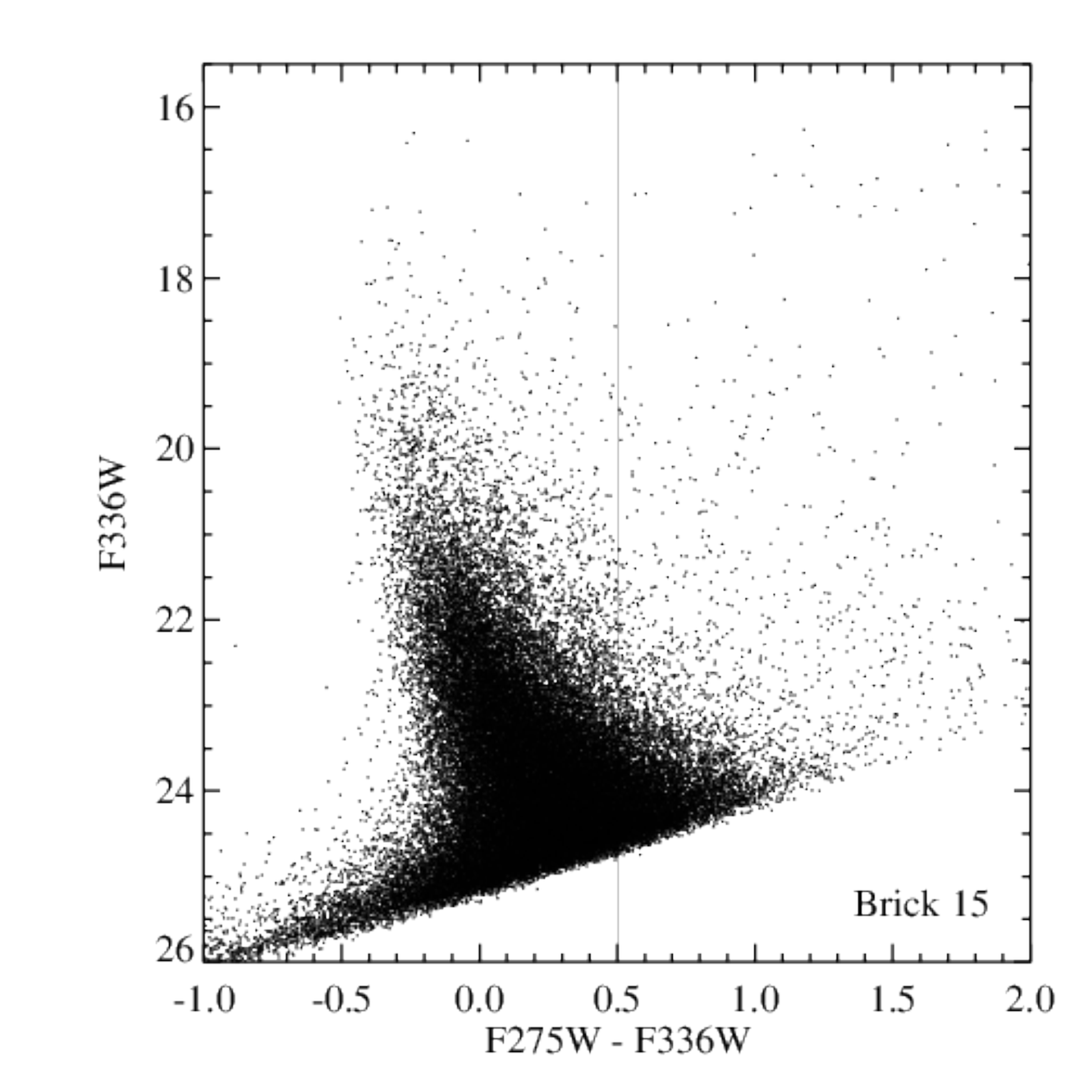}  
\includegraphics[width=3.25in]{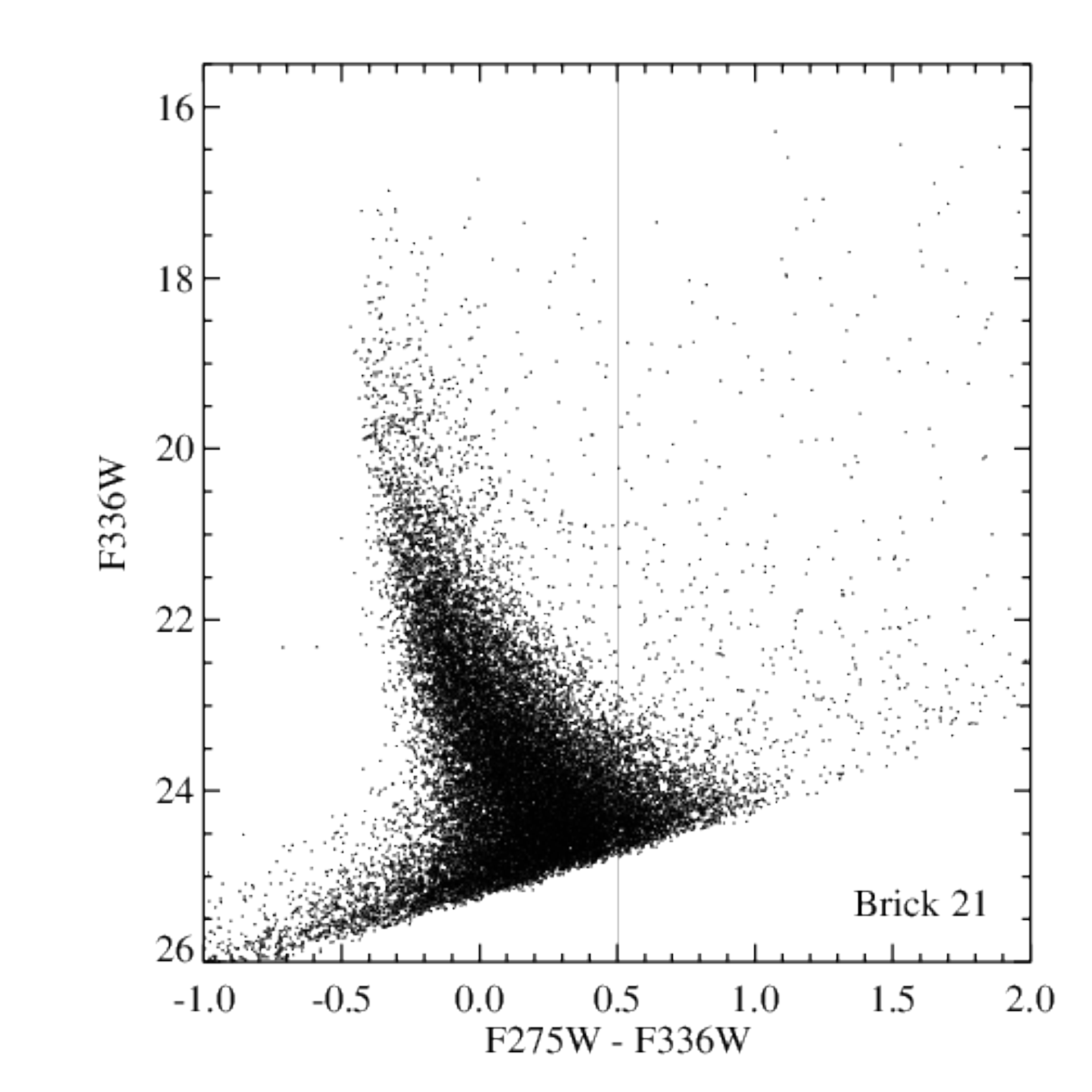}  
}
\caption{UV Hess diagrams for complete Bricks 1, 9, 15, and 21 (from
  upper left to lower right).  The Hess diagram is displayed with a
  square root scaling to highlight both high and low density features.
  Outside of Brick 1, the UV CMD morphology is dominated by main
  sequence stars.  In the bulge-dominated Brick 1, however, a much
  bluer population due to hot HP-HB stars is also present
  \citep{rosenfield2012}.  The UV data reach a consistent depth
  across all bricks, due to the low level of crowding.  The vertical
  line is given for reference, and matches those in
  Figures~\ref{fakeCMDfig} and \ref{fakeFGfig}.
  \label{hessbrickuvfig}}
\end{figure}
\vfill
\clearpage

\begin{figure}
\centerline{
\includegraphics[width=3.25in]{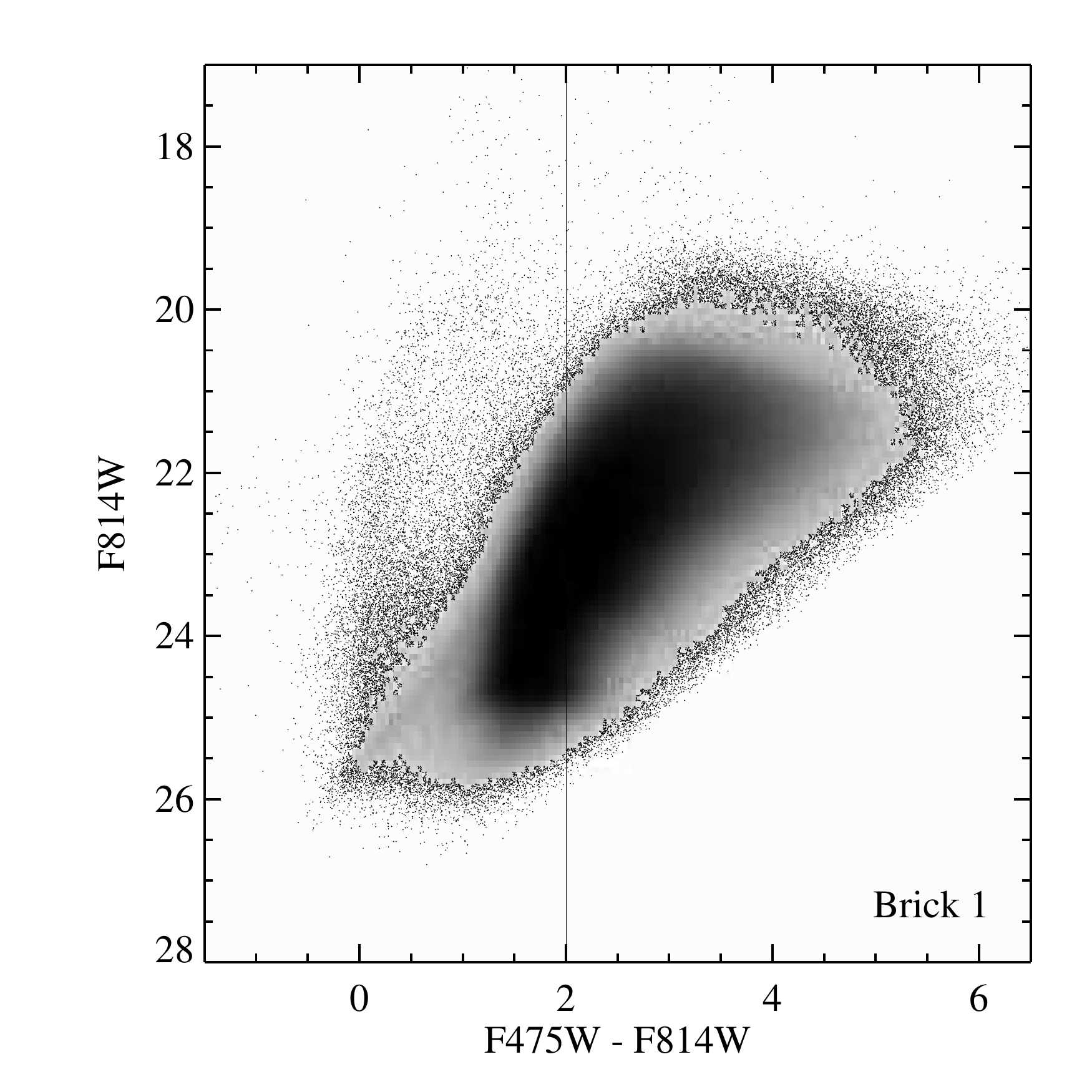}  
\includegraphics[width=3.25in]{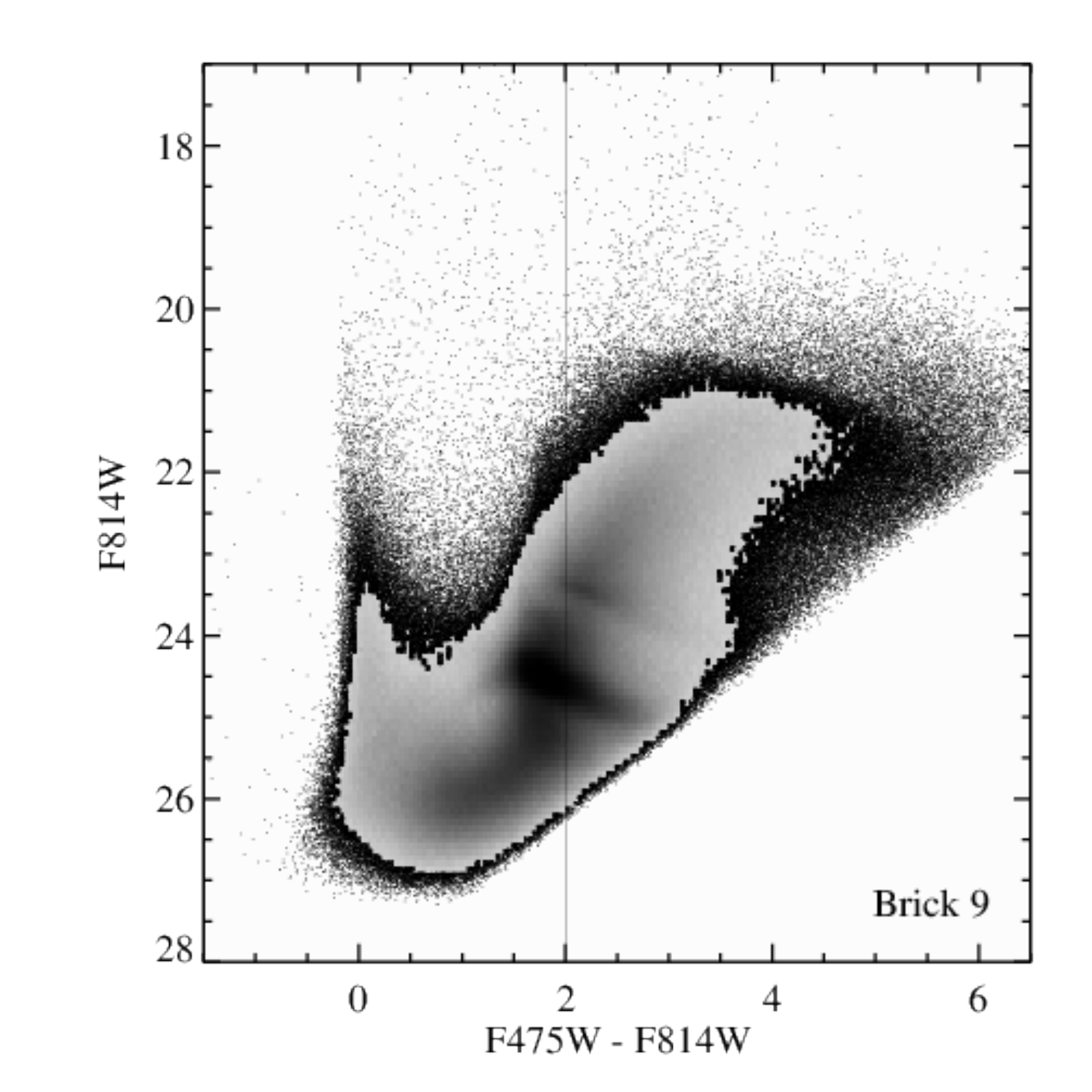}  
}
\centerline{
\includegraphics[width=3.25in]{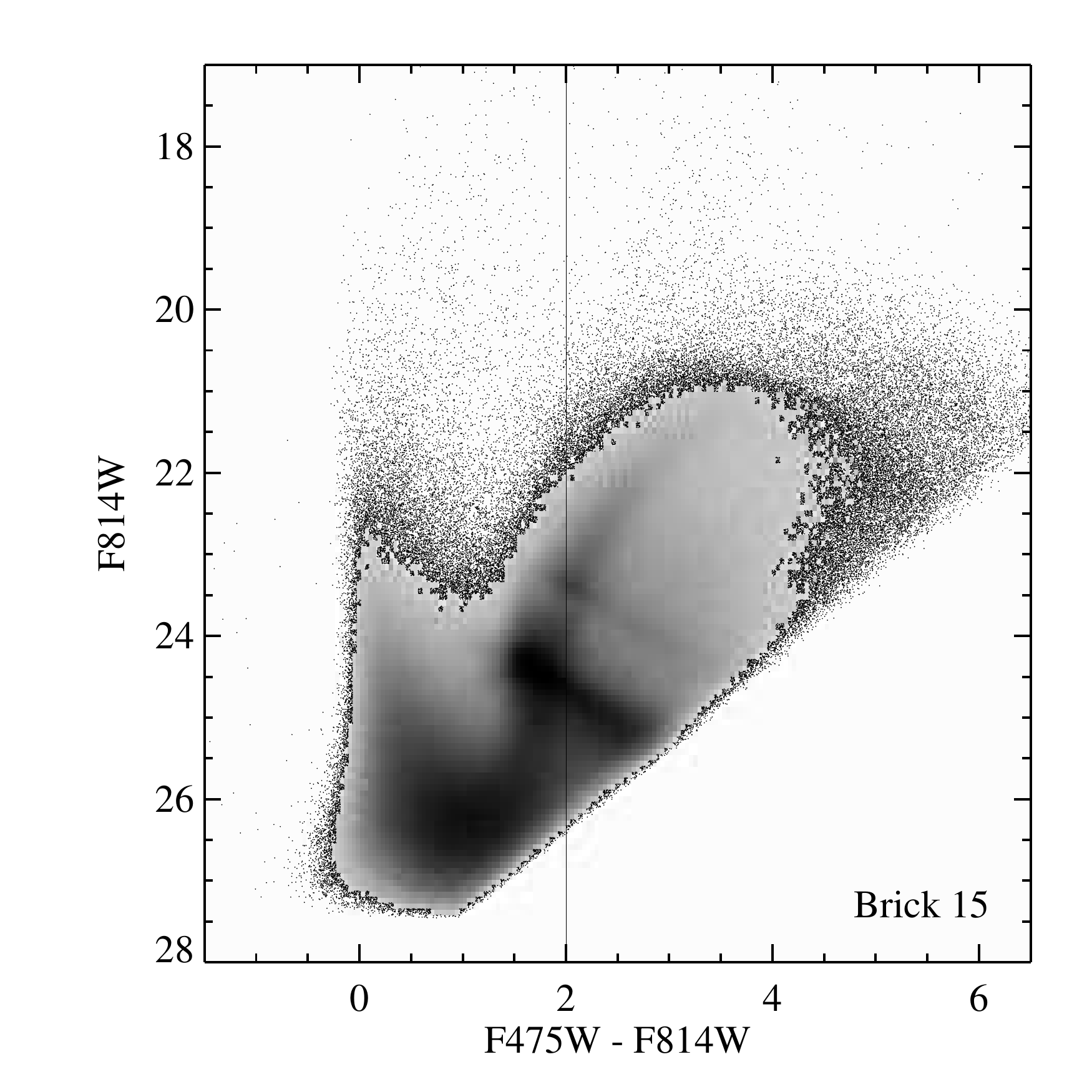}  
\includegraphics[width=3.25in]{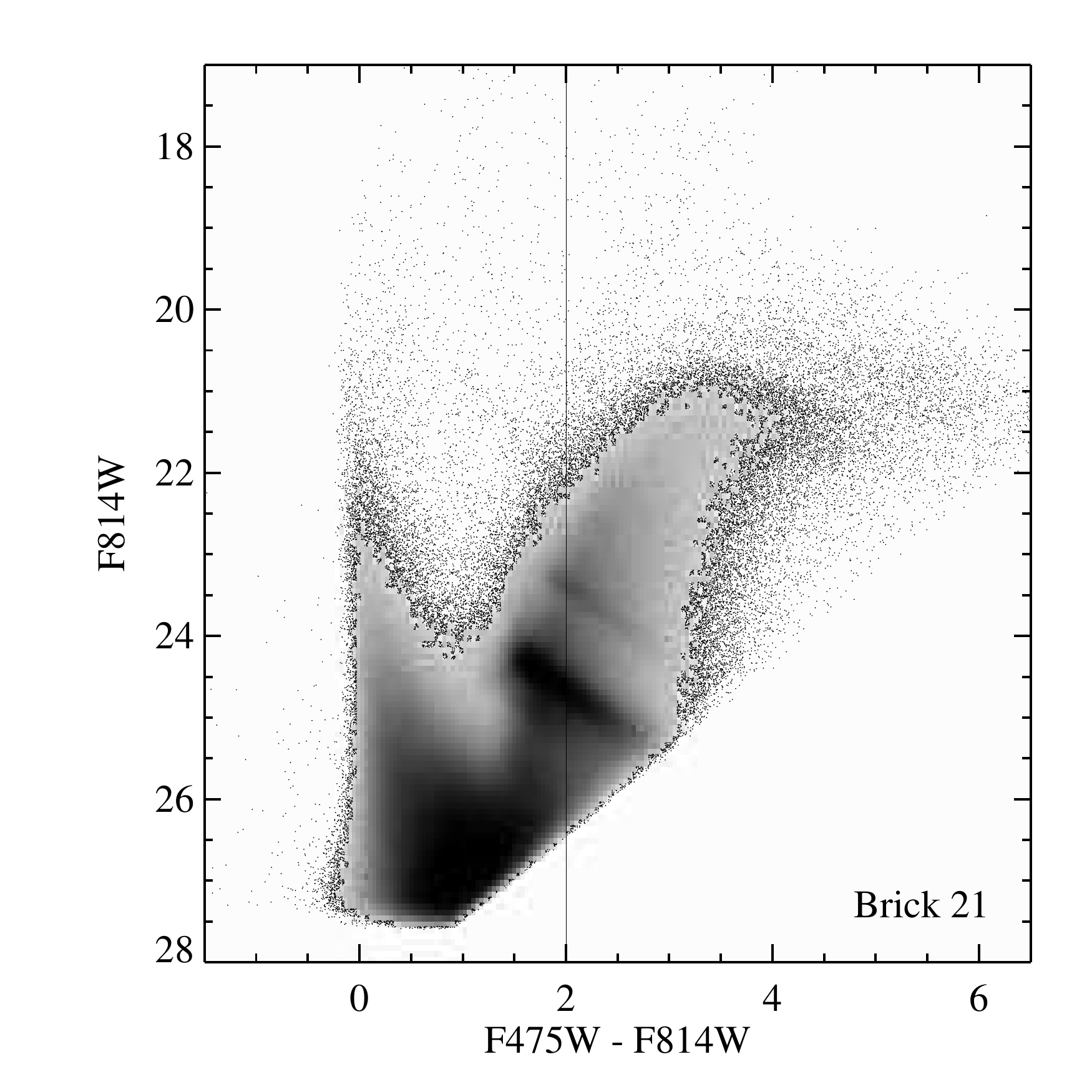}  
}
\caption{Optical Hess diagrams for complete Bricks 1, 9, 15, and 21
  (from upper left to lower right).  The Hess diagram is displayed
  with a square root scaling to highlight both high and low density
  features.  The optical CMDs show a varying mixture of red RGB stars
  and blue main sequence stars.  The RGB also shows a clear red clump
  at $\fw{814}\!\sim\!24.5$, and a brighter AGB bump at
  $\fw{814}\!\sim\!23.5$.  The depth in the optical depends strongly on
  radius because the ACS images are crowding-limited at most radii.
  The CMD for the bulge-dominated Brick 1 also shows a small diagonal
  sequence extending to $\fw{814}\!\sim\!26$, $\fw{475}-\fw{814}\!\sim\!-0.5$,
  due to blue horizontal branch stars \citep{williams2012}.  The vertical
  line is given for reference, and matches those in
  Figures~\ref{fakeCMDfig} and \ref{fakeFGfig}.
  \label{hessbrickoptfig}}
\end{figure}
\vfill
\clearpage

\begin{figure}
\centerline{
\includegraphics[width=6.0in]{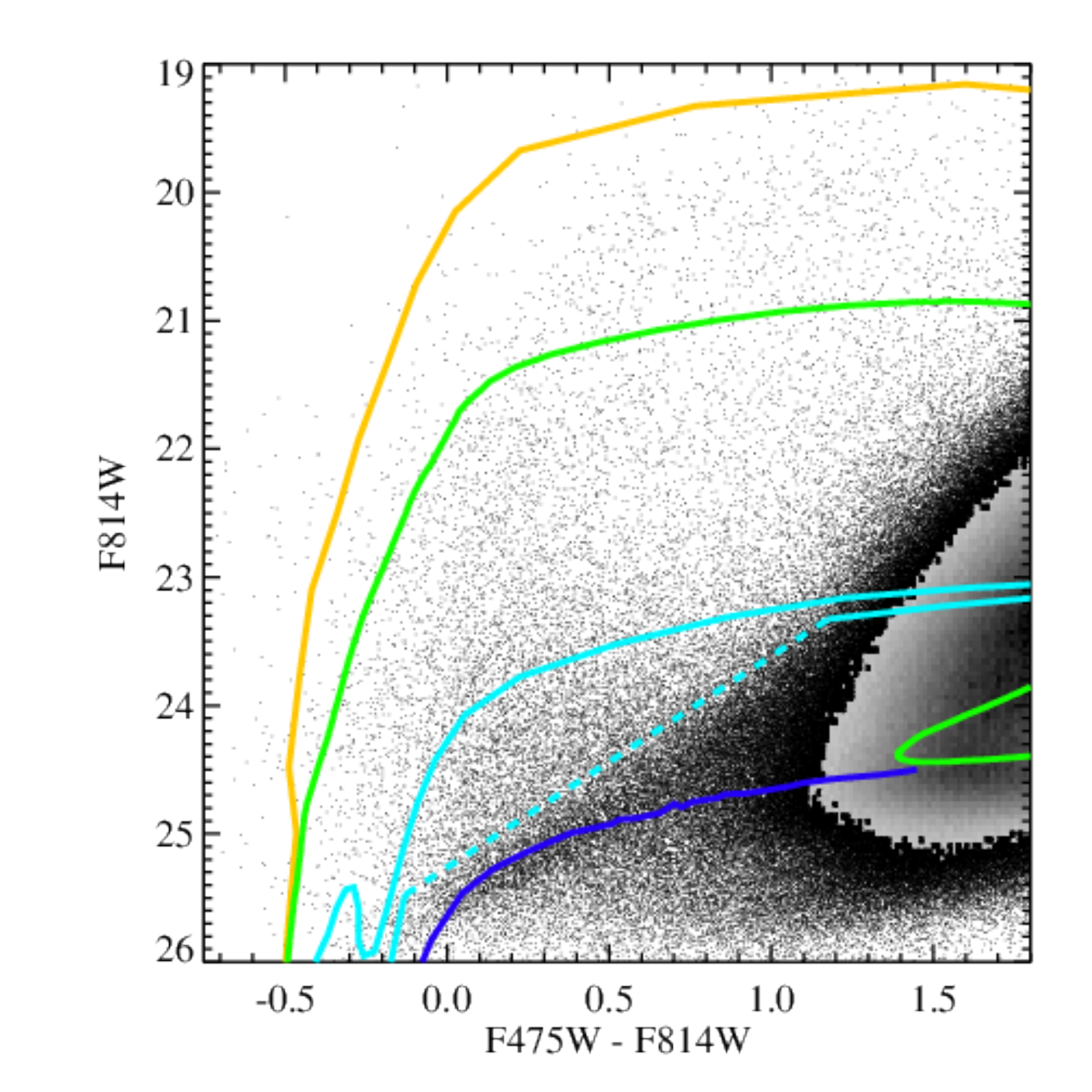}  
}
\caption{Optical CMD of Brick 1, focusing on the diagonal sequence of
  blue stars.  Overlayed are evolutionary tracks from Bressan et
  al.~(in prep) for post-AGB, post early AGB, AGB manqu\'e, and the
  zero age horizontal branch stars (plotted as orange, green, light
  blue, and dark blue, respectively), assuming zero foreground
  reddening with chemical composition $Z = 0.07$ and $Y = 0.389$ and
  an $\alpha$-enhanced composition typical of bulges
  \citep[e.g.,][]{Bensby10}. We converted the tracks to the WFC3/UVIS
  photometric system following the \citet{Girardi08} bolometric
  corrections updated with the latest WFC3/UVIS filter throughputs,
  and assumed $m-M=24.47$. Stars are plotted as a Hess diagram in
  regions of high density on the CMD; however, to highlight the
  diagonal feature, the scaling of the Hess diagram and the transition
  to plotting points has been changed from in
  Figure~\ref{hessbrickoptfig}.  The dashed portion of the light blue
  AGB manqu\'e track indicates a very rapid phase of evolution; while
  this track appears to go through the overdensity of stars, the
  evolution through this phase is sufficiently rapid that the actual
  contribution of AGB manqu\'e stars to this feature is likely to be
  small.  \label{bulgetrackfig}}
\end{figure}
\vfill
\clearpage

\begin{figure}
\centerline{
\includegraphics[width=3.25in]{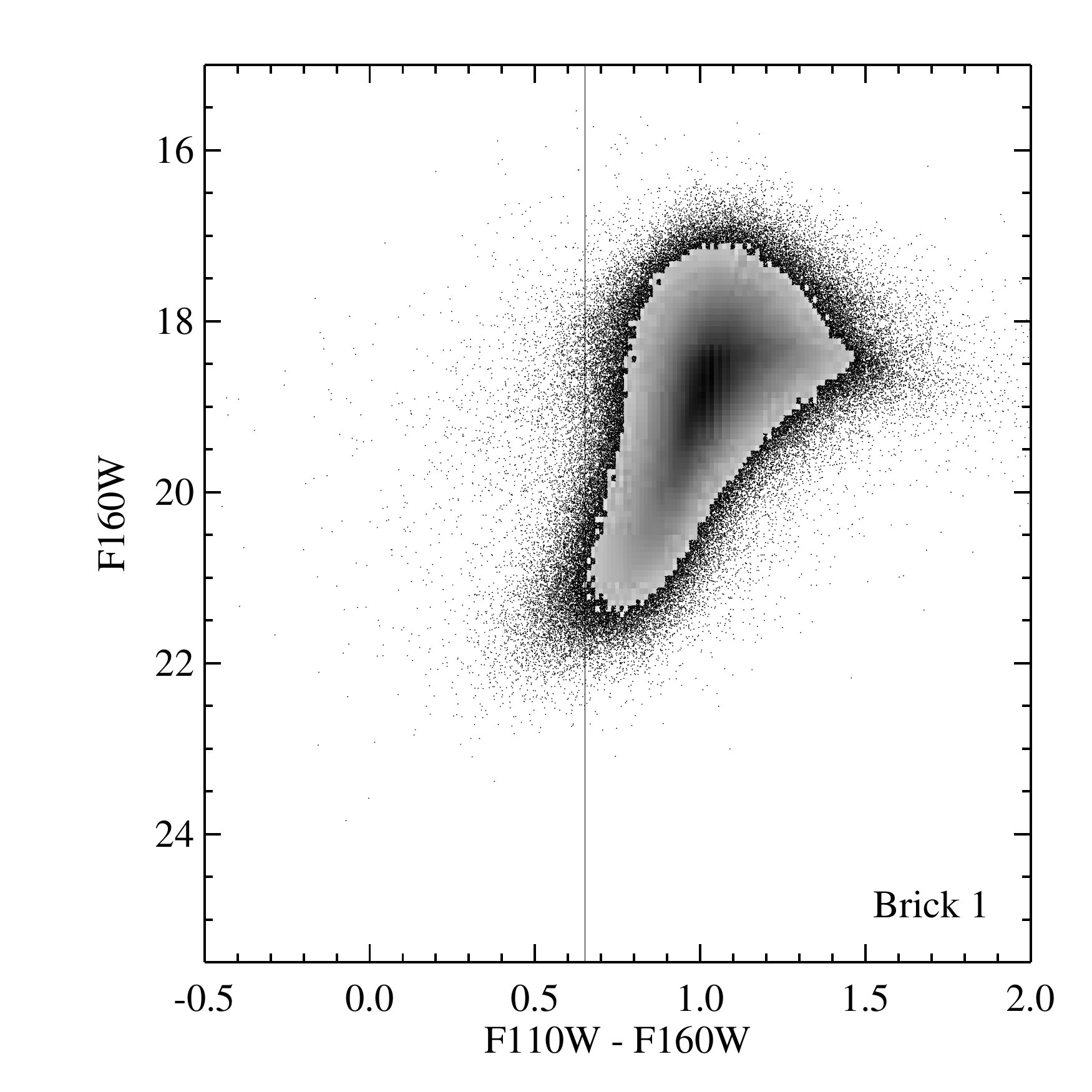}  
\includegraphics[width=3.25in]{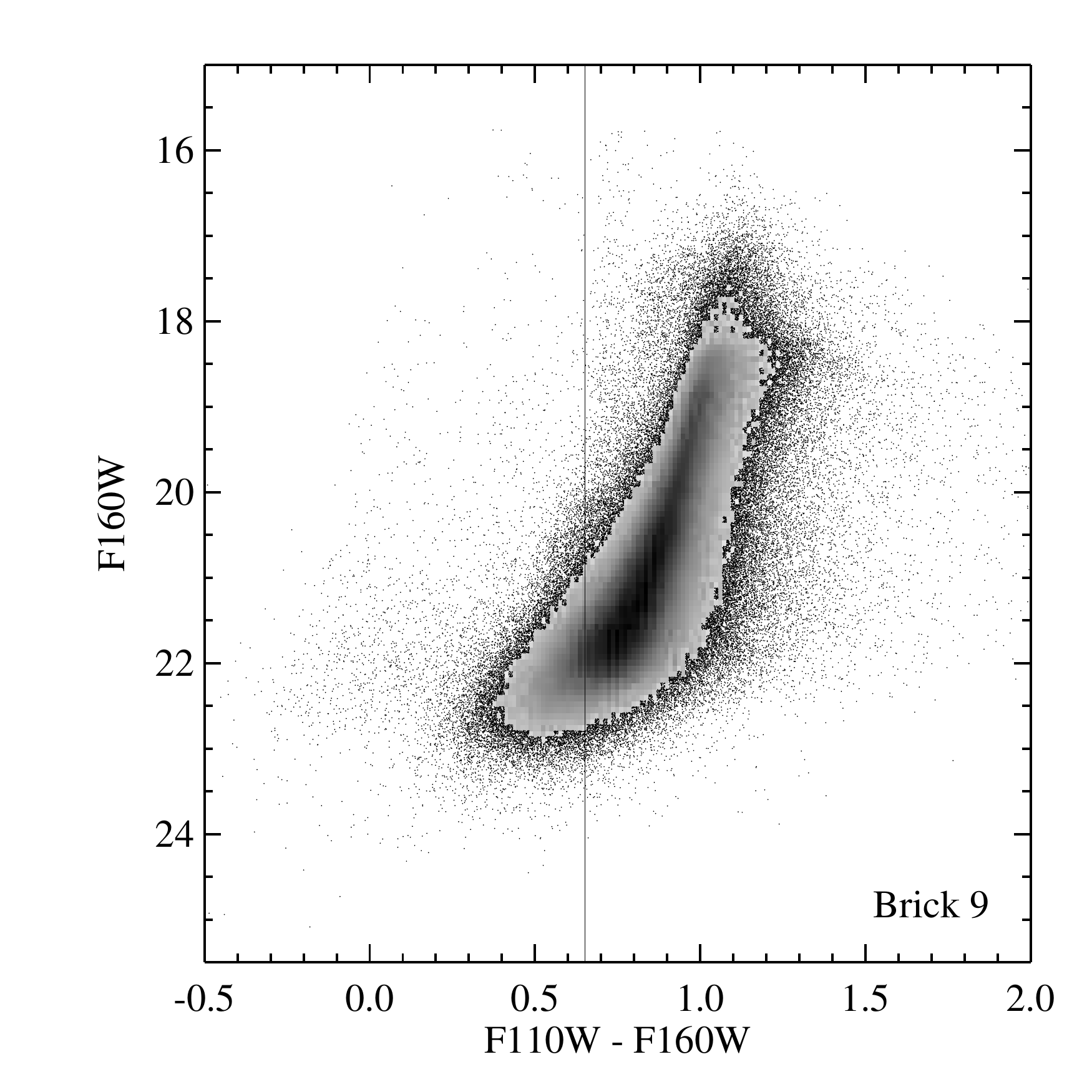}  
}
\centerline{
\includegraphics[width=3.25in]{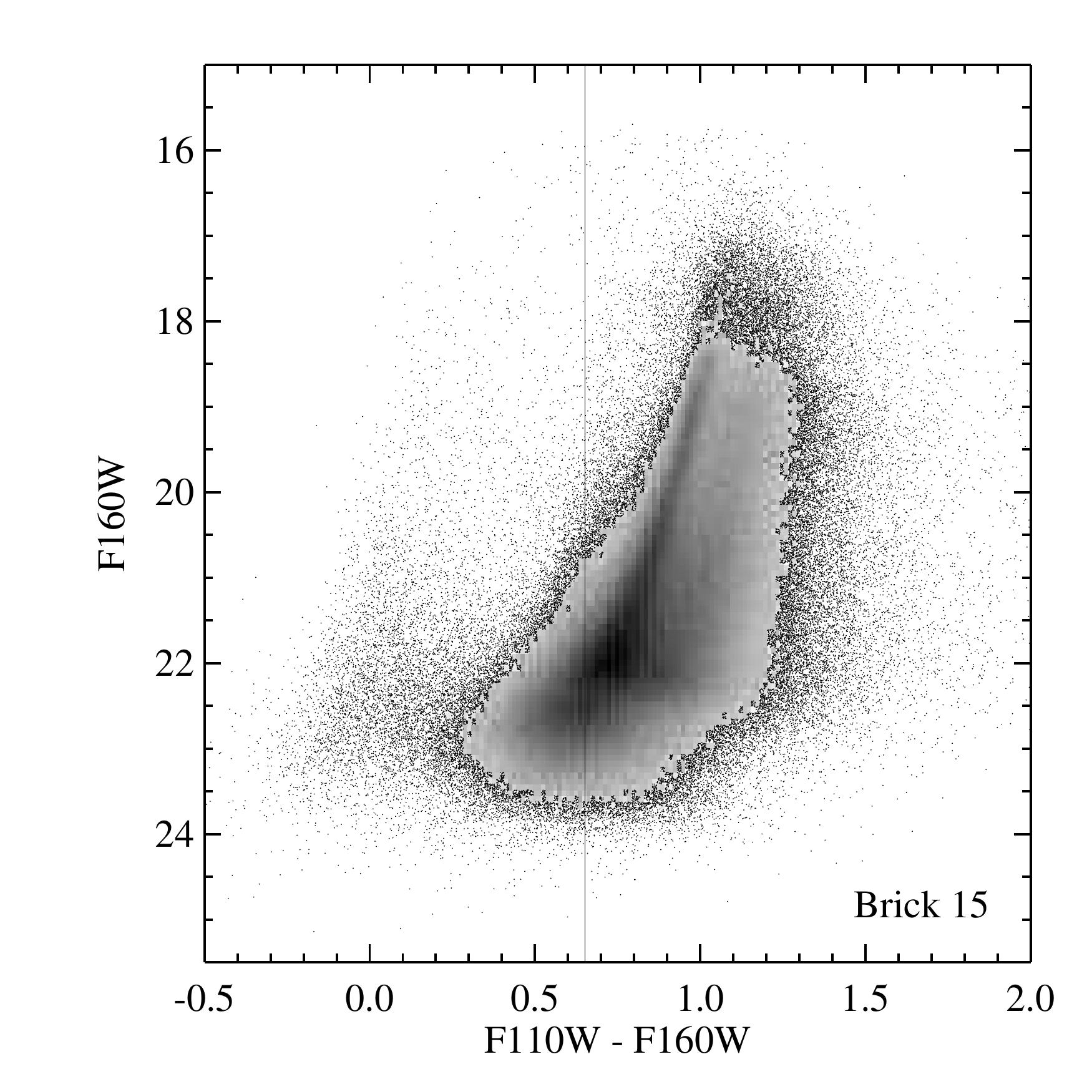}  
\includegraphics[width=3.25in]{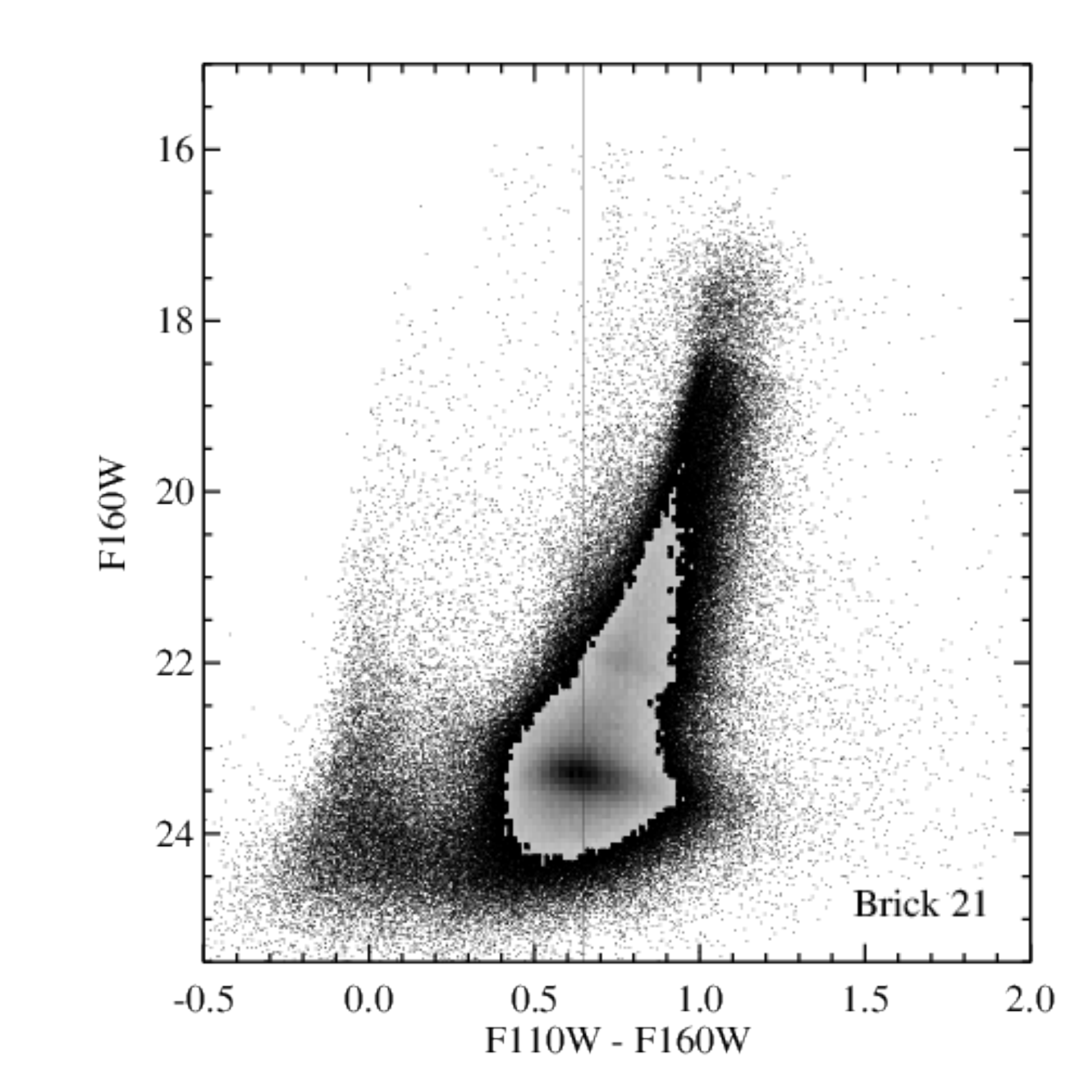}  
}
\caption{NIR Hess diagrams for complete Bricks 1, 9, 15, and 21 (from
  upper left to lower right).  The Hess diagram is displayed with a
  square root scaling to highlight both high and low density features.
  The NIR CMD is dominated by RGB stars, although AGB stars and MS are
  also present.  The depth in the NIR depends strongly on radius
  because the NIR images are crowding-limited at all radii.  In the outermost
  brick shown (Brick 21), the data reach below the NIR red clump (at
  $\fw{160}\!\sim\!23.5$). There is also a clear AGB bump at
  $\fw{160}\!\sim\!22$.  Reddening and extinction blur CMD features down
  and to the right.   The vertical
  line is given for reference, and matches those in
  Figures~\ref{fakeCMDfig} and \ref{fakeFGfig}.
  \label{hessbrickirfig}}
\end{figure}
\vfill
\clearpage

\begin{figure}
\centerline{
\includegraphics[width=6.25in]{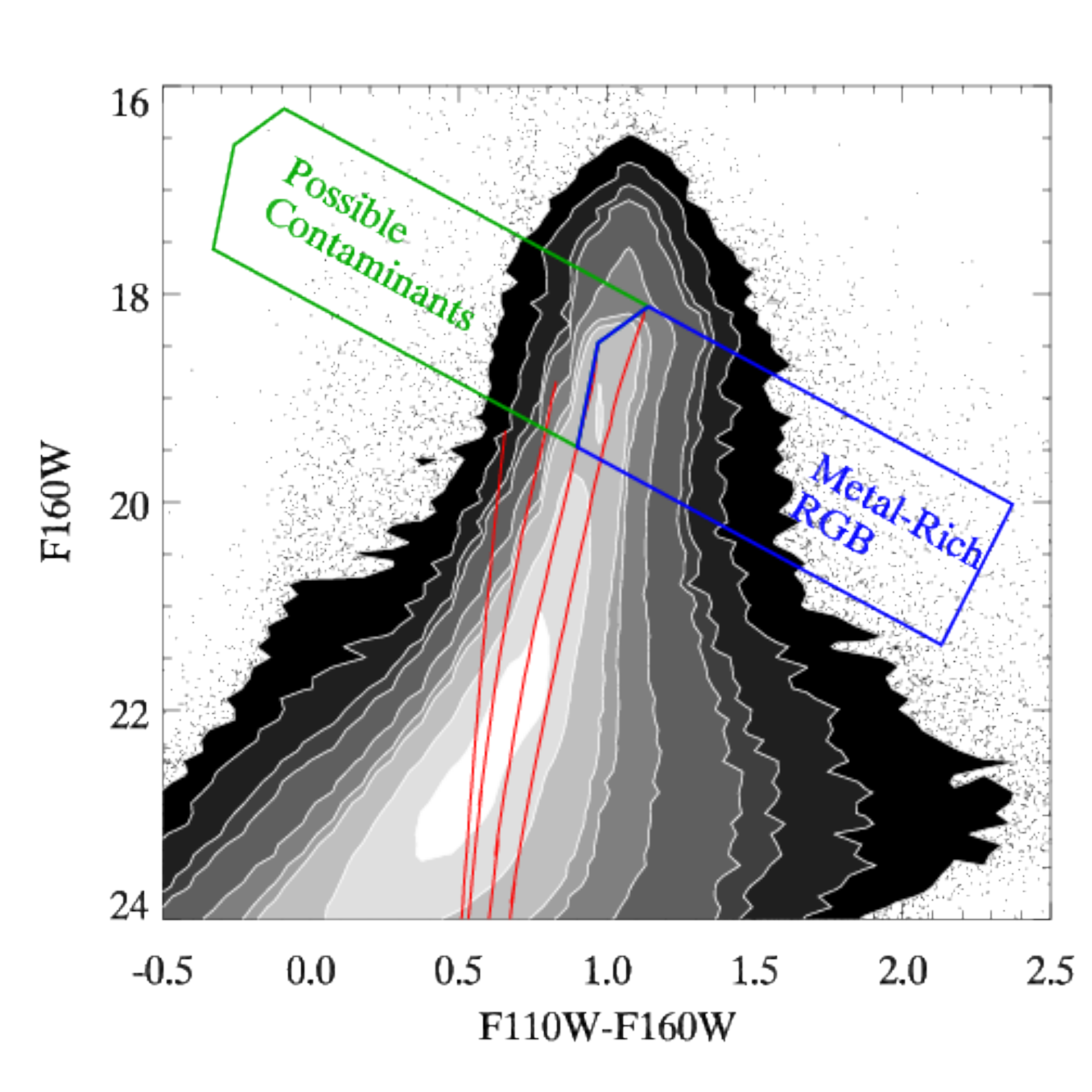}  
}
\caption{An example of selection of RGB stars from the NIR CMD.  The
  blue selection region selects RGB stars with [Fe/H] $>-0.7$ and
  reddenings up to $E(B-V)=3$.  This metal-rich RGB selection box only
  includes the brightest stars, to ensure uniform selection of RGB
  stars even in the crowded inner disk (i.e.,
  Figure~\ref{maglimradiusfig}).  The green box indicates possible
  contaminants that could be reddened into the RGB selection area.
  These contaminants are metal-poor RGB stars, red supergiants and AGB
  stars.  This box typically contains 25\% of the stars in the
  metal-rich selection box.  Red lines indicate RGB positions for
  [Fe/H] $= (-2.3, \, -1.3, \, -0.7, \, 0.0)$ with an age of 10 Gyr.
  The NIR data are drawn from the merged catalog of Bricks 8 and 9;
  identical selection criteria were used in all other regions.  A
  foreground extinction correction of $E(B-V) = 0.062$ was applied to
  the data before plotting.  \label{rgbselectionfig}}
\end{figure}
\vfill
\clearpage

\begin{figure}
\centerline{
\includegraphics[width=6.25in]{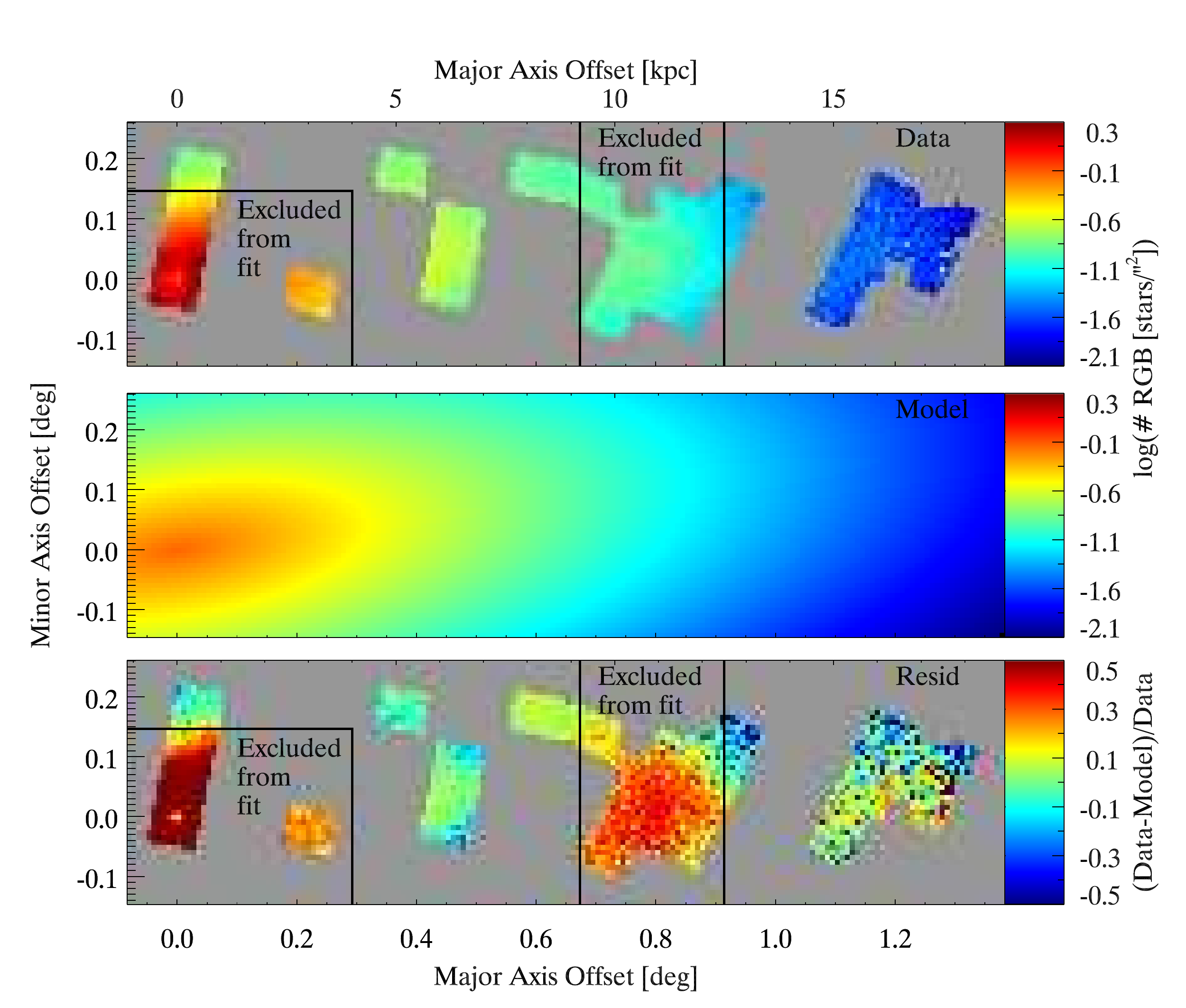}  
}
\caption{Top: Surface density of luminous RGB stars falling in the
  cyan selection region from Figure~\ref{rgbselectionfig}.  The
  overall RGB surface density distribution is quite smooth, with the
  exception of the inner region of the bulge, where incompleteness
  begins to affect even the brightest RGB stars.  Middle: The best-fit
  single inclined exponential disk model, which was derived by fitting
  the observed surface density in the upper panel, after excluding the
  central regions of the bulge and the region form 9.2--12.5~kpc along
  the major axis.  Bottom: Residual map showing the fractional
  deviation of the model from the data.  Typical excursions are of
  order 10\%, with the exception of the $10\kpc$ star forming ring,
  which appears as a true enhancement in the stellar surface density
  of $\sim$40\%, and is therefore a dynamical structure, rather than
  simply a localized enhancement in the rate of star formation.  A
  literature value for the position angle of 35$^\circ$ was assumed in
  rotating the axes to be roughly along the major and minor axis.  The
  fit along the major axis suggests a position angle of 43.2$^\circ$.
\label{densitymapfig}}
\end{figure}
\vfill
\clearpage

\end{document}